\documentclass[12pt,a4paper,twoside]{kkd}
\usepackage{epsfig}
\usepackage{amssymb}
\usepackage{setspace}
\usepackage{subfigure}

\usepackage[round]{natbib}

\usepackage{amsmath}
\usepackage{bm,psfrag}
\usepackage{color}
\usepackage{ragged2e} %Added 29/06/2009
\makeatletter
\begin{document}

\pagenumbering{roman}
\addcontentsline{toc}{chapter}{Title page}
 
%\newpage
%\begin{singlespacing}
\setlength{\textheight}{285.9 mm}
\oddsidemargin 1.2 cm 

\thispagestyle{empty}
\def\maketitle{%
  \null
  \thispagestyle{empty}%
%  \vfill
  \begin{center}%\leavevmode
   % \normalfont
    {\Large {\bf \@title\par}}%
    \vskip 1.0 cm
    %{\bf {\sl Thesis submitted to}}\\
    {\bf   {\em Thesis submitted to the}}\\
    %\vskip .2 cm
    {\textbf{\em Indian Institute of Technology Kharagpur}}\\
    %\vskip .2 cm
    {\bf{\em For the award of the degree}}\\
    \vspace{0.75cm}
    {\bf {\em of}}\\
    \vspace{0.35cm}
    {\Large {\bf Doctor of Philosophy }}\\
    \vspace{.25cm}   
    {\bf {\em by}}\\
    %\vspace{.25cm}
    {\Large {\bf\@author\par}}%
    \vspace{0.7cm}
    {\bf {\em Under the guidance of}}\\
    \vspace{.35cm}
    {\Large {\bf Prof. Somnath Bharadwaj\\ and \\ \vspace{.3cm} Dr. Sk. Saiyad Ali}}\\{\bf (Jadavpur University)}\\
       
\end{center}%
  %\vfill
  \null
}
\setlength{\topmargin}{-15. mm}

\begin{singlespacing}
\title{Visibility-based  Power Spectrum Estimation for Low-Frequency Radio Interferometric Observations}
\author{Samir Choudhuri}
\maketitle
\vspace{.1cm}
\begin{figure}[h]
\centerline{\psfig{figure=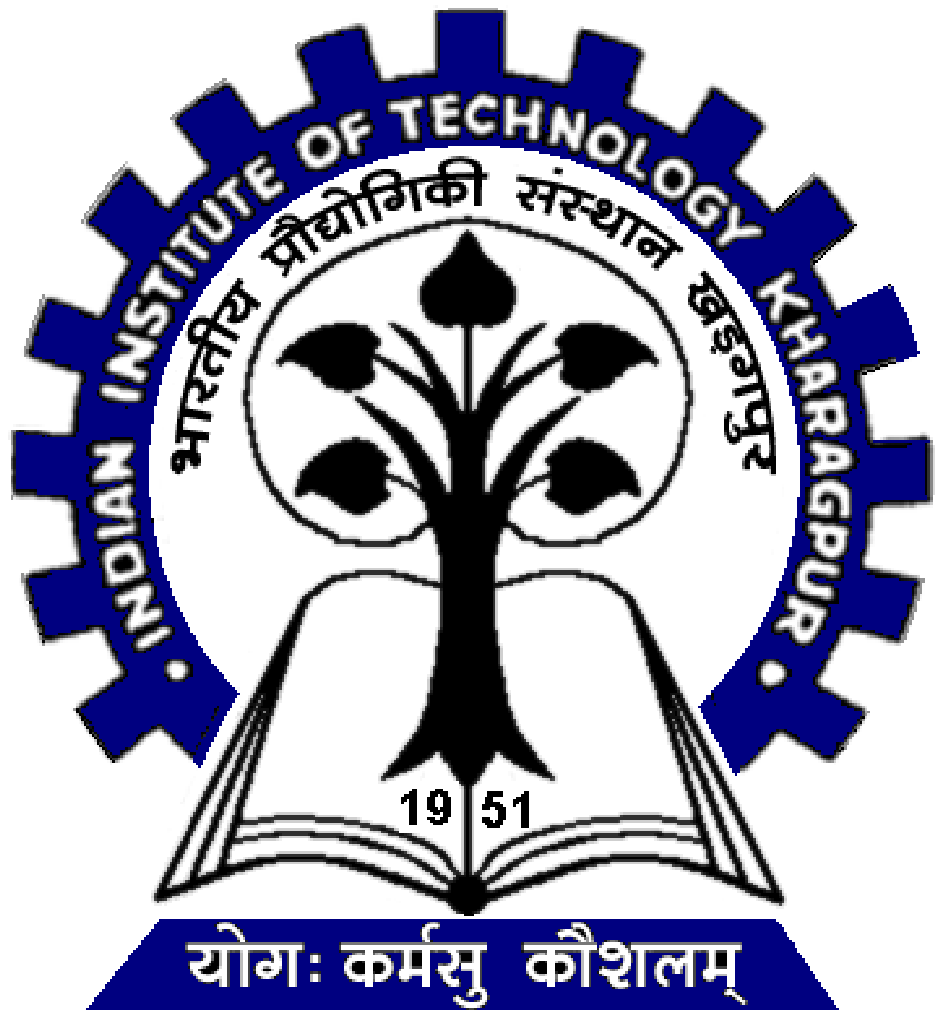,width=4.5cm,height=4.5cm}}
\end{figure}
\begin{center}
%\vspace{.3cm}
{\large {\bf DEPARTMENT OF PHYSICS}}\\
\vspace{.15cm}
{\large {\bf INDIAN INSTITUTE OF TECHNOLOGY KHARAGPUR}}\\
\vspace{.15cm}
{\large {\bf July 2017}}\\
\vspace{.25cm}
{\bf $\copyright$ 2017 Samir Choudhuri. All rights reserved.}

\end{center}

\end{singlespacing}

\setlength{\topmargin}{0.05cm} %\Added by Partha

\textheight 21.1 cm 
\textwidth 16.0 cm 
\pagestyle{headings}

\newcommand{\degr}{\ensuremath{^{\circ}}}

\renewcommand{\labelenumi}{(\roman{enumi})}
\renewcommand{\thefootnote}{\fnsymbol{footnote}}

\setlength{\headheight}{10.0 mm}
\setlength{\headsep}{10.0 mm}
\setlength{\topskip}{0.00 mm}
\setlength{\textheight}{215.9 mm}
\setlength{\textwidth}{152.4 mm}

\setlength{\parskip}{.75 mm}
\setlength{\parindent}{6.00 mm}
\setlength{\floatsep}{4 mm}
\setlength{\textfloatsep}{4 mm} \setlength{\intextsep}{4 mm}
\setcounter{secnumdepth}{3} \setcounter{tocdepth}{2}
\newtheorem{theorem}{Theorem}[section]
\newtheorem{lemma}{Lemma}[chapter]
\newtheorem{Remark}{Remark}[chapter]
\newtheorem{corollary}{Corollary}
\newtheorem{definition}[theorem]{Definition}
\newtheorem{observation}[theorem]{Observation}
\newtheorem{fact}{Fact}[theorem]
\newtheorem{proposition}[theorem]{Proposition}
\newtheorem{rule-def}[theorem]{Rule}
\newcommand{\ccnt}[1]{\multicolumn{1}{|c|}{#1}}
\newcommand{\fns}[1]{\footnotesize{#1}}

%\singlespacing
%\doublespacing
% defineing short form------

\newcommand{\sce}{\setcounter{equation}}
\newcommand{\nn}{\nonumber}
\newcommand{\lb}{\label}
\newcounter{saveeqn}
\newcommand{\alpheqn}{\setcounter{saveeqn}{\value{equation}}
\stepcounter{saveeqn}\setcounter{equation}{0}%
\renewcommand{\theequation}{\mbox{\arabic{chapter}.\arabic{saveeqn}\alph{equation}}}}
\newcommand{\reseteqn}{\setcounter{equation}{\value{saveeqn}}%
\renewcommand{\theequation}{\arabic{chapter}.\arabic{equation}}}
\newcounter{savecite}
\newcommand{\alphcite}{\setcounter{savecite}{\value{cite}}
\stepcounter{savecite}\setcounter{cite}{0}%
\renewcommand{\thecite}{\mbox{\arabic{savecite}\alph{cite}}}}
\newcommand{\resetcite}{\setcounter{cite}{\value{savecite}}%
\renewcommand{\thecite}{\arabic{cite}}}

\newcommand{\be}{\begin{equation}}
\newcommand{\e}{\end{equation}}
\newcommand{\bear}{\begin{eqnarray}}
\newcommand{\ear}{\end{eqnarray}}
\newcommand{\nline}{\nonumber \\}
\newcommand{\f}{\frac}
\newcommand{\de}{{\rm d}}
\newcommand{\del}{\partial}
\newcommand{\mathsc}[1]{{\normalfont\textsc{#1}}}
\def\apj{Astrophysical Journal}
\def\apjl{Astrophysical Journal Letters}
\def\apjs{Astrophysical Journal Supplement Series}
\def\mnras{Monthly Notices of Royal Astronomical Society}
\def\mnrasl{Monthly Notices of Royal Astronomical Society: Letters}
\def\aaps{Astronomy and Astrophysics Supplement Series}
\def\araa{Annual review of astronomy and astrophysics}
\def\prd{Physical Review D}
\def\prl{Physical Review Letters}
\def\phyrep{Physics Report}
\def\nat{Nature}
\def\aj{Astronomical Journal}
\def\aa{Astronomy \& Astrophysics}
\def\aap{Astronomy \& Astrophysics}
\def\japa{Journal of Astrophysics and Astronomy}
\def\jcap{Journal of Cosmology and Astroparticle Physics}
\def\memras{Memoirs of the Royal Astronomical Society}
\def\pasa{PASA}

\def\r{{\mathcal R}}
\def\a{{\mathcal A}}
\def\xh1{x_{{\rm H~{\textsc{i}}}}\,}
\def\xi{x^{i}_{{\rm H~{\textsc{i}}}}\,}
\def\xb{\bar{x}_{{\rm H~{\textsc{i}}}}}
\def\Ph1{P_{{\rm H~{\textsc{i}}}}}
\def\eh1{\eta_{{\rm H~{\textsc{i}}}}}
\def\Mh1{M_{{\rm H~{\textsc{i}}}}}
\def\te{\tilde{\eta}_{{\rm H~{\textsc{i}}}}}
\def\roh1{\rho_{{\rm H~{\textsc{i}}}}}
\def\np{(\dot{N}_{phs}/10^{57}\, {\rm sec^{-1}})}
\def\tq{(\tau_Q/10^7 \, {\rm yr})}
\def\u{{\vec U}}
\def\th{\vec{\theta}}
\def\rn{r_{\nu}}
\def\rnp{r'_{\nu}}
\def\newblock{}
\def\E{\hat{E}}
\def\k{{\bf k}}
\def\kp{k_\parallel}
\def\HI{H~{\textsc{i}}\,}
\def\HII{H~{\textsc{ii}}\,}
\def\gt{\textgreater}
\def\lt{\textless}
\def\newblock{}
\def\V{{\mathcal V}}
\def\h{{\rm H}}

%\newcommand{\wt}{\widetilde}

%%%%%%%%%%%%%%%%%%%%%%%%%%%%%%%%%%%
\newcommand{\eps}{\varepsilon}
\renewcommand{\l}{\ell}
%\renewcommand{\r}{r}

%%%%%%%%%%%%%%%%%%%%%%%%%%%%%%%%%%%%

\clearpage{\pagestyle{empty}\cleardoublepage}
\thispagestyle{empty}
\newcommand{\dead}[2]{ & & & & & & & {{\hspace{5cm}\bf #1}} &  {{\it #2}} \\}
\newcommand{\deads}{& & & & & & & & \\}
%\topskip 18cm
%\begin{center}
\begin{tabular}{lcccccccl}
\deads
\deads
\deads
\deads
\deads
\deads
\deads
\deads
\deads
\hline
\dead{\large {To}}{My parents}
%\dead{}{and the tax payers of INDIA}
\end{tabular}

\vspace{12.10cm}
{\bf ``The journey, Not the destination matters...''} by  T.S. Eliot
%\Large { ...In the infinite beauty we all join in one...}
%\topskip 0cm
\newpage

\clearpage{\pagestyle{empty}\cleardoublepage}
\addcontentsline{toc}{chapter}{Certificate of Approval}
%MAKE IT TOTALLY BLANK
%\thispagestyle{empty}
%\vspace{1cm}
 %\draftstring{}
%\watermarkgraphic{water.png} \watermark
\begin{center}
%{\Large \underline{\bf CERTIFICATE OF APPROVAL}}
{\Large \underline{\bf APPROVAL OF THE VIVA-VOCE BOARD}}
\end{center}

%\vspace{.5cm}
\hspace{11.6cm} Date: $24/07/2017$ 

\justifying

\hspace{-.55cm}

\noindent
Certified that the thesis entitled {\bf Visibility-based  Power Spectrum Estimation for Low-Frequency Radio Interferometric Observations} submitted by {\bf SAMIR CHOUDHURI} to the Indian Institute
of Technology Kharagpur, for the award of the degree of Doctor of
Philosophy has been accepted by the external examiners and that the
student has successfully defended the thesis in the viva-voce
examination held today.

\vspace{1.95cm}

\noindent
Signature\hspace{4.22cm}Signature\hspace{4.22cm}Signature

\vspace{0.1cm}
\noindent
(Member of the DSC)\hspace{2.cm}(Member of the
DSC)\hspace{2.cm}(Member of the DSC)

\vspace{1.95cm}

\noindent
Signature\hspace{10.22cm}Signature

\vspace{0.1cm}
\noindent
(Supervisor)\hspace{9.65cm}(Co Supervisor)

\vspace{1.95cm}

\noindent
Signature\hspace{10.22cm}Signature

\vspace{0.1cm}
\noindent
(External Examiner)\hspace{8.15cm}(Chairman)

\clearpage{\pagestyle{empty}\cleardoublepage}
\addcontentsline{toc}{chapter}{Certificate by the Supervisors}
%MAKE IT TOTALLY BLANK
%\thispagestyle{empty}
\vspace{3cm}
 %\draftstring{}
%\watermarkgraphic{water.png} \watermark
\begin{center}
{\Large \underline{\bf CERTIFICATE}}
\end{center}

\vspace{.5cm}

\noindent
This is to certify that the thesis entitled {\bf Visibility-based  Power Spectrum Estimation for Low-Frequency Radio Interferometric Observations}, submitted by {\bf Samir Choudhuri} to the Indian Institute
  of Technology Kharagpur, is a record of bona fide research work
  under our supervision and we consider it worthy of consideration for
  the award of the degree of Doctor of Philosophy of the Institute.

\vspace{5.0cm}
\noindent
\hspace{.5cm}\underline{\hspace{1.2cm} Supervisor \hspace{1.2cm}}
\hspace{5.4cm}\underline{\hspace{1.2cm} Co Supervisor \hspace{1.2cm}}

\noindent
\hspace{.5cm}Prof. Somnath Bharadwaj
\hspace{5.2cm}Dr. Sk. Saiyad Ali

\vspace{0.5cm}
\noindent
\hspace{.5cm}Date:
\hspace{9.0cm}Date:

\clearpage{\pagestyle{empty}\cleardoublepage} 
\addcontentsline{toc}{chapter}{Declaration}
\oddsidemargin 0.9 cm 
\evensidemargin 1.4 cm
\setlength{\textwidth}{152.4 mm}
 \vspace*{2cm}
 %\draftstring{}
%\watermarkgraphic{water.png} \watermark
\begin{singlespacing}

{\Large {\bf DECLARATION}}

\vspace{1cm} I certify that 

\begin{description}

\item 
\hspace{0.1cm} a. The work contained in the thesis is original and has 
been done by myself under the general supervision of my supervisor(s).

\item 
\hspace{0.1cm} b. The work has not been submitted to any other 
Institute for any degree or diploma.

\item
\hspace{0.1cm} c. I have followed the guidelines provided by the 
Institute in writing the thesis.

\item
\hspace{0.1cm} d. I have conformed to the norms and guidelines given 
in the Ethical Code of Conduct of the Institute.

\item
\hspace{0.1cm} e. Whenever I have used materials (data, theoretical 
analysis, and text) from other sources, I have given due credit to
them by citing them in the text of the thesis and giving their details
in the references.

\item
\hspace{0.1cm} f. Whenever I have quoted written materials from other
sources, I have put them under quotation marks and given due credit to
the sources by citing them and giving required details in the
references.

\end{description}

%\end{enumerate}

\vspace{2.5cm}
\hspace{9.5cm} Signature of the Student

%%\vspace{5.65cm}
%%\hspace{10.5cm} iv

\end{singlespacing}

\clearpage{\pagestyle{empty}\cleardoublepage} 
\addcontentsline{toc}{chapter}{Acknowledgements}
\vspace{-0.75cm}

\oddsidemargin 0.9 cm 
\evensidemargin -.4 cm
\setlength{\textwidth}{152.4 mm}

\begin{center}
{\Large \bf Acknowledgments}
\end{center}
%{\setlength{\baselineskip}{14pt} \setlength{\parskip}{2pt}

This is the end of my long Ph.D journey that I have started five years ago.
In this journey I met many people and I have been benefited form their 
contribution and support which make me complete to the thesis. Now, this 
is the time to acknowledge all of them.

It is really my pleasure to express my sincere gratitude to my
supervisors Prof.  Somnath Bharadwaj and Dr. Sk. Saiyad Ali. Without
their continuous help it would not be possible to complete my thesis.
Numerous ideas, intuition, insight and perfection towards research of
Prof. Bharadwaj have always inspired me to do well in the field of
research.  I am thankful to Dr. Sk. Saiyad Ali who helped me in many
ways during my research period.

Next, I would like to thank my collaborators who had contributed
directly to my Ph.D work. I would like to thank my seniors Suman Da
and Abhik Da for all the discussions that we had during my research
career.  I would like to thank Nirupam Da from whom I learned many
things that are directly related to my research. I thank Prasun Da for
his help to develop the simulation that I have used in this
thesis. Also, I would like to thank Huib Intema for providing me the
calibrated TGSS data sets that I have used in the $6^{th}$ chapter of
this thesis.

I would like to thank Prof Jayaram Chengalur for helping me to learn
the radio astronomy software packages (e.g. AIPS, CASA). I thank all
the instructors of the Radio Astronomy School (RAS-2013) at NCRA, Pune
from whom I learned the basics of Radio Astronomy techniques.

I am also grateful to my doctoral scrutiny committee Prof. Sayan Kar,
Prof. Sugata Pratik Khastgir, Prof. Anirban Dasgupta, Dr. Arghya
Taraphder for their encouragement.

Now, I would like to thank my friends and juniors at the Centre for
Theoretical studies (CTS) with whom I spent a major portion of time in
the last five years. I would like to thank Rajesh, Anjan, Debanjan,
Suman, Siddartha, Preetha, Abinash, Srijita, Soumaya (Kaka), Rajibul,
Soumya, Varatrajan with whom I spent many
exciting memorable moments.  I wish to thank Sasmita Di for her help
during my stay at Kharagpur.  We (Nirupam Da, Sasmita Di and me) spent
many memorable moments at kharagpur.

During my stay at B.C.Roy Hall I had found a nice friend Subha with
whom I spent many enjoyable moments. He helped me in many ways mainly,
preparing my slides at night before my presentation. Also, I
would like to thank Bikash with whom I spent a large fraction of time
during my stay at hostel.

I would to thank the CTS staff members, Ujal da, Subhabrata da and Gopal da
for their constant help and support. I am thankful to all the staff members of the Physics department for helping in the various official processes.

I would like to thank the NCRA, Pune staff for providing the
hospitality during my visit at NCRA. Also, I would like to thank all
the people associated with GMRT for helping me out during my GMRT
observations.  I would like to thank ORT staffs for their warm
hospitality during my academic visits there.

I would like to acknowledge the University Grants Commission (UGC),
India for financial support.

I express my gratitude to my parents for their continuous support and
 unconditional love and encouragement. I dedicate my thesis to
 them. Also, I want to thank my two elder sisters Bula and Mala for
 their endless love since my childhood. Finally, I would like to thank
 again all the people who helped me to finish my thesis.

%\vspace{2.5cm}
\hspace{9.5cm} Samir Choudhuri

\clearpage

%\newpage

 %I am really fortunate
%to get a supervisor such a person who help 

%\clearpage{\pagestyle{empty}\cleardoublepage} %Added by PSD
\addcontentsline{toc}{chapter}{List of Symbols}
%\thispagestyle{empty}
%\newpage

\newcommand{\sline}[2]{ {{\bf #1}} & & & &  {#2} \\}
\newcommand{\sskip}{& & & & \\}
%%%%%%%%%%%%%%%%% definitions %%%%%%%%%%%%%%%%%%%%%%%%%%%

\def\S{{\mathcal S}}
\def\V{\mathcal{V}}
\def\N{{\mathcal N}}
\def\A{{\bf A}\,}

\def\E{\mathcal{P}}
\def\B{\mathcal{B}}
\newcommand{\wt}{\widetilde}

\begin{singlespacing}
\vspace{-3.5cm}
%%%%%%%%%%%%%%%%%%%%%%%%%%%%%%%%%%%%%%%%%%%%%%%%%%%%%%%%%%
\chapter*{List of Symbols}
\vspace{-1.cm}
%\centering

\begin{center}
{\bf \Large{  Acronyms}}
\end{center}
\begin{center}
\begin{tabular}{lcccl}
\hline
\sskip
\sline{\large {\bf Acronym  }}{\large {\bf Full form }} 
\hline \hline

\sskip 
\sline{CMBR}{Cosmic Microwave Background Radiation}
\sskip
\sline{EoR}{Epoch of Reionization}
\sskip
\sline{FoV}{Field of View}
\sskip
\sline{FWHM}{Full Width at Half Maxima}
\sskip
\sline{GMRT}{Giant Metrewave Radio Telescope}
\sskip
\sline{ISM}{Inter-Steller Medium}
\sskip
\sline{HI}{Neutral hydrogen}
\sskip
\sline{CASA}{Common Astronomy Software Applications}
\sskip
\sline{WSRT}{Westerbork Synthesis Radio Telescope}
\sskip
\sline{LOFAR}{Low Frequency Array}
\sskip
\sline{MWA}{Murchison Widefield Array}
\sskip
\sline{PAPER}{Precision Array to Probe the Epoch of Reionization}
\sskip
\sline{SKA}{Square Kilometer Array}
\sskip
\sline{HERA}{Hydrogen Epoch of Reionization Array}
\sskip
\sline{OWFA}{Ooty Wide Field Array}
\sskip

\hline
\end{tabular}
\end{center}

\newpage

\begin{center}
\begin{tabular}{lcccl}
\hline

\sskip
\sline{CHIME}{Canadian Hydrogen Intensity Mapping Experiment}
\sskip
\sline{BAOBAB}{Baryon Acoustic Oscillation Broadband and Broad-beam Array}
\sskip
\sline{WMAP}{Wilkinson Microwave Anisotropy Probe}
\sskip
\sline{MAPS}{Multi-frequency Angular power spectrum}
\sskip
\sline{FFTW}{Fastest Fourier Transform in the West}
\sskip
\sline{RFI}{Radio Frequency Interference}
\sskip
\sline{TGE}{Tapered Gridded estimator}
\sskip
\sline{PB}{Primary Beam}
\sskip
\sline{MFS}{Multi Frequency Synthesis}
\sskip
\sline{MS-MFS}{Multi Scale Multi Frequency Synthesis}
\hline
\end{tabular}
\end{center}
\newpage
\begin{center}
{\bf \Large{Symbols}}
\end{center}
\begin{center}
\begin{tabular}{lcccl}
\hline
\sskip
\sline{\large {\bf Symbols  }}{\large {\bf Full form }} 
\hline \hline

\sskip
\sline{$u,v,w$}{Three components of Baseline vector}
\sskip
\sline{$\ell$}{Angular multipole}
\sskip
\sline{$C_{\ell}$}{Angular power spectrum}
\sskip
\sline{$C^M_{\ell}$}{Model angular power spectrum}
\sskip
\sline{$\nu$}{Observing frequency}
\sskip
\sline{$\vec{\theta}$}{Two dimensional vector in the sky plane}
\sskip
\sline{$I(\th, \nu)$}{Specific Intensity}
\sskip
\sline{$\bar{I}(\nu)$}{Mean Specific Intensity}
\sskip
\sline{$T(\th, \nu)$}{Brightness Temperature}
\sskip
\sline{$\u$}{Baseline vector}
\sskip
\sline{${\bf d}$}{Antenna pair separation}
\sskip
\sline{$\V(\u, \nu)$}{Visibility measured at baseline $\u$}
\sline{}{and frequency $\nu$ (Jy)}
\sskip
\sline{$\lambda$}{Observing wavelength}
\sskip
\sline{$\S(\u, \nu)$}{Sky signal component of measured Visibility at baseline $\u$}
\sline{}{and frequency $\nu$ (Jy)}
\sskip
\sline{$\N(\u, \nu)$}{Noise at baseline $\u$}
\sline{}{and frequency $\nu$ (Jy)}
\sskip
\sline{$F (\u, \nu)$}{Foregrounds Contribution of measured Visibility at baseline $\u$}
\sline{}{and frequency $\nu$ (Jy)}
\sskip

\hline
\end{tabular}
\end{center}

\newpage

\begin{center}
\begin{tabular}{lcccl}
\hline
\sskip
\sline{\large {\bf Symbols  }}{\large {\bf Full form }} 
\hline \hline

\sskip
\sline{${\cal A}(\th, \nu)$}{Antenna primary beam}
\sskip
\sline{$\delta I(\th, \nu)$}{Fluctuations in Specific Intensity}
\sskip
\sline{$\tilde{a}(\u, \nu)$}{Fourier transform of $A(\th, \nu)$}
\sskip
\sline{$\Delta \tilde{I}(\u, \nu)$}{Fourier transform of $\delta
  I(\th,   \nu)$}
\sskip
\sline{$\delta T(\th, \nu)$}{Fluctuations in Temperature}
\sskip
\sline{$\Delta \tilde{T}(\u, \nu)$}{Fourier transform of $\delta
  T(\th,   \nu)$}
\sskip
\sline{$J_1(x)$}{First order Bessel function}
\sskip
\sline{$P(\u, \nu)$}{Power spectrum at baseline $\u$ and frequency $\nu$}
\sskip
\sline{$(\del B_{\nu}/{\del T})$}{Conversion factor from temperature}
\sline{}{to specific intensity}
\sskip
\sline{$k_B$}{Boltzmann constant (Joule/K)}
\sskip
\sline{$V_{2}$}{Two visibility correlation}
\sskip
\sline{$S_{2}$}{Signal correlation}
\sskip
\sline{$N_{2}$}{Noise covariance}
\sskip
\sline{$\Omega$}{Solid Angle}
\sskip
\sline{${\cal W}(\theta)$}{Window function}
\sskip
\sline{${\tilde w(\u)}$}{Fourier transform of ${\cal W}(\theta)$}
\sskip
\sline{$B(\u)$}{Baseline sampling function}
\sskip
\sline{$N_r$}{Number of realizations}
\sskip

\hline
\end{tabular}
\end{center}

\newpage

\begin{center}
\begin{tabular}{lcccl}
\hline
\sskip
\sline{\large {\bf Symbols  }}{\large {\bf Full form }} 
\hline \hline

\sskip
\sline{$g_a$}{Gain error of antenna ``a''}
\sskip
\sline{$\sigma$}{Standard deviation}
\sskip
\sline{$l,m,n$}{Direction cosines in the sky plane}
\sskip
\sline{$dN/dS$}{Differential source count}
\sskip
\sline{$T_{sys}$}{System Temperature}
\sskip
\sline{$A_{eff}$}{Effective collecting area}
\sskip
\sline{$\Delta\nu$}{Channel width}
\sskip
\sline{$\Delta t$}{Integration time}
\sskip
\sline{$N_{ant}$}{Total number of antennas}
\sskip
\sline{$N_{chan}$}{Number of channels}
\sskip
\sline{$T_{obs}$}{Total observing time}
\sskip
\sline{$z$}{Redshift}
\sskip
\sline{$\theta_x,\,\theta_y$}{Two perpendicular components of $\vec{\theta}$}
\sskip
\sline{$\sigma_n$}{Noise rms in visibility}
\sskip
\sline{$\theta_{{\rm FWHM}}$}{FWHM of the antenna beam pattern}
\sskip
\sline{$\theta_0$}{$0.6\, \theta_{{\rm FWHM}}$}
\sskip
\sline{${\cal A}_W(\th, \nu)$}{Modified Antenna primary beam}
\sskip
\sline{$\tilde{a}_W(\u, \nu)$}{Fourier transform of ${\cal A}_W(\th, \nu)$}
\sskip
\sline{$P({\rm{\bf k}})$}{Power spectrum}
\sskip
\sline{$k_{\perp},k_{\parallel}$}{Two components of ${\bf k}$}
\sskip

\hline
\end{tabular}
\end{center}

\newpage

\begin{center}
\begin{tabular}{lcccl}
\hline
\sskip
\sline{\large {\bf Symbols  }}{\large {\bf Full form }} 
\hline \hline

\sskip
\sline{$P(k_{\perp},k_{\parallel})$}{Power spectrum at $(k_{\perp},k_{\parallel})$}
\sskip
\sline{$r$}{Comoving distance from the present day observer}
\sline{}{to the redshift $z = 1420/(\nu -1)$}
\sskip
\sline{$r^{\prime}_{\nu}$}{$d r_{\nu}/d \nu$}
\sskip
\sline{$\tau$}{Delay channel}
\sskip
\sline{$N_c$}{Number of channels}
\sskip
\sline{$B_{bw}$}{Observing Bandwidth}
\sskip

\hline
\end{tabular}
\end{center}
\end{singlespacing}
\thispagestyle{empty}

%\clearpage{\pagestyle{empty}\cleardoublepage} %Added by PSD
\addcontentsline{toc}{chapter}{List of Tables}
%\newpage

\listoftables

\clearpage
%\newpage

%\setcounter{page}{17}

\thispagestyle{empty}   %Due to this line there is problem in contents

\newpage

%Just add the line \setcounter{page}{17} complile then remove
%problem may be solved temporarily 

%\chaptermark{$\empty$}
%\clearpage{\pagestyle{empty}\cleardoublepage} %Added by PSD
\addcontentsline{toc}{chapter}{List of Figures}
%\listoffigures

\listoffigures

%\newpage

%\thispagestyle{empty}

%\newpage

%\clearpage
\clearpage{\pagestyle{empty}\cleardoublepage} 
\chaptermark{$\empty$}
\addcontentsline{toc}{chapter}{Abstract}

%\thispagestyle{empty}

%\pagenumbering{roman}

%\vspace{-0.5cm}
%\vspace{2cm}    %%%Partha

\begin{center}
{\Large \bf Abstract}
\end{center}
%\vspace{.2cm}

\begin{singlespacing}
Precise measurement of the power spectrum of the diffuse background
sky signal using low frequency radio interferometers is an important
topic of current research. The problem is particularly challenging due
to the presence of foregrounds and system noise. In this thesis we
present a visibility based estimator namely, the Tapered Gridded
Estimator (TGE) to estimate the power spectrum of the diffuse sky
signal. The TGE has three novel features.  First, the estimator uses
gridded visibilities to estimate the power spectrum which is
computationally much faster than individually correlating the
visibilities. Second, a positive noise bias is removed by subtracting
the auto-correlation of the visibilities which is responsible for the
noise bias. Third, the estimator allows us to taper the field of view
so as to suppress the contribution from the sources in the outer
regions and the sidelobes of the telescope's primary beam.

We first consider the two dimensional (2D) TGE to estimate the angular
power spectrum $C_{\ell}$. We validate the estimator and its
statistical error using realistic simulations of Giant Meterwave Radio
Telescope (GMRT) $150\, {\rm MHz}$ observations, which includes
diffuse synchrotron emission and system noise. We further developed
the simulation by including the discrete point sources. We use
different ``CLEANing'' strategies to investigate the accuracy of point
source subtraction from the central region of the primary beam, and to
identify the best ``CLEANing'' strategy. It is difficult to correctly
model and subtract the point sources from the periphery and the
sidelobes of the primary beam. We see that the TGE successfully
suppresses contributions from these unsubtracted sources and correctly
recovers the $C_{\ell}$ of the Galactic synchrotron emission.

Finally we have extended the TGE to estimate the three dimensional
(3D) power spectrum $P({\bf k})$ of the cosmological 21-cm signal.
Analytic formulas are presented for predicting the variance of the
binned power spectrum.  The estimator and its variance predictions are
validated using simulations of $150 \, {\rm MHz}$ GMRT observations.

We have used the 2D TGE to estimate $C_{\ell}$ using visibility data
for two of the fields observed by TIFR GMRT Sky Survey (TGSS). We find
that the sky signal, after subtracting the point sources, is dominated
by the diffuse Galactic synchrotron radiation across the angular
multipole range $200 \le \ell \le 500$.  We present a power law fit,
$C_{\ell}=A\times\big(\frac{1000}{l}\big)^{\beta}$, to the measured
$C_{\ell}$ over this $\ell$ range. We find that the values of $\beta$
are in the range of $2$ to $3$. In future, we plan to extend our analysis
for the whole sky using TGSS survey and to find out the variation of 
both $A$ and $\beta$ as a function 
of Galactic coordinate.
 
{\bf Keywords}: methods: statistical, data analysis,
 techniques: interferometric, cosmology: diffuse radiation

\end{singlespacing}

%\newpage
\clearpage

\thispagestyle{empty}

\newpage
%\newpage

%\clearpage {\pagestyle{empty}\cleardoublepage}
\addcontentsline{toc}{chapter}{Contents}
\tableofcontents
\clearpage{\thispagestyle{empty}}
%\cleardoublepage
\thispagestyle{empty}

\setcounter{chapter}{0}

\def\lsim{~\rlap{$<$}{\lower 1.0ex\hbox{$\sim$}}}
\setcounter{section}{0}
\setcounter{subsection}{0}
\setcounter{subsubsection}{2}
\setcounter{equation}{0}
%%\pagenumbering{arabic}
\pagenumbering{arabic} 
\setcounter{page}{0}    %%%%%KKD used \setcounter{page}{0}  
\oddsidemargin 0.9 cm 
\evensidemargin -.4 cm
\setlength{\textwidth}{152.4 mm}
\def\dt{\tilde{\eta}_{\rm HI}}
\def\u{{\bf U}}
\def\th{\vec{\theta}}
\def\x{{\bf x}}
\def\HI{{\rm HI}}
\def\H{{\rm H}}
\def\mK{\rm mK}
\def\d{\eta_{\HI}}
\def\chb{\bar{\chi}_{\rm HI}}
\def\dH{\Delta_{\rm HI}}
\def\n{{\bf n}}
\def\Tc{T_\gamma}
\def\Tb{T_b}
\def\Ts{T_s}
\def\Tg{T_g}
\def\nue{\nu_e}
\def\Bon{B_{10}}
\def\Bo{B_{01}}
\def\A10{A_{10}}
\def\hp{h_p}
\def\kb{k_B}
\def\rn{r_{\nu}}
\def\rnp{r_{\nu}^{'}}
\def\nH{n_{HI}}
\def\k{{\bf k}}
\def\kp{k_\parallel}
\def\kpr{{\bf k}_\perp}
\def\gsim{~\rlap{$>$}{\lower 1.0ex\hbox{$\sim$}}}

\chapter[Introduction]{{\bf Introduction}}
\label{Intro}
Low frequency radio astronomy has become a topic of intense research
during the last two decades. It promises to improve our current
understanding of a wide range of astrophysical phenomena spanning from
our own Galaxy to the high redshift universe.  There currently are
several low frequency interferometers in different parts of the world
which are operating in different frequency bands. For example, the
Giant Meter Wave Radio Telescope
(GMRT \footnote{http://www.gmrt.ncra.tifr.res.in}; \citealt{swarup})
currently operates in the frequency range $150$ to $1420 {\rm
MHz}$. The GMRT has 30 steerable antennas each of diameter $45~{\rm
m}$. A total 14 out of the 30 antennas are randomly distributed in a
central square $1.1~{\rm km}\times 1.1~{\rm km}$ in extent, while the
rest of the antennas are distributed approximately in a ’Y’ shaped
configuration. This configuration provides a good sensitivity for both
compact and extended sources. Other radio telescopes such as the
Donald C. Backer Precision Array to Probe the Epoch of Reionization
(PAPER{\footnote{http://astro.berkeley.edu/dbacker/eor}},
\citealt{parsons10}), the Low Frequency Array
(LOFAR{\footnote{http://www.lofar.org/}},
\citealt{haarlem,yata13}) and the Murchison Wide-field Array
(MWA{\footnote{http://www.mwatelescope.org}} \citealt{bowman13,tingay13})
are also targeted to observe the low frequency radio sky. Upcoming
instruments like the upgraded GMRT, the Square
Kilometer Array (SKA1
LOW{\footnote{http://www.skatelescope.org/}}, \citealt{koopmans15})
and the Hydrogen Epoch of Reionization Array
(HERA{\footnote{http://reionization.org/}}, \citealt{neben16}) are
planned to achieve even higher sensitivity by increasing the
instantaneous bandwidth and also the collecting area. Several other
upcoming interferometers like the Ooty Wide Field Array (OWFA;
\citealt{prasad,subrahmanya16,subrahmanya16a}) and the Canadian Hydrogen Intensity Mapping
Experiment
  (CHIME{\footnote{http://chime.phas.ubc.ca/}; \citealt{bandura}) are
  planned for 21-cm intensity mapping experiments. These currently
  functioning and future telescopes motivate the study presented in
  this thesis.

Hydrogen is the most abundant element of the baryonic content of the
Universe. The hyperfine transition in the ground state of neutral
hydrogen (HI) emits a photon of wavelength 21-cm or 1420 ${\rm MHz}$
which lies in the radio band. Observations of this radiation are one
of the most promising future probes of the high redshift Universe.
The redshifted 21-cm radiation from the cosmological HI distribution
appears as a diffuse background radiation in all low frequency
observations below 1420 ${\rm MHz}$ \citep{madau97}.
The power spectrum of the angular and frequency fluctuations of the
brightness temperature of this radiation provides us a useful tool to quantify the large
scale structures in the universe in the post-reionization era ($z <
6$) \citep{bharadwaj01,bharadwaj011,bharadwaj03,bharadwaj04}. This
radiation has been perceived as a important probe for studying the
epoch of reionization (EoR) in redshift range $20\ge z \ge6$ \citep{furlanetto04a,furlanetto04b}.  The properties of the
first stars and galaxies can be inferred by measuring the 21-cm
radiation during this era \citep{fan06,choudhury06}. The evolution of
the Universe during the dark ages, before the formation of any
luminous source, ($30< z<200$) can also be studied using the 21-cm
radiation \citep{loeb04,bharadwaj044}. In summary, the redshifted
21-cm line can be used as a tool to probe the evolution of the
Universe from the Dark Ages through the EoR to the present
epoch \citep{bharadwaj05,furlanetto06,morales10,pritchard12,mellema13}.

There are several observations towards detecting the redshifted
21-cm radiation. \citet{ali08} have used GMRT observation to
characterize the background radiation at 150 {\rm
MHz}. \citet{ghosh1,ghosh2} set an upper limit on the power spectrum
of the 21-cm fluctuations using GMRT 610 {\rm MHz} observations.
\citet{switzer13} have used observation using the Green Bank Telescope (GBT)
to constrain HI fluctuations at $z\sim0.8$. \citet{masui13} measure the 21 cm brightness fluctuations at $z\sim0.8$ using cross-correlation with large-scale structure traced by galaxies. \citet{paciga13} used GMRT {\rm 150 MHz}
observation to give an upper limit of 21-cm power spectrum which is about $(248 {\rm mK})^2$
for $k = 0.50 {\rm hMpc}^{-1}$ at
$z\sim8.6$. Recently, \citet{beardsley16} have set the upper limit
of $\Delta^2 \leq 2.7 \times 10^4$ {\rm mK}$^2$ at $k=0.27$ {\rm
hMpc}$^{-1}$ at $z=7.1$. The best upper limit of the 21-cm power
spectrum achieved till date is (22.4 ${\rm mK})^2$ in
the range $0.15\lt k\lt 0.5 ~{\rm hMpc}^{-1}$ at z =
8.4 \citep{ali15}.

\section{Observational Challenges}
The brightness temperature fluctuations of the 21-cm signal is expected to be 4-5 orders of magnitude lower than the  astrophysical foregrounds \citep{shaver99,ali08,paciga11,ghosh1,ghosh2}. Accurately modelling the foregrounds and subtracting them from the data are the biggest challenges for the detection of the cosmological 21-cm signal. Other strong component like radio frequency interference (RFI), system noise and the ionospheric distortion also corrupt the 21-cm signal in low frequency observations. The dominant factor in the system noise comes from the sky temperature $T_{sky}$. All the EoR fields are targeted at the position of sky where $T_{sky}$ is relatively low. As for foregrounds, the main contributions come from the (a) point sources (b) diffuse Galactic synchrotron emission (DGSE) (c) Extragalactic and (d) Galactic free-free emission. The last two components are much lower as compared with the others \citep{shaver99}.

Extragalactic point sources dominate the low frequency
sky \citep{ali08} at the angular scales $\le 4^{\circ}$ which are
relevant for telescopes like the GMRT, LOFAR and SKA. There are
currently several surveys which cover a large portion of the sky at
low frequency (e.g. 3C survey \citep{edge59}, 6C
survey \citep{hales88}, 3CR
survey \citep{bennet62}). Recently, \citet{intema16} present the
source catalogue for almost $90\%$ of the sky at $150 {\rm MHz}$ using
GMRT. The DGSE is the most dominant foreground component if the point
source are subtracted to a sufficiently low flux
level \citep{bernardi09,ghosh12,iacobelli13}. The measurement of the
diffuse Galactic synchrotron emission at $408 {\rm
MHz}$ \citep{haslam82}, $1.4 {\rm GHz}$ \citep{reich82,reich88} and
$2.3 {\rm GHz}$ \citep{jonas98} showed that it is the most dominant
foreground at angular scale larger than $\approx
1^{\circ}$. \citet{laporta08} have measured the angular power spectrum
of the Galactic synchrotron emission using single dish all-sky total intensity maps at $408 {\rm MHz}$ and $1420 {\rm MHz}$. They have reported that the angular power spectrum can be modeled  as $C_l\sim\ell^{\alpha}$ in the $\ell$ range  $10\le\ell\le300$, with $\alpha\in
[-3.0,-2.6]$. \citet{bernardi09} have analysed  Westerbork Synthesis Radio Telescope (WSRT) data observed at
$150 {\rm MHz}$ and found that the angular power shows a power law
with slope $-2.2$ ($C_{\ell}\sim\ell^{-2.2}$) at $\ell\le900$. Another
measurement using GMRT $150 {\rm MHz}$ observations showed the same
power law behaviour with a slope $-2.34$ for
$253\le\ell\le800$ \citep{ghosh12}. Recently, LOFAR observation at
$150 {\rm MHz}$ showed a slightly lower slope of $-1.8$ for
$100\le\ell\le1300$ \citep{iacobelli13}. A precise characterization
and a detailed understanding of the DGSE are needed to reliably remove
foregrounds in cosmological $21\, {\rm cm}$ experiments.  The study of the angular
power spectrum ($C_{\ell}$) of the DGSE is interesting in its own
right. This will shed light on the cosmic ray electron distribution,
the strength and structure of the Galactic magnetic field in the
turbulent interstellar medium (ISM) of our
Galaxy \citep{Waelkens,Lazarian}.

Foreground removal is an important issue for quantifying both the DGSE
and the cosmological 21-cm signal. Accurate subtraction of the point
sources is needed to study the DGSE in low frequency observations. For
21-cm signal, subtraction of both the point sources and the DGSE is
required. A large variety of techniques have been proposed to remove
the foregrounds from the low frequency data in the context of the 21-cm
signal.  The different approaches may be broadly divided into two
classes (1) Foreground Removal, and (2) Foreground Avoidance.  All the
foreground removal techniques rely on the fact that foregrounds behave
smoothly along the frequency direction. Various methodologies have been explored
for foreground subtraction and for detecting the underlying 21-cm
signal
\citep{ali08,jelic08,bowman09,paciga11,ghosh1,ghosh2,chapman12,parsons12,liu12,trott12,pober13,paciga13,parsons14,trott16}. Foreground avoidance is based
on the idea that the Cylindrical Power Spectrum
$P(k_{\perp},k_{\parallel})$ due to the foregrounds is expected to be
restricted within a wedge in the $(k_{\perp},k_{\parallel})$ space
(\citealt{adatta10}). The 21-cm power spectrum can be estimated using
the uncontaminated Fourier modes outside this
wedge \citep{vedantham12,thyag13,pober14,liu14a,liu14b,dillon14,dillon15,ali15}.
With their merits and demerits, these two approaches are considered
complementary \citep{chapman16}.

\section{Power Spectrum Estimation}
Several different estimators have been proposed and used in literature
to estimate the power spectrum of the diffuse sky
signal. \citet{seljak97} has proposed an image based estimator for the
angular power spectrum $C_{\ell}$. \citet{bernardi09}
and \citet{iacobelli13} have used this estimator to measure $C_{\ell}$
of the diffuse synchrotron emission using $150{\rm MHz}$ observations with WSRT and LOFAR
respectively. \citet{dillon15} have proposed an image based estimator
to measure the three dimensional (3D) power spectrum $P({\bf k})$ of
the cosmological 21-cm signal. Radio interferometers directly measure
the visibilities which are the Fourier transform of the sky signal. It
is convenient to directly estimate the power spectrum from the
measured visibilities. \citet{begum06} and \citet{dutta08} have used a
visibility based estimator to estimate the power spectrum of the 21-cm
signal from the ISM of external
galaxies. \citet{liu12} and \citet{trott16} have proposed visibility
based estimators for the three dimensional 21-cm $P({\bf k})$. In a
recent paper \citet{jacob16} have used multiple power spectrum
analysis pipelines to estimate $P({\bf k})$ and compare their outputs
using MWA data. \citet{shaw14} and \citet{liu16} present power spectrum
estimators that incorporate the spherical sky.

\section{Objective and Motivation}

In this thesis we present a visibility based estimator, the Tapered
Gridded Estimator (TGE) to estimate the fluctuations of the diffuse
sky signal.  The 2D TGE estimates the angular power spectrum
$C_{\ell}$ from the measured visibilities. This quantifies the two
dimensional (2D) brightness temperature fluctuations of the sky
signal at a fixed frequency. We have further extended the 2D TGE to
the 3D TGE to estimate the three dimensional (3D) power spectrum
$P({\bf k})$ of the brightness temperature fluctuations of the
redshifted 21-cm signal. The spatial fluctuations of the cosmological
HI distribution appear as brightness temperature fluctuations in frequency 
and angular position in the sky.

It is also possible to estimate the power spectrum from the images but
the noise properties of the visibilities are better understood than
the image pixel. The noise in the different visibilities is
uncorrelated, whereas the noise in the image pixels may be correlated
depending on the baseline $uv$ coverage. The noise bias in the estimated
power spectrum can be avoided by subtracting the self correlation term
which is responsible for the noise bias. The visibility based power spectrum
estimator also avoids the imaging artifact due to the error in the
deconvolution process. Another important factor for any estimator is
the total computation time required to estimate the power spectrum
from the visibilities. Current generation radio telescopes are
expected to generate huge volumes of visibility data in observations
spanning large bandwidth and collecting area. In such a situation any
estimator should have enough efficiency to handle such a huge data
volume.

The wide field foreground is an important issue for estimating the
power spectrum of the faint diffuse signal which mainly comes from the
central region of the primary beam. The bright point sources from
the outer region, if not removed properly, may have significant
contribution in the estimated power spectrum of the diffuse
signal. One possible solution is to make a large image and subtract
all the point sources from the whole region.  But it is
computationally very challenging to make such a large image and also
cumbersome to identify all the sources for subtraction. The primary
beam at the outer region is highly time and frequency dependent. The
deviation from the circular symmetry and rotation of the earth make it
more difficult to accurately model the point sources in the outer
region.

The TGE can solve the above mentioned problems to a large extent. The TGE
incorporates three novel features.  First, the estimator uses the
gridded visibilities to estimate the angular power spectrum
$(C_{\ell})$, this is computationally much faster than individually
correlating the visibilities. Second, a positive noise bias is removed
by subtracting the auto-correlation of the visibilities. Third, the
estimator allows us to taper the field of view (FoV) so as to restrict
the contribution from the sources in the outer regions and the
sidelobes of the telescope's primary beam. The mathematical formalism
of the TGE and its variances are presented in this thesis. The
estimator and its variance predictions are validated using realistic
simulations. Finally, we apply the 2D TGE to the real GMRT data and
quantify the $C_{\ell}$ of the diffuse Galactic synchrotron emission
over some range of angular scale.

\section{Outline of the thesis}
We present the brief summary of the work presented in this thesis

In {\bf Chapter 2} we present  two estimators  namely, the Bare Estimator and the TGE to quantify the angular 
power spectrum of the sky signal directly from the  visibilities measured
in radio interferometric observations.  This is relevant 
for  both the foregrounds and the cosmological $21$-cm 
signal buried therein. Also, the analytic prediction for the statistical error for these two estimators are presented in this chapter. Both the estimators and their statistical errors are validated using simulated visibilities for the GMRT.
The simulations include the diffuse Galactic synchrotron emission along with the system noise. We have also studied the effect of some of the real life problem like the gain error and the ``w-term'' effect in the estimated $C_{\ell}$.

In {\bf Chapter 3} we further developed the earlier simulations by including point sources. We use different ``CLEANing'' strategies to investigate the accuracy of point source subtraction from the simulated visibilities, and to identify the best ``CLEANing'' strategy. We apply the TGE to the residual data to measure the angular power spectrum of the diffuse emission. We also assess the impact of individual ``CLEANing'' procedures for point source subtraction in recovering the input power spectrum  $C_{\ell}$ of the  diffuse Galactic synchrotron emission.

In {\bf Chapter 4}  we show that by tapering  the sky response it is possible to suppress the contribution from the outer region of the primary beam where it is highly frequency dependent. Using simulated  $150 \, {\rm MHz}$ observations, 
we demonstrate that the TGE suppresses the contribution due to point sources from the outer parts to measure the angular power spectrum $C_{\ell}$
of the underlying diffuse signal. We also show from the simulation that this method can self-consistently compute the noise bias and accurately subtract it to provide an unbiased estimation of $C_{\ell}$.

In {\bf Chapter 5} we present an improved 2D TGE which resolves the overestimate (discussed in Chapter 2) due to the patchy $uv$ distribution. Next, the  2D TGE
 is  extended to the 3D TGE for the power spectrum $P(\k)$ of the 21-cm
brightness temperature fluctuations. Analytic formulas are also
presented for predicting the variance of the binned power spectrum.
The estimator and its variance predictions are validated using
simulations of $150 \, {\rm MHz}$ GMRT observations. We show that the
estimator accurately recovers the input model for the 1D Spherical
Power Spectrum $P(k)$ and the 2D Cylindrical Power Spectrum
$P(k_\perp,k_\parallel)$, and the predicted variance is also in
reasonably good agreement with the simulations. 

In {\bf Chapter 6} we apply the 2D TGE to estimate $C_{\ell}$ using
visibility data for two of the fields observed by the TIFR GMRT Sky
Survey (TGSS{\footnote{http://tgss.ncra.tifr.res.in}},\citealt{sirothia14}). We have used the data which was
calibrated and processed by \citet{intema16}. We find that the sky
signal, after subtracting the point sources, is dominated by the
diffuse Galactic synchrotron radiation across the angular multipole
range $200 \le \ell \le 500$.  We present power law fits to the
measured $C_{\ell}$ over this $\ell$ range.

In {\bf Chapter 7} we summarize our findings and highlight some of the
future scopes of the thesis.

%\clearpage{\pagestyle{empty}}\cleardoublepage} %%%%%%%%%%%%%%%%%%%%
%\newpage
\setcounter{section}{0}
\setcounter{subsection}{0}
\setcounter{subsubsection}{2}
\setcounter{equation}{0}
%\pagenumbering{arabic}

\def\u{\vec{U}}
\def\S{{\mathcal S}}
\def\V{\mathcal{V}}
\def\N{{\mathcal N}}
\def\A{{\bf A}\,}
\def\Sc{S_2}

%-------------------------------------------
\chapter[Visibility based angular power spectrum estimation]{{\bf Visibility based  angular power spectrum estimation in low frequency radio interferometric observations}\footnote{This chapter is adapted
   from the paper ``Visibility based  angular power spectrum estimation in low frequency radio interferometric observations''
   by \citet{samir14}}}
\label{chap:chap2}

\section{Introduction}
Observations of the redshifted 21-cm radiation from the large scale
distribution of neutral hydrogen (HI) is one of the most promising
probes to study the high redshift Universe
(recent reviews: \citealt{morales10, mellema13}). This radiation appears as a
very faint, diffuse background radiation in all low frequency radio
observations below $1420 \,{\rm MHz}$. At these frequencies
the sky signal is largely dominated by different foregrounds which are four
to five orders of
magnitude stronger than the redshifted 21 cm signal
(\citealt{ali08,bernardi09,ghosh12,pober13}). Foreground removal 
is possibly the most serious challenge for detecting the  cosmological 
21-cm signal.  Various methodologies  have been explored  for 
foreground subtraction and  for detecting the underlying $21 \,{\rm cm}$
signal \citep{jelic,ghosh2,mao3,liu12,cho,jacobs,parsons14,dillon14}.

The Galactic synchrotron emission is expected
to be the most dominant foreground 
at angular scale $>10^{'}$ after point source subtraction at $10 -
20\, {\rm mJy}$ level
\citep{bernardi09,ghosh12}. A precise characterization and a detailed
understanding of the Galactic synchrotron  emission is needed to reliably
remove foregrounds in  $21\, {\rm cm}$ experiments.
The study of the Galactic synchrotron emission is interesting in its own
right.  This will  shed light on the cosmic ray electron distribution, 
the  strength and structure of the Galactic magnetic field,  and the
magnetic turbulence \citep{Waelkens,Lazarian,iacobelli13}. 

\cite{bernardi09} and \cite{ghosh12} have respectively analyzed 
$150 \, {\rm MHz}$
WSRT and GMRT observations where they find that the measured angular
power spectrum  can be well fitted with a power law 
($C_{\ell} \propto \ell^{-\beta}$, $\beta=2.2 \pm 0.3$ for WSRT 
and $\beta=2.34 \pm
0.28$ for GMRT) upto $\ell \le 900$. At relatively higher
frequencies, \citet{giardino01} and \citet{giardino02} have analyzed
the fluctuations in the Galactic synchrotron radiation using the $2.3
\, {\rm GHz}$ Rhodes Survey and the $2.4 \, {\rm GHz}$ Parkes radio
continuum and polarization survey, where they find a slope $\beta=2.43
\pm 0.01$ ($2 \le \ell \le 100$) and $\beta=2.37 \pm 0.21$ ($40 \le
\ell \le 250$) respectively. At tens of GHz, \cite{bennett03} have
determined the angular power spectrum of the Galactic synchrotron
radiation using the Wilkinson Microwave Anisotropy Probe (WMAP) data
where they find a scaling $C_{\ell} \sim \ell^{-2}$ within $\ell \le
200$. The structure of the Galactic synchrotron emission is not well
quantified  at the frequencies and angular scales relevant for detecting
 the 
cosmological $21$-cm signal, and there is considerable scope for further 
work in this direction.

Radio interferometric observations measure the complex visibility. 
The measurement is done directly in Fourier
space which makes interferometers ideal instruments for measuring the
angular power spectrum of the sky signal. 
The visibility based power spectrum
estimator formalism has been extensively used for analyzing CMB data
from interferometers (\citealt{hobson1,white,hobson2,myers}). 
A  visibility based estimator has also been successfully
employed to study the power spectrum of the HI in the 
interstellar medium (ISM) of several nearby  galaxies ( eg. 
\citealt{begum06,  Dutta}).  A direct visibility based approach 
has been proposed for quantifying the power spectrum of the 
cosmological $21$-cm signal expected at the GMRT (\citealt{bharadwaj011,bharadwaj03,bharadwaj05})
and recently  for the ORT \citep{ali14}.  
Visibility based power spectrum estimators have been used to 
analyze   GMRT data in the context of HI observations
(\citealt{ali08,paciga11,ghosh1,ghosh2,ghosh12}). A recent
paper  \citep{paul} has proposed  visibility correlations to detect  the EoR signal 
using  drift scan observations with the  MWA.

It is possible to  estimate the angular power spectrum of the 
sky signal from the synthesized radio image 
(eg. \citealt{bernardi09,bernardi10,iacobelli13}). 
The noise properties of the visibilities are better understood
than those of the image pixels. The noise in the different 
visibilities is uncorrelated, whereas the noise in the image 
pixels may be correlated depending on the baseline $uv$ coverage. 
The visibility based power spectrum estimators have the added 
advantage that they avoid possible imaging artifacts due to the 
dirty beam, etc (\citealt{trott}).  

In this paper we consider two estimators which use the measured 
visibilities to quantify the angular power spectrum of the 
sky signal. The Bare Estimator, which has been utilized   in \citet{ali08}
and \citet{ghosh1}, directly uses pairwise correlations of the measured
visibilities. 
The Tapered Gridded Estimator, which has been utilized  in 
\citet{ghosh2} and \citet{ghosh12}, uses the visibilities after gridding
on a rectangular grid in the $uv$ plane.  The latter incorporates 
the feature that it allows a tapering of the sky response and thereby 
suppresses the sidelobes of the telescope's primary beam. Earlier work 
\citep{ghosh2} has shown this to be a useful ingredient in foreground
removal for detecting the cosmological $21$-cm signal. In this paper
we have carried out a somewhat detailed  investigation 
in order to place these two estimators on sound theoretical footing. 
The theoretical predictions are substantiated using simulations. 
As a testbed for the estimators, we consider a situation where 
the point sources have been identified and subtracted out so that 
the residual visibilities are dominated by the Galactic synchrotron 
radiation. We investigate how well the estimators are able to recover 
the angular power spectrum of the input model used to simulate
the Galactic synchrotron emission at $150 \, {\rm MHz}$. We have also analyzed the effects
of gain errors and the $w$-term. Most of our simulations are for the 
GMRT, but we also briefly consider simulations for  LOFAR. 
The estimators considered here can be generalized to the multi-frequency 
angular power spectrum  (MAPS, \citealt{kdatta07})  which can be used 
to quantify the cosmological $21$-cm signal. We plan to investigate this
 in a future study. 

A brief outline of the paper follows. In Section 2 we  establish the
relation between the visibility correlation and the angular power spectrum. 
In Section 3 we describe the simulations which we have used to validate 
the theoretical results of this paper.  In Sections 4 and 5 
we consider the Bare and the Tapered Gridded Estimators respectively. 
The theoretical analysis and the results from the simulations are 
all presented in these two sections. 
Section 6 presents a brief comparison between the two estimators,
and in Sections 7 and 8 we consider the effect of gain errors and the 
$w$-term respectively. Much of the analysis of the previous sections
is in the context of the GMRT. In Section 9 we apply the estimators
to simulated LOFAR data and present the results.  We present discussion 
 and conclusions in Section 10.

\section{Visibility Correlations and the angular power spectrum}
\label{v2ps}
In this section we discuss  the  relation between the two visibility 
correlation and the  angular power spectrum of the specific intensity
$I(\th,\,\nu)$ or equivalently the brightness temperature $T(\th,\,\nu)$
distribution  on the sky   under the flat-sky approximation.
 Here $\th$ is a two   dimensional vector on the
plane of the sky with origin at the center of the field of view
(FoV). It is useful to decompose the specific intensity 
as $ I(\th,\nu)=\bar{I}(\nu)+\delta I(\th,\,\nu)$ where the first
term $\bar{I}(\nu) $ is an uniform background brightness and the
second term $\delta I(\th,\,\nu)$ is the angular fluctuation in the
specific intensity. We assume that  $\delta I(\th,\,\nu) $ is a
particular realization of a statistically homogeneous and isotropic
Gaussian random process on the sky. In radio interferometric observations, the
fundamental observable quantity is a set of complex visibilities
$\V(\u,\nu)$ which are sensitive to only the angular fluctuations in
the sky signal. The baseline $\u$ quantifies the antenna pair separation
${\bf d}$ projected on the plane perpendicular to the line of sight in
units of the observing wavelength $\lambda$. The 
 measured visibilities   are a sum of  two contributions 
$\V(\u, \nu)=\S(\u, \nu)+\N(\u,\nu)$,
the  sky signal and system noise respectively.
We assume 
that the signal and the noise are both uncorrelated Gaussian random 
variables with zero mean.  The
visibility contribution $\S(\u,\nu)$ from  the sky signal records the
Fourier transform of the product of the primary beam pattern
${\mathcal A}(\th, \nu)$ and $\delta I(\th,\,\nu)$. The primary beam pattern
${\cal A}(\th, \nu)$ quantifies how the individual antenna responds to
signals from different directions in the sky.  Using the convolution
theorem, we then have
\begin{equation}
\S(\u,\nu)=  \int \, d^2 U{'}  \,
  \tilde{a}\left(\u - \u{'},\,\nu\right)\, 
 \, \Delta \tilde{I}(\u{'},\,\nu),   
\label{eq:a2}
\end{equation}
where $\Delta \tilde{ I}(\u,\,\nu)$ and $\tilde{a}\,(\u, \nu)$ are the
Fourier transforms of $\delta I(\th,\,\nu) $ and ${\cal A}(\th,\,\nu)$
respectively. Typically, the term arising from the uniform specific 
intensity distribution 
 $\bar{I}(\nu) \tilde{a}\,(\u, \nu)$ makes no contribution to
the measured visibilities, and we have dropped this.
 We refer to $\tilde{a}\,(\u, \nu)$ as the
aperture power pattern.  The individual antenna response ${\cal A}(\th, \nu)$ 
 for any telescope is usually quite complicated
depending on  the  telescope aperture, the reflector and 
the feed \citep{lfra}. It is beyond
the scope of the present paper to consider the actual single antenna 
response of any particular telescope. We make the simplifying assumption
that the telescope has an uniformly illuminated circular aperture 
of diameter $D$ 
whereby we have the primary beam pattern (Figure 1)
\begin{equation}
{\cal A}(\th,\,\nu) = 
\left[ \left(\frac{2 \lambda}{\pi\theta D} \right)
J_1\left(\frac{\pi\theta D}{\lambda}\right) \right]^2 
\label{eq:b1} 
\end{equation}
 where $J_1$ is the Bessel function of the first kind of order one, the
 primary  beam pattern  is normalized to unity  at the pointing center
$[{\cal A}(0)=1]$, 
and  the aperture power pattern is  
\begin{equation}
\tilde{a}(\u, \nu)=\frac{8 \lambda^4}{\pi^2 D^4}\bigg[\bigg(\frac{D}{\lambda}\bigg)^2 \cos^{-1}\bigg(\frac{\lambda U}{D}\bigg)-U\sqrt{\bigg(\frac{D}{\lambda}\bigg)^2-U^2}\bigg],
\label{eq:b2} 
\end{equation}

We note that  $\tilde{a}(\u, \nu)$ in  eq. (\ref{eq:b2}) peaks at $U=0$,
declines monotonically with increasing $U$,  and is zero for 
 $U \ge  D/{\lambda}$. The primary beam pattern (Figure 1) is 
well approximated by a circular Gaussian function 
\begin{equation}
{\cal A}_G(\th,\nu)=\exp[-\theta^2/\theta^2_0]
\label{eq:b1a} 
\end{equation} 
of  the same full width at half maxima (FWHM) as eq. (\ref{eq:b1}). 
The parameter $\theta_{0}$ here   is related to the
 full width half maxima  $\theta_{\rm FWHM}$ of the primary beam pattern
 ${\cal A}(\th,\nu)$ (eq. \ref{eq:b1})  as ${\theta}_{0} =0.6 \theta_{\rm FWHM}$,  
and 
\begin{equation}
\tilde{a}_G(\u, \nu)=\frac{1}{\pi U_0^2} \ e^{-U^2/U_0^2}
\label{eq:b2a} 
\end{equation}
where $U_0=(\pi \theta_0)^{-1}=0.53/ \theta_{\rm FWHM} $.  While the Gaussian
$\tilde{a}_G(\u, \nu)$  (eq. \ref{eq:b2a}) provides a  good
approximation  to $\tilde{a}(\u, \nu)$ (eq. \ref{eq:b2}) particularly 
in the vicinity of $U=0$, there is  however  a significant difference in 
that $\tilde{a}(\u, \nu)$ has a compact support and is exactly zero for all 
$U \ge  D/{\lambda}$ whereas $\tilde{a}_G(\u, \nu)$, though it has an 
extremely small value for $U \ge  D/{\lambda}$,  does not become 
zero anywhere.  In practice 
it is extremely difficult to experimentally determine the full primary 
beam pattern ${\cal A}(\th,\,\nu)$ for a telescope. However,  
the value of $\theta_{\rm FWHM}$ is typically well determined. 
This has motivated the Gaussian approximation to be used extensively 
for both theoretical predictions \citep{bharadwaj011,bharadwaj05} and analyzing observational 
data \citep{ali08,ghosh12}. 
The close
match between ${\cal A}(\th,\,\nu)$ (eq. \ref{eq:b1})  and 
${\cal A}_G(\th,\,\nu)$ (eq. \ref{eq:b1a}) indicates that we
 may also expect the 
Gaussian approximation to provide a good fit 
to the telescope's actual primary beam pattern, particularly within the main
lobe.  This, to some extent, justifies the use of the Gaussian approximation
 in the earlier works.  The Gaussian approximation  simplifies the calculations
rendering them amenable to analytic treatment, and we use it on several 
occasions as indicated later in this paper.  For much of the investigations
presented in this paper we have considered $D=45 \ {\rm m}$ and 
$\lambda=2 \ {\rm m}$  which corresponds to    GMRT  $150 \ {\rm MHz}$ 
observations. We have also considered $D=30.75  \ {\rm m}$ and  
$\lambda=2 \ {\rm m}$ which corresponds to  LOFAR $150 \ {\rm MHz}$ 
observations.  For both these telescopes, 
Table~\ref{tab:1} summarizes  the values of some of the relevant parameters. 
Note that these values correspond to the idealized telescope model
discussed above, and they are somewhat different from the values 
actually measured for the respective telescopes. For example, the GMRT
primary beam pattern has $\theta_{\rm FWHM}=186^{'}$ whereas we have used
$\theta_{\rm FWHM}=157^{'}$ based on  our idealized  model.  We discuss 
the observational consequence of this $\sim 16\%$ difference later in 
 Section~\ref{sum} of this paper. For the rest of this paper we focus on the 
GMRT , except in  Section~\ref{lofar} where we shift our attention to LOFAR. 
Our entire analysis is based on the idealized telescope model described
above and  the relevant  parameters are listed in 
Table~\ref{tab:1}  for both these telescopes.

\begin{table}
\begin{center}
\begin{tabular}{|l|cc|c|c|c|c|}
\hline
$150 \, {\rm MHz}$&$D$  &$\theta_{\rm FWHM}$ & $\theta_0$ & $U_0$ & $\sigma_0$ \\
\hline
& &$1.03  \lambda/D$  & $0.6  \theta_{\rm FWHM} $ & $0.53/ \theta_{\rm FWHM} $ & $0.76/  \theta_{\rm FWHM} $\\
\hline
GMRT& $45 \, {\rm m}$  & $157^{'}$& $95^{'}$ &11.54 & 16.6\\
\hline
LOFAR & $30.75\, {\rm m}$ & $230^{'}$& $139^{'}$& 7.88& 11.33\\
\hline
\end{tabular} 
\caption{This shows some  relevant parameters for the primary beam pattern calculated using the 
idealized telescope model (eqs. \ref{eq:b1},\ref{eq:b2}),
and the Gaussian approximation  (eqs. \ref{eq:b1a},\ref{eq:b2a}). The parameter $\sigma_0$ is defined
in eq.~(\ref{eq:dellu}).}  
\label{tab:1}
\end{center}
\end{table}

\begin{figure}
\begin{center}
\psfrag{theta}[c][c][1.][0]{$\theta$ $[{\rm arcmin}$]}
\psfrag{Pbeam}[c][c][1.][0]{${\mathcal A}(\theta,\nu)$}
\psfrag{Bessel}[r][r][1.][0]{Model}
\psfrag{Gaussian}[r][r][1.][0]{Gaussian}
\includegraphics[width=60mm,angle=-90]{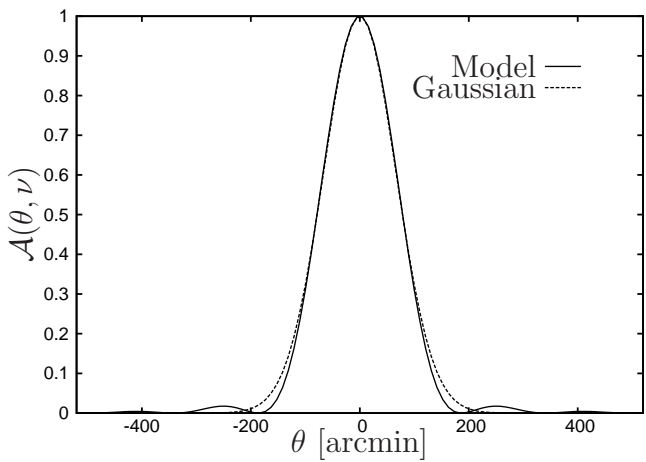}
\caption{The solid curve shows the $150\, {\rm  MHz}$   GMRT primary beam pattern ${\cal A}(\th,\,\nu)$ 
predicted by eq. (\ref{eq:b1}), and the dashed curve shows Gaussian approximation  (eq. \ref{eq:b1a}) with 
the same $\theta_{\rm FWHM}$.}
\label{fig:gauss}
\end{center}
\end{figure}

In the flat sky approximation 
the statistical properties of the background intensity fluctuations
$\delta I(\th,\,\nu)$ can be quantified through the two dimensional (2D)
 power spectrum  $P(U, \nu)$ defined as,
\begin{equation}
 \langle \Delta \tilde{I}(\u ,\nu) \Delta \tilde{I}^{*}(\u' ,\nu) \rangle = 
\delta_{D}^{2}(\u-\u') \, P(U, \nu),
\label{eq:v4}
\end{equation}where $\delta_{D}^{2}(\u-\u')$ is a two dimensional Dirac delta
function. The angular brackets $\langle ... \rangle$ here denote an
ensemble average over different realizations of the stochastic
intensity fluctuations on the sky. We also assume that the $P(U,
\nu)$ depends only on the magnitude $U=|\u|$  i.e. the fluctuations
are statistically isotropic. 
We note that $ P(U, \nu)$  is related to $C_{\ell}(\nu)$ the
angular power spectrum of the brightness temperature fluctuations
 through \citep{ali08} 
\begin{equation}
C_{\ell}(\nu) =\left( \frac{\partial B}{\partial T}\right)^{-2}
P(\ell/2 \pi, \nu) \,,
\label{eq:c2}
\end{equation}  
where the angular multipole $\ell$ corresponds to 
$U=\ell/2 \pi$, $B$ is the Planck function and $({\partial B}/{\partial T})=2
k_B/\lambda^2$ in the Raleigh-Jeans limit which is valid at the
frequencies of our interest. We will drop the $\nu$ dependence henceforth as 
the rest of the calculations are done at a fixed  frequency $ \nu = 150\, 
{\rm  MHz}$.

We now consider the two visibility correlation which is defined as 
\begin{equation}
V_2(\u, \u+\Delta \u)=\langle  \V(\u){\V}^*(\u+\Delta \u) \rangle \,,
\end{equation}
and which has the contribution 
\begin{equation}
S_2(\u, \u+\Delta \u) = \int d^2 U{'} \, \tilde{a}(\u-\u{'}) \, \tilde{a}^{*}(\u+\Delta \u-\u{'})\,P(U{'}) \,
\label{eq:v6}
\end{equation}
from the sky signal.

\begin{figure}
\begin{center}
\psfrag{1000Para}[r][r][1][0]{$\u=1,000$}
\psfrag{alpha}[Br][r][1][0]{$\beta=2.34$}
\psfrag{gaussian}[br][r][1.][0]{Gaussian}
\psfrag{x}[t][c][1][0]{$\Delta \u$}   %check***
\psfrag{y}[b][c][0.8][0]{$S_2(\u, \u+\Delta \u)/S_2(U)$}
\includegraphics[width=85mm,angle=0]{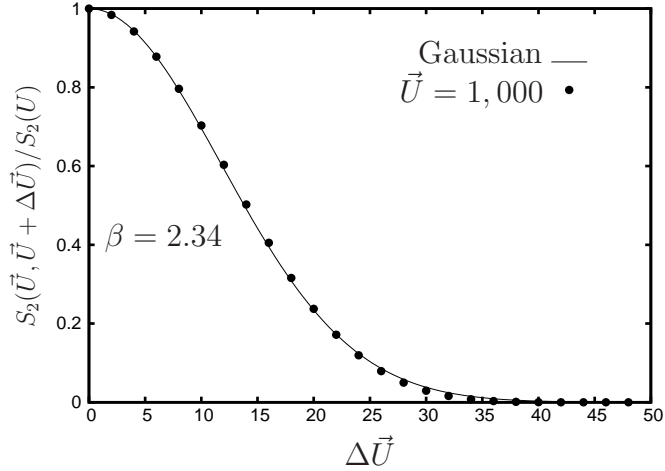}
\caption{This figure shows how the sky signal contribution to the two visibility correlation 
  varies with $\Delta\u$ for a fixed  value $U=1,000$. The points  show the 
results from eq. (\ref{eq:v6}) for $P(U)=A U^{-2.34}$,
and the solid line shows the  Gaussian fit  given in eq.  (\ref{eq:dellu}). }
\label{fig:deltau}
\end{center}
\end{figure}

The visibilities at the baselines $\u$ and $\u+\Delta\u$ are  correlated
 only if there is a significant overlap between 
 $\tilde{a}(\u-\u')$ and $\tilde{a}^{*}(\u+\Delta \u-\u')$.
The correlation  $S_2(\u,\u+\Delta\u)$  is strongest when $|\Delta \u| =0$,
declines rapidly with increasing $|\Delta \u|$,  and is zero for 
 $|\Delta \u| \ge 2D/\lambda$.  The correlation  $S_2(\u,\u+\Delta\u)$ 
depends on both, the magnitude of 
$\Delta \u$ as well as the angle between $\Delta \u$ and $\u$, and an 
earlier work \citep{bharadwaj03} has studied this in detail for the predicted 
post-reionization cosmological 21-cm signal.  In this work we have 
considered a power law power spectrum  $P(U)=A U^{-\beta}$ 
for different values of $\beta$ in the range $1.5$ to $3.5$,
and  we have used eq.~(\ref{eq:v6}) to study the  $\Delta \u$ dependence of 
$S_2(\u,\u+\Delta\u)$. We find that the  $\Delta\u$ dependence is isotropic 
to a great extent, and it can be well modelled using a Gaussian 
(Figure \ref{fig:deltau}) as 
\begin{equation} 
S_2(\u, \u+\Delta \u) = 
\exp\bigg[-\bigg(\frac{\mid \Delta \u \mid }{\sigma_0}\bigg)^2\bigg]
\, S_2(U),
\label{eq:dellu}
\end{equation}
where $\sigma_0=0.76/\theta_{\rm FWHM}$ (Table~\ref{tab:1}) and $S_2(U) \equiv 
S_2(\u,\u)$.  While the approximation in eq.~(\ref{eq:dellu}) matches the result of 
 eq.~(\ref{eq:v6}) quite well for small $\Delta \u$,  the approximation breaks down 
when  $\mid \Delta  \u \mid > 2 D/\lambda$ where $S_2(\u, \u+\Delta \u)=0$ contrary 
to the prediction of  eq~(\ref{eq:dellu}). This discrepancy, however, does not significantly 
affect the estimators (defined later) because the value of   $S_2(\u, \u+\Delta \u)$ 
predicted by eq~(\ref{eq:dellu})  is extremely small for  $\mid \Delta  \u \mid > 2 D/\lambda$.

A further simplification is possible  for $U \gg U_0$ where 
it is possible to approximate  $S_2(U)$ which is calculated using  eq.~(\ref{eq:v6}) by  assuming that 
the value of $P(U{'})$ does not  change much within 
the width of the function $|\tilde{a}\left(\u -  {\u}{'} \right)|^2 $. 
We then obtain  
\begin{equation}
S_2(U) = \left[\int d^2 U{'} \, \mid \tilde{a}(\u-\u{'}) \mid^2 \right]
P(U) \,.    
\label{eq:v6a}
\end{equation}
The integral in the square brackets has a constant value 
$\frac{\pi\theta_0^2}{2}\,$ in the Gaussian approximation which  yields the 
value $1.19\times10^{-3}$,  whereas we have $1.15\times10^{-3}$
if we use eq.~(\ref{eq:b2}) and numerically evaluate the integral in the 
square brackets. We see that the  Gaussian approximation is adequate for 
 the integral in  eq.~(\ref{eq:v6a}), and we adopt the value $\pi \theta_0^2/2$ 
for the entire subsequent analysis. 
We have calculated  $S_2(U)$  (Figure.\ref{fig:psconv})
using the convolution  in eq.~(\ref{eq:v6}), and compared this with the  
approximation in eq.~(\ref{eq:v6a}). We find that the approximation 
in eq.~(\ref{eq:v6a})  matches quite well with the convolution  (eq. \ref{eq:v6})
for baselines $U \ge 4 U_0 \sim 45$. Throughout  the subsequent analysis we have restricted 
the baselines to this range, and  we have used  eq.~(\ref{eq:v6a}) to evaluate  $S_2(U)$,
the  sky signal contribution to the visibility correlation.

\begin{figure}
\begin{center}
\psfrag{cl}[b][t][1][0]{$S_2(U)$}
\psfrag{U}[r][cB][1][0]{U}
\psfrag{a2.3}[r][c][1][0]{$\beta=2.34$}
\psfrag{a1.8}[l][c][1][0]{$\beta=1.8$}
\includegraphics[width=85mm,angle=0]{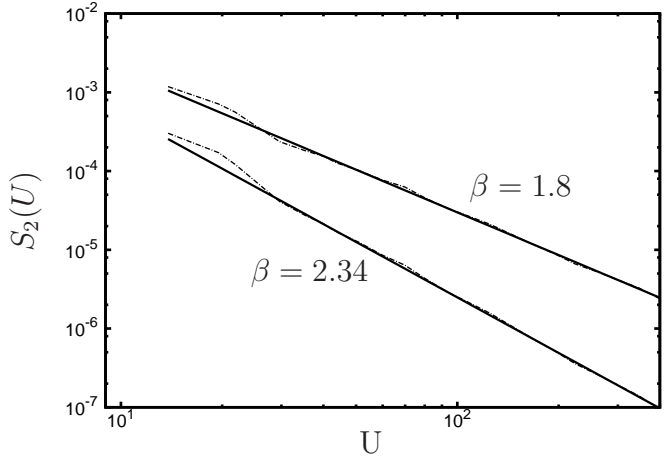}
\caption{This  shows the sky signal contribution to the visibility correlation ($S_2(U)$) 
for  two different power spectra with  slopes $\beta=1.8$ and $2.34$ respectively. 
The dash-dot curve shows the result of the convolution  in  eq.(\ref{eq:v6}) 
with $\Delta \u=0$   whereas the solid curve shows the result of  approximating this 
with eq. (\ref{eq:v6a}). We see that the approximation of eq. (\ref{eq:v6a})  matches 
the convolution reasonable well at large baselines $U \ge 4 U_0 \sim 45$.}
\label{fig:psconv}
\end{center}
\end{figure}

We finally have the approximate relation between the sky signal contribution 
to the two visibility 
correlation and the angular power spectrum 
\begin{equation}
\Sc(\u, \u+\Delta \u) = \frac{\pi \theta_0^2}{2} 
\left( \frac{\partial B}{\partial T}\right)^{2}
\exp\bigg[-\bigg(\frac{\Delta U}{\sigma_0}\bigg)^2\bigg] 
\, C_{\ell}
\label{eq:vcf}
\end{equation}
where $\ell= 2 \pi U$.  We thus see that the visibilities at two different
baselines $\u$ and $\u + \Delta \u$ are correlated only if the separation
is small $(\mid \Delta U \mid \le \sigma_0)$, and there is negligible correlation
if the separation is beyond a disk of radius $\sigma_0$. Further, the visibility
correlation $S_2(\u, \u+\Delta \u)$  
gives a direct estimate of the angular power spectrum $C_{\ell}$
at the angular 
multipole $\ell = 2 \pi U$. In addition to the sky signal $\S(\u)$, 
each  visibility also contains  a system noise contribution $\N(\u)$. 
For each visibility measurement, the real and imaginary parts of $\N(\u)$  are both   
random variables of zero mean and rms. $\sigma_n$. Further, the noise
in any two different visibilities is uncorrelated. We can then write the total
visibility correlation as 
\begin{equation}
V_{2ij} \equiv  \langle \V_i \V^{*}_j \rangle = V_0 \, e^{-\mid \Delta \u_{ij} \mid^2/
\sigma_0^2} 
\, C_{\ell_i} + \delta_{ij} 2 \sigma_n^2
\label{eq:vcorr}
\end{equation}
where $ [\V_i,\V_j] \equiv [\V(\u_i),\V(\u_j)]$, $V_0= \frac{\pi \theta_0^2}{2}
\left( \frac{\partial B}{\partial T}\right)^{2}$,  $\Delta \u_{ij}=
\u_{i}-\u_{j}$ and the Kronecker delta $\delta_{ij}$ is nonzero only if 
we correlate a visibility with itself. Equation (\ref{eq:vcorr}) relates
the two visibility correlation $V_{2ij}$ to $C_{\ell_i}$ the angular power 
spectrum  of the sky signal at the angular multipole  $\ell_i=2 \pi U_i$ 
and $\sigma_n^2$ the mean square system  noise,  
and we use this extensively in connection with 
 the estimators that we consider in the subsequent sections. 
  
\section{Simulating the sky signal}
\label{ps_simu}
We have used simulations of radio-interferometric observations to validate
the angular power spectrum estimators that we introduce in subsequent
sections of this paper.  In this section  we first describe  the simulations    
of the sky signal,  and  then describe how these were used to simulate
the expected visibilities. For the sky model, we assume that all 
point sources 
with flux above a sufficiently low threshold have been identified and removed
from the data so that the $150 \, {\rm MHz}$ radio sky is dominated 
by the diffuse Galactic Synchrotron radiation. 

The slope $\beta$ of  the angular power spectrum of diffuse Galactic synchrotron emission  is within the
 range $1.5$ to $3$ as found by all the previous measurements at frequencies $0.15 -94 \, {\rm GHz}$ 
(eg. \citealt{laporta08,bernardi09}). 
For the purpose of this paper we assume that the 
fluctuations in the 
diffuse Galactic Synchrotron radiation are a statistically homogeneous
and isotropic Gaussian random field  whose  statistical properties 
are completely specified by the  angular power spectrum. Further, we
assume that the angular power spectrum of brightness temperature 
fluctuations is well described by a single 
power law  over the entire range of angular scales of our interest. 
In this work we have adapted the angular power spectrum 
\begin{equation}
C^M_{\ell}=A_{\rm 150}  \times \left(\frac{1000}{\ell} \right)^{\beta},
\label{eq:cl150}
\end{equation}
where  $A_{\rm 150}=513 \, {\rm mK}^2$ and $\beta=2.34 $.
from \citet{ghosh12}. This is the input model for all our simulations.  

\begin{figure}
\begin{center}
\includegraphics[width=100mm,angle=0]{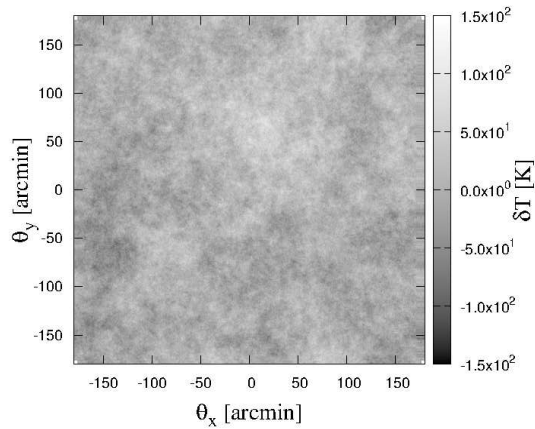}
\caption{This shows a single realization of the simulated $150 \, {\rm MHz}$ radio sky 
under the assumption that the bright point sources have been removed so that it is 
dominated by the diffuse Galactic synchrotron radiation. We have simulated a 
$5.8^{\circ} \times 5.8^{\circ}$ FoV with $\sim 10.2''$ resolution. }
\label{fig:grf}
\end{center}
\end{figure}

\begin{figure}
\begin{center}
\includegraphics[width=90mm,angle=0]{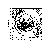}
\caption{This shows the $uv$  coverage  for $8 \, {\rm hr}$  GMRT $150 \, {\rm
    MHz}$  observations centered  on a field at a declination of $\delta=+60^{\circ}$. 
Only baselines with $\mid u \mid, \mid v \mid \le 1,000$ have been shown. 
Note that $u$ and $v$ are antenna separations measured in units of the observing wavelength,
and hence they are dimensionless.}
\label{fig:gmrtbasl}
\end{center}
\end{figure}

We have considered a  $5.8^\circ \times 5.8^\circ$  FoV  
for the GMRT simulations. This has been 
represented using  a $2048\times2048$ grid  with an angular resolution  of 
$\sim10.2$ arc-second .  We have first generated the  Fourier components of 
the brightness temperature fluctuations on the grid using ,
\begin{equation}
\Delta \tilde{T}(\u)=\sqrt{\frac{\Omega \, C_{\ell}}{2}}[x(\u)+iy(\u)],
\label{eq:ran}
\end{equation}
where $\Omega$ is the total solid angle of the simulation, and 
$x(\u)$  and $y(\u)$   are  independent Gaussian random variables with 
zero mean and unit variance.  We then use  a Fourier transform
to generate the brightness temperature fluctuations  $\delta T(\th)$ or 
equivalently the specific intensity fluctuations $\delta I(\th)$
on the grid.   Figure~\ref{fig:grf} shows one  realization
of the brightness temperature  fluctuations generated using the procedure
outlined above. We have generated $20$ different independent realizations 
of the sky by considering different sets of random numbers in 
eq.~(\ref{eq:ran}).

To simulate GMRT observations we consider $8 \, {\rm hr}$  observations 
targeted on a field located at $+60^{\circ}$ DEC  for which the $uv$ tracks 
for baselines within $\mid u \mid, \mid v \mid  \le 1,000$ are shown in
 Figure \ref{fig:gmrtbasl}.
We assume $16 {\rm s}$  integration time for each sampled visibility data
which gives us  $2,17,457$ visibility points. To calculate 
the visibilities, we have multiplied the simulated $\delta \tilde{I}(\theta)$  
with the primary beam pattern ${\mathcal A}(\th)$ (eq.~\ref{eq:b1})
and evaluated  the Fourier transform  of the product for each sampled
baseline $\u$ on the $uv$ track. In addition to the sky signal, each 
measured visibility will also have a system noise contribution. We have 
included this by adding  independent Gaussian  random noise  contributions
to both the real and imaginary parts of each visibility. This noise
is predicted to have an rms. of $\sigma_n=1.03 \, {\rm Jy}$ for
a single polarization at the GMRT.  

It is clearly visible in Figure \ref{fig:gmrtbasl} that the GMRT has a 
rather  sparse $uv$ coverage. The fact that we  have data for only a  
limited number  of the Fourier modes is expected to play an important role.
 This is  particularly  important for the cosmic variance 
which crucially depends on the number of independent Fourier modes. 
 In order to assess the impact  of the  sparse $uv$ 
coverage we have also considered a situation where exactly the same 
number of
visibility measurements  ($2,17,457$) are randomly  distributed 
within the region $\mid u \mid, \mid v \mid  \le 1,000$
on the  $uv$ plane.  

In the subsequent sections of this paper we have analyzed $20$ independent 
realizations of the sky signal,  with visibilities points that 
correspond to the $uv$ tracks shown in  Figure \ref{fig:gmrtbasl}. We 
refer to this ensemble of 20 simulated data sets 
as ``GMRT''.  We have also considered a random baseline
distribution and calculated the visibilities for the same $20$ realizations of the sky signal, 
and we  refer to this as ``Random''.  Finally, we have also carried out simulations 
for LOFAR which has a more uniform $uv$ coverage as compared to the GMRT.
These simulations are separately discussed in Section \ref{lofar}.

Finally, we note that the simulated baselines lying in the lower half of the $uv$ plane 
(e.g.  Figure \ref{fig:gmrtbasl}.)  are all folded  to the upper half  using the property 
$\V(\u)=\V^{*}(-\u)$.  The simulated baseline distribution that  we finally use for analysis 
is entirely restricted to the upper half of the $uv$ plane.

%1st est. 40 - 1000 202594
%                   170339 
%
%                  204165
%                  170616

\section{The Bare Estimator}
\label{sec:bare}
The Bare Estimator  directly uses  the individual visibilities to estimate the 
angular power spectrum. Each measured visibility corresponds to a Fourier mode of the 
sky signal, and  the visibility squared  $\mid \V \V^{*} \mid$  straight away gives the angular power  
spectrum. This  simple estimator, however, has a severe drawback because the noise contribution 
$ 2 \sigma^2_n$ is usually much larger than the  sky signal 
$ V_0 \, e^{-\mid \Delta \u_{ij} \mid^2/ \sigma_0^2} \, C_{\ell}$ in eq.~(\ref{eq:vcorr}). 
Any estimator that  includes the correlation of a visibility with itself suffers from a very large 
positive noise bias. It is, in principle, possible to model the constant noise bias and subtract it out.
This however   is extremely difficult in practice because  small calibration errors 
(discussed later in Section~\ref{sec:gerr})  would  introduce fluctuations in the noise bias 
 resulting  in  residuals that could exceed the sky signal. It is therefore desirable to avoid the 
noise bias by considering estimators which  do not include  the  contribution from the correlation of a
visibility with itself. 

The   Bare Estimator $\hat E_B(a)$ is defined as 
\begin{equation}
\hat E_B(a)=\frac{\sum_{i,j} \, w_{ij} \, \V_{i} \,  \V^{*}_{j} }{\sum_{i,j} w_{ij} V_0  
e^{-\mid \Delta \u_{ij} \mid^2/\sigma_0^2} } \,, 
\label{eq:be1}
\end{equation}
where we have  assumed that the  baselines have been divided into bins such that all the baselines
$U$ in the range $U_1 \le U < U_2$ are in  bin $1$, those in the range $U_2 \le U < U_3$ 
are in  bin $2$  etc., and $\hat E_B(a)$ refers to  a particular  bin  $a$.  The sum $i,j$ is  over all pairs of 
visibilities $\V_i,\V_j$  with  baselines $\u_i,\u_j$  in  bin $a$. 
We have restricted the sum to pairs within $\mid \u_i - \u_j \mid \le \sigma_0$
as the pairs with larger separations do not contribute much to the estimator. 
 The weight 
$w_{ij}=(1-\delta_{ij}) K_{ij}$ is chosen such that 
it is zero when we correlate a visibility with itself, thereby avoiding the positive noise bias. 

We now show that  $\hat E_B(a)$ gives an unbiased estimate of the angular 
power spectrum $C_{\ell}$ for bin $a$. 
The expectation value of the estimator can be expressed using eq. (\ref{eq:vcorr}) as 
\begin{equation}
\langle \hat E_B(a) \rangle 
=\frac{\sum_{i,j}  \, w_{ij} \, V_{2ij} }{\sum_{i,j}  w_{ij} V_0  
e^{-\mid \Delta \u_{ij} \mid^2/\sigma_0^2} }  
=\frac{\sum_{i,j} \, w_{ij} \,e^{-\mid \Delta \u_{ij} \mid^2/\sigma_0^2} C_{\ell_i}  }{\sum_{i,j}  w_{ij}  
e^{-\mid \Delta \u_{ij} \mid^2/\sigma_0^2} }  
\label{eq:be2}
\end{equation}
which can be written as 
\begin{equation}
\langle \hat E_B(a) \rangle = \bar{C}_{\bar{\ell}_a} 
\label{eq:be3}
\end{equation}
where $ \bar{C}_{\bar{\ell}_a}$ is the average  angular power spectrum  at 
 \begin{equation}
\bar{\ell}_a
=\frac{\sum_{i,j} \, w_{ij} \,e^{-\mid \Delta \u_{ij} \mid^2/\sigma_0^2} \ell_i  }{\sum_{i,j}  w_{ij}  
e^{-\mid \Delta \u_{ij} \mid^2/\sigma_0^2} }  \,.
\label{eq:be4}
\end{equation}
which is the   effective angular multipole  for bin $a$. 

We note that it is possible to express eq.~(\ref{eq:be2})  using matrix notation 
as
\begin{equation}
\langle \hat E_B(a) \rangle 
=\frac{Tr({\bf w} {\bf V}_2)}{Tr({\bf w} {\bf I}_2)}
\label{eq:be4b}
\end{equation}
where we have the matrices ${\bf w} \equiv w_{ij}$, ${\bf V}_2 \equiv V_{2ij}$, ${\bf I}_2= V_0  
e^{-\mid \Delta \u_{ij} \mid^2/\sigma_0^2}$ and $Tr({\bf A})$ denotes the trace of a matrix ${\bf A}$. 

We next evaluate  $\sigma^2_{E_B}(a)$  the variance of $\hat E_B(a)$. This  gives
$\delta C_{\ell_a}$  which is 
an estimate of the error in the  angular power spectrum measured from the data. 
We have 
\begin{equation}
[\delta C_{\ell_a}]^2 \equiv \sigma^2_{E_B}(a)=\langle \hat E^2_B(a) 
\rangle - \langle \hat E_B(a) \rangle^2
\label{eq:be5}
\end{equation} 
which can be simplified to 
\begin{equation}
\sigma^2_{E_B}(a)=\frac{\sum_{i,j,k,l} w_{ij} w_{kl} V_{2il} V_{2kj}}{[Tr({\bf w} 
{\bf I}_2)]^2}
=\frac{Tr({\bf w} {\bf V}_2 {\bf w} {\bf V}_2)}{[Tr({\bf w} {\bf I}_2)]^2}
\label{eq:be6}
\end{equation} 
under the assumptions  that  ${\bf w}$ is symmetric and  the measured 
visibilities are Gaussian 
random variables.

\begin{figure}
\begin{center}
\psfrag{cl}[b][t][1.][0]{$\ell (\ell+1) C_{\ell}/2 \pi \, [mK^2]$}
\psfrag{U}[c][c][1.][0]{$\ell$}
\psfrag{Ghosh}[cr][tr][1.][0]{$C^M_{\ell}$}
\psfrag{gmrt}[r][r][1.][0]{GMRT}
%\psfrag{Random}[r][r][1.2][0]{Random}
\includegraphics[width=80mm,angle=0]{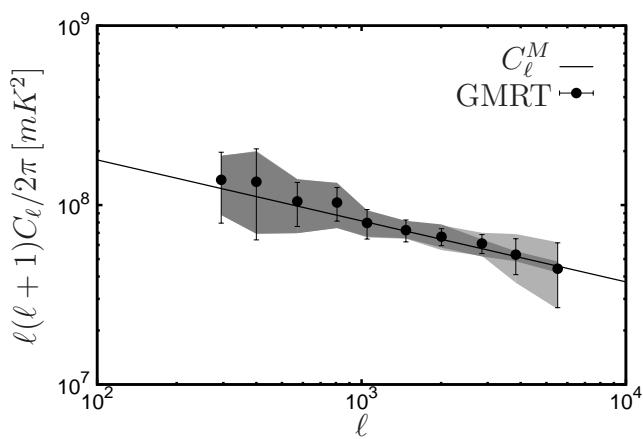}
\caption{This shows $C_{\ell}$ multiplied with $\ell (\ell + 1)/2 \pi$, 
plotted as a function of $\ell$. 
The  solid  line shows the input model (eq.~\ref{eq:cl150}) used for the 
simulations, and the points show the values recovered  by the  Bare
Estimator (eq.~\ref{eq:be1}).  The points show the mean and  the light  shaded 
region shows the $1\sigma$ variation measured from $20$ realizations of
the GMRT simulations. The dark shaded region shows the cosmic variance
which has been calculated by setting the system noise $\sigma_n=0$ in the simulation, 
 and the error bars show $1\sigma$ error bars 
predicted using eq.~(\ref{eq:be6}). The errors are dominated by the cosmic variance 
at $\ell \le 2,500$ where the dark and faint shaded regions coincide. 
We see that the Bare Estimator correctly 
recovers the input model, and the predicted error bars are consisted with 
the errors measured from the simulations.
}
\label{fig:ps1st}
\end{center}
\end{figure}

The system noise only appears in the diagonal elements of the visibility correlation matrix ${\bf V}_2$,
whereas the sky signal contributes to both the diagonal and the off-diagonal elements. Further, the diagonal
elements of the weight matrix ${\bf w}$ are all zero. Consequently the trace $Tr({\bf w}{\bf V}_2)$
in eq.~(\ref{eq:be3})  does not
pick up any contribution from the diagonal elements of  ${\bf V}_2$, and the expectation value of the 
estimator is not affected by the system noise. The variance $\sigma^2_{E_B}(a)$ however has contributions
from both diagonal and off-diagonal elements of ${\bf V}_2$. The diagonal elements  are dominated by the 
system noise, whereas the off-diagonal elements contribute to the cosmic variance. 

The weights $w_{ij}$ should, in principle, be chosen so as to maximize the signal 
to noise ratio ${\rm SNR}=\langle \hat E_B(a) \rangle/\sigma_{E_B}(a)$.  The optimal 
weights depend on the baseline distribution and $V_0 C_{\ell}/\sigma_n^2\,$, 
the relative amplitude of   the  signal to the noise  in the individual visibilities.  
Here we have made the simplifying assumption that all the visibility  pairs 
contribute equally to $\sigma^2_{E_B}(a)$. Each visibility pair is assigned 
the weight $w_{ij}=(1-\delta_{ij}) e^{-\mid \Delta \u_{ij} \mid^2/\sigma_0^2}$ which 
is proportional to its contribution to   $\langle \hat E_B(a) \rangle$. 

To test the  Bare Estimator we have used it to estimate $C_{\ell}$ from 
the simulated GMRT and Random data. For this analysis 
the visibilities with baselines $U$ in the range $40 \le U \le 1,000$ were 
divided in $20$ equally spaced logarithmic bins. Figure \ref{fig:ps1st}  
shows the mean and the rms. variation  of 
$\ell(\ell+1) C_{\ell}/2 \pi$ 
measured from the  $20$ independent realizations of the data. We find that 
the angular power spectrum estimated from the simulated GMRT data is in
good agreement  with the model (eq.~\ref{eq:cl150}) that was used 
to simulate the data. 
We next test  the predicted error estimate  $\delta C_{\ell}$ 
given by eq.~(\ref{eq:be6}).  To do this 
we have evaluated  $\sigma^2_{E_B}(a)$  by explicitly carrying out the 
sum $\sum_{ijkl}$  where the indices each runs over all the baselines
in bin $a$. For ${\bf V}_2$ (eq.~\ref{eq:vcorr}) we have used the mean 
$C_{\ell}$ estimated from the $20$ realizations and the value of 
$\sigma_n$ that was used  for the system noise in the simulation. We
find that $\delta C_{\ell}$ predicted by the analytic error estimate 
(eq.~\ref{eq:be6}) is in reasonably good agreement with the rms.
obtained from the $20$ independent realizations of the data.  
The results for the Random data are very similar to those for GMRT, 
and we have not shown these separately here. 

In conclusion of this section we find that the Bare Estimator 
(eq.~\ref{eq:be1}) is able to successfully extract the angular power 
spectrum directly from the measured visibilities. We further show that 
(eq.~\ref{eq:be6}) provides  a reasonably good estimate of the statistical 
errors for the measured angular power spectrum. The errors depend 
on the choice of the weights $w_{ij}$, the baseline distribution, the 
magnitude of the signal  and the system noise.   In  Figure \ref{fig:ps1st} 
we  see that the error decreases
with increasing $\ell$ until $\ell \sim 2,500$ beyond which the error 
increases again. We find that this feature does not change significantly
between the GMRT and the Random simulations. Based on this we 
conclude that this behaviour of the error is largely determined 
by the relative contributions from the signal whose magnitude 
falls with $\ell$ and the system noise which has been  assumed to be 
constant across all baselines. 
The errors at $\ell \le 2,500$ are cosmic variance dominated, 
whereas the errors are dominated by the system noise at larger 
$\ell$.  

\section{The Tapered  Gridded Estimator}
\label{sec:grid}
The telescope primary beam 
is usually not very well quantified at large angles 
where we have the frequency dependent pattern of nulls and sidelobes
(Figure~\ref{fig:gauss}). Point sources
located near the nulls and the sidelobes are a problem for estimating 
the angular power spectrum of the diffuse background radiation. 
Further, point sources located far away from the pointing center, 
particularly those located near the nulls, introduce ripples along the 
frequency 
direction in the multi-frequency angular power spectrum.
This poses a severe problem for separating the foregrounds 
from the cosmological 21-cm signal. As pointed out in \citet{ghosh2}, 
it is possible to avoid these problems  by tapering the sky response
 through 
a frequency independent window function ${\cal W}(\theta)$.  In this work we
choose a Gaussian ${\cal W}(\theta)=e^{-\theta^2/\theta^2_w}$  
such that $\theta_w = f \theta_0$
with $f \le 1$ so that the window function  cuts off the sky response 
well  before the first null. This tapering is achieved by convolving 
the measured visibilities 
\begin{equation}
\V_c(\u)=\tilde{w}(\u)\otimes \V(\u)
\label{eq:ge1}
\end{equation}
where $\tilde{w}(\u)=\pi\theta_w^2e^{-\pi^2U^2\theta_w^2}$ is the Fourier 
transform of ${\cal W}(\theta)$. 
The convolved visibilities $\V_c(\u)$ are the Fourier transform 
of the product ${\cal W}(\theta)\,  {\cal A}(\theta) \, \delta I(\th)$
whose  sky response  can be well controlled through the window 
function ${\cal W}(\theta)$. 

Current  radio interferometers are expected to produce considerably 
large volumes of visibility data in observations spanning many 
frequency channels and large observing times.  Given the potentially 
large computational requirement,  it is  useful to compress the 
visibility data by gridding it.  We choose a rectangular grid
in the $uv$ plane and consider the convolved visibilities  
\begin{equation}
\V_{cg} = \sum_{i}\tilde{w}(\u_g-\u_i) \, \V_i
\label{eq:ge2}
\end{equation}
where $\u_g$ refers to the different grid points and $\V_i$ 
refers to the measured visibilities. We now focus our attention 
on $\S_{cg}=\sum_{i}\tilde{w}(\u_g-\u_i) \, \S_i$ which is the sky
signal contribution to $\V_{cg}$. This can be written as 
\begin{equation}
\S_{cg}=\int d^2 U \, \tilde{w}(\u_g-\u) B(\u) \S(\u)
\label{eq:ge3}
\end{equation}
where $B(\u)=\sum_i \delta^2_D(\u-\u_i)$ is the  baseline sampling function
of the measured visibilities and $\delta^2_D(\u)$ is the 2D Dirac delta
function. The integral in eq.~(\ref{eq:ge3}) is dominated by the contribution
from within a disk of radius $\sim (\pi \theta_w)^{-1}$ centered
around $\u_g$. Assuming that the sampling function  $B(\u)$ is nearly uniform 
within this disk  we can  replace $B(\u)$ in eq.~(\ref{eq:ge3}) by its 
average value 
\begin{equation}
\bar{B}(\u_g)=\left[\frac{\int d^2 U \, \tilde{w}(\u_g-\u) B(\u)}
{\int d^2 U \, \tilde{w}(\u_g-\u)} \right]
\label{eq:ge4a}
\end{equation}
 evaluated at the grid point $\u_g$.  We  then have the approximate equation 
\begin{equation}
\S_{cg}= \bar{B}(\u_g) \int d^2 U \, \tilde{w}(\u_g-\u)  \S(\u) \,.
\label{eq:ge4}
\end{equation}
Considering eq.~(\ref{eq:ge4a}) for $\bar{B}(\u_g)$, the denominator has value ${\cal W}(0)=1$  whereby
$\bar{B}(\u_g)=\sum_i  \tilde{w}(\u_g-\u_i)$ and we have 
\begin{equation}
\S_{cg}=\left[\sum_i  \tilde{w}(\u_g-\u_i)  \right] \int d^2 U \, \tilde{w}(\u_g-\u)  \S(\u) \,. 
\label{eq:ge5}
\end{equation}
We note that eq.~(\ref{eq:ge5})  holds only if we have an uniform  and sufficiently dense 
baseline distribution in the vicinity of the
grid point $\u_g$. This breaks down if we have a patchy and sparse baseline distribution, and it is 
then necessary to use
 \begin{equation}
\S_{cg}=\sum_i  \tilde{w}(\u_g-\u_i)   \S(\u_i) \,. 
\label{eq:ge5b}
\end{equation}
 In such a situation  it is necessary to take the exact patchy $uv$ distribution  into account, and  it is 
difficult to make generic analytic predictions. Here  we have assumed an uniform 
baseline distribution,  and we have used eq.~(\ref{eq:ge5})  extensively in the subsequent calculations,

The integral in eq.~(\ref{eq:ge5})  is the Fourier  transform of the product 
${\cal W}(\theta)\,  {\cal A}(\theta) \, \delta I(\th) \equiv 
 {\cal A_W}(\theta) \, \delta I(\th)$. We may think of $ {\cal A_W}(\theta)$
as a modified primary beam pattern which has a new $\theta_{\rm FWHM}$
which is a factor $f/\sqrt{1+f^2}$ smaller than  $\theta_{\rm FWHM}$
given in Table~\ref{tab:1} and whose sidelobes are strongly suppressed. 
We can approximate the modified primary beam pattern  as a Gaussian 
$ {\cal A_W}(\theta)=e^{-\theta^2/\theta_1^2}$ with $\theta_1=f (1+f^2)^{-1/2} \theta_0$.
Using this, we can generalize eq.~(\ref{eq:vcorr}) to calculate  the correlation 
of the gridded visibilities 
$V_{c2g g^{'}} = \langle \V_{c g}  \V^{*}_{c g^{'}} \rangle$. 
The crucial point is that we have to replace $V_0$ and $\sigma_0$ in  eq.~(\ref{eq:vcorr}) 
with $V_1= \frac{\pi \theta_1^2}{2} \left( \frac{\partial B}{\partial T}\right)^{2}$ and
$\sigma_1 =  f^{-1} \sqrt{1+f^2} \sigma_0$ in order to account for the modified primary beam
pattern ${\cal A_W}(\theta)$.  We then have 

\begin{equation}
V_{c2g g^{'}} = K_{1g} K^{*}_{1 g^{'}} V_1 e^{-\mid \Delta \u_{g g^{'}} \mid^2/\sigma_1^2} C_{\ell_g} +
2 \sigma_n^2 K_{2 g g^{'}} 
\label{eq:ge7}
\end{equation}

where $\ell_g=2 \pi U_g$, $K_{1g}=\sum_i  \tilde{w}(\u_g-\u_i)$, 
$K_{2g g^{'} }=\sum_i  \tilde{w}(\u_g-\u_i) \tilde{w}^{*}(\u_{g^{'}}-\u_i)$
and  $\Delta \u_{g g^{'}}= \u_{g}-\u_{g^{'}}$.

We now define the estimator ${\hat E}_g$ for the angular power spectrum at a single grid point 
$g$ as 
\begin{equation}
{\hat E}_g=\frac{(\V_{cg} \V^{*}_{cg} - \sum_i \mid  \tilde{w}(\u_g-\u_i) \mid^2 \,  
\mid \V_i \mid^2)}{(\mid K_{1g} \mid^2 V_1 - K_{2gg} V_0)} \,.
\label{eq:ge8}
\end{equation}
Using eq.~(\ref{eq:ge7}) and eq.~(\ref{eq:vcorr})   respectively to evaluate the  expectation values 
\begin{equation}
\langle \V_{cg} \V^{*}_{cg}  \rangle =  \mid K_{1g} \mid^2  V_1 C_{\ell_g} + 2 \sigma_n^2 K_{2 g g} 
\end{equation}
and 
\begin{equation}
 \sum_i \mid  \tilde{w}(\u_g-\u_i) \mid^2 \,  \langle \mid \V_i \mid^2 \rangle 
=  V_0 \sum_i \mid  \tilde{w}(\u_g-\u_i)\mid^2  C_{\ell_i} + 2 \sigma_n^2  K_{2gg} 
\end{equation}
we see that the system noise contributions to these two terms are exactly equal and it exactly 
cancels out in  $\langle {\hat E}_g \rangle$. Further, assuming that 
$ \sum_i \mid  \tilde{w}(\u_g-\u_i) \mid^2 C_{\ell_i}   \approx C_{\ell_g} K_{2gg}$ we have 
\begin{equation}
\langle {\hat E}_g \rangle = C_{\ell_g} \,.
\end{equation}
We see that ${\hat E}_g$ defined in eq.~(\ref{eq:ge8}) gives an unbiased  estimate of the angular
power spectrum $C_{\ell}$ avoiding the positive noise bias caused by the  system noise. 

The terms $K_{1g}$ and $K_{2gg}$ in eq.~(\ref{eq:ge8}) are both proportional to $N_g$ the number of visibilities 
that contribute to the grid point $g$.  For large  $N_g$   it is reasonable to assume that   $\mid K_{1g} \mid^2 \gg K_{2gg}$
and  we thereby  simplify   eq.~(\ref{eq:ge8}) to obtain 
\begin{equation}
{\hat E}_g=\frac{(\V_{cg} \V^{*}_{cg} - \sum_i \mid  \tilde{w}(\u_g-\u_i) \mid^2 \,  
\mid \V_i \mid^2)}{\mid K_{1g} \mid^2V_1}
\label{eq:ge9}
\end{equation}
for the estimator. 

 We  use this to define the binned Tapered Gridded Estimator 
\begin{equation}
{\hat E}_G(a) = \frac{\sum_g w_g  {\hat E}_g}{\sum_g w_g } \,.
\end{equation}
where $w_g$ refers to the weight assigned to the contribution from any particular 
grid point. This has an expectation value 
\begin{equation}
\langle {\hat E}_G(a) \rangle = \frac{ \sum_g w_g C_{\ell_g}}{ \sum_g w_g}
\label{eq:ge10}
\end{equation}
which can be written as 
\begin{equation}
\langle \hat E_G(a) \rangle = \bar{C}_{\bar{\ell}_a} 
\label{eq:ge11a}
\end{equation}
where $ \bar{C}_{\bar{\ell}_a}$ is the average  angular power spectrum  at 
 \begin{equation}
\bar{\ell}_a =
\frac{ \sum_g w_g \ell_g}{ \sum_g w_g}
\label{eq:ge11}
\end{equation}
which is the   effective angular multipole  for bin $a$. 

We next calculate the variance of $\hat E_G(a)$ defined as 
\begin{equation}
[\delta C_{\ell_a}]^2 \equiv \sigma^2_{E_G}(a)=\langle \hat E^2_G(a) 
\rangle - \langle \hat E_G(a) \rangle^2\,.
\label{eq:ge12}
\end{equation} 
 Explicitly using eq.~(\ref{eq:ge9}) yields a rather unwieldy expression 
which is not very useful for making analytic predictions  for the variance. 
The first term   in the  numerator  of eq.~(\ref{eq:ge9})   which is of  order $N_g^2$ 
 makes a much larger  contribution to the variance than the second term  
$\sum_i \mid  \tilde{w}(\u_g-\u_i) \mid^2 \,   \mid \V_i \mid^2$ which is of order $N_g$.
In our analysis  we make the simplifying assumption  that we can drop the 
second term which yields 
\begin{equation}  
\sigma^2_{E_G}(a) = \frac{\sum_{g g^{'}} w_g w_g^{'}  \mid K_{1g}^{-1}   K_{1g^{'}}^{*-1} 
 V_{c2g g^{'}} \mid^2}{V_1^2[\sum_{g } w_g]^2} \,.
\label{eq:ge13}
\end{equation}
We further approximate $K_{2g g^{'}}= e^{-\mid \Delta \u_{g g^{'}} \mid^2/\sigma_1^2} K_{2gg}$
which allows us to write the variance as 
\begin{equation}  
\sigma^2_{E_G}(a) = \frac{\sum_{g g^{'}} w_g w_g^{'} e^{-2 \mid \Delta \u_{g g^{'}} \mid^2/\sigma_1^2}
\mid  C_{\ell_g} + \frac{2 K_2g g^{'} \sigma_n^2}{K_{1g} K^{*}_{1 g^{'}} V_1}  \mid^2}
{[\sum_{g } w_g]^2}
\label{eq:ge14}
\end{equation}
using eq.~(\ref{eq:ge7}).

\begin{figure*}
\begin{center}
\psfrag{cl}[b][b][1.][0]{$\ell (\ell+1) C_{\ell}/2 \pi \, [mK^2]$}
\psfrag{U}[c][c][1.][0]{$\ell$}
\psfrag{Ghosh}[cr][tr][1.][0]{$C^M_{\ell}$}
\psfrag{gmrt}[r][r][1.][0]{GMRT}
\psfrag{rndm}[r][r][1.][0]{Random}
\includegraphics[width=75mm,angle=0]{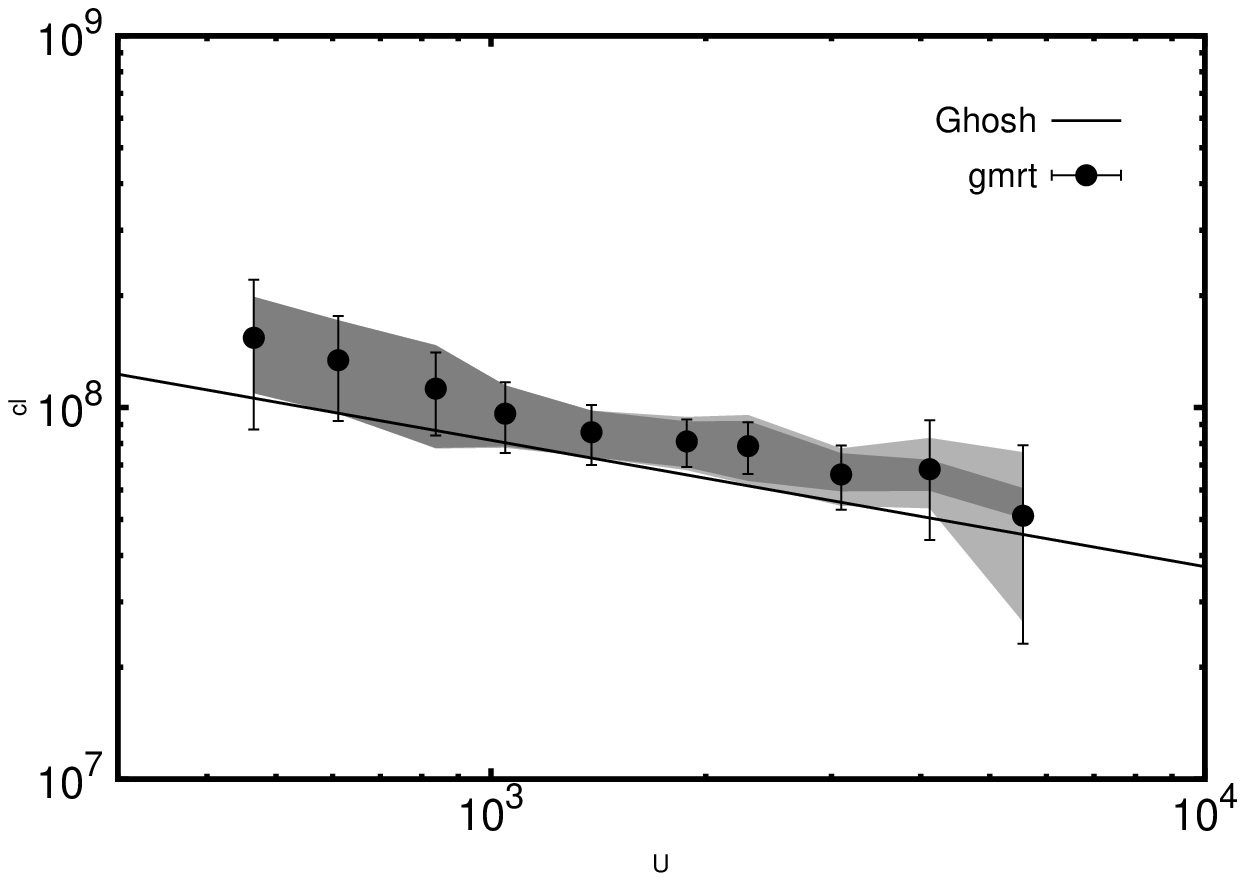}
\includegraphics[width=75mm,angle=0]{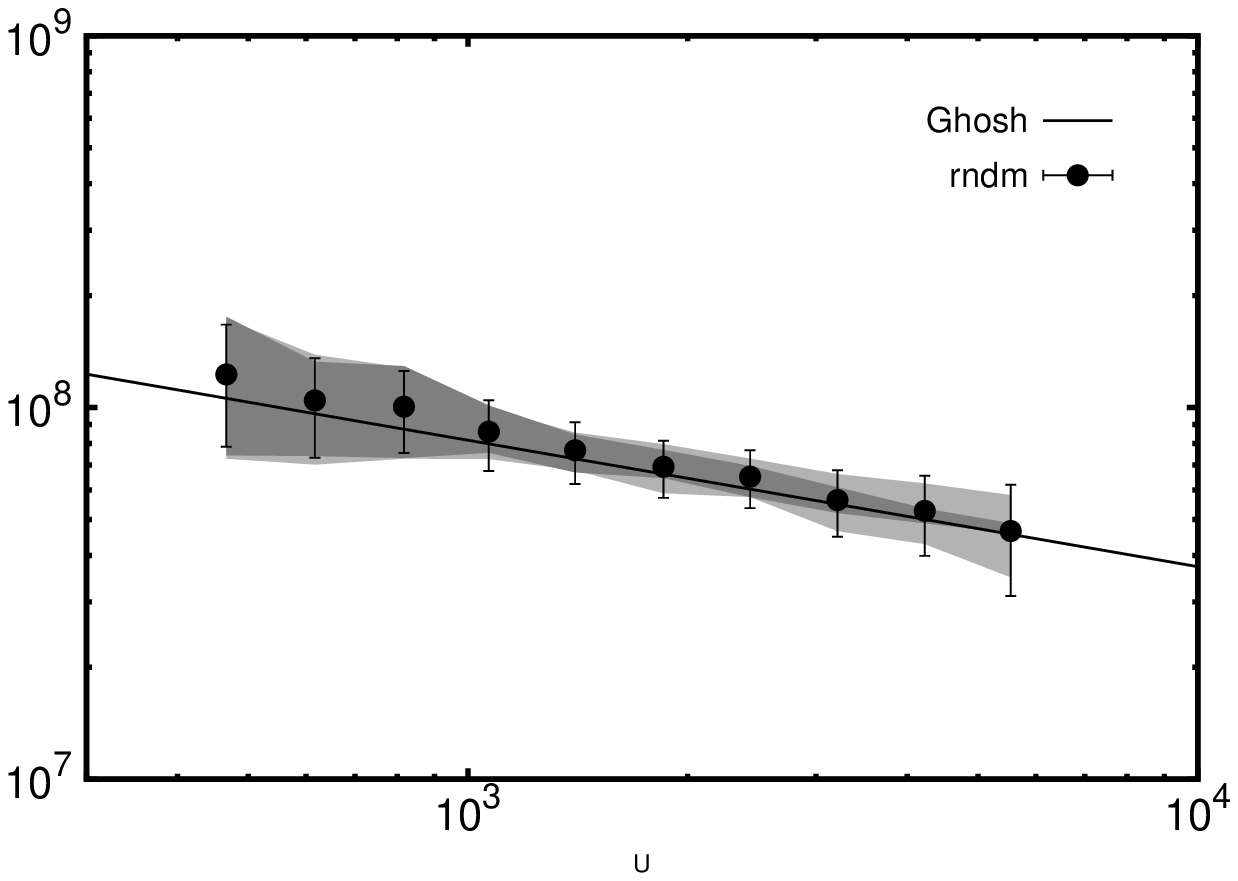}
\caption{Same as Figure \ref{fig:ps1st}, but for the Tapered Gridded Estimator.}
\label{fig:ps2nd}
\end{center}
\end{figure*}

We have applied the Tapered Gridded Estimator to the simulated GMRT and Random data.
The $20$ realizations were used to calculate the mean and the variance of the estimated $C_{\ell}$.
We have considered the values $f=1.0,0.8, 0.65$ and $0.4$ for the tapering window, and have 
also tried two different weight schemes $w_g=1$ and $w_g=K^2_{1g}$ respectively. The former assigns
 equal weight to every grid point that has same data, this is expected to minimize the cosmic variance.
 The latter scheme assigns a larger weight to grid points which have a denser visibility sampling
 relative to the grid points with sparser sampling. This is expected to minimize the system noise
 contribution.  The grid spacing $\Delta U$ in the $uv$ plane
is chosen based on two considerations. A very small value of $\Delta U$ results in a very large number
of grid points which do not contain independent signal contributions. This also unnecessarily increases 
the computation time. In contrast, a large value of $\Delta U$ implies that the signal in many visibilities 
is very poorly represented in the gridded data, resulting in  a loss of signal.   We have chosen a  grid spacing
 $\Delta u=\sqrt{\ln2}/(2\pi\theta_w)$ which corresponds to one fourth of the FWHM of $\tilde{w}(\u)$ as an 
optimum value. For any fixed grid position $\u_g$, we have restricted the contribution to baselines $\u_i$ 
 within $\mid \u_g - \u_i \mid \le 6 \Delta U$. The weight function  $\tilde{w}( \u_g - \u_i)$ falls considerably
and we do not expect a significant contribution from the visibilities beyond this baseline  separation. 
The tapering also modifies the smallest baseline where the approximation of eq.~(\ref{eq:v6a}) is valid, 
and the grid points $\u_g$ in the range $U_{min}=\sqrt{1+f^2} f^{-1} 40$ to $1,000$ were binned into 
$10$ equally spaced logarithmic bins for this analysis.

Figure \ref{fig:ps2nd} shows the results for $f=0.8$ and $w_g=\mid K_{1g}\mid^2$. We see that for  both 
GMRT and Random 
the estimated $C_{\ell}$ are roughly within the $1\sigma$ region of the input model angular power spectrum 
$C^{M}_{\ell}$. For GMRT, 
however, the estimated $C_{\ell}$ values all appear to be somewhat in excess of  $C^M_{\ell}$ indicating 
that we have an overestimate of  the angular power spectrum relative to $C^{M}_{\ell}$. In 
comparison, the $C_{\ell}$ values are in better agreement with $C^{M}_{\ell}$ for the Random simulation. 
For both GMRT and Random the error estimates predicted by
eq.~(\ref{eq:ge14}) are in good agreement with the rms. fluctuation
estimated from the 20 realizations.  We note  that the rms. fluctuation of $C_{\ell}$ is more for GMRT in 
comparison to Random.

The  Tapered Gridded Estimator is  expected to give  an unbiased estimate of  $C_{\ell}$ 
provided  we have a uniform and sufficiently  dense  baseline distribution. We  test this 
using  the Random simulations which have a uniform baseline distribution.  
In such a  situation we expect the  deviation $C_{\ell} -C_{\ell}^M $ to arise purely from 
 statistical fluctuations. The deviation  is expected to have values  around  $~\sigma/\sqrt{N_r}$  
and   converge to $0$ as $N_r$, the number of  realizations,  is increased. 
For this purpose we have studied (Figure \ref{fig:dev})
how  the  fractional deviation $(C_{\ell}-C_{\ell}^M)/C_{\ell}^M$  
  varies if we increase the number of  realizations from  $N_r=10$ to  $100$. We find that it is 
more convenient to use $20$ equally spaced logarithmic bins in $\ell$ to highlight the convergence
of the fractional deviation with increasing $N_r$. Note that we have used $10$ bins (as mentioned earlier)
everywhere except in (Figure \ref{fig:dev}). 
For the Random simulation (right panel), 
 we find that as the number of realizations is increased  the convergence of the  fractional 
deviation to $0$ is clearly visible  for $\ell \ge 1.2 \times 10^3$ $(U \ge 200)$.
Further, the fractional deviation is also found to be consistent with $\sigma/(\sqrt{20} \, C_{\ell}^M)$
and  $\sigma/(10 \, C_{\ell}^M)$ expected for $N_r=20$ and $100$ respectively. 
At smaller baselines, however, the behaviour is not so clear.  The approximation 
eq.~(\ref{eq:v6a}) for the  convolution and the 
approximation for the primary beam pattern each introduce around $2-5 \%$ errors 
in the estimated $C_{\ell}$ at small baselines. Further, for a uniform baseline distribution 
the bins at the smallest  $\ell$ values contain  fewer  baselines and also fewer grid points, and are 
susceptible to larger  fluctuations. The discrete $uv$ sampling due to the finite number of baselines 
is also expected to introduce some errors at all values of $\ell$. To test this effect, 
we have considered a situation where $N_r=100$ and the total number of baselines is 
increase to  $869,828$ which is  a factor of $4$ larger compared to the other simulations. 
We find that for  $\ell \ge 3 \times 10^3$ the  fractional deviation  falls from $\sim 5 \%$
to $\sim 2 \%$ when the baseline density is increased, this difference is not seen at smaller
baselines. In summary, the tests clearly show that for a uniform baseline distribution
the estimator is unbiased for  $\ell \ge 1.2 \times 10^3$.  In contrast, for the GMRT (left panel)
the fractional  deviation does not converge to $0$ as $N_r$ is increased. We see that $C_{\ell}$ 
is overestimated at all values of $\ell$. As mentioned earlier, the GMRT has a patchy $uv$ coverage  
for which eq.(\ref{eq:ge4}),  which assumes a uniform baseline distribution, breaks down.  
The overestimate is a consequence of GMRT's patchy $uv$ coverage, and is not inherent  to the  Tapered 
Gridded Estimator. The rms. fluctuations also are larger for GMRT in comparison to the Random simulations
(Figure \ref{fig:ps2nd}).  This too  is a consequence of GMRT's patchy $uv$ coverage.

\begin{figure}
\begin{center}
\psfrag{dev}[c][c][1.][0]{$(C_{\ell}-C^M_{\ell}) /C^M_{\ell}$}
\psfrag{U}[c][c][1.][0]{$\ell$}
\psfrag{gmrt}[c][c][1.][0]{GMRT}
\psfrag{rndm}[c][c][1.][0]{Random}
%\psfrag{100}[r][r][1.][0]{100}
%\psfrag{20}[r][r][1.0][0]{20}
%\psfrag{10}[r][r][1.0][0]{10}
\psfrag{a100}[r][r][0.8][0]{100}
\psfrag{a10}[r][r][0.8][0]{10}
\psfrag{a20}[r][r][0.8][0]{20}
\psfrag{b100}[r][r][0.8][0]{100a}
\includegraphics[width=75mm,angle=0]{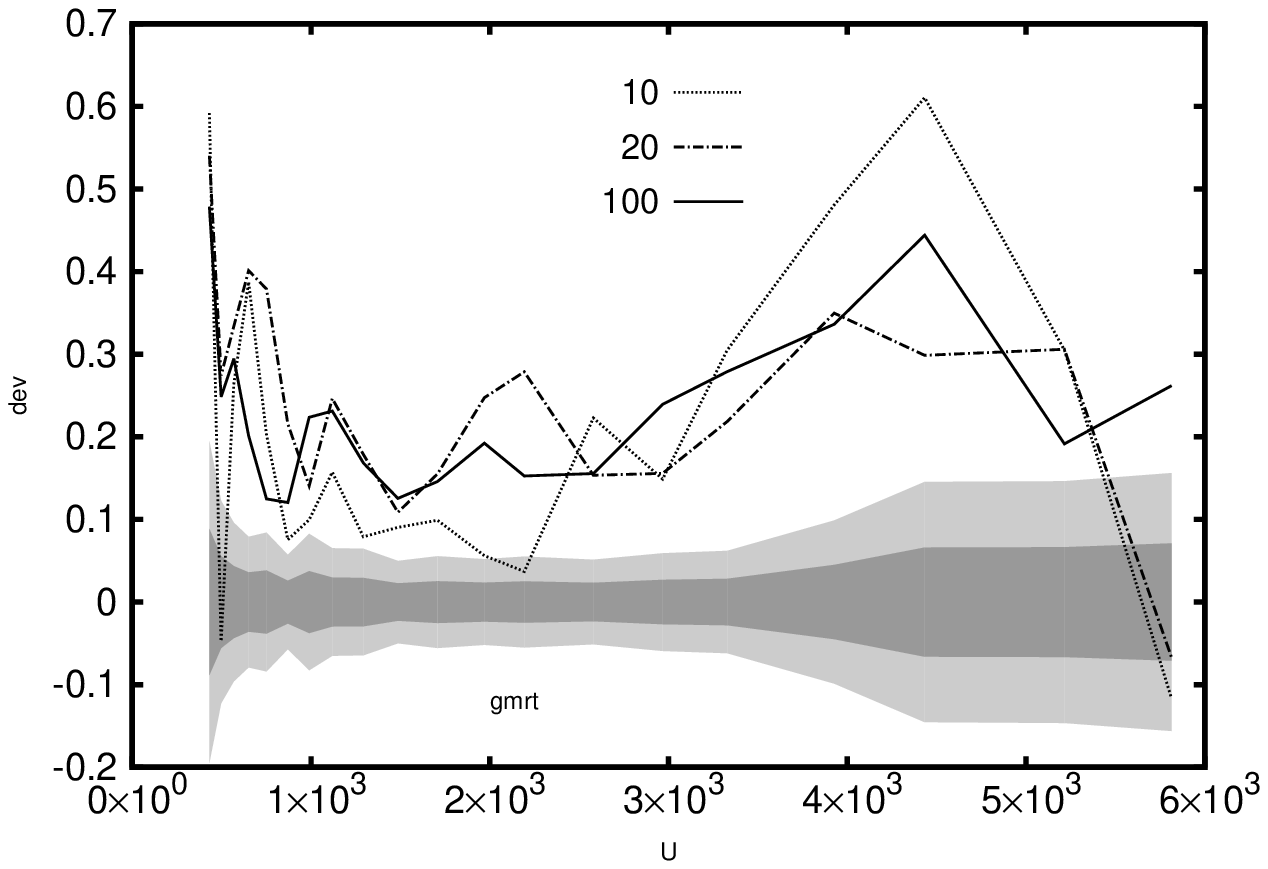}
\includegraphics[width=75mm,angle=0]{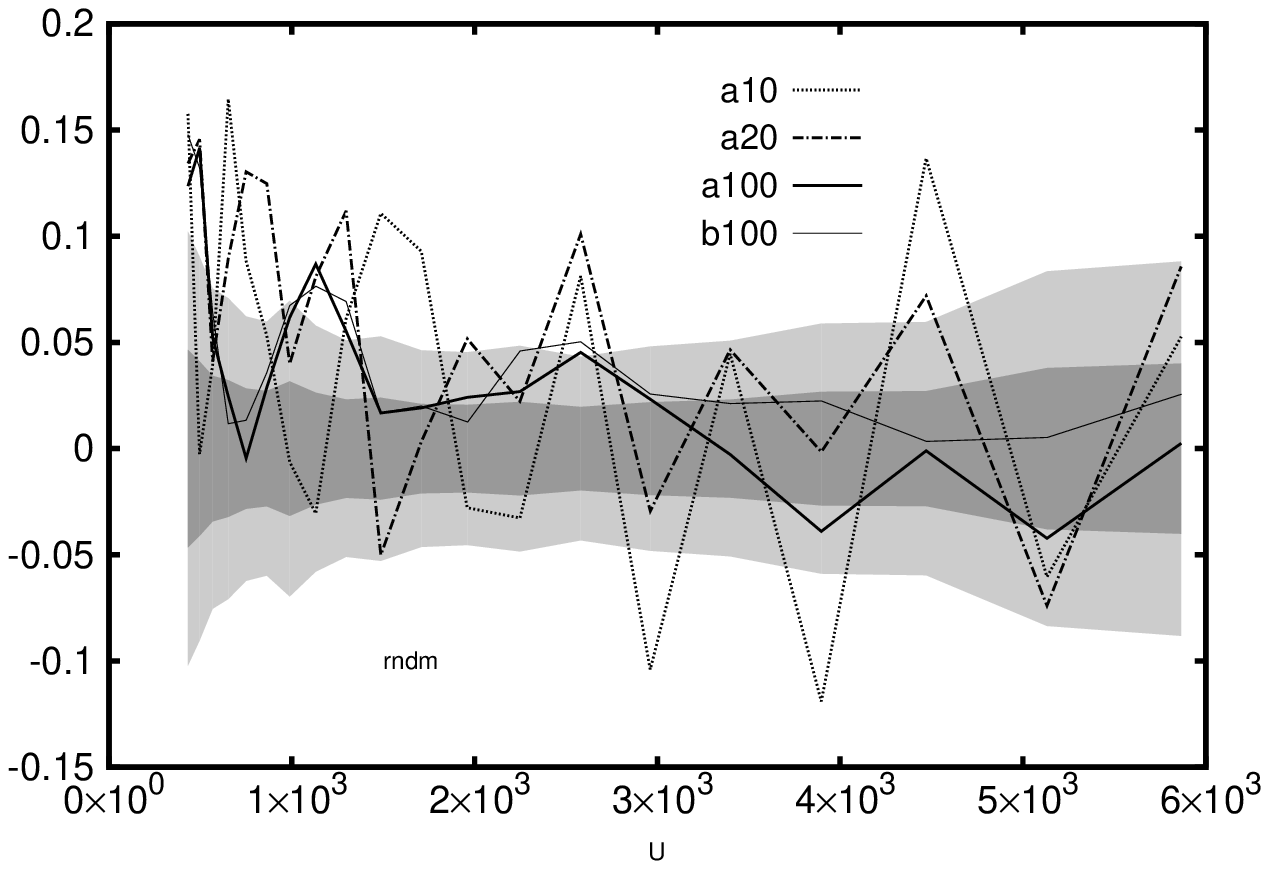}
\caption{The different curves  show  the fractional deviation  $(C_{\ell}-C^M_{\ell})/C^M_{\ell}$
for the  different numbers of realizations $(N_r)$ shown in the figure.  The  curve  100a corresponds to 
$N_r=100$ with  $869,828$ baselines,   which is  $4$ times 
the number of baselines in the other simulations. The two shaded region show  $\sigma/(\sqrt{Nr} \, C_{\ell}^M)$
for $N_r=20$ and $100$ respectively. We have used $f=0.8$ and $w_g=\mid K_{1g} \mid^2$, with $20$ equally spaced
logarithmic bins in $\ell$.}
\label{fig:dev}
\end{center}
\end{figure}

We now study how the estimator behaves for different values of
$f$. Figure \ref{fig:dev2ndwt2} and Figure \ref{fig:sig2ndwt2}
respectively show the relative deviation $(C_{\ell}-C^M_{\ell})
/C^M_{\ell}$ and the relative error $\sigma/C^M_{\ell}$ for different
values of $f$ with $w_g=\mid K_{1g} \mid^2$. Here, $C_{\ell}$ and
$\sigma$ refer to the mean and rms. estimated from the 20
realizations. We find that the deviations are roughly within the
$1\sigma$ errors for all the cases that we have considered.  For
GMRT, the deviation increases with decreasing $f$. This effect is only
visible at low $\ell$ for Random. The error $\sigma$, increases with
$f$ for both GMRT and Random. In all cases, the error is found to
decrease until $\ell\sim2000$ and then increase subsequently. As
mentioned earlier for the Bare Estimator, we interpret this as a
transition from cosmic variance to system noise dominated errors as
$\ell$ is increased. The sky coverage of the modified primary beam
${\cal A_W}(\theta)$ falls with a decrease in $f$. This explains the
behaviour of the cosmic variance contribution which increases as $f$
is reduced. We further see that the system noise contribution also
increases as $f$ is reduced. This can be attributed to the term
$V_1=\frac{\pi\theta_1^2}{2}$ which appears in
eq.~(\ref{eq:ge14}). This effectively increases the system noise
contribution relative to $C_{\ell}$ as $f$ is reduced.

\begin{figure}
\begin{center}
\psfrag{cl}[b][b][1.][0]{$ (C_{\ell}-C^M_{\ell}) /C^M_{\ell}$}
\psfrag{U}[c][c][1.][0]{$\ell$}
\psfrag{f1wt2}[c][c][0.8][0]{f=1}
\psfrag{f2wt2}[r][r][0.8][0]{f=0.8}
\psfrag{f3wt2}[r][r][0.8][0]{f=0.65}
\psfrag{f4wt2}[r][r][0.8][0]{f=0.4}
\psfrag{gmrt}[c][c][1.][0]{GMRT}
\psfrag{rndm}[c][c][1.][0]{Random}
\includegraphics[width=75mm,angle=0]{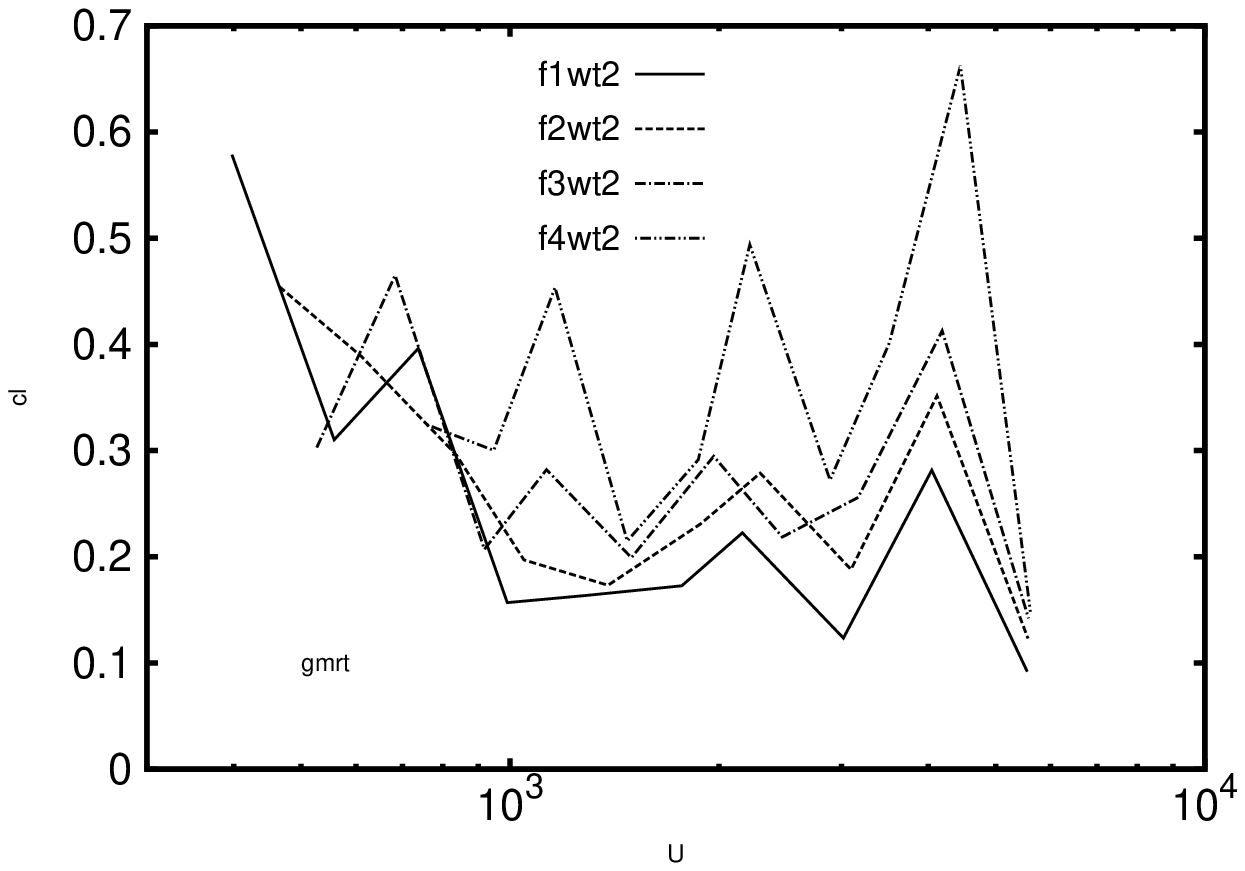}
\includegraphics[width=75mm,angle=0]{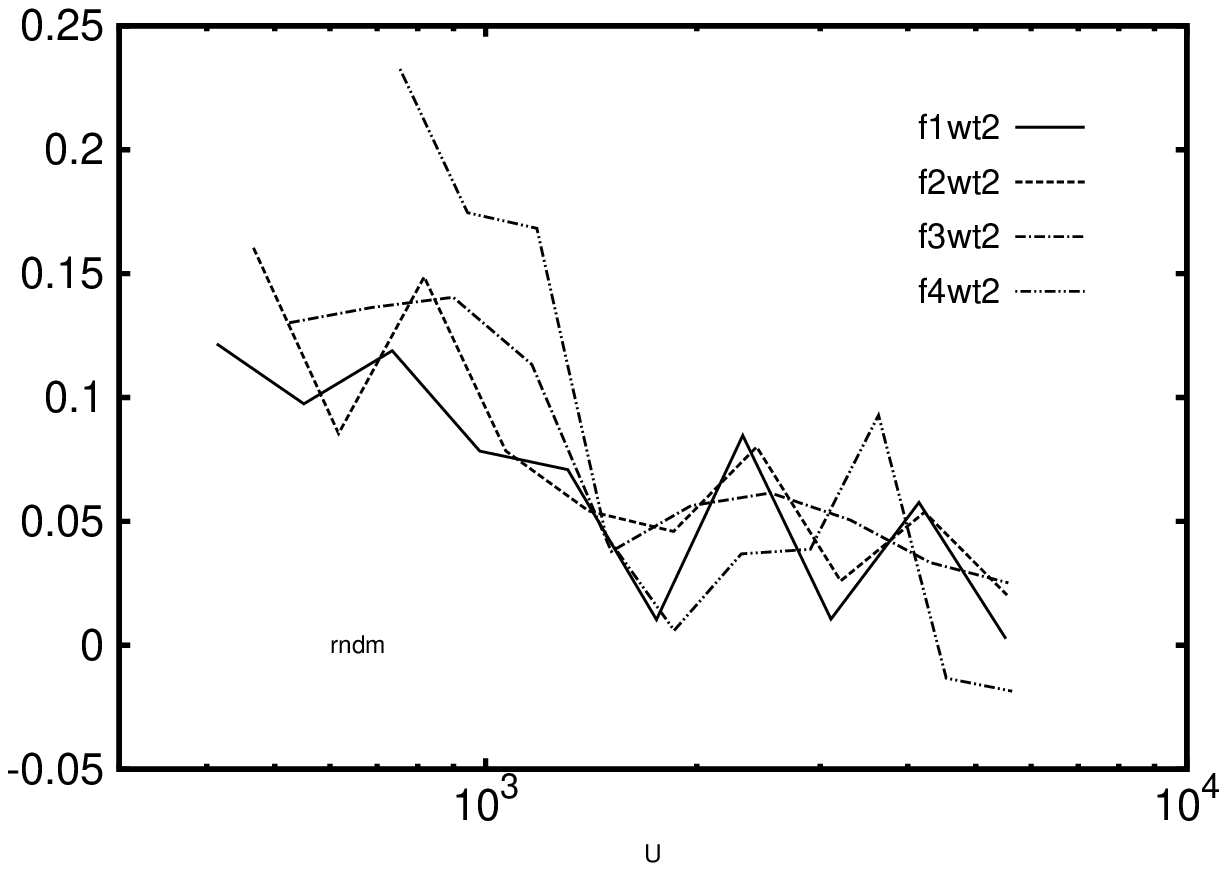}
\caption{This shows the fractional deviation of the estimated $ C_{\ell}$ from the input model $ C^M_{\ell}$. Here, we have used $w_g=\mid K_{1g} \mid^2$ and the  different $f$ values  shown in this figure.}
\label{fig:dev2ndwt2}
\end{center}
\end{figure}

\begin{figure}
\begin{center}
\psfrag{cl}[b][b][1.][0]{$\sigma/C^M_{\ell}$}
\psfrag{U}[c][c][1.][0]{$\ell$}
\psfrag{f1wt2}[c][c][0.8][0]{f=1}
\psfrag{f2wt2}[r][r][0.8][0]{f=0.8}
\psfrag{f3wt2}[r][r][0.8][0]{f=0.65}
\psfrag{f4wt2}[r][r][0.8][0]{f=0.4}
\psfrag{gmrt}[c][c][1.][0]{GMRT}
\psfrag{rndm}[c][c][1.][0]{Random}
\includegraphics[width=75mm,angle=0]{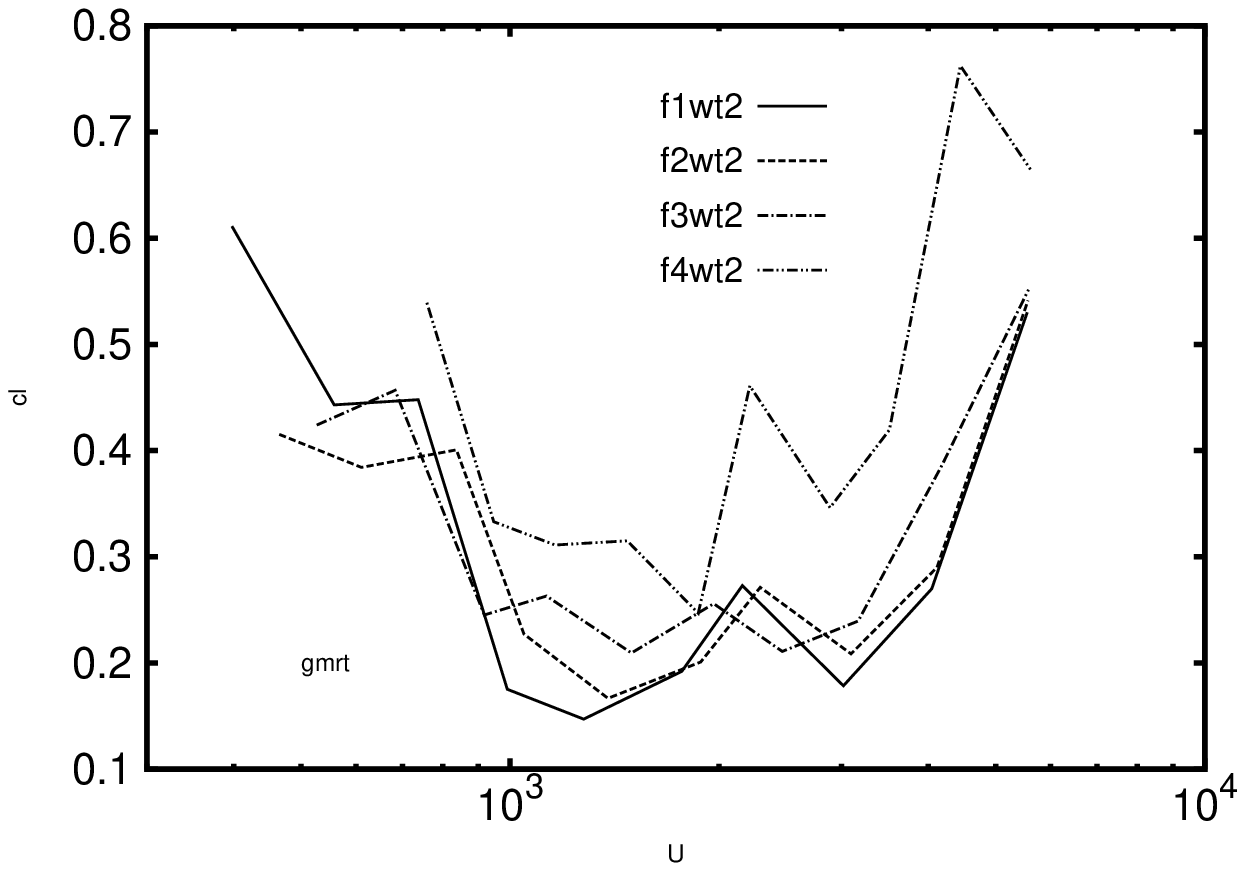}
\includegraphics[width=75mm,angle=0]{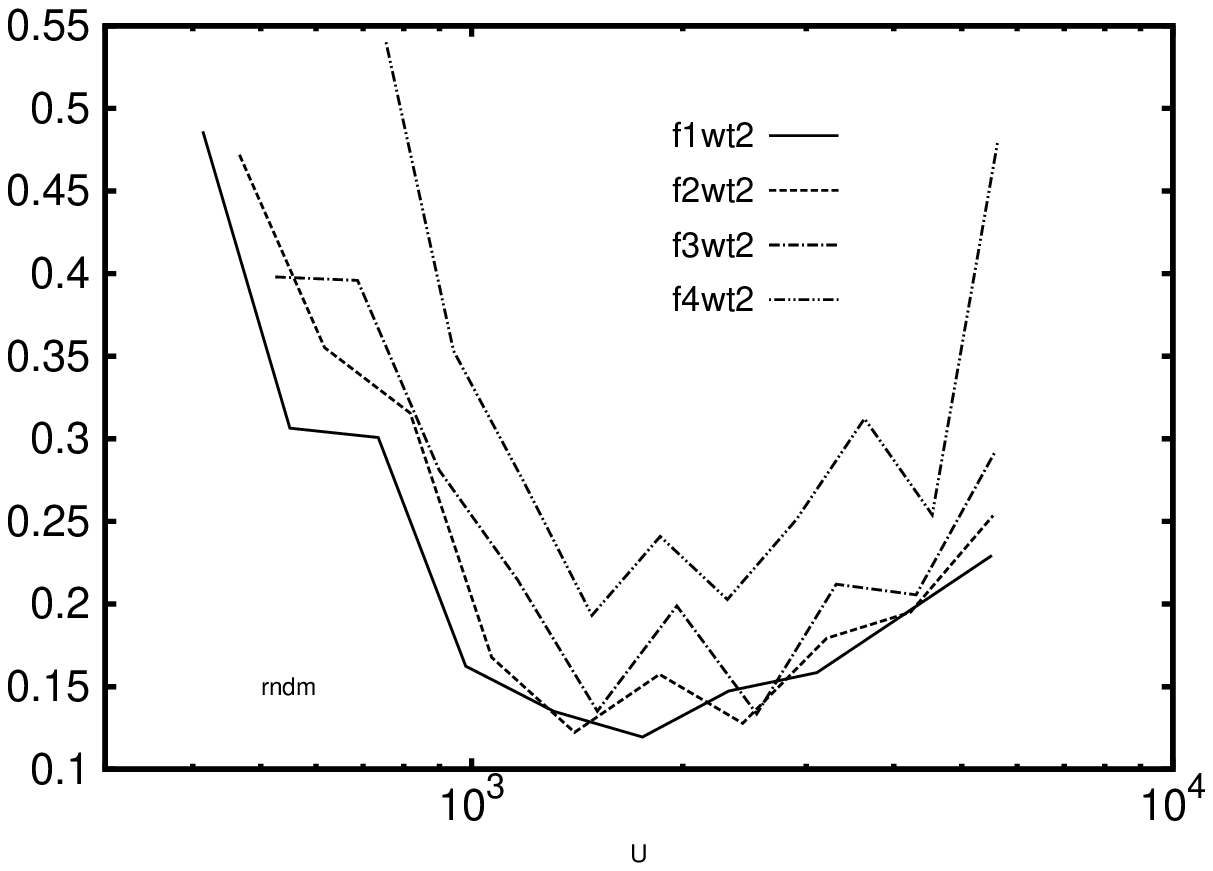}
\caption{This figure shows the relative error ($\sigma/C^M_{\ell}$) estimated from 20 realization of the simulation. Here, we have used $w_g=\mid K_{1g} \mid^2$ and the  different $f$ values shown in this figure.}
\label{fig:sig2ndwt2}
\end{center}
\end{figure}

We have studied the relative performance of the two weight scheme
mentioned earlier. Figure {\ref{fig:sig2ndf2}} shows the relative
deviation and the relative error for both $w_g=1$ and $w_g=\mid K_{1g}
\mid^2$ for $f=0.8$. As expected, the first scheme performs better in
the cosmic variance dominated regime. The difference between the two
weight scheme, however, is not very large in this regime. The second
weight scheme performs significantly better in the system noise
dominated region. In this region the errors are nearly doubled if we
use $w_g=1$ instead of $w_g=\mid K_{1g} \mid^2.$ 

In summary, we have introduced a Gridded Estimator for the angular
power spectrum where it is possible to avoid the positive noise bias
which arises due to the contribution from the correlation of a visibility with itself.
Further, the estimator allows the possibility to taper the sky response
and thereby implement  sidelobe suppression. We have used simulated visibility data
to validate the estimator. We find that the estimator provides an unbiased estimate
of $C_{\ell}$ for $\ell \ge 1.2 \times 10^{3}$ if we have a sufficiently 
dense, uniform baseline distribution. We also find that eq.~(\ref{eq:ge14}) provides 
a good analytic estimate of the errors in the measured $C_{\ell}$.
The estimator is found to be sensitive to the telescope's $uv$ coverage,  
and we have somewhat of an overestimate for the GMRT which has a patchy
$uv$ coverage. This deviation, however, is roughly within the $1\sigma$ 
error bars and is not expected to be a serious issue. It is possible to carry
out  simulations with the actual observational $uv$ coverage and use these
to compensate for the overestimate. The new telescopes like LOFAR 
(discussed later) have a denser and more uniform $uv$ coverage, and we do not 
expect this issue to be of concern there. The $1\sigma$ errors, we find, increase as the 
tapering is increased. The choice of $f$, however, is decided by issues
related to point source removal not considered here. We find that the weight
scheme $w_g=\mid K_{1g} \mid^2$ performs better than $w_g=1$, and we use the former
for the subsequent analysis.

\section{A comparison of the two  estimators}
Comparing the Bare Estimator with the Tapered Gridded Estimator we see (left panel of Figure~\ref{fig:sig2ndf2}) that the former is more successful in recovering the input sky model. The statistical errors also (right panel of Figure~\ref{fig:sig2ndf2}) , we find, are somewhat smaller for the Bare Estimator. The Bare Estimator deals directly with the measured visibilities, and in a sense we expect it to outperform any other estimator which deals with gridded visibilities. What then is the motivation to consider a Gridded Estimator which is not able to recover the input model with as much accuracy as the Bare Estimator The Bare Estimator deals directly with the visibilities and the computational time for the pairwise correlation in eq.~(\ref{eq:be1}) scales proportional to $N^2$, where $N$ is the total number of visibilities in the data. Further, the error calculation in eq.~(\ref{eq:be6}) is expected to scale as $N^4$. In contrast, the computation time is expected to scale as N for Tapered Gridded Estimator. This N dependence arises in the process of gridding the visibilities, the correlation eq.~(\ref{eq:ge9}) and the error estimate eq.~(\ref{eq:ge14}) are both independent of $N$.

Figure~\ref{fig:time} show the computation time for the two estimators as the number of visibilities varied. We see that the computation time shows the expected $N$ dependence for large values of $N(>1000)$. The Bare Estimator takes less computation time when $N$ is small ($N\le10^4$). However, the computation time for the Bare Estimator and its error estimate are larger than that for the Tapered Gridded Estimator for $N\ge10^5$. The Bare Estimator is extremely computation extensive for a large $N$ and it is preferable to use the Gridded Estimator when $N\ge10^5$. Based on this we focus on the Tapered Gridded Estimator for most of the subsequent discussion.

\begin{figure}
\begin{center}
\psfrag{cl}[c][c][0.8][0]{$\sigma/C^M_{\ell}$}
\psfrag{dev}[c][c][0.8][0]{$(C_{\ell}-C^M_{\ell}) /C^M_{\ell}$}
\psfrag{U}[c][c][0.8][0]{$\ell$}
\psfrag{f2}[r][l][0.8][0]{f=0.8}
\psfrag{f2wt1}[r][c][0.8][0]{$w_g=1$}
\psfrag{f2wt2}[r][r][0.8][0]{$w_g=\mid K_{1g}\mid^2$}
\psfrag{bare}[r][l][0.8][0]{Bare}
\psfrag{gmrt}[c][c][0.8][0]{GMRT}
\includegraphics[width=75mm,angle=0]{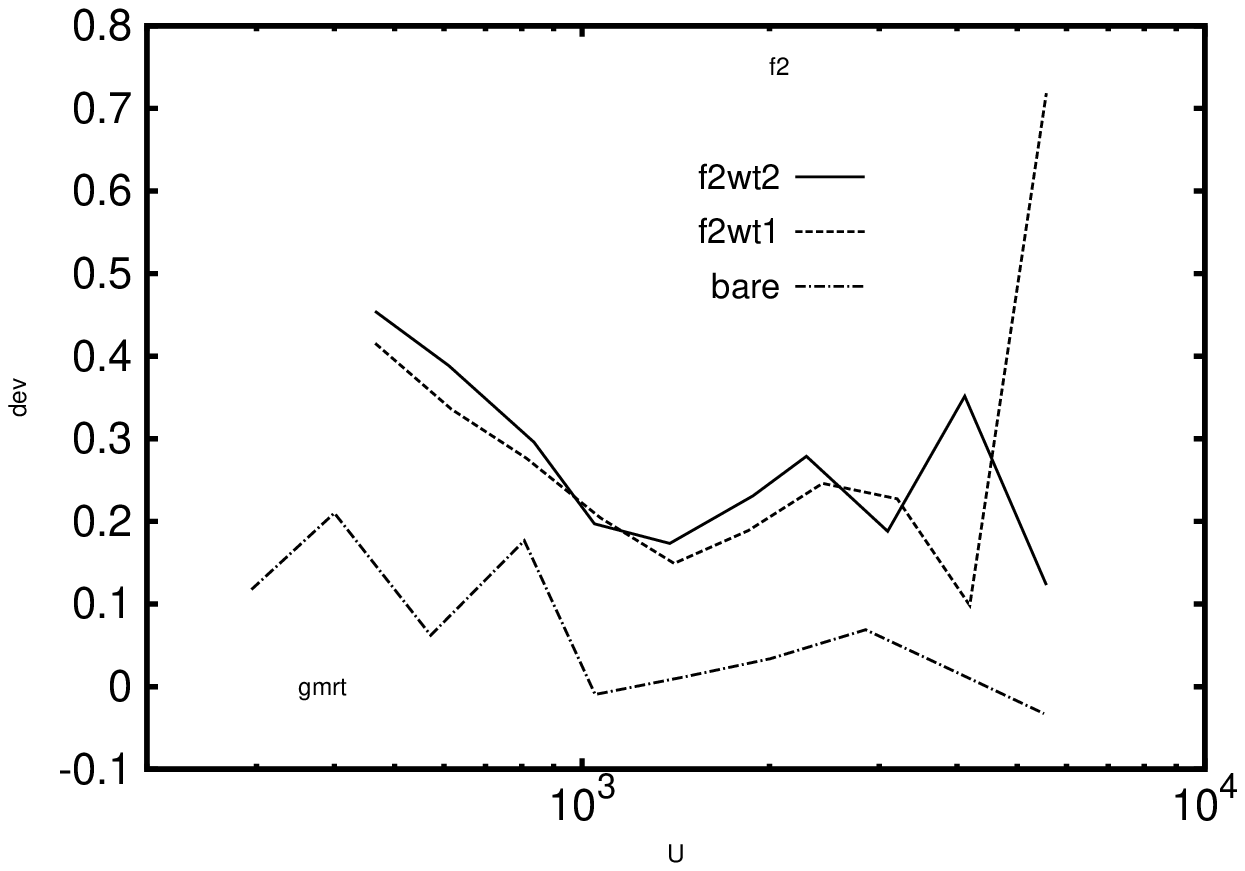}
\includegraphics[width=75mm,angle=0]{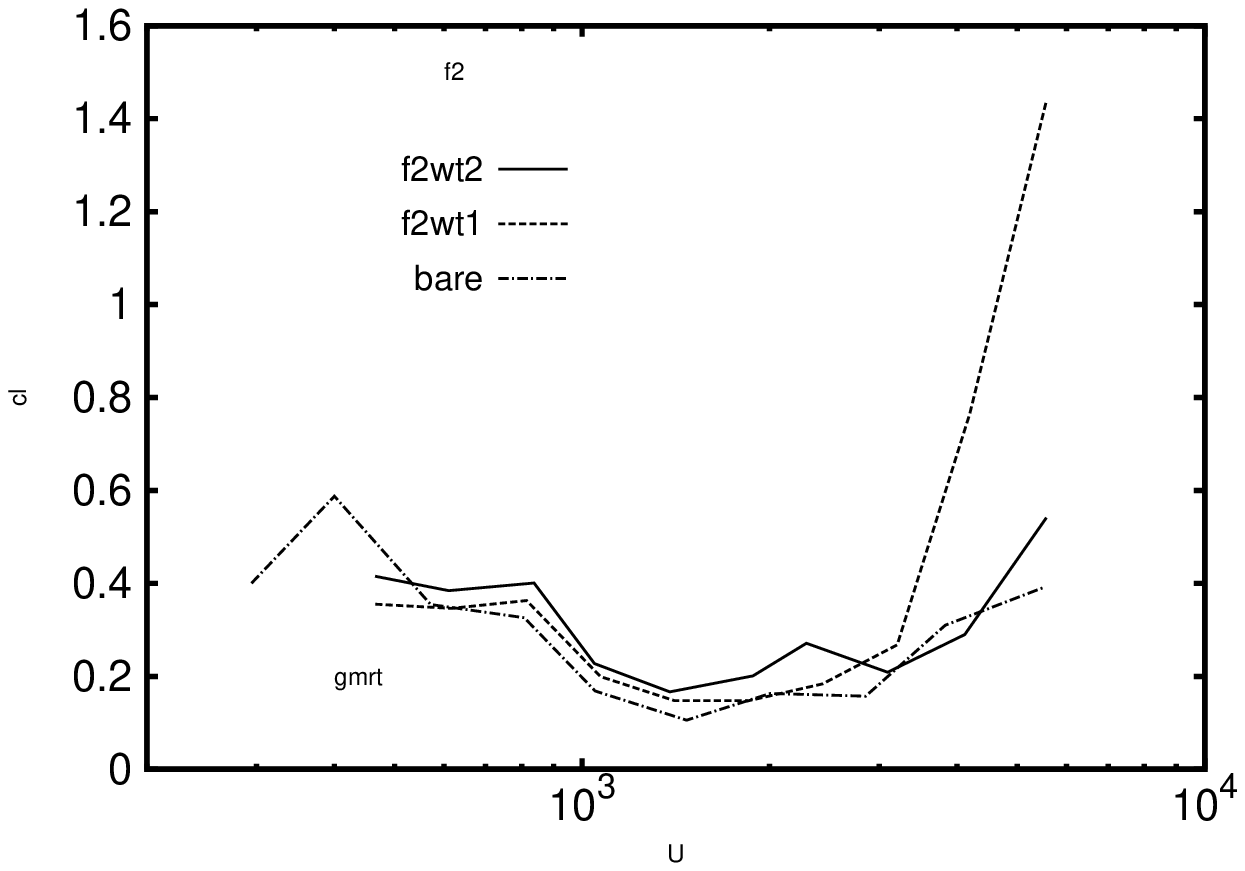}
\caption{ The left (right) panel shows the the fractional deviation (error) for the two weight schemes 
 $w_g=1$ and $\mid K_{1g}\mid^2$ respectively, both with $f=0.8$.  The results for the Bare estimator have also 
been shown for comparison.}
\label{fig:sig2ndf2}
\end{center}
\end{figure}

\begin{figure}
\begin{center}
\psfrag{bline}[c][c][0.8][0]{No. of visibility}
\psfrag{time}[b][c][1.][0]{Time [sec]}
\psfrag{1st}[r][r][0.8][0]{$C_{\ell}$ Bare}
\psfrag{2nd}[r][r][0.8][0]{$C_{\ell}$ Gridded}
\psfrag{1stvar}[r][r][0.8][0]{$\sigma$ Bare}
\includegraphics[width=70mm,angle=-90]{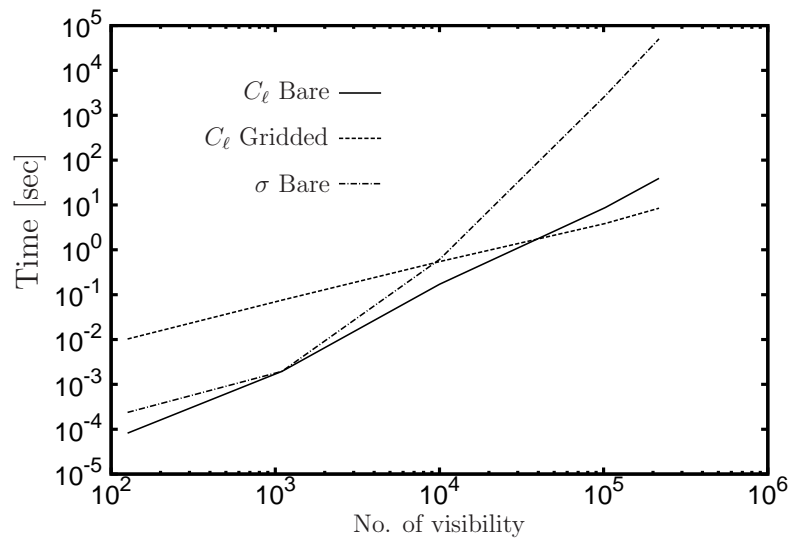}
\caption{This shows how the computation time varies with the number of visibility data  for the two different
 estimators.  The computation time for  analytically predicting  the error (eq.~\ref{eq:be6}) 
for the Bare Estimator is also shown.} 
\label{fig:time}
\end{center}
\end{figure}

\section{Gain Errors}
\label{sec:gerr}
The measured visibilities have undetermined time varying gains which arise due to  the 
atmosphere, receiver system, etc. The calibration procedure attempts to determine these
gains and correct for them, but this generally leaves unknown residual gain errors in the data.
\citet{adatta,adatta10} have studied the impact of the residual gain errors on  bright source subtraction 
and place a tolerance limit  for  detecting the reionization $21 \,{\rm cm}$ signal. Here we study the 
effect of gain errors on the estimators that we have defined earlier. For this work we assume antenna
 dependent  gain errors whereby the calibrated visibilities can be written as 

\begin{equation}
\V(\u_{ab})=g_ag_b^*[\S(\u_{ab})+\N(\u_{ab})]
\end{equation}
where $a,b$ refer to the two antennas corresponding to the baseline $\u_{ab}$, and
  $g_a=(1+\alpha_a)e^{i\phi_a}$
and $g_b=(1+\alpha_b)e^{i\phi_b}$ are the respective antenna gains.
Here the $\alpha_a$s and the $\phi_a$s are 
respectively the amplitude and the phase errors of the individual 
antenna gains.
We have assumed that both $\alpha_a$ and $\phi_a$ are Gaussian random variables of zero mean and
 variance $\sigma_{\alpha}^2$ and $\sigma_{\phi}^2$ respectively. The errors are assumed to be independent 
in different antennas  and at different time instants . 

The two visibility correlation can  be
written as,
\begin{equation}
\langle \V(\u_{ab})\V^{*}(\u_{cd})\rangle=\langle g_ag_b^*g_c^*g_d\rangle 
[S_2(\u_{ab},\u_{cd})+N_2(\u_{ab},\u_{cd})]
\label{eqn:gerrcorr}
\end{equation}
where the product of the gains is to be averaged over different realizations of the gain errors $\alpha$ and $\phi$. We now have three
different possibilities which we discussed below.

{\bf Case I:} The two visibilities $\V(\u_{ab})$ and $\V(\u_{cd})$ are 
at two different time instants or they have no antenna 
in common. In this situation we have

\begin{equation}
\langle g_ag_b^*g_c^*g_d\rangle=e^{-2\sigma_{\phi}^2}.
\label{eq:gerrcs1}
\end{equation}
 
 {\bf Case II:} The two visibilities $\V(\u_{ab})$ and $\V(\u_{cd})$ are at the same time instant and have only one antenna in common. In this situation we have

\begin{equation}
\langle g_ag_b^*g_c^*g_d\rangle=(1+\sigma_{\alpha}^2)e^{-\sigma_{\phi}^2}.
\label{eq:gerrcs2}
\end{equation}

{\bf Case III:} Both  $\V(\u_{ab})$ and $\V(\u_{cd})$ referred
the same measured visibility. In this situation we have
\begin{equation}
\langle g_ag_b^*g_c^*g_d\rangle=(1+\sigma_{\alpha}^2)^2.
\label{eq:gerrcs3}
\end{equation}

The signal contribution to both the  estimators defined 
earlier is dominated by Case I, whereas the noise is dominated by
Case III. Based on this it is possible to generalize eq.~(\ref{eq:vcorr}) to 
obtain the approximate relation 
\begin{equation}
V_{2ij} = e^{-2\sigma_{\phi}^2}V_0 \, e^{-\mid \Delta \u_{ij} \mid^2/
\sigma_0^2} 
\, C_{\ell_i} + (1+\sigma_{\alpha}^2)^2\delta_{ij} 2 \sigma_n^2
\label{eq:vcorrm}
\end{equation}
which takes into account the effect of gain errors. It is also
possible to generalize eq.~(\ref{eq:ge7}) for the gridded visibilities
in a similar fashion. Using these  to calculate the effect of gain errors
 on the  estimators defined earlier,  we have  
\begin{equation}
\langle \hat E(a) \rangle = e^{-2 \sigma^2_{\phi}} \bar{C}_{\bar{\ell}_a} \,.
\label{eq:gerr1}
\end{equation}
for both the Bare and the Tapered Gridded Estimators. 
We see that  both the  estimators  are  unaffected by the error in the gain amplitude, 
 however the  phase errors cause the expectation value  of the estimator to decrease by a factor 
$e^{-2 \sigma^2_{\phi}}$.  It is quite straightforward to generalize eq.~(\ref{eq:be6}) and 
eq.~(\ref{eq:ge13}) to incorporate the effect of the gain errors in  the  variance  of the  Bare 
 and the Tapered Gridded Estimators   respectively. The main effect is that the signal contribution 
is suppressed by a factor  $e^{-2 \sigma^2_{\phi}}$ whereas the system noise contribution is 
jacked up by a factor $(1+\sigma_{\alpha}^2)^2$ (eq.~\ref{eq:vcorrm}). We consequently expect 
the ${\rm SNR}$ to remain unchanged in the cosmic variance dominated regime at low $\ell$, 
whereas  we  expect the ${\rm SNR}$ to fall  in the system noise dominated regime (large $\ell$). 
Further, we also expect the transition from the cosmic variance to the system noise dominated 
regime to shift to smaller $\ell$ values if the gain errors increase.

\begin{figure}
\begin{center}
\psfrag{cl}[b][b][0.8][0]{$\ell (\ell+1) C_{\ell}/2 \pi \, [mK^2]$}
\psfrag{U}[c][c][0.8][0]{$\ell$}
\psfrag{snr}[c][c][0.8][0]{${\rm SNR}$}
\psfrag{model}[c][l][0.8][0]{$C^M_{\ell}$}
\psfrag{model60}[r][r][0.8][0]{$C^M_{\ell}\times e^{-\sigma^2_{\phi}}$}
\psfrag{a50}[c][r][0.8][0]{$\sigma_{\alpha}=0.5$}   
\psfrag{p60}[r][r][0.8][0]{$\sigma_{\phi}=60^{\circ}$}
\psfrag{p10}[r][r][0.8][0]{$\sigma_{\phi}=10^{\circ}$}
\psfrag{gmrt}[c][c][0.8][0]{GMRT}
\psfrag{00}[r][r][0.8][0]{Uncorrupted}
\psfrag{a10p10}[r][r][0.7][0]{$\sigma_{\alpha}=0.1,\sigma_{\phi}=10^{\circ}$}
\psfrag{a10p60}[r][r][0.7][0]{$\sigma_{\alpha}=0.1,\sigma_{\phi}=60^{\circ}$}
\psfrag{a50p10}[r][r][0.7][0]{$\sigma_{\alpha}=0.5,\sigma_{\phi}=10^{\circ}$}
\psfrag{a50p60}[r][r][0.7][0]{$\sigma_{\alpha}=0.5,\sigma_{\phi}=60^{\circ}$}
\includegraphics[width=75mm,angle=0]{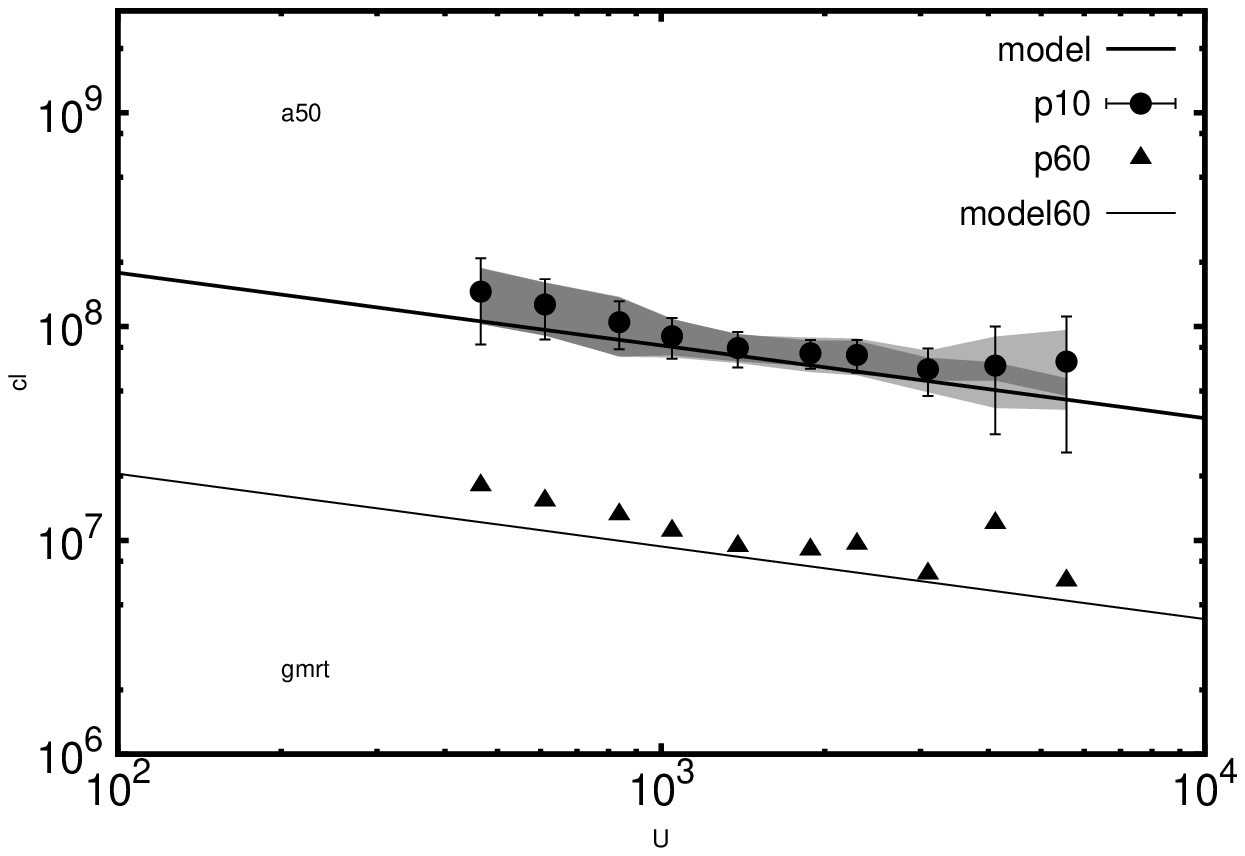}
\includegraphics[width=75mm,angle=0]{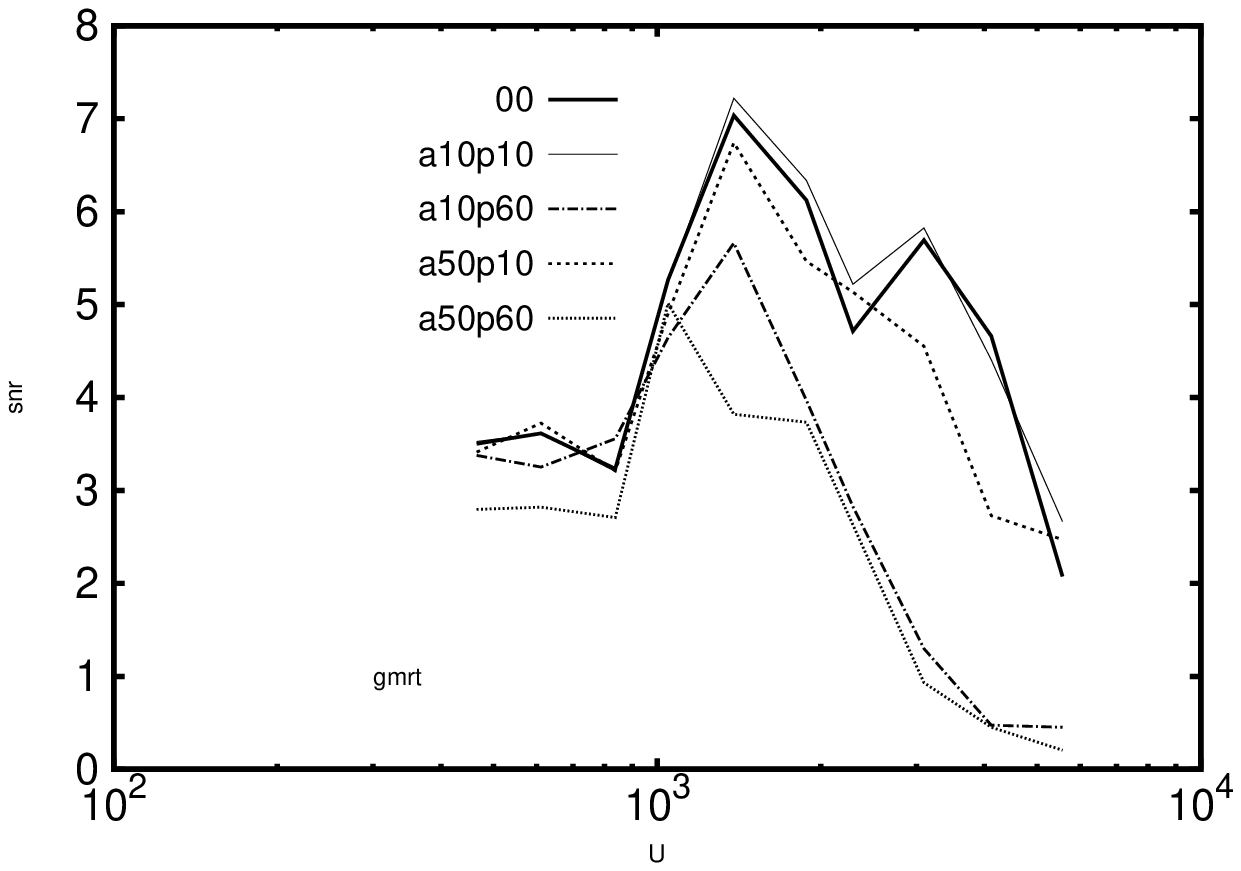}
\caption{The left panel shows the same as 
Figure~\ref{fig:ps1st} for the Tapered Gridded Estimator
 using corrupted visibilities
with the $\sigma_{\alpha}$ and $\sigma_{\phi}$ values shown in the figure.   We have also 
shown $e^{-\sigma^2_{\phi}} \times C^M_{\ell}$ with $\sigma_{\phi}=60^{\circ}$ for comparison. 
The right panel shows the ${\rm SNR}$ for different values of  $\sigma_{\alpha}$ and $\sigma_{\phi}$.}
\label{fig:gerr}
\end{center}
\end{figure}

We have carried out simulations to test the effect of gain errors on the angular power spectrum
estimators. For this we have generated $20$ different realizations of the random gain errors  and 
used these to corrupt the simulated  visibilities described in Section~\ref{ps_simu}. The  simulations were
carried out for different values of $\sigma_{\alpha}$ and $\sigma_{\phi}$. We have applied both the Bare and the 
Tapered Gridded Estimators on  the corrupted visibilities. Both the estimators show very similar behaviour 
under gain errors, and we show the results for only the Tapered Gridded Estimator.  

 We have considered two values  $\sigma_{\alpha}=0.1$ and $0.5$ which respectively  correspond  to $10 \%$ and
 $50 \%$ errors  in the gain amplitude. The left panel of  Figure~\ref{fig:gerr} shows the results for 
 $\sigma_{\alpha}=0.5$.  We see  that the expectation value of the estimator 
is unaffected by the errors in the  gain amplitude. For the phase errors, we have considered the values
$\sigma_{\phi}=10^{\circ}$ and   $60^{\circ}$ for which $e^{-2\sigma^2_{\phi}}$ have values $0.94$ and $0.11$ 
respectively. The left panel of  Figure~\ref{fig:gerr} shows   that  eq.~(\ref{eq:gerr1}) provides a good
 description for the effect of the gain errors on the 
angular power spectrum estimator. We see the net result of the phase errors  is that the estimated  angular power
spectrum  is reduced by a factor   $e^{-2\sigma^2_{\phi}}$ relative to the input model.

 The right  panel of  Figure~\ref{fig:gerr} shows the ${\rm SNR}$ for 
 different values of  $\sigma_{\alpha}$ and $\sigma_{\phi}$.  The rms. fluctuation  $\sigma_{E_G}$ of  the
 estimator  is expected to  depend exponentially  as $e^{-2\sigma^2_{\phi}}$ on the phase errors and have 
a  $(1+\sigma_{\alpha}^2)^2$ dependence on the amplitude errors (eq.~\ref{eq:vcorrm}). We find that the 
simulated ${\rm SNR}$ are more sensitive to  the phase errors in comparison to the amplitude errors. 
 The   ${\rm SNR}$ is  nearly invariant  to  gain errors
in the cosmic variance dominated regime (low $\ell$) where $\sigma_{E_G}$ 
is  reduced by the same factor $e^{-2\sigma^2_{\phi}}$ as the expectation value of the estimator. 
However, the transition from the cosmic variance dominated to the system noise dominated regime 
(approximately the peak of the  ${\rm SNR}$ curves) shifts to smaller $\ell$ if the gain errors are 
increased.  The amplitude errors, we see, reduces the  ${\rm SNR}$ at large $\ell$ where the error is 
dominated by the system noise.

\section{The  $w$-term}
The entire analysis, until now, has been  based on the assumption that 
the visibility contribution $\S(\u)$ from  the sky signal is  the
Fourier transform of the product  of ${\mathcal A}(\th)$ and $\delta I(\th)$.
This is only an approximate relation which is valid   only if the filed of 
view is sufficiently small. The actual relation is 
\begin{align}
\S(u,v,w)=  \int \, dl dm \frac{\delta I(l,m) {\mathcal A}(l,m)}{\sqrt{1-l^2-m^2}}
e^{-2\pi i[ul+vm+w(\sqrt{1-l^2-m^2}-1)]}\,\,,
\label{eq:weq}
\end{align}
where the $w$-term, which we have ignored until now, is the  baseline component along the line of sight to the 
phase 
center and $l,m$ are the direction cosines corresponding to any point  on the sky.  
In a situation where the primary beam pattern falls of within a small angle from the phase center, 
it is adequate to treat the region of sky under observation as a 2D plane and use $(l,m)=(\theta_x,\theta_y)$. 
For example,  the GMRT  has a FWHM of $186^{'}$   for which  $\sqrt{1-l^2-m^2} \approx 0.997$. The term 
$\sqrt{1-l^2-m^2}$ which appears in the denominator of  eq.~(\ref{eq:weq}) incorporates the curvature 
of the sky.   We see that  this makes an insignificant  contribution at the small angles of our interest, 
and hence  may be ignored. The term $w(\sqrt{1-l^2-m^2}-1)$ which appears in the phase in eq.~(\ref{eq:weq}) 
has a value $\sim 10^{-3} \times w$ for the angle mentioned earlier, and this is not  necessarily small.  
The value of $w$ depends on the telescope configuration and the observing direction, and may be quite large 
$(> 10^3)$.  It is therefore necessary to assess the impact of the $w$-term on the angular power spectrum 
estimators defined earlier.

We have simulated  GMRT visibilities  using  eq.~(\ref{eq:weq}) keeping the $w$-term.  The $20$ realizations 
of the sky signal and the baseline tracks are the same as described in section~\ref{ps_simu}, and we have used 
the flat sky approximation ({\it ie.} we have dropped $\sqrt{1-l^2-m^2}$ from the denominator).  
We have applied both the Bare and the Tapered Gridded Estimator to this simulated visibility data. 
We show results  for only the  Tapered Gridded Estimator, the results are very similar for the  Bare Estimator
and we have not shown these separately.   Figure~\ref{fig:wterm} shows the relative change in the estimated 
angular power spectrum if we include the $w$-term. We find that  the  change due to the $w$-term is less than
 $3 \%$ for all values of $\ell$ barring the largest $\ell$ value where there is a $~9 \%$ change. 
The  $w$-term has a larger effect at the large baselines which also correspond to a larger value of 
$w$.  We find that the change caused by the $w$-term is less than $10 \%$ of the statistical fluctuations 
for most values of $\ell$. In summary, for angular power spectrum estimation  it is adequate to ignore the 
$w$-term at the angular scales of our  interest for the GMRT.

\begin{figure} 
\begin{center}
\psfrag{u}[t][c][1.][0]{$\ell$}
\psfrag{err}[b][c][1.][0]{relative error}
\psfrag{sigma}[r][r][0.8][0]{$0.1 \times$ statistical}
\psfrag{1stw}[r][r][0.8][0]{$w$-term}
\includegraphics[width=85mm,angle=0]{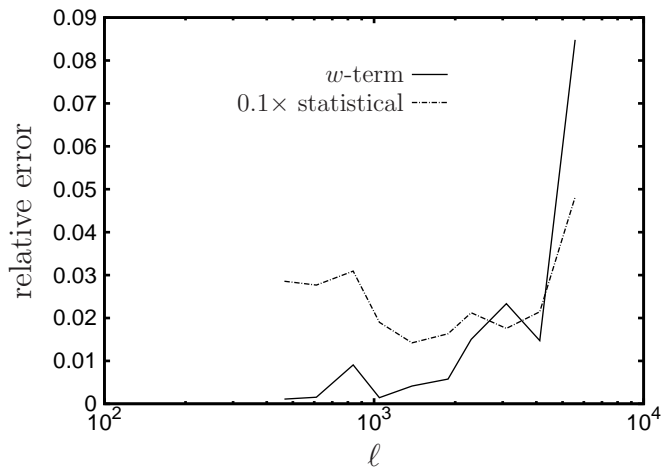}
\caption{This  shows the relative change in the estimated angular power spectrum using Tapered Gridded Estimator due to the $w$-term.  For comparison we have also shown $0.1 \times \delta C_{\ell}/C_{\ell}$ which corresponds to $10 \%$ 
of the relative statistical error in $C_{\ell}$. }
\label{fig:wterm}
\end{center}
\end{figure}

\section{LOFAR}
\label{lofar}
LOFAR, the Low Frequency Array, is an  innovative  new radio
telescope which operates at the lowest radio  frequencies ($10-240
\, {\rm MHz}$) \citep{haarlem}. It
consists of an interferometric array of dipole antenna stations
distributed throughout the Netherlands and Europe. The individual
stations perform the same basic functions as the dishes of a
traditional interferometric radio telescope. Hence, the station beam
which is analogous to the primary beam ultimately determines the FoV
for a given observation. In the High Band Antennas (HBAs, $110 - 240 \,{\rm MHz}$), groups of dipole
pairs are combined into HBA tiles and the station beam is formed from
the combined signal from the tiles. The HBA tiles are sensitive to two orthogonal 
linear polarizations. Close to the phase centre, the
LOFAR station beam can be well modeled with a circular Gaussian and the FWHM  of the Gaussian varies
approximately from $3.0^{\circ}$  to  $5.0^{\circ}$ in the frequency range  $115 \, - 185 \, {\rm MHz}$ 
with  $\theta_{\rm FWHM}=3.8^{\circ}$ at $150 \,  {\rm MHz}$. 

%%%%%%%%%%%%%%%%%%%%%%%%%%%%%%%%%%%
In this section we consider the possibility of using LOFAR to estimate the angular power spectrum 
of the  $150 \,  {\rm MHz}$ sky signal after point source subtraction. 
The LOFAR has a wider field of view compared to the GMRT and 
we have simulated a $\sim \, 8^{\circ}\times 8^{\circ}$ region of the sky 
with an  angular resolution of $14^{''} \times 14^{''}$.  Here again we have 
generated $20$ independent realizations of the sky signal.  The simulations 
were carried out in exactly the same way as described in Section~\ref{ps_simu} 
using the LOFAR parameters given in Table~\ref{tab:1}. 
We have generated the  LOFAR baseline distribution for the 62 
antennas   in the central core region for  $8$ hrs of observing time. 
 Visibilities were generated with a time  interval
of $40$s and  we obtain a total of $669,809$ visibilities 
in the  baseline range $30\le \u\le800$. We have included 
the $w$-term for calculating the LOFAR visibilities. 
 The LOFAR has a denser 
$uv$ coverage compared to the GMRT, and the simulated baseline 
range is nearly uniformly covered.   We have used 
$\sigma_n=2.2$Jy \citep{haarlem} for the system  noise   in the simulations. 
Given the large volume of data, we have only used the Tapered Gridded 
Estimator with $f=0.8$ and $w_g=K_{1g}^2$.

\begin{figure}
\begin{center}
\psfrag{cl}[b][b][0.8][0]{$\ell (\ell+1) C_{\ell}/2 \pi \, [mK^2]$}
\psfrag{U}[c][c][0.8][0]{$\ell$}
\psfrag{model}[cr][tr][0.8][0]{$C^M_{\ell}$}
\psfrag{lofar}[r][r][0.8][0]{LOFAR}
\psfrag{fwt}[r][r][0.8][0]{$f=0.8,w_g=\mid K_{1g}\mid^2$}
\includegraphics[width=80mm,angle=0]{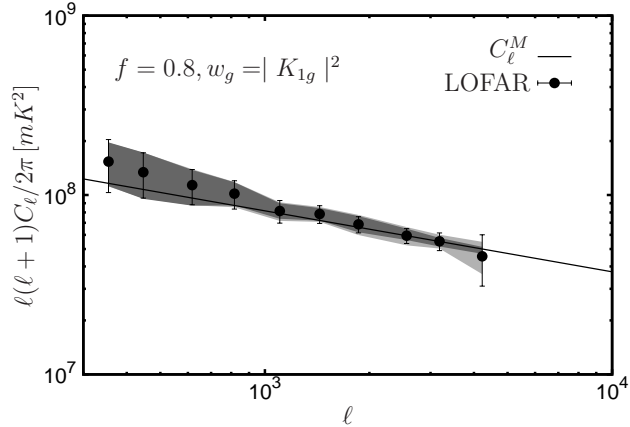}
\caption{Same as Figure~\ref{fig:ps1st} for the Tapered Gridded Estimator and the 
simulated LOFAR data.}
\label{fig:ps2ndlofar}
\end{center}
\end{figure}

Figure  \ref{fig:ps2ndlofar}  shows the angular power spectrum estimated from our simulations. 
We see that  the estimated $C_{\ell}$ values are all  within the $1\sigma$ region of the input model angular
power spectrum $C^{M}_{\ell}$. The estimated $C_{\ell}$ values, however,  are somewhat in excess of 
 $C^M_{\ell}$ at small $\ell$ $(< 1,000)$. The fractional deviation  
$(C_{\ell}-C^M_{\ell})/C^M_{\ell}$ is around $\sim 30 \%$ at the smallest $\ell$ bin, and 
it is  $\sim 15 \%$ at $\ell\sim 800$.
The excess  is not seen   at larger  $\ell$ 
 where the estimated values are in excellent agreement with  $C^M_{\ell}$.
We also see that  the error estimates predicted by
eq.~(\ref{eq:ge14}) are in good agreement with the rms. fluctuation
estimated from the 20 realizations. The transition from cosmic variance dominated 
errors to  system noise dominated errors occurs at $\ell \sim 2,000$ similar to the GMRT. 
The LOFAR has considerably more baselines compared to the GMRT, and 
the  errors in the estimated  angular power spectrum   are  smaller for LOFAR 
in comparison to GMRT.  

As mentioned earlier in the context of the GMRT, the excess in the estimated
 $C_{\ell}$ may be a consequence of  patchy $uv$ coverage at   
 small baselines  ($U < 160$). The   average baseline density in the region  $U < 160$
is several times larger than the average within  $U < 800$, however this 
does not guaranty that the former is less patchier  than the latter. Further, it
is not possible to say anything definite from a visual inspection of the 
baseline distribution. The convolution with the primary beam pattern and 
the window function introduces a $\sim 8 \%$  deviation between $C_{\ell}$ and 
$C_{\ell}^M$  at $U < 160$. The exact cause of the excess at small $\ell$
is at present not fully understood.

\section{Discussion  and conclusions}
\label{sum}
 In this paper  we have introduced two estimators for quantifying the angular power spectrum of the sky 
brightness temperature. Both of these estimators use the visibilities measured in radio
interferometric observations. The Bare Estimator works directly with the measured visibilities, and 
uses pairwise visibility correlations to estimate the angular power spectrum.  The Tapered Gridded Estimator 
uses the visibility data after gridding on a rectangular grid in the $uv$ plane. Here it is possible to 
taper the sky response so as to suppress the sidelobes and reduce the filed of view.  Earlier work 
\citep{ghosh2} shows tapering  to be an important ingredient in  foreground removal for detecting the 
cosmological $21$-cm signal. We have investigated the properties of the estimators, and present analytic 
formulae for the expectation value (eqs.~\ref{eq:be3} and \ref{eq:ge11a}) and the 
variance (eqs.~\ref{eq:be6} and \ref{eq:ge14}).  The expectation value of both the estimators is free
from the positive system noise bias which arises due to  the correlation of a visibility with itself. The 
system noise  affects only the variance. 

We have carried out simulations to validate the estimators. The simulated sky signal  assumes that the 
point sources have been removed and the residuals are dominated by the diffuse Galactic synchrotron 
radiation which is modelled as   a homogeneous and isotropic Gaussian random field with a power law
 angular power spectrum. We consider GMRT observations for most of the analysis. We find that the Bare 
Estimator is able to recover the input model to a good level of precision.  The computation time is 
found to scale as $N^2$ with the number of visibility data. Further, the scaling is $N^4$ for the variance. 

We find that the Tapered Gridded Estimator is able to recover 
the input model $C_{\ell}^M$ to a high level of precision provided the baselines have a uniform $uv$ coverage. For the GMRT which has a patchy $uv$ coverage,  
 the $C_{\ell}$ estimated from the Tapered Gridded Estimator is largely within the $1\sigma$ errors from  the 
input model $C^M_{\ell}$.  There is, however, indication that  the angular power spectrum 
is  overestimated to some extent. Comparing the results to a situation with  a uniform 
random baseline distribution, 
we conclude that the overestimate is a consequence of GMRT's patchy $uv$ coverage and is not inherent 
to the  Tapered Gridded Estimator which is unbiased by construction.  It is possible to use simulations to quantify this overestimate and 
correct for  this in a real observation.  We do not anticipate this overestimate to be a very 
major obstacle  for the Tapered Gridded  Estimator. The computation time for this estimator and its 
variance both scale as $N$. Long observations spanning many frequency channels will produce large volumes of
visibility data. The Bare Estimator is computationally very expensive for large $N$, and  a 
Gridded  Estimator is the only feasible alternative. Consequently, we have focused on the Tapered Gridded 
Estimator  for much of the analysis in the later part of this paper. 

Residual gain errors corrupt the measured visibilities, and this is a potential 
difficulty for estimating the angular power spectrum. We have analyzed the effect
of gain errors on the two estimators introduced in this paper. Our analysis, 
validated by simulations, shows that the expectation value of the estimators is 
unaffected by  amplitude errors. The phase errors cause a decrement by the factor 
$e^{-2 \sigma^2_{\phi}}$ in  the expectation value. The statistical errors in the estimated
$C_{\ell}$ are affected by both the amplitude and the phase errors, however this is more 
sensitive to the phase errors relative to the amplitude errors. We have also investigated 
the effect of the $w$-term. We find that the $w$-term does not cause a very big change in the 
estimated $C_{\ell}$ at the scales of our interest here.  Our analysis here shows that 
the residual phase errors can lead to the angular power spectrum being underestimated 
by a factor $e^{-2 \sigma^2_{\phi}}$ which has  a value $\sim 0.1$ for $ \sigma_{\phi}=60^{\circ}$. 
It is therefore  imperative to independently quantify the magnitude of the residual 
phase errors for a correct estimate of the angular power spectrum. 

In addition to GMRT, we have also applied the estimators to  simulated LOFAR data. 
We find that the $C_{\ell}$ estimated using the Tapered Gridded Estimator is within 
the $1\sigma$ errors of the input model. There is, however, indication that there is 
some overestimation $(15 - 30 \%)$  at low $\ell$ $(< 1,000)$.   The exact cause of this excess at small $\ell$ is at present not fully understood.

The two estimators considered here both avoid  the positive noise bias  which 
arises due to the system noise contribution in the visibilities. This is 
achieved by not including the contribution from the correlation of a visibility
with itself.  As an  alternative one could  consider an estimator which 
straight away squared the measured or the gridded visibilities. In this 
situation it is necessary  to separately identify the noise bias contribution 
and subtract it out. The noise bias contribution is expected to be independent  
of frequency  and $\ell$. It is, in principle, possible to 
identify a  frequency and $\ell$ independent component and subtract it out. 
However, our analysis in this paper shows that the errors in the amplitude 
of the calibrated gains  affect the noise bias. Frequency and baseline 
dependent gain errors would manifest themselves as the frequency and $\ell$ 
dependence of the noise bias. This is a major obstacle which is bypassed
by our estimators. 

The multi-frequency angular power spectrum (MAPS, \citealt{kdatta07}) 
jointly quantifies the angular and frequency dependence of the fluctuations 
in the  sky signal.   This can be estimated directly from the measured
visibilities (eg. \citealt{ali08}), and it can be  used to  detect the 
cosmological $21$-cm signal \citep{ghosh2}.  In  future work we plan to
generalize the  analysis of this paper to the multi-frequency angular
power spectrum and address various issues, including point source removal, 
which are  relevant  for  detecting the cosmological $21$-cm signal.

%\clearpage{\pagestyle{empty}\cleardoublepage} %%%%%%%%%%%%%%%%%%%%
%\newpage
\setcounter{section}{0}
\setcounter{subsection}{0}
\setcounter{subsubsection}{2}
\setcounter{equation}{0}
%\pagenumbering{arabic}

%-------------------------------------------
\chapter[Point source removal for power spectrum estimation]{{\bf Point source subtraction for angular power spectrum estimation from low-frequency radio-interferometric data}\footnote{This chapter is adapted
   from the paper ``Point source subtraction for angular power spectrum estimation from low-frequency radio-interferometric data''
   by \citet{samir16c}}}
\label{chap3}

\section{Introduction}

Observations of redshifted $21\, {\rm cm}$ radiation from neutral hydrogen (HI) 
hold the potential of tracing the large scale structure of the Universe over a 
redshift range of $200 \ge z \ge 0$. Accurate cosmological HI tomography and 
power spectrum measurement, particularly from the Epoch of Reionization (EoR), 
by ongoing or future low-frequency experiments will provide us a significant 
amount of information about various astrophysical and cosmological phenomena to 
enhance our present understanding of the Universe. However, a major challenge 
in statistical detection of the redshifted $21\, {\rm cm}$ signal arises from 
the contamination by Galactic and extragalactic ``foregrounds'' 
\citep{shaver99,dmat1,santos05}.

The two major foreground components for cosmological HI studies are (1) the 
bright compact (``point'') sources and (2) the diffuse Galactic synchrotron 
emission \citep{ali08,paciga11,bernardi09,ghosh12,iacobelli13}. An accurate 
and precise subtraction of the bright point sources is a primary step for 
measurement of the redshifted $21\, {\rm cm}$ signal. \citet{bowman09} and 
\citet{liu09}, for example, have reported that point sources should be 
subtracted down to a $10-100\,{\rm mJy}$ threshold in order to detect the $21\, {\rm 
cm}$ signal from the EoR. It has been recently demonstrated also using both 
simulated and observed data from MWA that foreground (particularly the point 
sources) must be considered as a wide-field contaminant to measure the  $21\, 
{\rm cm}$ power spectrum \citep{pober16}. Detection of the weak cosmological 
signal will also require a proper removal of the diffuse component of the 
foreground. However, detecting and characterizing the diffuse emission itself 
also require removal of the point sources properly. Thus, understanding the 
impact of point source subtraction on the diffuse emission (either foreground 
Galactic synchrotron or cosmological HI signal) is an important step for
all such experiments. A detailed investigation and analysis of the
Galactic synchrotron emission power spectrum can be used to study the
distribution of cosmic ray electrons and the magnetic fields in the
ISM of our own Galaxy and also interesting in its own right \citep{Waelkens,Lazarian,iacobelli13}.

Keeping aside calibration errors, the problem of subtracting point sources 
ultimately reduces to a problem of deconvolution of point sources, in presence 
of diffuse (foreground and/or cosmological HI signal) emission, to fit their 
positions and flux densities as accurately as the instrumental noise permits. 
The optimum strategy of modeling and subtracting point sources in presence of 
diffuse emission is an open question in the general context of interferometric 
radio frequency data analysis. A comparatively large field of view as well as a large number of strong point sources and bright Galactic synchrotron emission 
make it more relevant at low radio frequency. Hence, for EoR and post-EoR 
cosmological HI studies at low frequencies, particularly due to the weakness of 
the desired signal compared to the foregrounds and the improved sensitivity of 
the current and future telescopes (e.g. the Giant Metrewave Radio Telescope, the Low Frequency Array, the Murchison Wide-field Array, the Precision Array to Probe the Epoch of Reionization, the Primeval Structure Telescope {\footnote{PaST; http://web.phys.cmu.edu/~past}}, the Hydrogen Epoch of Reionization Array, the Square Kilometer Array etc.), this is 
one of the major and important issue to be taken care of. 

Naturally, a significant amount of effort has gone into addressing the problem 
of foreground removal for detecting the $21\, {\rm cm}$ power spectrum from EoR \citep{wang06,mac06,morales06,jelic08,geil08,gleser08,liu09a,liu09,harker10,petrovic11,bernardi11,mao3,liu12,chapman12,paciga13}. In contrast, foreground 
avoidance \citep{parsons12,trott12,morales12,vedantham12,hazelton13,pober13,dillon13,thyag13,pober14,parsons14,dillon14,liu14a,liu14b,ali15,jacobs15,trott16} 
is an alternative approach based on the idea that contamination from any 
foreground with smooth spectral behaviour is confined only to a wedge in 
cylindrical $(k_{\perp}, k_{\parallel})$ space due to chromatic coupling of an 
interferometer with the foregrounds. The HI power spectrum can be estimated from 
the uncontaminated modes outside the wedge region termed as the $EoR \,\, 
window$ where the HI signal is dominant over the foregrounds. With their merits 
and demerits, these two approaches are considered complementary \citep{chapman16}. 

Here we have considered the issue of accurate modeling and subtraction of point 
sources in presence of diffuse emission using simulated radio interferometric 
data. This is part of a coherent effort of end-to-end simulation of realistic 
EoR signal and foreground components, including instrumental effects, and 
finally using suitable power spectrum estimator to recover the signal. In this 
endeavor, we have developed a novel and fast estimator of angular power spectrum that consistently avoid the noise bias, and tested it with simulated diffuse 
Galactic synchrotron emission \citep{samir14}. Here, we have further developed the simulations to include the point sources in the model (as well as 
instrumental noise) to investigate the effectiveness of various point source 
subtraction strategies. This paper describe the details of the simulations and 
analysis, including the adopted point source modeling and subtraction 
strategies, and the effects on the residual diffuse emission (in terms of both 
first and second order statistics). A companion paper has reported the 
usefulness of the new estimator in recovering the diffuse emission power 
spectrum from the residual data in such situation \citep{samir16a}. There it is 
demonstrated that the contribution due to point sources from the outer parts of 
the main lobe of the primary beam can be suppressed by tapering the sky 
response using this newly developed Tapered Gridded Estimator (hereafter, TGE). 
The same estimator is used for the analysis presented in this paper. A further 
generalization of the estimator to deal with spherical and cylindrical power 
spectrum is presented in \citet{samir16b}. Please note, even 
though these exercises are in the context of EoR experiments, for the sake of 
simplicity, we have so far not included the weak cosmological signal in the 
model. We leave that, and also more complicated instrumental effects, for 
future studies. Here we only establish the ability of the developed estimator 
to recover the diffuse emission power spectrum accurately after point source 
subtraction. Thus, apart from EoR experiments, these results are also relevant 
in more general situation, e.g. detailed study of Galactic synchrotron emission 
\citep{samir16c}.

The current paper is organized as follows. In Section 2, we discuss the details 
of the foreground point source and diffuse emission simulation, and Section 3 
discusses the method of analysis using different CLEANing options. Section 4 
and 5 highlights the result of point source subtraction in the images and in 
the recovered power spectrum. Finally, we present summary and conclusions in 
section 6.

\section{Multi-frequency Foreground Simulation}
\label{sec:simu}

In this section we describe the details of the foreground simulation to produce the sky model for generating visibilities for low radio frequency observation 
with an interferometer. Even if the simulation, described in this paper, is 
carried out specifically for $150 \ {\rm MHz}$ observation with GMRT, it is 
generic and can easily be extended to other frequency and other similar 
telescopes (including the SKA). Earlier studies \citep{ali08,paciga11} 
have found that, for $150\, {\rm MHz}$ GMRT small field observations, the 
bright compact sources are the dominating foreground component for EoR signal 
at the angular scales $\le  4^{\circ}$, the other major component being the 
Galactic diffuse synchrotron emission \citep{bernardi09,ghosh12,iacobelli13}. 
We build our foreground sky model keeping close to the existing observational 
findings. The sky model includes the main two foreground components (i) 
discrete radio point sources and (ii) diffuse Galactic synchrotron emissions. 
The contributions from these two foregrounds dominate in low frequency radio 
observations and their strength is $\sim 4-5$ orders of magnitude larger than 
the $\sim 20-30\, {\rm mK}$ cosmological $21$-cm signal \citep{ali08,ghosh12}. 
Galactic and extragalactic free-free diffuse emissions are not included as a 
part of the sky model, though each of these is individually larger than the 
HI signal.

\subsection{Radio Point Sources}
\label{sec:ptsrc}

Most of the earlier exercise of numerical simulation conducted so far have not 
included the bright point source foreground component in the multi-frequency 
model. In such analysis, it is generally assumed that the brightest point 
sources are perfectly subtracted from the data before the main analysis, and 
the simulated data contains only faint point sources and other diffuse 
foreground components, HI signal and noise. We, however, simulate the point 
source distribution for sky model using the following differential source 
counts obtained from the GMRT $150 \ {\rm MHz}$ observation \citep{ghosh12}:
\begin{equation}
\frac{dN}{dS} = \frac{10^{3.75}}{\rm Jy.Sr}\,\left(\frac{S}{\rm Jy}\right)^{-1.6} \,.
\label{eq:dnds}
\end{equation}
The Full Width Half Maxima (hereafter FWHM) of the GMRT primary beam (PB) at 
$150 \ {\rm MHz}$ is $\approx3.1^{\circ}$. To understand and quantify how the 
bright point sources outside the FWHM of the PB affect our results, we consider 
here a larger region ($7^{\circ} \times 7^{\circ}$) for point source 
simulation. Initially, $2215$ simulated point sources, with flux density in 
the range $9\,{\rm mJy}$ to $1\,{\rm Jy}$ following the above mentioned 
distribution, are randomly distributed over this larger region. Out of those 
sources, $353$ are within ${95}^{'}$ from the phase centre (where the PB 
response falls by a factor of $e$). We note that the antenna response falls 
sharply after this radius. For example, the primary beam response is $\lesssim$ 
0.01 in the first sidelobe. Hence, outside this ``inner'' region, only sources 
with flux density greater than $100 \, {\rm mJy}$ are retained for the next 
step of the simulation. In the outer region, any source fainter than this will 
be below the threshold of point source subtraction due to primary beam 
attenuation. With $343$ sources from the ``outer'' region, we finally include 
total $696$ sources in our simulation. Figure~\ref{fig:ptsrc} shows the angular 
positions of all $2215$ sources over this region, as well as of the $696$ 
sources after the flux density restriction. Note that, we have assumed all the 
sources are unresolved at the angular resolution of our simulation. In reality, 
there will also be extended sources in the filed. Some of the extended sources 
can be modelled reasonably well as collection of multiple unresolved sources. 
However, other complex structures will probably need more careful modelling or 
masking, and such sources are not included in this simulation for simplicity.

The flux density of point sources changes across the frequency band of 
observation. We scale the flux density of the sources at different frequencies 
using the following relation,
\begin{equation}
S_{\nu}=S_{\nu_0}(\frac{\nu}{\nu_0})^{-\alpha_{\rm ps}} \,
\label{eq:fluxscale}
\end{equation}
where $\nu_0=150\,{\rm MHz}$ is the central frequency of the band, $\nu$ 
changes across the bandwidth of $16{\rm MHz}$ and $\alpha_{\rm ps}$ is the 
spectral index of point sources. The point sources are allocated a randomly 
selected spectral index uniform in the of range $0.7$ to $0.8$ 
\citep{jackson08,randall12}. Please note that the subsequent point source 
modeling and subtraction are carried out in such a way that the final outcomes 
do not depend on the exact distribution function of the spectral index. Before 
calculating the visibilities, the flux density of the point sources are 
rescaled, according to their angular separation from the phase centre, by 
multiplying with the PB response. We model the PB of GMRT assuming that the 
telescope has an uniformly illuminated circular aperture of $45\,{\rm m}$ 
diameter (D) whereby the primary beam pattern is given by,
\begin{equation}
{\cal A}(\th,\,\nu) = 
\left[ \left(\frac{2 \lambda}{\pi\theta D} \right)
J_1\left(\frac{\pi\theta D}{\lambda}\right) \right]^2 
\label{eq:b1} 
\end{equation}
where $J_1$ is the Bessel function of the first kind of order one. The primary 
beam pattern is normalized to unity at the pointing center $[{\cal A}(0)=1]$. The central part of the model PB (eq.~\ref{eq:b1}) is a reasonably good approximation to the actual PB of the GMRT antenna, whereby, it may vary at the outer region. In our analysis, we taper the outer region through a window function for which the results by using model PB do not change significantly.

\begin{figure*}
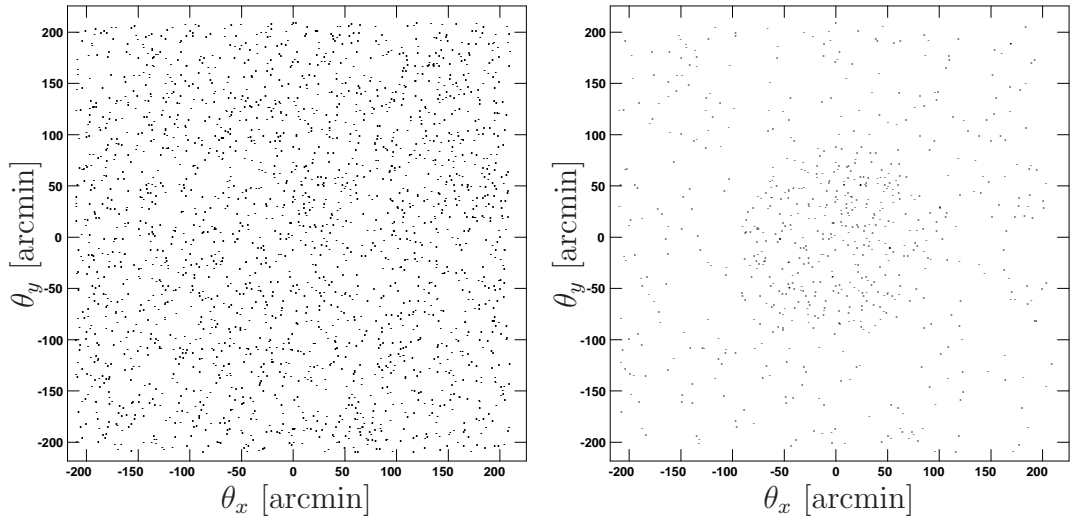

\begin{center}
\psfrag{thetax}[t][c][1][0]{$\theta_x$ $[{\rm arcmin}$]}
\psfrag{thetay}[c][c][1][0]{$\theta_y$ $[{\rm arcmin}$]}
\includegraphics[width=70mm,angle=0]{chapter3/fig/ptsrc1small1.ps}
\includegraphics[width=70mm,angle=0]{chapter3/fig/ptsrc696.ps}
\caption{The angular position of the simulated point sources over a $7^{\circ} \times 7^{\circ}$ region. The left panel shows positions of all $2215$ sources over this whole field, and the right panel shows $696$ sources after applying a flux density cutoff. The number of point sources in the flux density range $9\,{\rm mJy}$ to $1\,{\rm Jy}$ inside the FWHM of the primary beam is $N_{in}=353$ and outside of the FWHM with flux density more than 100 {\rm mJy} is $N_{out}=343$.}
\label{fig:ptsrc}
\end{center}
\end{figure*}

\subsection{Diffuse Synchrotron Emission}
\label{sec:diff}

In this section, we first describe the simulation of the diffuse Galactic 
synchrotron emission which are used to generate the visibilities. The angular 
slope $\beta$ of the angular power spectrum of diffuse Galactic synchrotron 
emission is within the range $1.5$ to $3$ as found by all the previous 
measurements at frequency range $0.15 -94 \, {\rm GHz}$ (e.g. 
\citealt{tegmark96,tegmark2000,giardino02,bennett03,laporta08,bernardi09,ghosh12,iacobelli13}). For the purpose of this paper, we assume that the fluctuations 
in the diffuse Galactic synchrotron radiation are a statistically homogeneous 
and isotropic Gaussian random field whose statistical properties are completely 
specified by the angular power spectrum. We construct our sky model of the 
diffuse Galactic synchrotron using the measured angular power spectrum at 
$150\,{\rm MHz}$ \citep{ghosh12}
\begin{equation}
C^M_{\ell}(\nu)=A_{\rm 150}\times\left(\frac{1000}{\ell} \right)^{\beta}\times\left(\frac{\nu}{150{\rm MHz}}\right)^{-2\alpha_{\rm syn}}   \,,
\label{eq:cl150}
\end{equation}
where $\nu$ is the frequency in ${\rm MHz}$, $A_{\rm 150}=513 \, {\rm mK}^2$ 
and $\beta=2.34 $ adopted from \citet{ghosh12} and $\alpha_{\rm syn}=2.8$ 
from \citet{platinia98}. The diffuse emissions are generated in a $1024 \times 
1024$ grid with angular grid size of $\sim 0.5^{'}$, covering a region of 
$\ 8.7^{\circ} \times 8.7^{\circ}$. This axis dimension is $\approx 2.8$ times 
larger than the FWHM of the GMRT primary beam.

\begin{figure*}
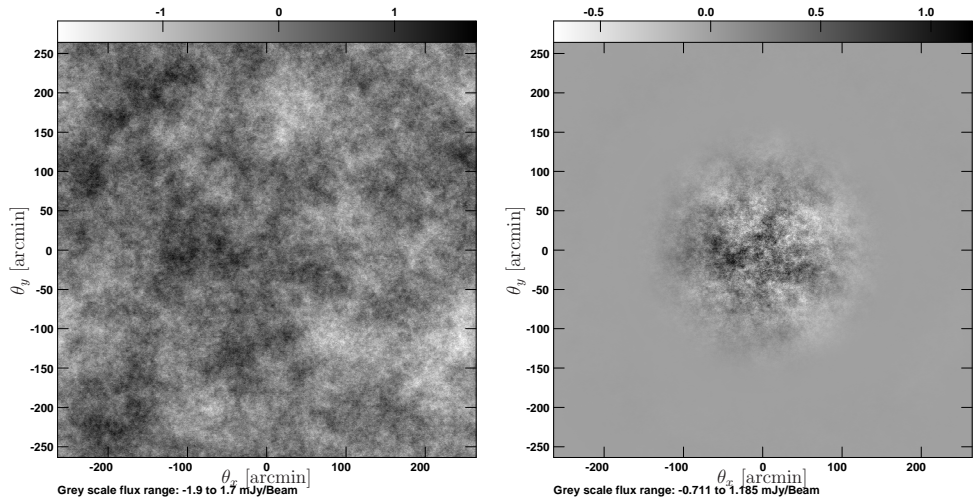

\begin{center}
\psfrag{thetax}[c][c][0.6][0]{$\theta_x$ $[{\rm arcmin}$]}
\psfrag{thetay}[c][c][0.6][0]{$\theta_y$ $[{\rm arcmin}$]}
\psfrag{cl}[b][t][1][0]{$C_{\ell} [mK^2]$}
\psfrag{U}[c][c][1][0]{$\ell$}
\psfrag{model}[c][l][1][0]{Model}
\psfrag{beam}[r][r][1][0]{with PB}
\psfrag{nobeam}[r][r][1][0]{without PB}
\includegraphics[height=70mm,angle=0]{chapter3/fig/diffmulti65nopb.ps}
\includegraphics[height=70mm,angle=0]{chapter3/fig/diffmulti65.ps}
\caption{The simulated  intensity map for the diffuse synchrotron radiation at $150 \, {\rm MHz}$ before (left panel) and after (right panel) multiplying the GMRT primary beam. The total angular size of each map is $8.7^{\circ} \times 8.7^{\circ}$ with a grid size $\sim 0.5^{'}$. Here, the grey scale is in units of ${\rm mJy/Beam}$.}
\label{fig:diff}
\end{center}
\end{figure*}

To simulate the diffuse emission, we mainly followed the same procedure as discussed in \citet{samir14}. We first create the Fourier components of the 
temperature fluctuations on a grid using
\begin{equation}
\Delta \tilde{T}(\u,\nu_0)=\sqrt{\frac{\Omega \, C^M_{\ell}(\nu_0)}{2}}[x(\u)+iy(\u)],
\label{eq:ran}
\end{equation}
where $\Omega$ is the total solid angle of the simulated area, and $x(\u)$ and 
$y(\u)$ are independent Gaussian random variables with zero mean and unit 
variance. Then, we use the Fastest Fourier Transform in the West (hereafter 
FFTW) algorithm \citep{frigo05} to convert $\Delta\tilde{T}(\u,\nu_0)$ to 
$\delta T(\th,\nu_0)$, the brightness temperature fluctuations or equivalently 
the intensity fluctuations $\delta I(\th,\nu_0)$ on the grid. The intensity 
fluctuations $\delta I(\th,\nu) = (2 k_B/\lambda^2)\, \delta T(\th,\nu) $ can 
be calculated using the Raleigh-Jeans approximation which is valid at the 
frequency of our interest. Figure~\ref{fig:diff} shows one realization of the 
intensity fluctuations $\delta I(\th,\nu_0)$ map at the central frequency 
$\nu_0=150\,{\rm MHz}$ with and without multiplication of the GMRT primary 
beam. The multiplication of the primary beam with intensity fluctuations in 
the sky plane is equivalent to the convolution of the Fourier transform of the 
both quantities in the $uv$ plane. The recovered angular power spectrum is 
affected due to the convolution of the primary beam only at large angular 
scales ($\lesssim 45~\lambda$). This affect has been shown already in Figure 3 
of \citet{samir14}. Based on a large number of realizations of the simulated 
diffuse intensity map, we find that the estimated angular power spectrum 
without multiplication of PB is in good agreement with the input model power 
spectrum (eq. \ref{eq:cl150}) at the scales of our interest $(\ell\sim300-2\times10^4)$.  

Finally, we generate the specific intensity fluctuations at any other frequency 
in the observation frequency band from that of the reference frequency using 
the scaling relation 
\begin{equation}
\delta I(\th,\nu)= (2 k_B/\lambda^2)\, \delta T(\th,\nu_0)(\frac{\nu}{\nu_0})^{-\alpha_{\rm syn}} \,.
\label{eq:tempscale}
\end{equation}
In general, the spectral index $\alpha_{\rm syn}$ of the diffuse emission may 
have a spatial variation and the synchrotron power spectrum may be different at 
different frequencies. However, the effect of this on point source subtraction 
is expected to be negligible, and the final results do not depend on the 
constancy of the synchrotron power spectrum slope. Here, we assume that the 
value of $\alpha_{\rm syn}$ is fixed over the whole region and across the 
observation band in the multi-frequency simulation. 

\subsection{Simulated GMRT Observation}

The simulations are generated keeping realistic GMRT specifications in mind, 
though these parameters are quite general, and similar mock data for any other 
telescope can be generated easily. The GMRT has $30$ antennas. The diameter of 
each  antenna is $45\rm m$. The projected shortest baseline at the GMRT can be  
$60\rm m$, and the longest baseline is $\rm 26 \,km$. The instantaneous 
bandwidth is $16 \, {\rm MHz}$, divided into $128$ channels, centered at $150 
\, {\rm MHz}$. We consider all antennas targeted on a arbitrary field located 
at R.A.=$10{\rm h} 46{\rm m} 00{\rm s}$ Dec=$59^{\circ} 00^{'} 59^{''}$ for a 
total of $8 \, {\rm hr}$ observation. The visibility integration time was 
chosen as $16 \,{\rm s}$. The mock observation produces $783000$ samples per 
channels in the whole $uv$ range. Each baseline generates $128$ visibilities 
because of $128$ spectral channels in the observation frequency-band. 
Figure~\ref{fig:uvtrack} shows the full $uv$ coverage at central frequency for 
the simulated GMRT Observation. Table~\ref{tab:1} summarizes the GMRT 
parameters used in this work.

The angular power spectrum of the diffuse synchrotron emission (eq. 
\ref{eq:cl150}) declines with increasing baseline $U=\mid \u \mid$ ($\ell = 2 
\pi U$), and drops significantly at the available longest baseline. Hence, for 
our simulation, the contributions of the diffuse emission have been taken from 
only baselines $U \le 3,000~\lambda$ to reduce the computation time. To calculate the 
visibilities, we multiply the simulated intensity fluctuations $\delta I(\th, 
\,\nu)$ with the PB (eq.~\ref{eq:b1}), and we use 2-D FFTW of the product in a 
grid. For each sampled baseline $U \le 3,000~\lambda$, we interpolate the gridded 
visibilities to the nearest baseline of the $uv$ track in 
Figure~\ref{fig:uvtrack}. We notice that the $w$-term does not have significant 
impact on the estimated angular power spectrum of diffuse synchrotron emission 
\citep{samir14}. But, to make the image properly and also to reduce the point 
source sidelobes, it is necessary to retain the $w$-term information. The 
$w$-term also improves the dynamic range of the image and enhances the 
precision of point source subtraction. We use the full baseline range to 
calculate the contribution from the point sources. The sky model for the point 
sources is multiplied with PB ${\cal A}(\th,\,\nu)$ before calculating the 
visibilities. Using the small field of view approximation, the visibilities for 
point sources are computed at each baseline by incorporating the $w$ term:
\begin{equation}
V(\u,\nu) \approx \int d^2 \theta  {\cal A}(\th,\,\nu) \,  \delta I(\th,\nu) \, e^{- 2 \pi i\big(u\theta_x+v\theta_y+w\big(\sqrt{1-\theta_x^2-\theta_y^2}-1\big)\big)} \, .
\label{eq:vis}
\end{equation}
\begin{figure}
\begin{center}
\hspace*{-1cm}
\includegraphics[width=70mm,angle=-90]{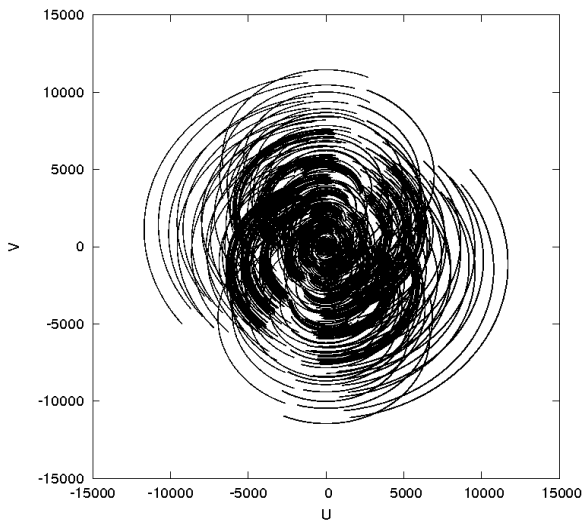}
\caption{The GMRT $uv$ coverage with phase centre at R.A.=$10{\rm h} 46{\rm m} 00{\rm s}$ Dec=$59^{\circ} 00^{'} 59^{''}$ for total observation time $8{\rm hr}$. Note that $u$ and $v$ are antenna separations measured in units of wavelength at the central frequency $150\, {\rm MHz}$.}
\label{fig:uvtrack}
\end{center}
\end{figure}
\begin{table}
\begin{center}
\begin{tabular}{|l|l|}
\hline
{\rm Parameter}&{\rm Value (GMRT)}  \\
\hline
${\rm R.A.}$& $10{\rm h} 46{\rm m} 00{\rm s}$  \\
\hline
${\rm Dec}$&$59^{\circ} 00^{'} 59^{''}$ \\
\hline
$N_{\rm ant}$& $30$  \\
\hline
${\rm Bandwidth}$ & $16\,{\rm MHz}$\\
\hline
$N_{\rm chan}$ & $128$\\
\hline
$\Delta\nu$ & $125\,{\rm kHz}$ per channel\\
\hline
$\Delta t$ & $16\,{\rm sec}$\\
\hline
$T_{\rm obs}$ & $8\,{\rm hr}$\\
\hline
\end{tabular} 
\caption{The GMRT parameters used to generate mock visibility data for the simulated sky model described in Section~\ref{sec:simu}.}
\label{tab:1}
\end{center}
\end{table}
The system noise of the interferometer is considered to be independent
at different baselines and channels, and is modelled as Gaussian
random variable. We add independent Gaussian random noise to both the
real and imaginary parts of each visibility contribution. For a single
polarization, the rms noise in the real or imaginary part of a
visibility is predicted to be \citep{thompson},
\begin{equation}
\sigma
=\frac{\sqrt2k_BT_{sys}}{A_{eff}\sqrt{\Delta \nu
      \Delta t}}
\label{eq:rms}
\end{equation}
where $T_{sys}$ is the total system temperature, $k_B$ is the Boltzmann 
constant, $A_{eff}$ is the effective collecting area of each antenna, $\Delta 
\nu$ is the channel width and $\Delta t$ is correlator integration time. For a 
channel width of $\Delta\nu=125\,{\rm kHz}$ and integration time $\Delta t = 
16\,{\rm sec}$, the rms noise comes out to be $\sigma_n=1.03 \, {\rm Jy}$ for GMRT at single polarization. The two polarizations are assumed to have identical sky 
signals but independent noise contribution.

In summary, our simulated visibilities for the GMRT observation are sum of two 
independent components namely the sky signal and the system noise. As outlined 
above, the realistic sky signal contains the contribution of the extragalactic 
point sources and the diffuse synchrotron emission from our own Galaxy. The 
visibility data does not contain any calibration errors, ionospheric effects 
and radio-frequency interference (RFI). We leave a detailed investigation of 
these effects for future work.
\section{Data Analysis}
\label{sec:data}
\begin{figure*}[h]
\begin{center}
\psfrag{cl}[b][t][1.5][0]{$C_{\ell} [mK^2]$}
\psfrag{U}[c][c][1.5][0]{$\ell$}
\psfrag{model}[r][r][1][0]{Model}
\psfrag{allnotaper}[r][r][1][0]{Total}
\includegraphics[width=68mm,angle=0]{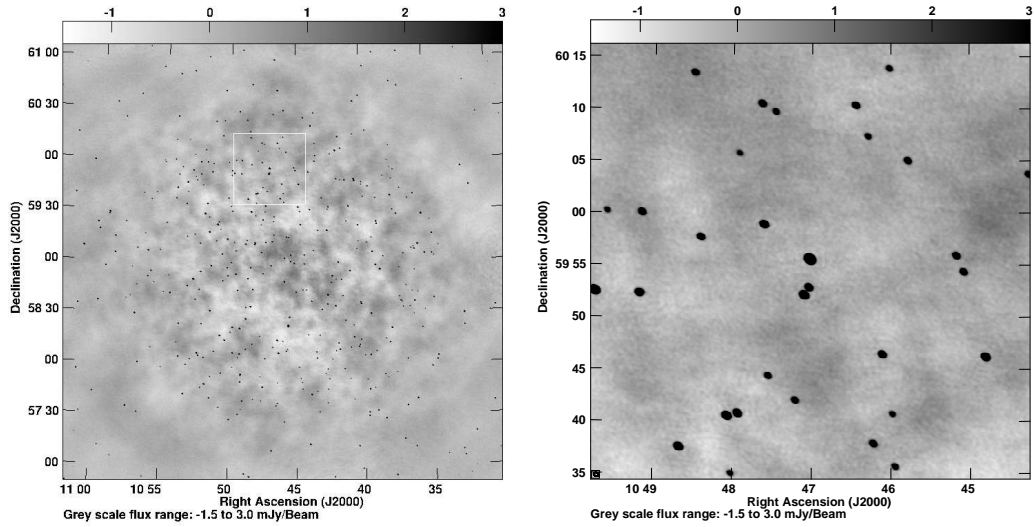}
\includegraphics[width=68mm,angle=0]{chapter3/fig/img_cleansmall.ps}
\caption{The left panel shows the CLEANed image ($4.2^{\circ}\times4.2^{\circ}$) of the simulated sky centered at R.A.=$10{\rm h} 46{\rm m} 00{\rm s}$ Dec=$59^{\circ} 00^{'} 59^{''}$. The synthesized beam has a ${\rm FWHM} \sim 20^{''}$. A zoom of the square region, $42^{'}\times42^{'}$ in size, marked in the left panel is shown in the right panel. This representative region is used in Figure~\ref{fig:rescompare} for comparison of ``residual'' images. In the central region the  ``off-source'' rms  noise is $\approx 0.3\, {\rm mJy/Beam }$. Here, the grey scale is in units of ${\rm mJy/Beam}$.}
\label{fig:imgclean}
\end{center}
\end{figure*}

The simulated visibility data described above is generated using sky emission 
model containing a combination of point sources and Galactic diffuse 
synchrotron emission (along with instrumental noise). Our next goal is to 
analyse these simulated data to recover the statistical properties of the 
diffuse emission, and compare those with the known input model parameters. As 
mentioned earlier, to estimate the power spectrum of the diffuse emission, our 
approach is to first remove the point source foreground accurately. This 
requires imaging and deconvolution to model the point sources, and then 
subtracting them from the data. In reality, there are many issues which make an 
accurate subtraction of point sources from radio interferometric wide-field 
synthesis images challenging. These include residual gain calibration errors 
\citep{adatta,adatta10}, direction dependence of the calibration due to 
instrumental or ionospheric/atmospheric conditions \citep{intema09,yata12}, 
the effect of spectral index of the sources \citep{rau11}, frequency dependence 
and asymmetry of the primary beam response, varying point spread function 
(synthesized beam) of the telescope \citep{bowman09,liu09a,morales12,ghosh12}, 
high computational expenses of imaging a large field of view, and CLEANing a 
large number of point sources \citep[particularly severe at low radio frequency 
(e.g. $150 \, {\rm MHz}$) images,][]{pindor11} etc. Earlier, \citet{adatta, 
adatta10} have studied the effect of calibration errors in bright point source 
subtraction. They have concluded that, to detect the EoR signal, sources 
brighter than $1\, {\rm Jy}$ should be subtracted with a positional accuracy 
better than 0.1 arcsec if calibration errors remain correlated for a minimum 
time $\sim$ 6 hours of observation. The polarized galactic synchrotron emission 
is expected to be Faraday-rotated along the path, and it may acquire additional 
spectral structure through polarization leakage at the telescope. This is a 
potential complication for detecting the HI signal \citep{jelic,moore13}. To 
cope with the capabilities of current and forthcoming radio telescopes, 
recently there have been a significant progress in developing calibration, 
imaging and deconvolution algorithms \citep{bhat13,cornwell08} which can now 
handle some of the above-mentioned complications.

In this paper, we take up a study of the effect of incomplete spectral modeling 
and of different deconvolution strategies to model and subtract point sources 
using simulated data at $150\,{\rm MHz}$. The power spectrum estimator that we 
have used takes care of, at least to a large extent, issues like asymmetry of 
the primary beam, direction dependence of the calibration for the outer region 
of the field of view and high computational expenses of imaging and removing 
point sources from a large field of view etc. We leave studying the other 
calibration related issues for future work.

For our analysis, we use the Common Astronomy Software Applications (CASA) 
\footnote{http://http://casa.nrao.edu/} to produce the sky images from the 
simulated visibility data. To make a CLEAN intensity image, we use the 
Cotton-Schwab CLEANing algorithm \citep{schwab84} with Briggs weighting and 
robust parameter 0.5, and with different CLEANing thresholds and CLEANing boxes 
around point sources. The CLEANing is done also with or without multifrequency 
synthesis (MFS; \citealt{sault94,conway90,rau11}). During deconvolution, MFS, if used, takes 
into account the spectral variation of the point sources using Taylor series 
coefficients as spectral basis functions. In a recent paper \citet{offringa16} 
suggest that CASA's MS-MFS algorithm can be used for better spectral modelling 
of the point sources. The large field of view ($\theta_{FWHM}= 3.1^{\circ}$) of 
the GMRT at $150\, {\rm MHz}$ lead to significant amount of errors if the 
non-planar nature of the GMRT antenna distribution is not taken into account. 
For this purpose we use $w-$projection algorithm \citep{cornwell08} implemented in 
CLEAN task within the CASA. For different CLEANing strategies, we assess the 
impact of point sources removal in recovering the input angular power spectrum 
$C_{\ell}$ of diffuse Galactic synchrotron emission from residual $uv$ data. 
Effectively, by CLEANing with these different options, we identify the optimum 
approach to produce the best model for point source subtraction and $C_{\ell}$ 
estimation. We investigate the CLEANing effects both in the image domain by 
directly inspecting the ``residual images'' after the point source subtraction, 
and also in the Fourier domain by comparing the power spectrum of the residual 
data with the input power spectrum of the simulated diffuse emission. For 
discussion on some of the relevant methods and an outline of the power 
spectrum estimation, please see \citet{samir14} and references therein.

The left panel of Figure~\ref{fig:imgclean} shows the resultant CLEANed image 
of the simulated sky of the target field with angular size $4.2^{\circ} \times 
4.2^{\circ}$. The synthesized beam has a ${\rm FWHM} \sim 20^{''}$. The image, 
as mentioned earlier, contains two different emission components (i.e. point 
sources, diffuse synchrotron emission) and noise. The grey scale flux density 
range in Figure~\ref{fig:imgclean} is saturated at ${\rm 3\, mJy}$ to clearly 
show the diffuse emission. The inner part ($\approx 1.0^{\circ} \times 
1.0^{\circ}$) of CLEANed image has rms noise $\approx 0.3\, {\rm mJy/Beam}$, 
and it drops to $\approx 0.15\,{\rm mJy/Beam }$ at the outer part of the image 
where the response of the GMRT primary beam attenuates quite a bit compared to 
the phase centre. In the right panel of Figure~\ref{fig:imgclean}, we also show 
a small portion (marked as a square box in the left panel) of the image with an 
angular size $42^{'}\times42^{'}$. We note that there is a strong point source 
at the centre of this small image with a flux density of ${\rm 676\, mJy/Beam}$ 
and spectral index of $0.77$. The intensity fluctuations of the diffuse 
emission are also clearly visible in both the panels of 
Figure~\ref{fig:imgclean}.

Figure~\ref{fig:imgcleanps} shows the angular power spectrum $C_{\ell}$ 
estimated from the simulated visibilities before any point source subtraction. 
We find that the estimated  power spectrum, as expected, is almost flat across 
all angular scales. This is the Poisson contribution from the randomly 
distributed point sources which dominate $C_{\ell}$ at all angular multipoles 
$\ell$ in our simulation. In this paper, we do not include the clustering 
component of the point sources which becomes dominant only at large angular 
scales ($\ell\le900$) \citep{ali08} where it introduces a power law $\ell$ 
dependence in the angular power spectrum. We also note that the convolution 
with the primary beam affects the estimated angular power spectrum at small 
$\ell$ values (Figure 3, \citealt{samir14}),  and it will be difficult to 
individually distinguish the Poisson and the clustered part of the point source 
components with the GMRT. The total simulated power spectrum $C_{\ell}$ 
(Figure~\ref{fig:imgcleanps}) is consistent with the previous GMRT ${\rm 
150MHz}$ observations (\citealt{ali08,ghosh12}). In 
Figure~\ref{fig:imgcleanps} we also show the input model angular power spectrum 
of the diffuse emission along with 1-$\sigma$ error bar (shaded region) 
estimated from 100 realizations of the diffuse emission map. Note that the 
angular power spectrum of the diffuse emission is buried deep under the point 
source contribution which dominates at all the angular scales accessible to the 
GMRT. We would like to emphasis that, in this paper, our aim is to study how 
well we can recover this diffuse power spectrum from the residual visibility 
data after point source subtraction is carried out to the desired level.

It is quite difficult to model and subtract out the point sources from the 
sidelobes and the outer parts of the main lobe of the primary beam. Our recent paper 
\citep{samir16a} contains a detailed discussion of the real life problems for 
modelling and subtracting point sources from these regions. In this paper we 
have restricted the point source subtraction to the central region of the 
primary beam (as detailed in the next section). Here we have used the TGE to 
estimate the angular power spectrum $C_{\ell}$ from the visibilities (both 
before and after point source subtraction). The TGE tapers the sky response to 
suppress the effect of the point sources outside the FWHM of the primary beam. 
This is achieved by convolving the visibilities with a window function whose 
width can be varied. It is also devised in such a way that it calculates the 
noise bias internally, and subtracts its contribution to extract only the 
desired signal. The TGE is an unbiased estimator for the angular power spectrum 
$C_{\ell}$, and a detailed description has been presented in our earlier paper 
\citep{samir14}. Here we have applied the TGE to the simulated visibility data 
to estimate $C_{\ell}$ in logarithmic intervals of $\ell$ after averaging all the frequency channels. The same estimator 
may also be extended to quantify the cosmological $21\,{\rm cm}$ signal, we 
plan to address this in future.

\begin{figure}[h]
\begin{center}
\psfrag{cl}[b][t][0.8][0]{$C_{\ell} [mK^2]$}
\psfrag{U}[c][c][0.8][0]{$\ell$}
\psfrag{model}[r][r][0.8][0]{Model}
\psfrag{allnotaper}[r][r][0.8][0]{Total}
\hspace*{-1.8cm}
\includegraphics[width=120mm,height=60mm,angle=0]{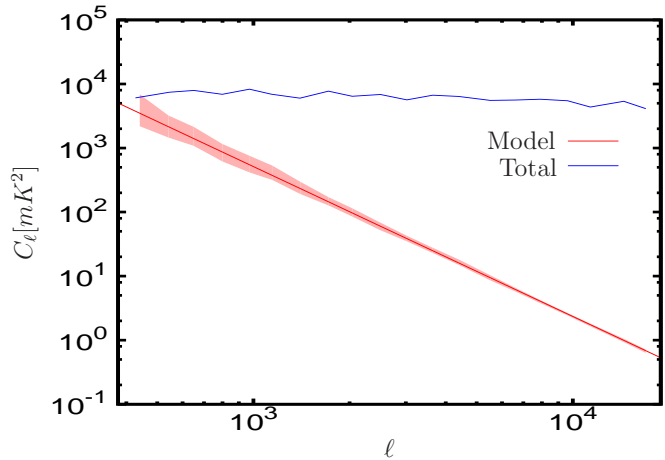}
\caption{The angular power spectrum $C_{\ell}$ estimated from the initial visibility data which contains two foreground components, point sources and diffuse synchrotron emission. For comparison, we show the model synchrotron power spectrum (lower curve) with 1-$\sigma$ error estimated from 100 realizations of the diffuse emission map. The total power spectrum (upper curve), dominated by the point sources, is flat in nature due to the Poisson distribution of positions of the discrete point sources in our simulation.}
\label{fig:imgcleanps}
\end{center}
\end{figure}

\section{Point Source Subtraction}
\label{sec:pssub}

The bright discrete point sources are the most dominant foreground component 
for detecting the redshifted HI 21-cm EoR signal. As shown in 
(Figure~\ref{fig:imgcleanps}), the $150\, {\rm MHz}$ radio sky is dominated by 
them at the angular scales  $\le  4^{\circ}$ \citep{ali08}. Therefore, it is 
very crucial to identify all point sources precisely from the image, and remove 
their contribution from the visibility data in order to estimate the power 
spectrum of background diffuse emission. In this section, we discuss the point 
source modeling and the effect of different CLEANing strategies on the 
``residual'' images, made from the point source subtracted visibility data.

We use different sets of parameter shown in Table~\ref{tab:2} for different 
CLEANing strategies. Pixels with flux density above a threshold value in the 
image are identified as point sources which are used to build the ``clean 
component'' model. The model visibilities corresponding to these clean 
components are subtracted from the original multi-frequency $uv$ data using 
the standard CASA task UVSUB. This should remove the point source contribution 
from the data to a large extent. The residual images, hence, are expected to 
be dominated by the diffuse emissions and the system noise. After point source 
subtraction we make residual images of size  $4.2^{\circ}\times4.2^{\circ}$.
Figure~\ref{fig:rescompare} shows a representative region of angular size 
$42^{'}\times42^{'}$, to illustrate the  effect of different cleaning schemes. 
 The different residual images (Image(a) to Image(f)) in 
Figure~\ref{fig:rescompare} correspond to the different CLEANing strategies in 
Table~\ref{tab:2} (Run(a) to Run(f)).

First we investigate the effect of spectral modelling of the clean components 
in the residual image. This is done by changing the parameter ``nterms'' where 
nterms=1 does not include any spectral correction, while nterms=2 builds the 
model by including spectral index during multi-frequency CLEANing. A more 
detailed discussions of these parameters can be found in \citet{rau11}. For point source subtraction with a CLEANing threshold of 
$1\, {\rm mJy}$ $(\approx 3\sigma_{im})$ and $\rm nterms=1$ and $2$, the 
``dirty'' images of the residual UVSUB data are shown in 
Figure~\ref{fig:rescompare} top row (left and right panel for $\rm nterms=1$ 
and $2$ respectively). The strong sidelobe patterns appear around the central 
bright source in Image(a) for incorrect spectral modelling. Most of these 
disappear in Image(b) where the spectral property of the bright source has been 
taken into account during CLEANing and continuum subtraction. 

In the middle row  of Figure~\ref{fig:rescompare}, we compare the residual 
images for two different CLEANing threshold $0.5 \,{\rm mJy}$ and $2.0 
\,{\rm mJy}$ (left and right panel respectively) while keeping $\rm nterms=2$ 
fixed for both. We notice that for the CLEANing threshold of $0.5 
\,{\rm mJy}$, part of the diffuse structure is also CLEANed and subtracted out 
form the data. On the other hand, all the diffuse structures (but also some 
residual from the point sources) are still present in the residual Image(d) 
where we use a higher threshold of $2.0 \,{\rm mJy}$ ($\approx 
6.0\,\sigma_{im}$). The overlayed contours in Figure~\ref{fig:rescompare} make 
the comparison more clear. For the panels in the top and the middle row, we 
CLEANed the {\it whole} image upto the specified threshold without making any 
CLEAN box around the point sources. This is more computation expensive as well 
as inadequate to handle the diffuse structure, and will remove positive and 
negative peaks of the diffuse signal. For EoR experiments, a part of the 
desired diffuse $21\,{\rm cm}$ signal, if present, may also be removed by such 
deep CLEANing without making boxes.

\begin{table*}
\begin{center}
\begin{tabular}{|l|c|l|l|}
\hline
{\rm Name}&{\rm nterms}&{\rm Threshold flux density}& {\rm CLEANing Box}\\
\hline
{\rm Run(a)}& $1$ &1.0\,{\rm mJy}& Single $4.2^{\circ}\times4.2^{\circ}$ {\rm Box}\\
\hline
{\rm Run(b)}& $2$ &1.0\,{\rm mJy}&Single $4.2^{\circ}\times4.2^{\circ}$ {\rm Box}\\
\hline
{\rm Run(c)}& $2$ &0.5\,{\rm mJy}&Single $4.2^{\circ}\times4.2^{\circ}$ {\rm Box}\\
\hline
{\rm Run(d)}& $2$ &2.0\,{\rm mJy}&Single $4.2^{\circ}\times4.2^{\circ}$ {\rm Box}\\
\hline
{\rm Run(e)}& $2$ &0.5\,{\rm mJy}&Circular region with radius $50^{''}$ \\
& & & around all sources in the image\\
\hline
{\rm Run(f)}& $2$ &2.0\,{\rm mJy}&Single $4.2^{\circ}\times4.2^{\circ}$ {\rm Box}\\
&&0.5\,{\rm mJy}&$1.6^{'}\times1.6^{'}$ {\rm Box} around\\
& & &each visible residual sources\\
\hline
\end{tabular} 
\caption{The set of parameters used for point source imaging with different CLEANing strategies.}
\label{tab:2}
\end{center}
\end{table*}

\begin{figure*}[h]
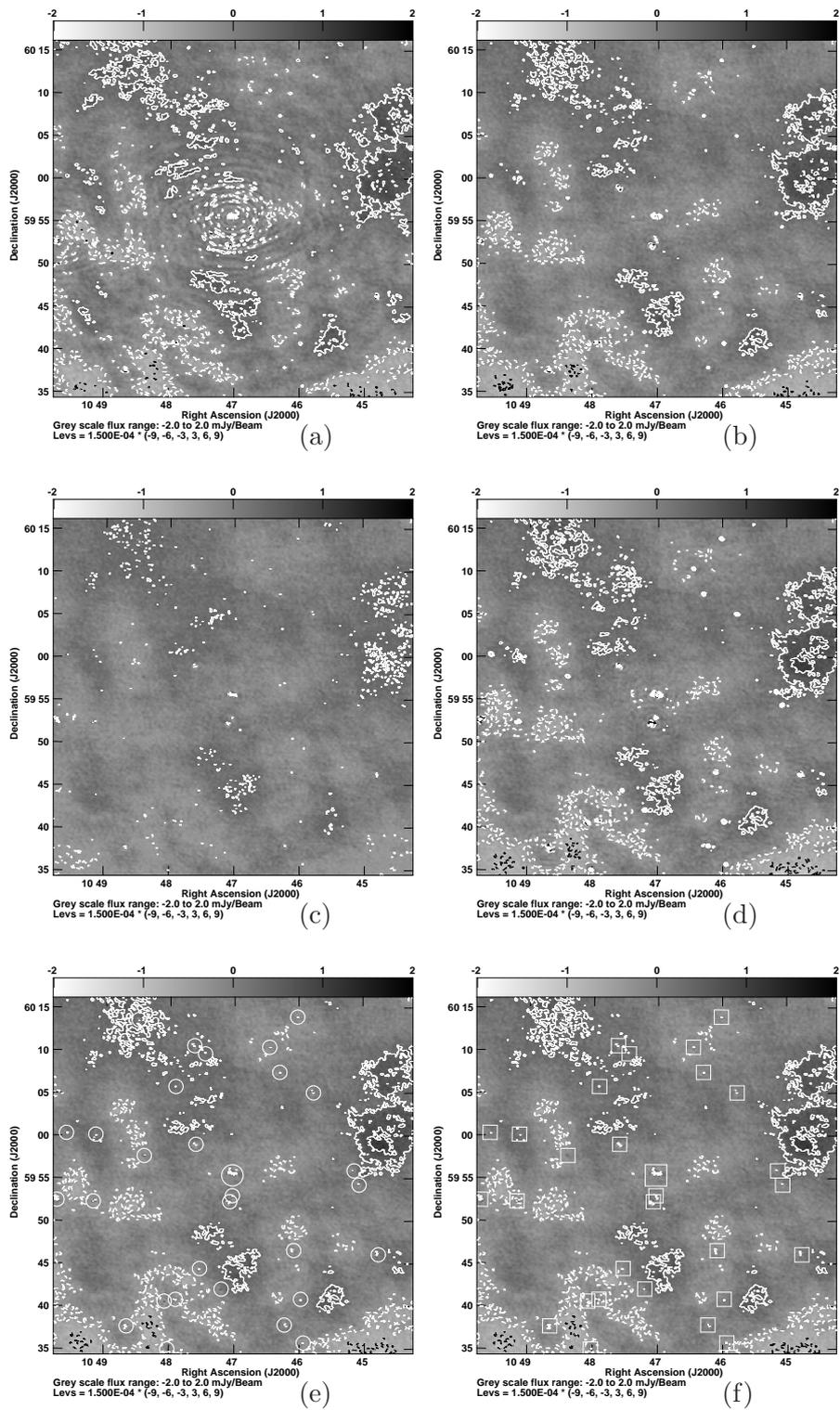

\begin{center}
\includegraphics[width=60mm,angle=0]{chapter3/fig/img1mjynterm12_2.ps}
\put(-50,5){\small (a)}
\includegraphics[width=60mm,angle=0]{chapter3/fig/img1mjy2_2.ps}
\put(-50,5){\small (b)}

\includegraphics[width=60mm,angle=0]{chapter3/fig/img0.5mjy2_2.ps}
\put(-50,5){\small (c)}
\includegraphics[width=60mm,angle=0]{chapter3/fig/img2mjy2_2.ps}
\put(-50,5){\small (d)}

\includegraphics[width=60mm,angle=0]{chapter3/fig/img0.5mjybox2_2star.ps}
\put(-50,5){\small (e)}
\includegraphics[width=60mm,angle=0]{chapter3/fig/img2mjyboxhand0.52_2star.ps}
\put(-50,5){\small (f)}
\caption{Residual images of the $42^{'}\times42^{'}$ representative region for various CLEANing strategies listed in Table~\ref{tab:2}, i.e  the residual images Image(a), Image(b), Image(c), ..., and Image(f) correspond to Run(a), Run(b), Run(c), ..., and Run(f) respectively. Here, the grey scale is in units of ${\rm mJy/Beam}$. Different contours with levels $(-9,-6,-3,3,6,9)\times0.15{\rm mJy/Beam}$ are also shown in these figures.}
\label{fig:rescompare}
\end{center}
\end{figure*}

Next we use CLEAN boxes to create the model for point source subtraction. This 
will ensure that the clean components are picked up only from the restricted 
regions defined by the shape of the box as highlighted in the bottom row of 
Figure~\ref{fig:rescompare}. Here, we select the boxes in two ways (see 
Table~\ref{tab:2}). In the first case, we use the mask file (circular box of radius $50^{''}$) from the catalogue sources which are used to generate the simulated data, and CLEANed upto 
$0.5\,{\rm mJy}$ threshold. For the second case, we first CLEANed the whole 
image upto a conservative limit of $2\,{\rm mJy}$. Then, by visually inspecting 
the image, we identified residual point sources which are not cleaned due to 
higher threshold, and placed rectangular boxes of size $1.6^{'}\times 1.6^{'}$ 
around each of them. These selected regions are then CLEANed upto a limit of 
$0.5\,{\rm mJy}$. The residual images for these two cases are shown in the 
bottom row of Figure~\ref{fig:rescompare} (Image(e) and Image(f)). We notice 
that there is no significant difference in the residuals for these two cases. 
In the next section, we assess impact of the different CLEAN strategies on the 
statistics such as distribution of visibilities and estimated angular power 
spectrum from different residual data sets.

\section{Results}
\label{sec:result}

\begin{figure*}
\begin{center}
\psfrag{jy}[b][t][0.8][0]{Flux Density[mJy]}
\psfrag{vjy}[b][t][0.8][0]{Re [Jy]}
\psfrag{number}[b][t][0.8][0]{Number}
\psfrag{tot}[r][r][0.8][0]{Total}
\psfrag{2}[r][r][0.8][0]{{\rm Run(d)}}
\psfrag{1}[r][r][0.8][0]{{\rm Run(b)}}
\psfrag{0.5}[r][r][0.8][0]{{\rm Run(c)}}
\psfrag{0.5bx}[r][r][0.8][0]{{\rm Run(e)}}
\psfrag{0.5bxhnd}[r][r][0.8][0]{{\rm Run(f)}}
\psfrag{gauss}[rb][rb][0.8][0]{Gaussian}
\includegraphics[height=50mm,angle=-90]{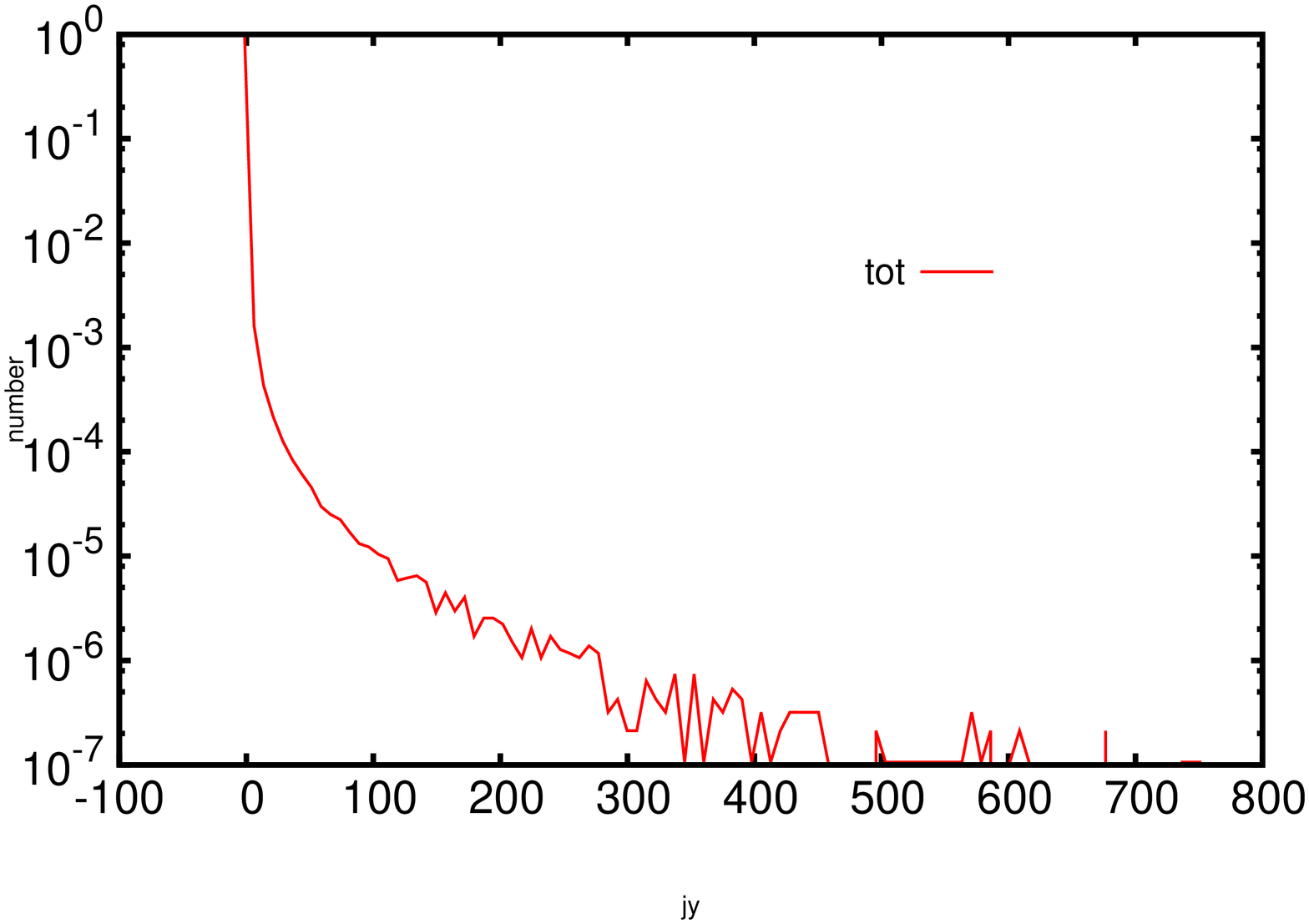}
\includegraphics[height=50mm,angle=-90]{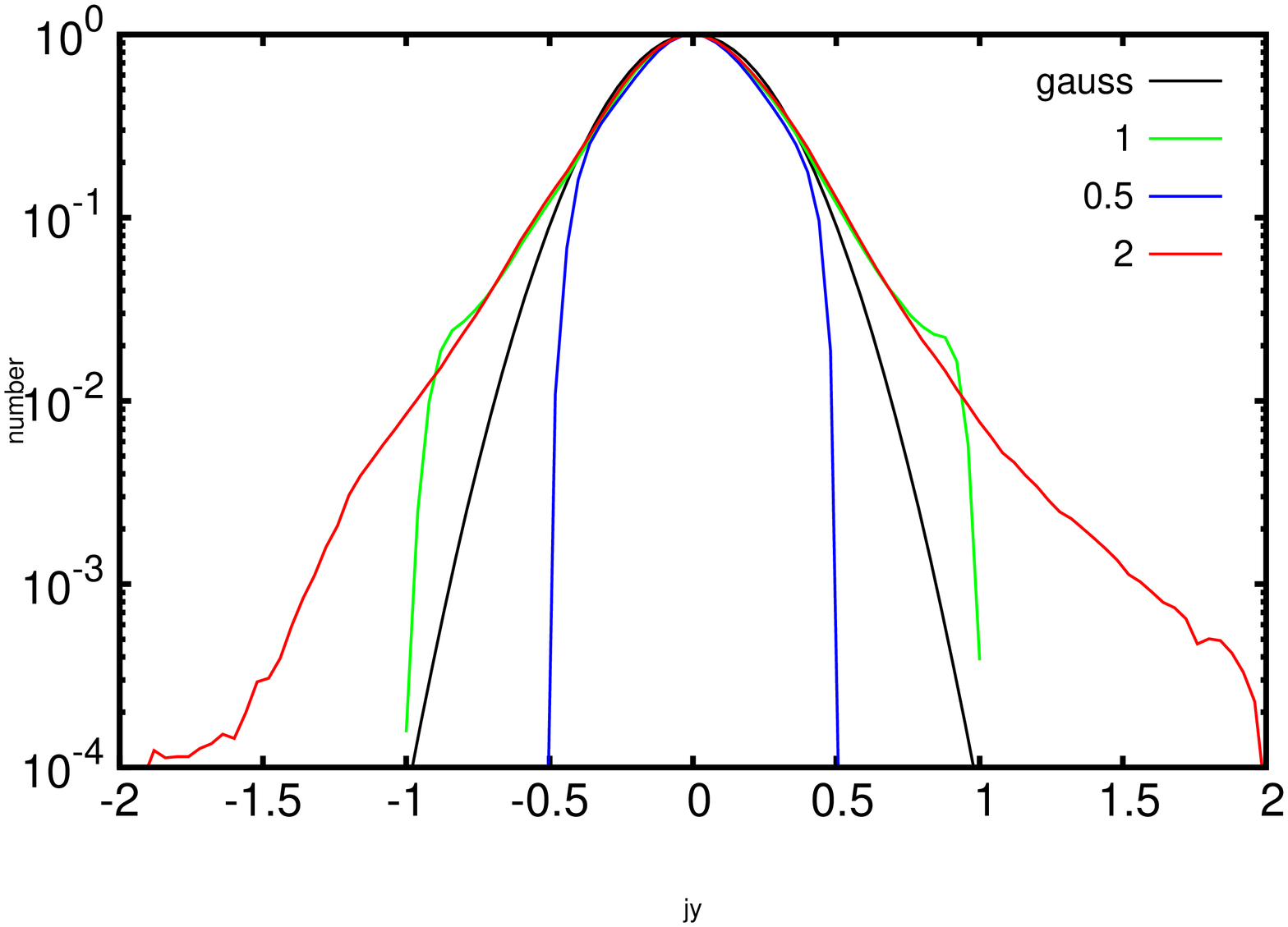}
\includegraphics[height=50mm,angle=-90]{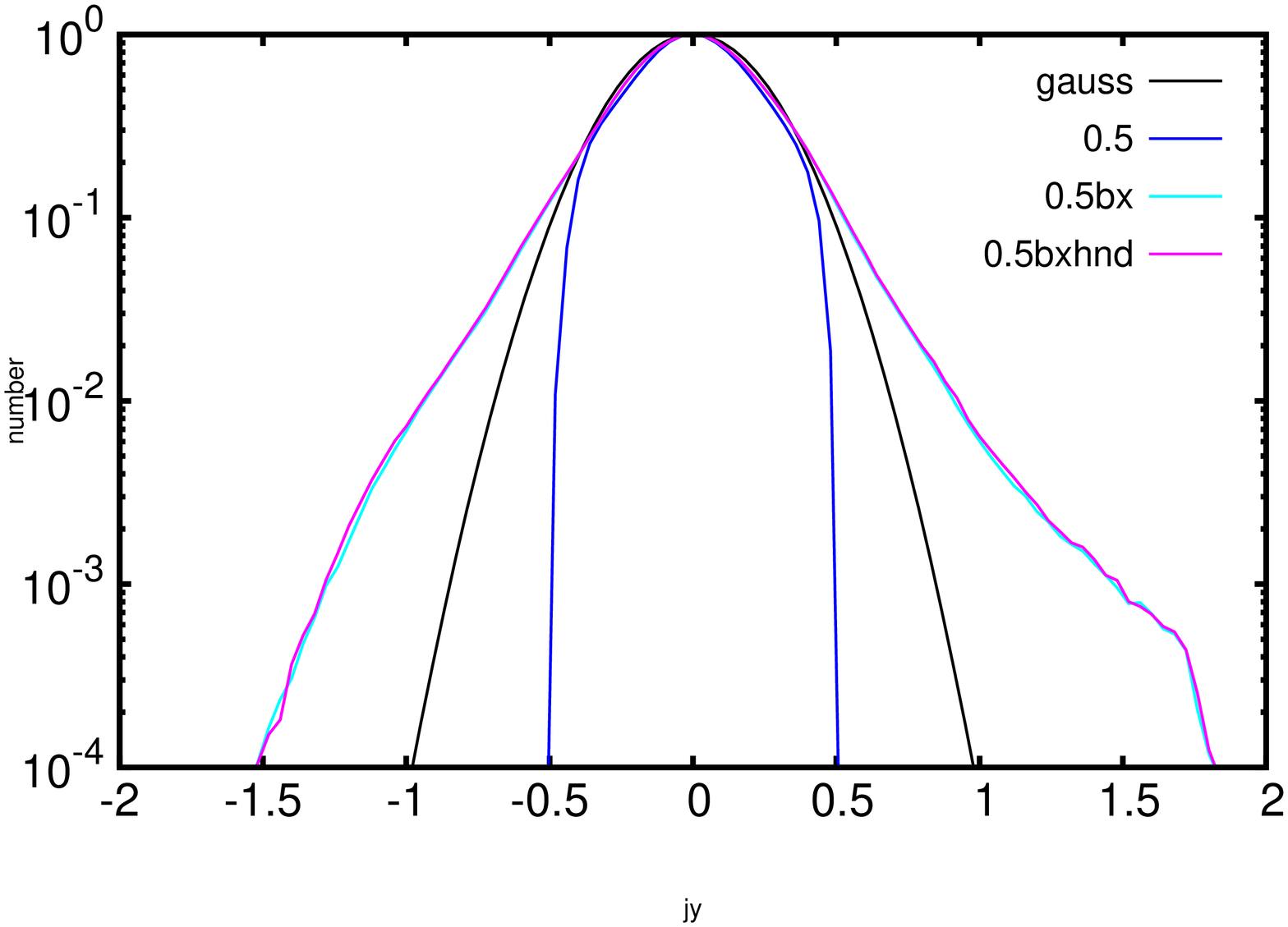}
\includegraphics[height=50mm,angle=-90]{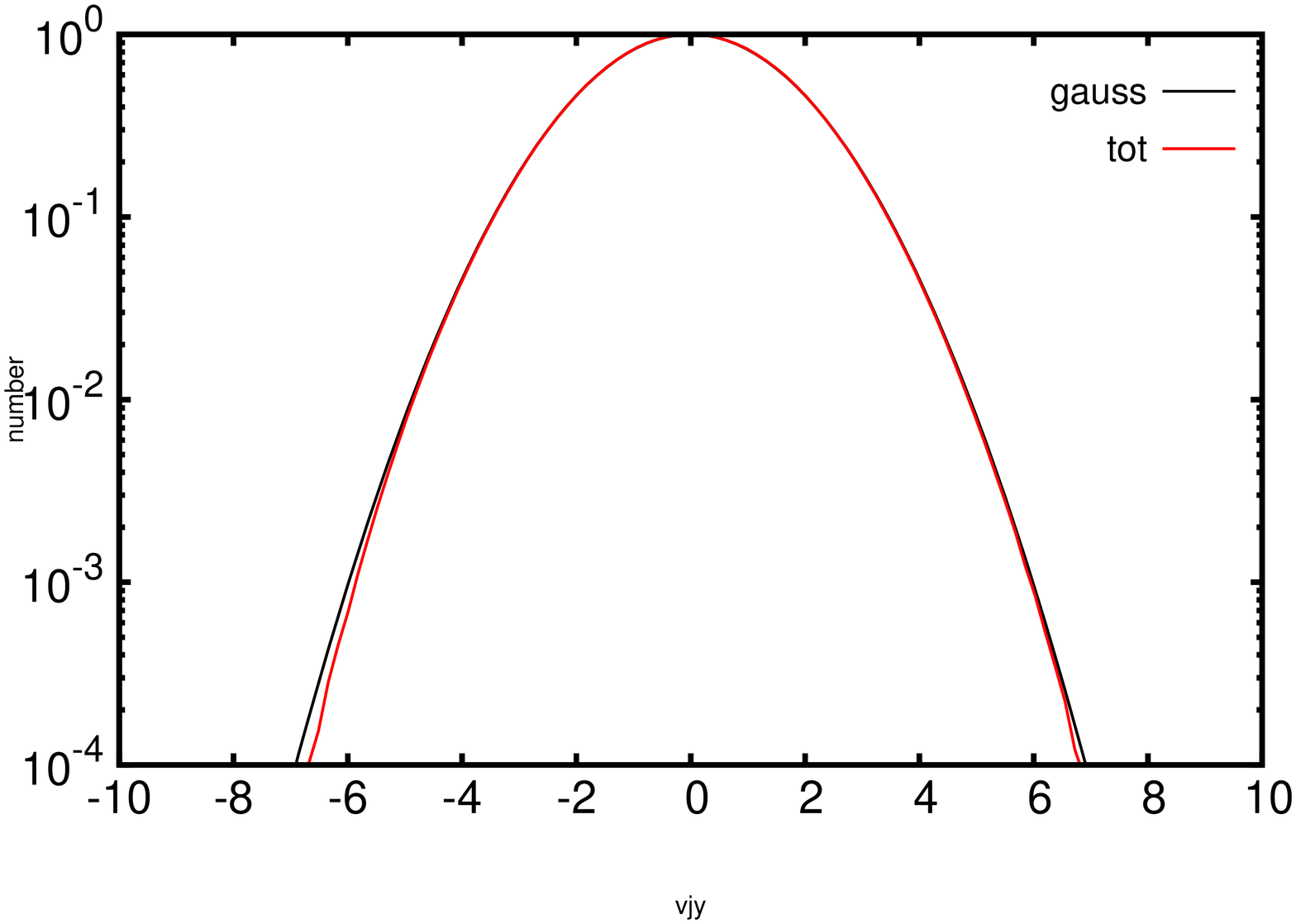}
\includegraphics[height=50mm,angle=-90]{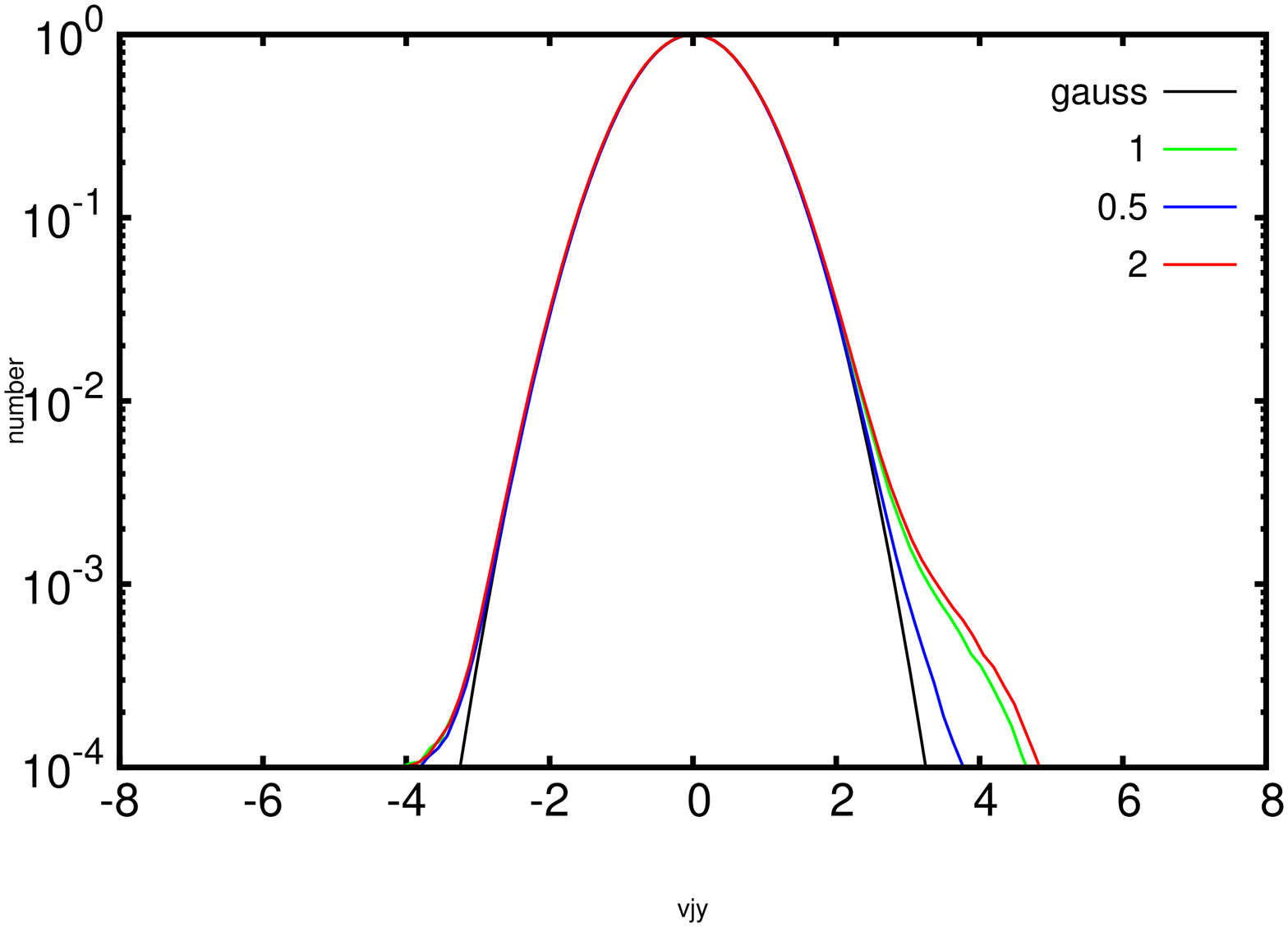}
\includegraphics[height=50mm,angle=-90]{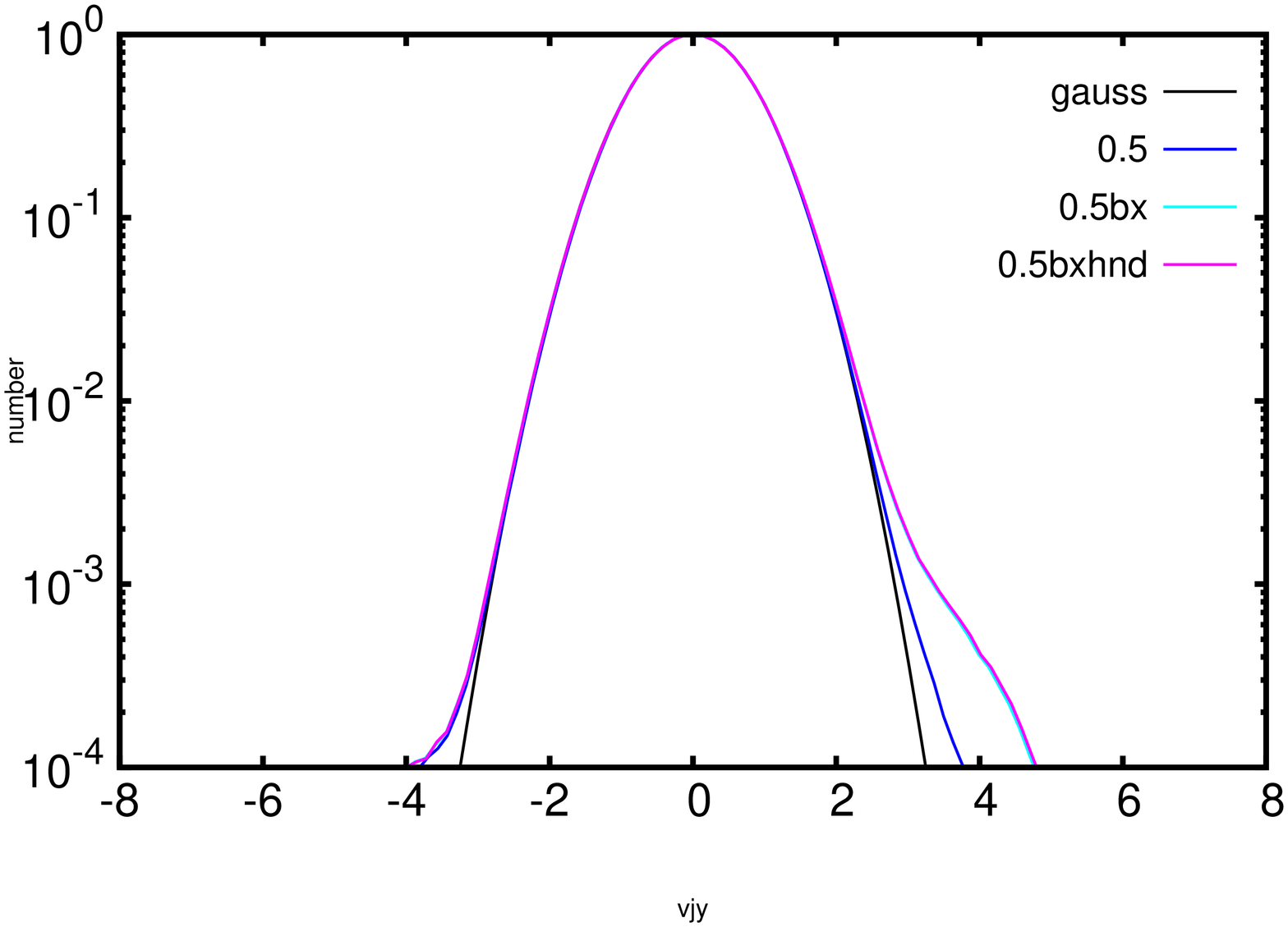}
\caption{The distribution of image plane pixel values (upper row) and the real part of visibilities (lower row) before point source subtraction (left panels) and after point source subtraction (middle and right panels) with different runs mentioned in  Table~\ref{tab:2}. The numbers in the y-axis are in logarithmic scale. The best fit Gaussian function for the distributions are also shown in the respective panels.}
\label{fig:hist}
\end{center}
\end{figure*}

We use different CLEANing strategies to subtract point sources from a 
$4.2^{\circ} \times 4.2^{\circ}$ region of the sky from simulated visibility 
data discussed in Section \ref{sec:pssub}. To compare the outcome of these 
strategies, we check the statistics of the visibilities as well as of the 
images. In Figure~\ref{fig:hist} we show the normalized histogram plots from 
images (top row) and from the visibility data (bottom row). First we consider 
the CLEANed and the residual images. The top-left panel of 
Figure~\ref{fig:hist} shows the distribution of the pixel values from the initial 
CLEANed map (Figure~\ref{fig:imgclean}). This plot shows a small number of 
pixels with high flux density values (due to the bright point sources). The 
distribution is, however, dominated by the diffuse foreground component with 
relatively small values ($\le 5.0\, {\rm mJy}$) over a large fraction of 
pixels. The top-middle and right panel show the histogram of the residual 
images from different CLEANing runs discussed in Section~\ref{sec:pssub}. A 
Gaussian with $\sigma=0.228\,{\rm mJy}$ is a fairly good fit to the distribution of 
the residuals upto a flux density limit of $\pm 0.5\, {\rm mJy}$. However, as 
evident from the top central panel, CLEANing with lower threshold (see 
Table~\ref{tab:2}) makes residual images more non-Gaussian. The histogram for 
Run(c), for example, is confined to lower flux density range, because ``blind'' 
CLEANing with very low threshold removes a part of diffuse structure. In the 
top right panel, we show the impact of choosing CLEAN boxes in different ways 
(Run(e) and (f) in Table~\ref{tab:2}), keeping a fixed threshold flux density 
of $ 0.5\, {\rm mJy}$. We find that there is no difference in the distribution 
of the residual images for {\rm Run(e)} and {\rm Run(f)}. Also, in all the 
cases, they follow the same Gaussian function upto $\pm 0.5\, {\rm mJy}$. 

Next, we consider the statistics of the visibilities. The corresponding 
visibility distribution functions are shown in the bottom row of 
Figure~\ref{fig:hist}. We use the real part the complex visibilities for the 
purpose of this comparison in the plots, but the imaginary parts also have a 
similar distribution. We find that the initial and residual visibility data 
both mostly follow a Gaussian distribution, but with different standard 
deviation. The initial visibility data (bottom row, left panel of 
Figure~\ref{fig:hist}) follows a Gaussian distribution with $\sigma=1.61$ Jy. 
The residual visibility data, however, can be fitted with a Gaussian function 
of $\sigma=0.76$ Jy upto a flux density limit of $\mid \rm Re(V)\mid <3{\rm 
Jy}$ containing the bulk of the data. The counts significantly deviate from a 
Gaussian at large visibility values most likely due to incomplete CLEANing.

The angular power spectrum $C_{\ell}$ have been estimated from the residual 
visibility data with the different CLEANing strategies. As mentioned earlier, 
this estimation is done using TGE. Here, we have used Gaussian window function to taper the sky response. The tapering is introduced through a 
parameter $f$, where $f$ is preferably $ \le 1$ so that modified window 
function inside the TGE cuts off the sky response well before the first null 
of the primary beam (see for details, Figure 1 of \citealt{samir16a}). The reduced field of view results in a larger cosmic variance for the angular modes which are within the tapered field of view. So, the tapering parameter $f$ will possibly be determined by optimizing between the reduced field of view and the cosmic variance. In this work we use $f=0.8$. It is expected that the estimated power spectrum from the residual data will be consistent with the input power spectrum if the point source subtraction is perfect and precise. Through angular power spectrum estimation 
from the different residual data sets, we try to find out the optimum approach 
for CLEANing to recover the underlying diffuse synchrotron emission power 
spectrum. Figure~\ref{fig:compnterm} shows the estimated $C_{\ell}$ from the 
residual visibility data for Run(a) and Run(b), that is for fixed CLEANing 
threshold of $1.0\, {\rm mJy}$ but $\rm nterms=1$ and $2$ respectively. We 
note that the residual sidelobes around the bright sources in the image with 
$\rm nterms=1$ (see Figure~\ref{fig:rescompare}a) introduced an excess power 
at large angular multipoles (small angular scales) $\ell \ge 6\times 10^3$ in 
the estimated angular power spectrum. On the other hand CLEANing with $\rm 
nterms=2$ reduces the residual sidelobes in the image after point source 
subtraction (see Figure~\ref{fig:rescompare}b). Hence, in this case the 
estimated $C_{\ell}$, as shown in Figure~\ref{fig:compnterm}, recover the 
input power spectrum better at large $\ell$ values as well. 

Figure~\ref{fig:compcutoff} shows the angular power spectra $C_{\ell}$ 
estimated from the residual visibility data obtained under the different 
CLEANing strategies Run(b), Run(c) and Run(d) with different threshold but 
fixed value of $\rm nterms=2$ (see Table \ref{tab:2}). For Run(d), which 
cleans upto $2.0\, {\rm mJy}$ ($\sim 6\,\sigma_{im}$), the angular power 
spectrum below $\ell \sim 7\times 10^3$ is properly recovered. However, due to 
insufficient CLEANing, it retains some extra residual power at large $\ell \ge 
7\times 10^3$. In contrast, as already noted earlier, Run(b) with CLEANing 
threshold of $1.0\, {\rm mJy}$ recovers the power spectrum for a larger range 
of $\ell$. The estimated angular power spectrum for Run(c), on the other hand, 
falls off by a factor $\sim 5$ compared to the input model power spectrum at 
all angular scales. This is due to the fact that Run(c) removes a part of 
diffuse structure from the map by CLEANing upto $1.5\sigma_{im}$.

\begin{figure}[h]
\begin{center}
\psfrag{cl}[b][t][1][0]{$C_{\ell} [mK^2]$}
\psfrag{U}[c][c][1][0]{$\ell$}
\psfrag{model}[r][r][0.8][0]{Model}
\psfrag{nterm1}[r][r][0.8][0]{{\rm Run(a)}}
\psfrag{nterm2}[r][r][0.8][0]{{\rm Run(b)}}
\includegraphics[width=80mm,angle=0]{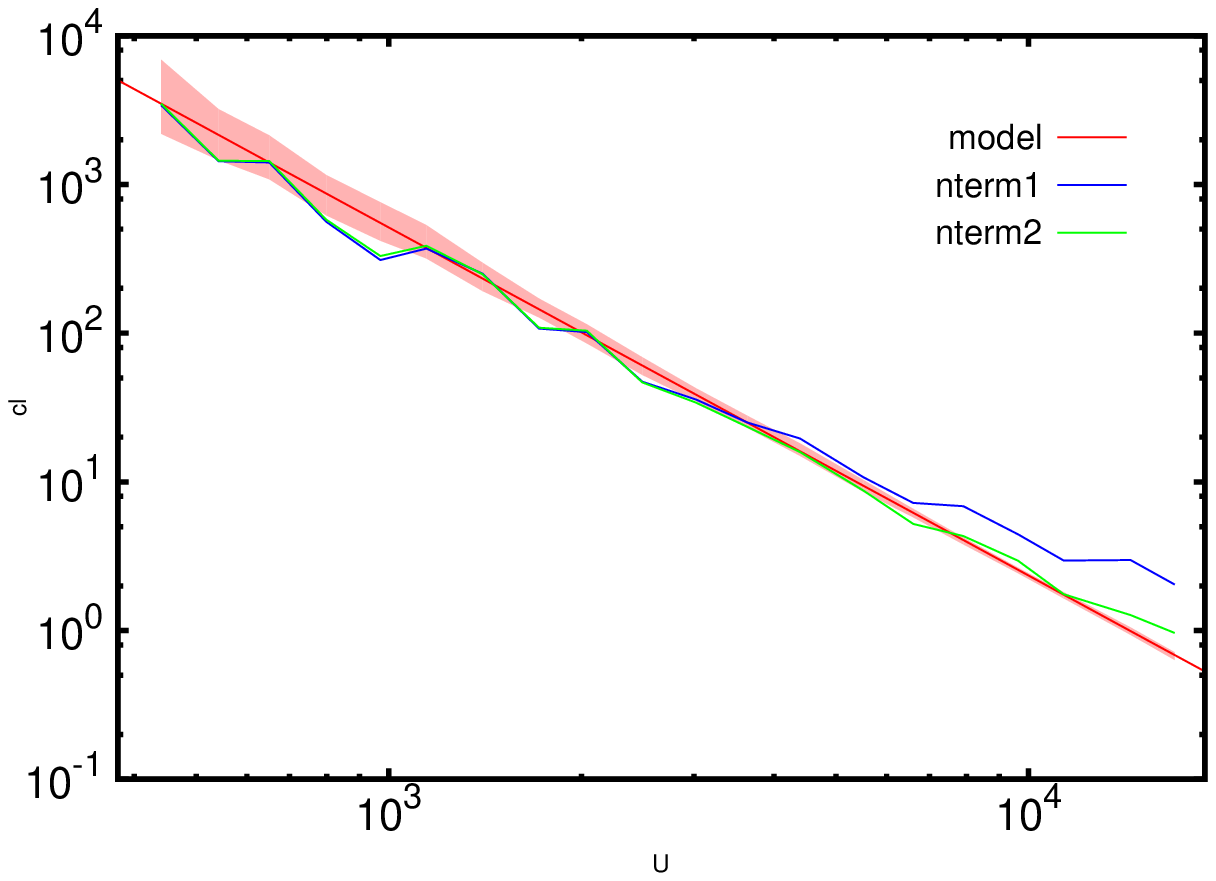}
\caption{The estimated power spectra from residual visibility data for Run(a) and Run(b) corresponding to threshold flux density of $1\,{\rm mJy}$ with $\rm nterms=1$ and $2$ respectively. The solid line shows the input model (eq. \ref{eq:cl150}) with 1-$\sigma$ error estimated from 100 realizations of the diffuse emission map.}
\label{fig:compnterm}
\end{center}
\end{figure}

\begin{figure}
\begin{center}
\psfrag{cl}[b][t][1][0]{$C_{\ell} [mK^2]$}
\psfrag{U}[c][c][1][0]{$\ell$}
\psfrag{model}[r][r][0.8][0]{Model}
\psfrag{2mjy}[r][r][0.8][0]{{\rm Run(d)}}
\psfrag{1mjy}[r][r][0.8][0]{{\rm Run(b)}}
\psfrag{0.5mjy}[r][r][0.8][0]{{\rm Run(c)}}
\includegraphics[width=80mm,angle=0]{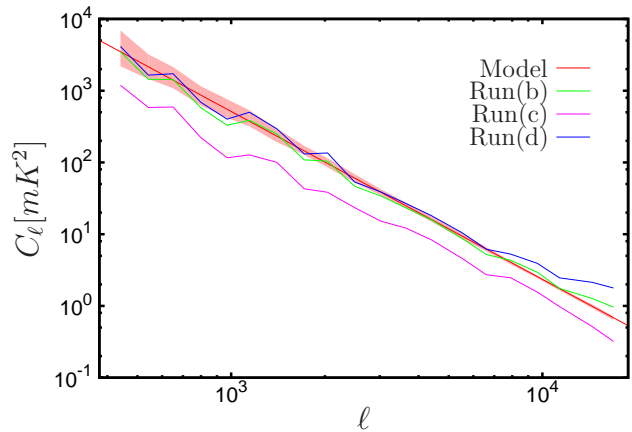}
\caption{The estimated power spectra for different CLEANing strategies, Run(b),(c) and (d) with different CLEANing threshold but fixed value of $\rm nterms=2$ (details in Table \ref{tab:2}).}
\label{fig:compcutoff}
\end{center}
\end{figure}

\begin{figure}
\begin{center}
\psfrag{cl}[b][t][1][0]{$C_{\ell} [mK^2]$}
\psfrag{U}[c][c][1][0]{$\ell$}
\psfrag{model}[r][r][0.8][0]{Model}
\psfrag{0.5mjy}[r][r][0.8][0]{{\rm Run(c)}}
\psfrag{o.5mjybox}[r][r][0.8][0]{{\rm Run(e)}}
\psfrag{0.5mjyboxhand}[r][r][0.8][0]{{\rm Run(f)}}
\includegraphics[width=80mm,angle=0]{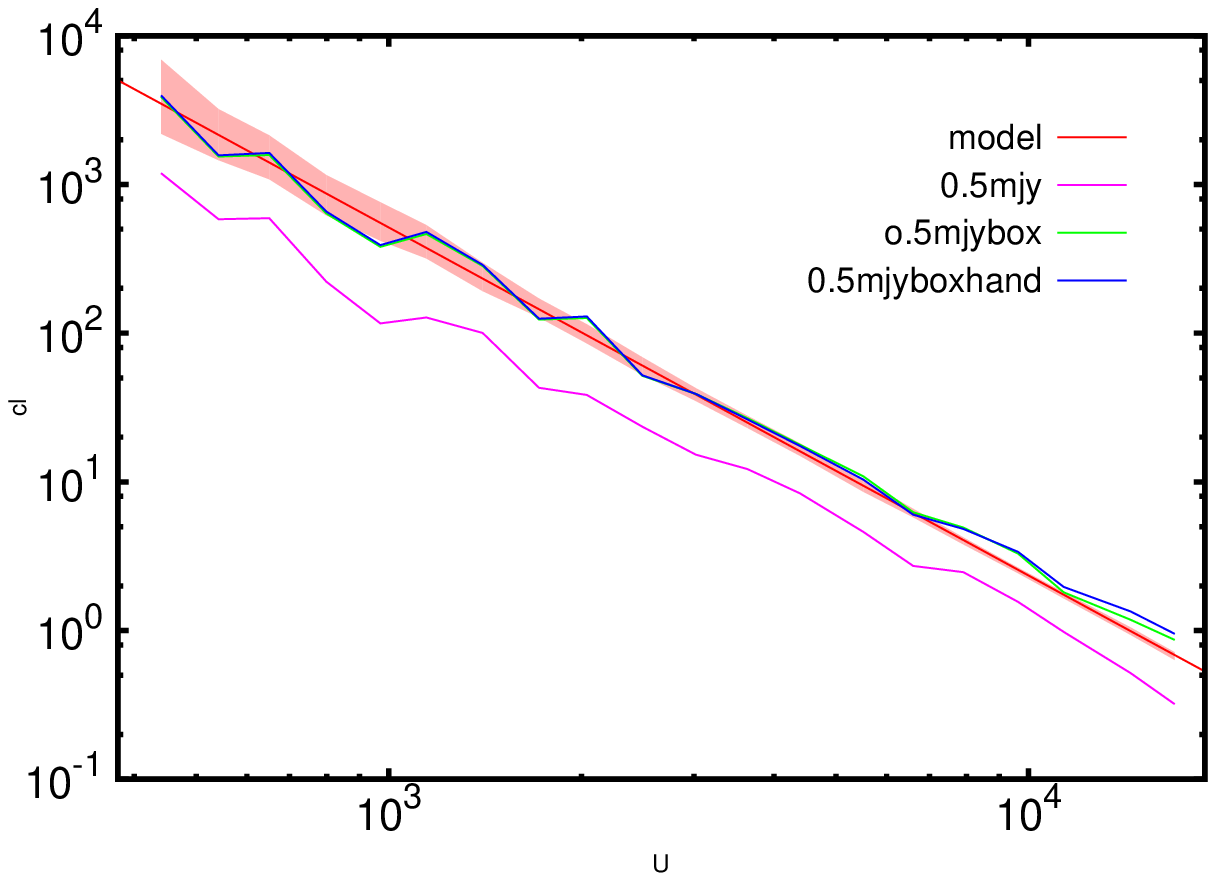}
\caption{The estimated power spectra for different CLEAN box options corresponding to Run(c),(e) and (f) described in Table \ref{tab:2}. For details see Section~\ref{sec:pssub} and Section~\ref{sec:result}.}
\label{fig:compbox}
\end{center}
\end{figure}

The effect of using different CLEANing box options (discussed in 
Section~\ref{sec:pssub}) in recovering $C_{\ell}$ is shown in 
Figure~\ref{fig:compbox}. Here we keep the other two parameters fixed at 
$\rm nterms=2$ and CLEANing threshold of $0.5\,{\rm mJy}$. It is clear from 
this figure that there is no significant change in the estimated power spectra 
for the two different CLEANing box strategies used in Run(e) and (f). In both of
these cases the estimated $C_{\ell}$ agree very well with the input power 
spectrum over the full range of $\ell$ probed here. For comparison, we also 
show the estimated power spectrum for Run(c) where the full image is CLEANed 
upto $0.5\, {\rm mJy}$ without selecting any CLEAN region around the point 
sources. As already shown, this partly removes the underlying diffuse emission 
from the image. Thus, the estimated $C_{\ell}$ in this case is a factor $\sim 
5$ lower compared to the input model power spectrum at all angular scales.

\section{Summary and conclusions}
\label{summa}

Precise subtraction of point sources from wide-field interferometric data is 
one of the primary challenges in studying the diffuse foreground emission as 
well as the weak redshifted HI 21-cm signal. In this paper, we consider the 
method of studying and characterizing the Galactic synchrotron emission using 
simulated $150\, {\rm MHz}$ GMRT observation in presence of point sources. The 
angular power spectrum $C_{\ell}$ of the diffuse emission is estimated from the 
residual visibility data using TGE after subtracting the point sources from 
only the inner part of the field of view of size $4.2^{\circ}\times 4.2^{\circ}$. We assess the impact of imperfect 
point source removal for different CLEANing strategies in recovering both the 
flux density distribution and the input $C_{\ell}$ of the diffuse Galactic 
synchrotron emission for the angular scale range probed by the GMRT.

The simulations are carried out for GMRT $150 \, {\rm MHz}$ observation for a 
sky model consisting of point sources and diffuse synchrotron emission. The sky 
model is multiplied with the model PB ${\cal A}(\th,\,\nu)$, before computing 
the visibilities for the frequency and the $uv$ coverage of the simulated GMRT 
observation. Finally, we add independent Gaussian random noise to both the real 
and imaginary parts of each visibility contribution. The standard analysis 
package CASA has been used to make images and to subtract point source model 
from the simulated visibility data. We use various CLEANing strategies as 
outlined in Section~\ref{sec:pssub} with different CLEANing boxes, threshold 
flux and spectral correction options. The residual data were then used for 
estimating $C_{\ell}$ of the diffuse component. We check the effect of point 
source subtraction by comparing image histograms, visibility distribution 
function as well as $C_{\ell}$ from the residual data. 

We find that all the different CLEANing strategies introduce some degree of 
non-Gaussianity in the residual data both in image and in visibility domain. 
The less precise point source subtraction generates more non-Gaussianity in the 
distribution of image-pixels beyond the CLEANing threshold. Equivalently, the 
visibility distributions also deviate significantly from a Gaussian. Comparing 
the recovered and the input power spectra, we find that both shallow CLEANing 
and incorrect spectral modelling of the point sources leave artifacts in the 
residual image near the position of bright point sources, and also results in 
excess power at the large angular multipoles ($\ell \sim 6\times 10^3$). On the 
other hand, very deep ``blind'' CLEANing removes part of the diffuse structure 
and reduces the amplitude of the power spectrum at all angular scale. The best 
possible situation is when, for a given region, source catalogue is available 
from other observations (even at a different frequency). Naturally, the optimum 
option there is to use the existing source catalogue to choose CLEANing regions 
for deep CLEANing (with threshold $\sim 1.5\sigma_{im}$) along with spectral 
correction for the point source model ($\rm nterms=2$ or higher). If a point 
source catalogue is not available, then one may use a moderate CLEANing 
threshold ($\sim 3\sigma_{im}$) for the whole image (which may still remove 
some of the diffuse signal). Alternatively, one may use a more conservative 
initial CLEANing threshold ($\sim 5 - 6\sigma_{im}$) for the whole image, and 
then choose CLEANing box around residual sources by visual inspection for a 
deeper ($\sim 1.5\sigma_{im}$) CLEANing. The latter strategy is useful only 
when one needs to remove the point sources from a relatively smaller region. 
Please note that, for the TGE, effect of the residual point sources from the 
outer region of the field is insignificant due to the tapering. Hence, we need 
to accurately subtract point sources only from the inner region, which makes 
it a viable option in the present case. We find that both this strategy and 
deep CLEANing based on source catalogue give a comparably good $C_{\ell}$ 
estimation for these simulated data.

The accurate removal of all the point sources from the wide-field image is 
complicated and difficult task in presence of instrumental systematics, 
calibration errors, RFI and ionospheric effects. In absence of the mentioned 
real-world obstacles, we subtract out all the point sources from the image with 
high level of accuracy, and the TGE successfully recovers the angular power 
spectrum $C_{\ell}$ of diffuse Galactic synchrotron emission from the residual 
visibility data at the angular scales probed by the GMRT. As a next step, we 
plan to incorporate some of the above mentioned ``real world" observational 
effects in our simulation, and investigate how precisely we can remove the 
point sources and estimate the angular power spectrum from the residual data. 
We leave this issue for future studies.

%\clearpage{\pagestyle{empty}\cleardoublepage} %%%%%%%%%%%%%%%%%%%%
%\newpage
\setcounter{section}{0}
\setcounter{subsection}{0}
\setcounter{subsubsection}{2}
\setcounter{equation}{0}
%\pagenumbering{arabic}

\def\u{\vec{U}}
\def\S{{\mathcal S}}
\def\V{\mathcal{V}}
\def\N{{\mathcal N}}
\def\A{{\bf A}\,}
\def\Sc{S_2}

%-------------------------------------------
\chapter[Tapering the sky response for $C_{\ell}$ estimation]{{\bf Tapering the  sky response for angular power spectrum estimation 
from low-frequency radio-interferometric data}\footnote{This chapter is adapted
   from the paper ``Tapering the  sky response for angular power spectrum estimation 
from low-frequency radio-interferometric data''
   by \citet{samir16a}}}
\label{chap:chap4}

\section{Introduction}
Foreground removal for detecting the Epoch of
Reionization (EoR) 21-cm signal is a topic of intense current research 
\citep{jelic08,bowman09,paciga11,chapman12,liu12,mao3,paciga13}.
Foreground avoidance
\citep{adatta10,parsons12,trott12,vedantham12,pober13,thyag13,parsons14,dillon14,pober14,liu14a,liu14b,ali15} 
is an alternate strategy based on the proposal  that the 
foreground contamination is restricted to a wedge  in
$(k_{\perp},k_{\parallel})$ space, and the signal can be estimated from the
uncontaminated modes outside the wedge. 
Point sources dominate the $150 \, {\rm MHz}$ sky  at the
angular scales $\le  4^{\circ}$ \citep{ali08} which are relevant 
for  telescopes like the Giant Metrewave Radio Telescope
\citep[GMRT;][]{swarup}, Low-Frequency Array
\citep[LOFAR;][]{haarlem} and the  upcoming Square Kilometre Array (SKA). 
It is difficult to model and subtract the point sources at 
the periphery of the telescope's field of view. The difficulties include the
fact that the antenna response  is highly frequency dependent near the nulls
of the primary beam,  and  the calibration differs from that of the
phase center due to ionospheric fluctuations.
Point source subtraction is also
important for measuring the angular power spectrum of the diffuse Galactic
synchrotron radiation \citep{bernardi09,ghosh12,iacobelli13}   which, apart 
from being an important foreground component for the EoR 21-cm signal,
is interesting in its own right. 

Most of the foreground subtraction techniques use the property of  smoothness
 along frequency for the various foreground components. \citet{ghosh1,ghosh2}
 found  that  residual point sources 
 located away from the phase center introduce oscillations  along frequency direction. 
The oscillation  are more rapid if the distance of the source from the phase 
center increases, and also with increasing baseline. Equivalently, the dominant
contribution to the width of the foreground wedge arises from the sources
located at the periphery of the field of view \citep{thyag13}.  
Using GMRT \citet{ghosh2,ghosh12} have shown that these oscillations can be reduced by tapering the sky response. In a recent paper \citet{pober16} showed that correctly modelling and subtracting the sidelobe foreground contamination is important for detecting the redshifted 21-cm signal.

In a recent paper \citet{samir14} have introduced the Tapered Gridded Estimator 
(TGE) for estimating the angular power spectrum $C_{\ell}$ directly from
radio-interferometric visibility data.  In this paper we use simulated
$150 \, {\rm MHz}$ GMRT data which incorporates point sources and the diffuse
Galactic synchrotron radiation to demonstrate that it is possible to suppress
the contribution from residual point sources in the sidelobes and the outer parts of the primary 
beam in estimating $C_{\ell}$ using the TGE. 

Noise bias  is an important issue for any estimator. For example, the image based estimator
\citep{seljak97} for $C_{\ell}$ and the visibility based estimator \citep{liu12}
for $P(k_{\perp},k_{\parallel})$ rely on modelling the noise properties of the
data and subtracting out the expected noise bias. However, the actual noise in
the observations could have baseline, frequency and time dependent  variations
 which  are  very difficult to model and there is the  risk of residual noise
 bias being mistaken as the signal. \citet{paciga11} have
avoided the noise bias by cross-correlating observations made on different
days. Another visibility based estimator \citep{begum06,dutta08}
individually correlates pairs of visibilities avoiding
the self correlation that is responsible for the noise bias. This, however, is
computationally  very expensive when the data volume is large. In this paper, we have
demonstrated that TGE, by
construction, estimates the actual noise bias internally from the data and
exactly subtracts this out to give an unbiased estimate of $C_{\ell}$. The entire discussion here is in the context of estimating  $C_{\ell}$ for the
diffuse Galactic synchrotron radiation. As mentioned earlier,  the same issues
are also relevant for measuring the EoR 21-cm power spectrum not considered here. 

In Section~\ref{sec:imgproblem} we discuss the conventional problem in standard imaging techniques. Simulation and data analysis processes are briefly discussed in Section~\ref{simu}.  Section~\ref{sec:est} discusses the estimator (TGE) that we used to suppress the outer region of the primary beam and the results are presented in Section~\ref{result}. Finally, we present summary and conclusion in Section~\ref{sum}.

\section{Problems in conventional Imaging}
\label{sec:imgproblem}
The contribution to the signal in radio frequency observations from the outer region of the primary beam and from the sidelobes is generally very small as compared to the inner region of the primary beam. In particular, the expected 21-cm signal, which itself is very faint, contributes mainly from the central part of the primary beam, and attenuated to a great extent in the outer region. Only the bright point sources from the outer region, if not accurately removed, may have significant impact on the statistical estimation of the diffuse signal. Thus, it is necessary to remove the effect of point sources from the outer region before estimating the residual power spectrum. However,  we will not be benefitted in terms of signal by including highly attenuated diffuse emission from the outer region.

Imaging a large enough region to model and subtract all the point sources before dealing with the diffuse emission may seems to be a direct solution of the above problem. But, in reality there are many issues which make this approach impractical. First of all, the field of view at low radio frequencies is large, and making larger images is computationally more expensive. In addition to that, non-coplaner nature of the baselines prevents us from making wide-field image without considering the effect of the ``w-term''. There are algorithms e.g. faceting \citep{cornwell92}, w-projection \citep{cornwell08}, WB-A projection \citep{bhat13} etc. to tackle this problem partly for radio interferometric observations. However, these algorithms still require significant computation to make an image of such a large region of the sky. Secondly, the number of bright point sources is quite large at low frequency. While imaging a very large region, selecting CLEANing region around each source is a tedious job. On the other hand, CLEANing without selecting regions removes a non-negligible part of the diffuse signal of our interest \citep[see][for details]{samir16c}. 

The next challenge is to accurately characterize the time and frequency dependence of the wide-field primary beam for effective point source subtraction from the periphery of the telescope's field of view \citep[e.g.][]{neben}. Both the frequency dependence and the deviation from circular symmetry are more prominent at the outer part of the primary beam. These, along with the rotation of primary beam on the sky, cause a strong time and frequency variation of the primary beam for point sources in the outer region. They create problem in accurately model the point sources that we want to subtract from the data. In fact, some of the variations are intractable in nature and  it is extremely difficult, if not impossible, to make accurate modelling and subtraction of the point sources from the outer part of  the primary beam.

Though we have not considered instrumental gains and ionospheric effects in this study, in real life any directional dependence of these quantities will also severely limit our ability to subtract point sources accurately from a large region. One can overcome this difficulty to some extent by going into complicated and messy procedure of direction dependent calibration (e.g. peeling) \citep{bhat08,intema09,kazemi11}. Again, (a) it is computationally more expensive, (b) part of the variation may be intractable, and (c) there is hardly any gain in terms of recovering the diffuse signal which is too weak in outer region. 

The future generation low frequency telescopes (e.g. SKA) that will presumably be used to carry out redshifted diffuse H~{\sc i} observation, will have larger field of view, large bandwidth, longer baseline and higher sensitivity. Hence the above issues will be even more relevant. Moreover, the expected huge data volume from observations with those telescopes will make it more challenging to address these problems by imaging a larger region for subtracting the point sources.
The following two sections outline a technique to overcome these problems by subtracting point sources only from the central region and using the TGE to recover the power spectrum of the diffuse emission in a more efficient way.

\section{Simulation and Data Analysis}
\label{simu}
The details of the simulation and data analysis,  including point source
subtraction, are presented in a companion paper \citep{samir16c} and we only
present a brief discussion here. Our  model of the $150 \, {\rm MHz}$ sky has
two components, the first being the diffuse Galactic synchrotron radiation
which is the signal that we want to detect. We use the measured 
angular power spectrum  \citep{ghosh12} 
\begin{equation}
C^M_{\ell}(\nu)=A_{\rm 150}\times\left(\frac{1000}{\ell} \right)^{\beta}\times\left(\frac{\nu}{150{\rm MHz}}\right)^{-2\alpha}   \,.
\label{eq:cl150}
\end{equation}
as the input model to generate  the brightness temperature fluctuations  on
the sky. Here  $\nu$ is the frequency in ${\rm MHz}$, $A_{\rm 150}=513 \, {\rm
  mK}^2$, $\beta=2.34 $  \citep{ghosh12} and $\alpha=2.8$ \citep{platinia98}. 
  The simulation covers a $\sim 8^{\circ} \times \sim 8^{\circ}$
region of the sky and a $16 \, {\rm MHz}$ bandwidth, centered at $150 \, {\rm MHz}$,  over $128$ spectral channels. The diffuse signal was simulated on a grid of
resolution $\sim 0.5^{'}$. 

The Poisson fluctuation of the extragalactic point sources dominates the $150
\, {\rm MHz}$ sky at the angular scales of our interest \citep{ali08}, and it is necessary
to subtract these or suppress their contribution in order to detect any 
  diffuse component like the Galactic synchrotron radiation which we consider here or
  the redshifted 21-cm cosmological signal which is much fainter and is not
  considered here. 
We use the $150\, {\rm MHz}$ differential source count measured  using GMRT 
\citep{ghosh12}
\begin{equation}
\frac{dN}{dS} = \frac{10^{3.75}}{Jy \cdot Sr}\cdot\,\left(\frac{S}{1
  Jy}\right)^{-1.6} \,.
\label{eq:b1}
\end{equation}
 to generate point sources in the flux range  $9 {\rm mJy}$ to 
$1 {\rm Jy}$ whose angular positions are randomly distributed 
within the $3.1^{\circ} \times 3.1^{\circ}$ Full Width Half Maxima (hereafter FWHM)  of the primary beam. 
The antenna response falls off beyond the FWHM, and we only include the
bright sources ($S \ge   100 {\rm mJy}$) outside the FWHM. We have $353$
and $343$ sources in the inner and outer regions respectively, and the sources
were assigned a randomly chosen spectral index $\alpha$ ($S_{\nu} \propto
\nu^{-\alpha}$) in the range $0.7$ to $0.8$.

We consider the mock GMRT observations 
targeted on a arbitrarily selected  field located at RA=$10{\rm h} \, 46{\rm
  m} \, 00{\rm s}$ and DEC=$59^{\circ} \, 00^{'} \, 59^{''}$. The GMRT has $30$
 antennas which for 
a total  $8 \, {\rm hr}$ of observation with  $16 {\rm s}$ integration time 
results in $783,000$ baselines $\u_i$ with $128$ visibilities $\V(\u_i,\nu)$
 (one per frequency channel) for  each baseline. The resolution of GMRT at $150 \, {\rm MHz}$ is $20^{''}$. 
The diffuse signal (eq. \ref {eq:cl150}) falls off with  increasing $U=\mid \u \mid$ 
($\ell = 2 \pi U$), and we  include this contribution for only the small baselines $U \le 3,000$
for which the visibility contribution is calculated using a 2 dimensional Fourier transform. 
We note that the $w$ term does not significantly affect the diffuse signal \citep{samir14}, 
however this is very important for correctly imaging and subtracting the point sources. 
We have included the point source contribution  for all the baselines in the simulation, and 
the visibilities are  calculated by individually summing over each point source  and including 
the $w$ term. We have modelled the GMRT primary beam pattern ${\cal A}(\th,\nu)$ 
 with the  square of a Bessel function (Figure \ref{fig:taper}) corresponding to the telescope's $45 \, {\rm m}$
diameter circular aperture. The simulated sky is multiplied with ${\cal A}(\th,\nu)$  
before calculating  the visibilities. Finally, we add the system noise contribution which is modelled 
a Gaussian random variable 
with standard deviation  $\sigma_n=1.03{\rm Jy}$ for  the real and imaginary parts of each visibility. 
We note that the GMRT has two polarizations which have identical sky signals but independent noise. 
 
\begin{figure}
\begin{center}
\psfrag{theta}[c][c][0.8][0]{$\theta$ $[{\rm Degrees}$]}
\psfrag{Pbeam}[c][c][0.8][0]{${\mathcal A}(\th)$}
\psfrag{Bessel}[r][r][0.8][0]{Primary Beam}
\psfrag{taper1}[r][r][0.7][0]{${\cal A_W}(\th)$,f=2.0}
\psfrag{taper2}[r][r][0.7][0]{${\cal A_W}(\th)$,f=0.8}
\psfrag{taper3}[r][r][0.7][0]{${\cal A_W}(\th)$,f=0.6}
\includegraphics[width=85mm,angle=0]{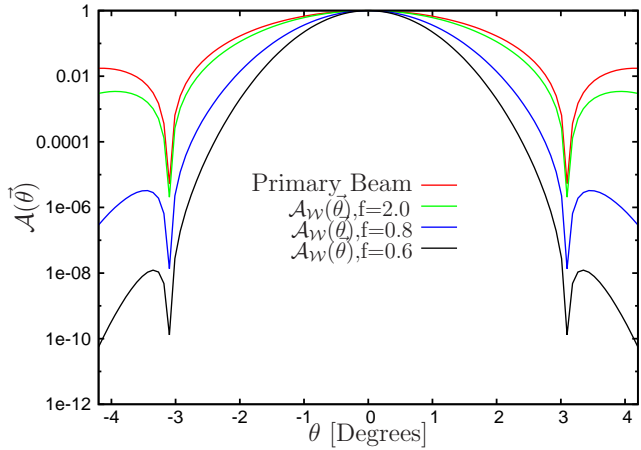}
\caption{The GMRT $150 \, {\rm MHz}$ primary beam  ${\cal A}(\th)$  
which has been modelled as the  square of a Bessel function.  The effective primary beam 
${\cal A_W}(\th)$, obtained after tapering the sky response for the 
different values of $f$ is also shown in the figure.} 
\label{fig:taper}
\end{center}
\end{figure}

We have used the  Common Astronomy Software Applications (CASA)
package to image and analyze our simulated data. The standard tasks  CLEAN and UVSUB were used to model 
and subtract out the point sources from a $4.2^{\circ}\times4.2^{\circ}$ region which covers an extent 
that is   approximately $1.5$  times the FWHM of the primary beam. We have tried different CLEAN strategies
for which the details are presented in our companion paper \citep{samir16c}, and for this work we adopt 
the most optimum parameter values which correspond to Run(e) of the companion paper.  
Figure~\ref{fig:img} shows the  ``dirty'' image of the entire simulation  region 
made from the residual visibility data after point source subtraction. The
central square box ($4.2^{\circ}\times4.2^{\circ}$) shows the region from which  the point sources have been  
subtracted. The features visible in this region correspond to the Galactic synchrotron radiation. 
It is difficult to model and subtract point sources from the periphery where the antenna response is 
highly frequency dependent. It also needs creating and cleaning a huge image that is computationally more expensive. Further, in real observations, any direction dependent gain away from the phase center will make it even more difficult. We have not attempted to subtract the point sources
from the  region outside the central box and  the residual point sources are
visible in this region of the  image.

Figure~\ref{fig:comtaper}  shows the angular power spectrum $C_{\ell}$ before and after point source subtraction;
the input  model for the diffuse radiation is also shown for comparison. 
Before subtraction, the point sources dominate $C_{\ell}$ at all angular multipoles $\ell$.  
After subtraction, we are able to recover the diffuse component at low angular multipoles
 $\ell\le3\times10^{3}$. However, the residual point sources still dominate at the  large 
$\ell$ values. The goal is to suppress the contribution from the residual point sources 
located at the periphery of the beam so that we can recover the input model over the entire $\ell$ 
range.  We show that it is possible to achieve this with the Tapered Gridded
Estimator discussed in the next section.

\begin{figure}
\begin{center}
\psfrag{a}[c][c][1.5][0]{{$\hspace{5mm}$ }}
\psfrag{b}[c][c][1.5][0]{{$\hspace{5mm}$ }}
\psfrag{x}[c][c][0.6][0]{$\theta_x$ $[{\rm Degrees}$]}
\psfrag{y}[c][c][0.6][0]{$\theta_y$ $[{\rm Degrees}$]}
\includegraphics[width=75mm]{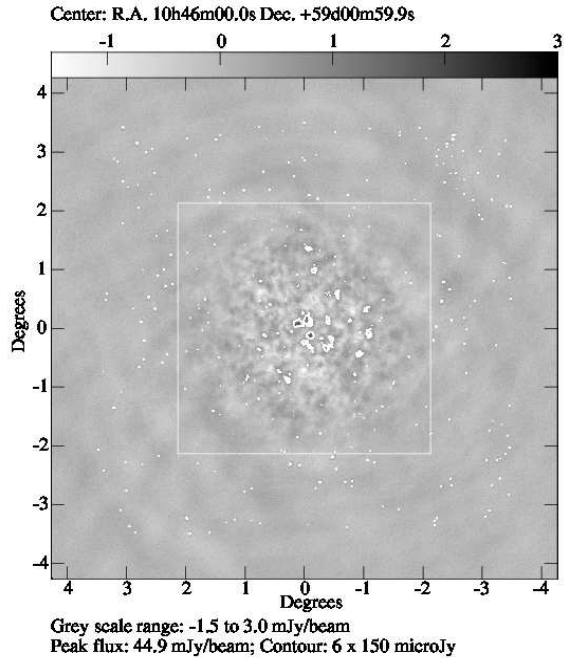}
\caption{``Dirty'' image of the entire simulation region made with
  the residual visibility data after point source subtraction. Point sources
  were subtracted from a central region (shown with a box, $4.2^{\circ}\times4.2^{\circ}$) whose extent is
  $\sim 1.3$  times the FWHM of the primary beam. The features visible inside
  the box all correspond to the diffuse radiation.  Residual point sources are
  visible outside the box, however the diffuse radiation is not visible  in
  this region.}
\label{fig:img}
\end{center}
\end{figure}

\section{The Tapered Gridded Estimator}
\label{sec:est}
The observed  visibilities are  a sum of  two independent parts  namely the  sky signal and the system noise 
\begin{equation}
\V(\u, \nu)=\S(\u, \nu)+\N(\u,\nu) \,.
\label{eq:c1}
\end{equation}
The signal $\S(\u, \nu)$ and the noise $\N(\u,\nu)$ are considered to be
  independent random variables, further the noise in the different
  visibilities are uncorrelated.
 The signal  contribution $\S(\u,\nu)$ records the Fourier transform of the product of 
$\delta I(\th,\,\nu)$, the fluctuation in specific  intensity of the sky signal, and 
the telescope's primary beam pattern ${\mathcal A}(\theta, \nu)$ shown in Figure \ref{fig:taper}. 
As mentioned earlier, it is difficult to model  and subtract point sources from the outer
region of the primary beam and the sidelobes. The residual point sources in the periphery 
of the telescope's field of view pose a problem for estimating the power spectrum of 
the diffuse radiation.  
In this section we discuss the Tapered  Gridded Estimator (TGE) 
which is  a technique for estimating the angular power spectrum from the visibility data.  
This technique suppresses the
contribution from  the sidelobes and the outer part of the primary beam by tapering the sky response. 
\citet{samir14} presents a detailed discussion of this estimator, and we only present a brief outline
 here. 

We  taper the sky response by multiplying the  field of view with a frequency independent 
Gaussian window function 
${\cal W}(\theta)=e^{-\theta^2/\theta^2_w}$. Here we parameterize $\theta_w=f \theta_0$ where 
$\theta_0=0.6 \times \theta_{FWHM}$ and $\theta_{FWHM}$ is the FWHM of the 
telescope's primary beam at the central frequency, and preferably $f \le 1$ so that ${\cal W}(\theta)$
  cuts off the sky response well before the first null of the  primary beam. 
We implement the tapering by convolving the measured visibilities with $\tilde{w}(\u)$ 
the Fourier transform of ${\cal W}(\theta)$. The convolved visibilities are evaluated 
on a grid in $uv$ space using 
\begin{equation}
\V_{cg} = \sum_{i}\tilde{w}(\u_g-\u_i) \, \V_i
\label{eq:c5}
\end{equation}
where $\u_g$ refers to the different grid points and $\V_i$ 
refers to the measured visibilities at baseline $\u_i$. The gridding significantly reduces
the data volume and the computation time required to estimate the power spectrum \citep{samir14}. 
The convolved visibilities are calculated separately  for each frequency channel. 
Then, for the purpose of this work, convolved visibilities for a grid are averaged over all frequencies. 

The  signal component of the convolved visibility is the Fourier transform of the product of a  
 modified   primary beam pattern ${\mathcal A_W}(\th, \nu)={\cal W}(\theta)\,  {\cal A}(\th, \nu)$
and $\delta I(\th, \nu)$
\begin{equation}
\S_c(\u,\nu)=  \int \, d^2 \th  \,
     {\mathcal A_W}(\theta, \nu)\delta I(\th,\,\nu)e^{2\pi i \u.\th} \,.
\label{eq:c3}
\end{equation}
It is clear that the convolved visibilities respond to the signal from a smaller region of the sky 
as compared to the measured visibilities. It may be noted that 
the tapering is effective only if the window function $\tilde{w}(\u_g-\u_i)$
in eq. (\ref{eq:c5})  is well sampled by  the   baseline distribution. 
The results of this paper, presented later, indeed justify this assumption for
the GMRT.

The correlation of the gridded visibilities $\langle \V_{c g}  \V^{*}_{c g} \rangle$ 
gives a direct  estimate of the angular power spectrum $C_{\ell_g}$ through 
\begin{equation}
\langle \V_{c g}  \V^{*}_{c g} \rangle
 = \mid K_{1g} \mid^2V_1 C_{\ell_g} + \sum_i  \mid \tilde{w}(\u_g-\u_i)
 \mid^2 \langle  \mid \N_i \mid^2 \rangle  
\label{eq:c6}
\end{equation}
where the angular multipole $\ell_g$ is related to the baseline $U_g$ as 
$\ell_g=2 \pi U_g$, $K_{1g}=\sum_i  \tilde{w}(\u_g-\u_i)$, 
$V_1= \left( \frac{\partial B}{\partial T}\right)^{2}\left[\int d^2 U{'} \,  
\mid \tilde{a}_W(\u-\u{'}) \mid^2 \right]$,
$\tilde{a}_W$ is the Fourier transform of ${\mathcal A_W}$
and $\left( \frac{\partial B}{\partial T}\right)$ is the conversion factor from 
brightness temperature to specific intensity. We see that the visibility correlation 
also has a term  involving $ \langle  \mid \N_i \mid^2 \rangle$
which is the variance of the noise contribution  present in the 
measured visibilities (eq. \ref{eq:c1}). 
This term, which   is independent of $C_{\ell}$, introduces a positive
definite noise bias.    The visibility correlation (eq. \ref{eq:c6}) provides
an estimate of $C_{\ell}$ except for the  additive noise bias. 
The TGE uses the same visibility data to obtain an  internal estimate of the
noise bias and subtract it from the visibility correlation. 
 We  consider the self-correlation term $B_{cg}=\sum_i \mid \tilde{w}(\u_g-\u_i) \mid^2 \,
\mid  \V_i \mid^2 $ for which 
\begin{equation}
 \langle B_{cg} \rangle
=   \sum_i \mid  \tilde{w}(\u_g-\u_i)\mid^2  (V_0  C_{\ell_i} 
+ \langle  \mid \N_i \mid^2 \rangle ) \,.
\end{equation}
where $V_0= \left( \frac{\partial B}{\partial T}\right)^{2}\left[\int d^2 U{'} \,  
\mid \tilde{a}(\u-\u{'}) \mid^2 \right]$,
$\tilde{a}$ is the Fourier transform of the primary beam pattern ${\mathcal
  A}$.  
 The term  $ \langle B_{cg} \rangle$, by construction, has  exactly the same
 noise bias as the visibility 
 correlation in eq. (\ref{eq:c6}). We use this to define the 
TGE estimator  
\begin{equation}
{\hat E}_g= (\mid K_{1g} \mid^2 V_1)^{-1} [ \V_{c g}  \V^{*}_{c g} - B_{cg}]
\end{equation}
which gives an unbiased estimate of the angular power spectrum at a grid point $g$.
A part of the signal also gets subtracted out with the noise bias. This loss is proportional to 
$N$ (the number of visibility data) whereas the visibility correlation is
proportional to $N^2$, 
and this loss is insignificant when the data size is large \citep{samir14}.   
The $C_{\ell_g}$ values estimated at each grid point  are  binned  in
logarithmic intervals of $\ell$,  
and we consider the bin-averaged $C_{\ell}$ as a function of the bin-averaged
angular multipole $\ell$.  

Tapering reduces the sky coverage which, in addition to suppressing the
point sources in  the periphery of the main lobe and the sidelobes, also
affects the diffuse signal. The reduced sky coverage causes the cosmic
variance of the estimated  $C_{\ell}$ to increase as $f$ is reduced (Figure
10, \citealt{samir14}). Further, the reduced sky coverage also restricts the 
 $\ell$ range ($\ell_{min} - \ell_{max}$) where it is possible to estimate
$C_{\ell}$,  and the value  of $\ell_{min}$ increases as $f$ is decreased.
\section{Results}
\label{result}
We have applied the Tapered Gridded Estimator (TGE) to the residual visibility data  after subtracting out the point sources. 
As mentioned earlier, the point sources have been identified and subtracted from  a $4.2^{\circ}\times4.2^{\circ}$ region  
(Figure~\ref{fig:img}) which covers an extent that is $\approx 1.3$ times
the FWHM   of the primary beam. However, the  point sources still remain 
at the periphery of the primary beam  and in the part of the sidelobe which has been included in the simulation.   The TGE tapers
the sky response which results in  an  effective primary beam ${\cal A_W}(\th)$  that  is  considerably narrower than the 
actual primary beam of the telescope ${\cal A}(\th)$.  Figure \ref{fig:taper} shows ${\cal A_W}(\th)$ for three different values of $f$ 
($2.0,0.8$ and $0.6$). For $f=2.0$  we see that ${\cal A_W}(\th)$ is not very significantly different from ${\cal A}(\th)$ in the region 
within  the first  null,  the difference however increases in the first sidelobe  and the sidelobe response is suppressed by a factor
 of $10$  at $\mid \th \mid \approx 4^{\circ}$. We see that the effective primary beam  gets narrower as the value of $f$ is reduced. 
The value of ${\cal A_W}(\th)$ is a factor of $\approx 10$ ($100$) lower compared to ${\cal A}(\th)$ for $f=0.8$ ($0.6$)
at  $\mid \th \mid = 2^{\circ}$ which corresponds to  the boundary of the region  within which the point sources have been subtracted.
We see that, for $f=0.8$ ($0.6$), tapering suppresses the first side lobe of ${\cal A_W}(\th)$ by a factor of $\approx 10^{5}$ ($10^{8}$) compared to
${\cal A}(\th)$  at  $\mid \th \mid = 4^{\circ}$. We expect the residual point source contribution to reduce by at least a factor of $10$ and $100$ for $f=0.8$ and $0.6$ respectively. 

\begin{figure}
\begin{center}
\psfrag{cl}[b][t][1][0]{$C_{\ell} [mK^2]$}
\psfrag{U}[c][c][1][0]{$\ell$}
\psfrag{model}[r][r][0.8][0]{Model}
\psfrag{allnotaper}[r][r][0.8][0]{Total}
\psfrag{notaper}[r][r][0.8][0]{Residual}
\psfrag{tap2}[r][r][0.8][0]{f=2.}
\psfrag{tap0.8}[r][r][0.8][0]{f=0.8}
\psfrag{tap0.6}[r][r][0.8][0]{f=0.6}
\includegraphics[width=85mm,angle=0]{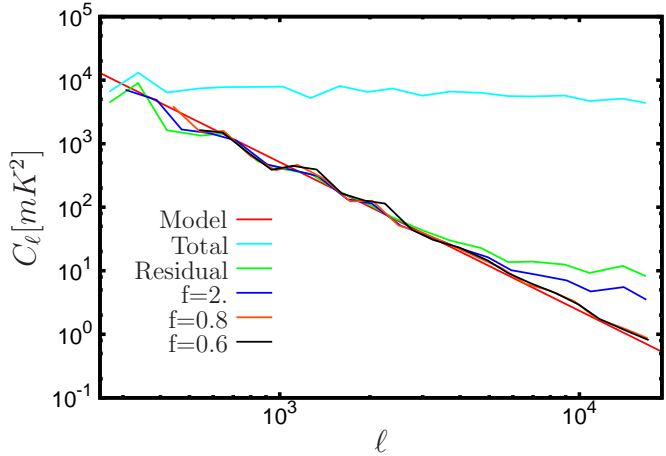}
\caption{Angular power spectrum $C_{\ell}$ of total and residual data. It also shows the estimated $C_{\ell}$ using  the TGE for the different values of $f$ are also shown in the figure. In this figure the curves for $f=0.6$ and $0.8$ overlaps with each other.}
\label{fig:comtaper}
\end{center}
\end{figure}

Figure \ref{fig:comtaper} shows the angular power spectrum ($C_{\ell}$) estimated from the residual visibility 
data using TGE with the $f$ values ($2.0,0.8$ and $0.6$) discussed earlier.  We see that in the absence of tapering 
we are able to recover the  angular power spectrum of the diffuse synchrotron radiation at  the low 
angular multipoles (large angular scales) $\ell < 3\times10^3$. The residual point source 
contribution is nearly independent of $\ell$ and has a value $C_{\ell} \approx 10$ ${\rm mK}^2$ which 
dominates the estimated $C_{\ell}$ at the large angular multipoles (small angular scales)  
$\ell \ge 10^4$. We have a gradual transition from the diffuse synchrotron dominated to a point 
source dominated $C_{\ell}$ in the interval $ 3\times10^3 \le \ell < 10^4$. The point source 
contribution comes down by a factor of  more than  $2$  if we use the TGE with $f=2.0$. 
We are now able to recover the  angular power spectrum of the diffuse synchrotron 
radiation to larger $\ell$ values ($\ell < 5\times10^3$) as compared to the situation 
without tapering.  The point source contribution, however, still dominates at larger $\ell$ 
values. We find that the point source contribution to $C_{\ell}$ is suppressed by more than a 
factor of $10$ if we use TGE with $f=0.8$ or $0.6$.  We are able to recover the 
angular power spectrum of the diffuse synchrotron radiation  over the entire $\ell$ range 
using either value of $f$. The fact that there is no noticeable change in  $C_{\ell}$
if the value of  $f$ is reduced from $0.8$ to $0.6$ indicates that a tapered
sky response with $f=0.8$ is adequate 
to detect the angular power spectrum of 
the diffuse synchrotron radiation  over the entire $\ell$ range of our interest here. 

The noise bias is an important issue in estimating the angular power spectrum, we illustrate 
this in Figure \ref{fig:comnse}. For this purpose we have used a smaller frequency bandwidth 
of $8 \ {\rm MHz}$ which increases the noise r.m.s. compared to the 
$16 \ {\rm MHz}$ bandwidth used throughout the rest of the paper. Figure \ref{fig:comnse} shows 
$C_{\ell}$ estimated with the TGE with $f=0.8$. We expect to recover the 
angular power spectrum of the diffuse synchrotron radiation  over the entire $\ell$ range  provided the
 noise bias is correctly estimated and subtracted out. 
 Figure \ref{fig:comnse} shows the estimated $C_{\ell}$  in the situation where the noise bias 
is not subtracted. We see that the noise bias makes a nearly constant contribution of 
$C_{\ell} \approx 7.5$ ${\rm mK}^2$ which dominates the estimated $C_{\ell}$ at large $\ell$. 
It is necessary to subtract the noise bias in order  to recover the $C_{\ell}$
of the diffuse  radiation at large $\ell$.   Figure \ref{fig:comnse}  demonstrates 
that the  TGE correctly subtracts out  the noise bias so that we are  able to recover the 
$C_{\ell}$  of the diffuse  radiation  over the entire $\ell$ range.

\begin{figure}
\begin{center}
\psfrag{cl}[b][t][1.5][0]{$C_{\ell} [mK^2]$}
\psfrag{U}[c][c][1.5][0]{$\ell$}
\psfrag{model}[r][r][1][0]{Model}
\psfrag{0.8nonse}[r][r][1][0]{No Noise Bias}
\psfrag{0.8nse}[r][r][1][0]{Noise Bias}
\includegraphics[width=80mm,angle=0]{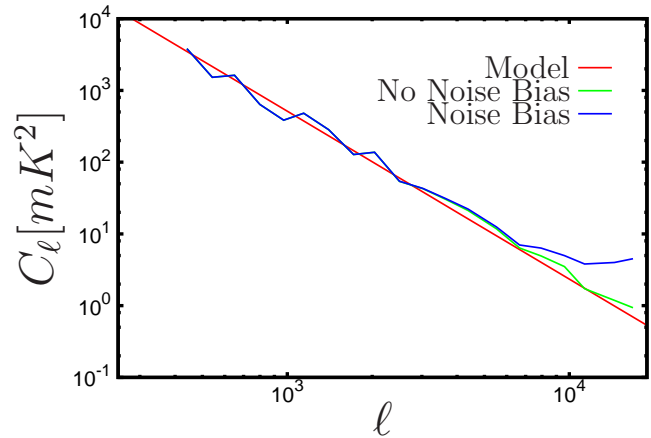}
\caption{Angular power spectrum $C_{\ell}$ estimated using  the TGE with $f=0.8$.
Results with the noise bias being present and with the noise bias subtracted are both shown here. }
\label{fig:comnse}
\end{center}
\end{figure}

\section{Summary and Conclusion}
\label{sum}
It is difficult to model and subtract point sources located at the periphery of the 
telescope's field of view. These residual point sources pose a problem for estimating 
the power spectrum of the diffuse background radiation if all visible point sources are removed with high level of accuracy from inside the main lobe of the primary beam. For example, \citet{pober16} have recently shown the effect of the residual point sources outside the main lobe on estimating the power spectrum for MWA observation. This issue is discussed here in 
the context of measuring the angular power spectrum of the diffuse Galactic synchrotron radiation 
using simulated $150 \, {\rm MHz}$ GMRT observations. However, the same  issue is also very 
important for  detecting the EoR 21-cm power spectrum which is a much 
fainter  diffuse signal that is not considered here. 

It is possible to suppress the contribution from the residual
point sources located at the periphery of the telescope's field of view through  a 
frequency independent window function  which  restricts or tapers the sky response. 
The Tapered Gridded Estimator(TGE)  achieves this tapering by convolving the measured visibilities 
with the Fourier transform of the window function. This estimator for the angular 
power spectrum has the added advantage that it internally estimates the noise bias
from the measured visibilities and accurately subtracts this out to provide an unbiased
estimate of  $C_{\ell}$.  In this paper we demonstrate, using simulated data, that the 
 TGE very effectively suppresses the contribution of the residual
point sources located at the periphery of the telescope's field of view. We also
 demonstrates that the TGE correctly estimates the noise bias from the input 
visibilities  and subtracts this out to give an unbiased estimate of $C_{\ell}$. 

The issues considered here are particularly important in the context of measuring the 
 EoR 21-cm power spectrum. While all the different frequencies have been collapsed for the 
present analysis, it is necessary to consider the multi-frequency angular power spectrum 
$C_{\ell}(\nu_1,\nu_2)$ or equivalently the three dimensional power spectrum $P(k_{\parallel},k_{\perp})$ 
to quantify the 21-cm signal. We plan to generalize the TGE for this context in future work.

%\clearpage{\pagestyle{empty}\cleardoublepage} %%%%%%%%%%%%%%%%%%%%
%\newpage
\setcounter{section}{0}
\setcounter{subsection}{0}
\setcounter{subsubsection}{2}
\setcounter{equation}{0}
%\pagenumbering{arabic}

\def\sige{\sigma_{P_G}}
\def\kpm{{k}_\perp}
\def\kp{k_\parallel}
\def\u{\vec{U}}
\def\S{{\mathcal S}}
\def\V{\mathcal{V}}
\def\N{{\mathcal N}}
\def\A{{\bf A}\,}
\def\Sc{S_2}

%-------------------------------------------
\chapter[21-cm Power spectrum estimator]{{\bf The  visibility based Tapered Gridded
  Estimator (TGE) for the redshifted 21-cm power spectrum}\footnote{This chapter is adapted
   from the paper ``The  visibility based Tapered Gridded
  Estimator (TGE) for the redshifted 21-cm power spectrum''
   by \citet{samir16b}}}
\label{chap:chap5}

\section{Introduction}
Observations of the redshifted neutral hydrogen (HI) 21-cm radiation 
hold the potential of probing a wide range of cosmological and
 astrophysical phenomena over a large  redshift range $0 < z \lsim 200$
\citep{bharadwaj05,furlanetto06,morales10,pritchard12,mellema13}. There now are several 
ongoing experiments  such as the Donald
C. Backer Precision Array to Probe the Epoch of Reionization
(PAPER,
\citealt{parsons10}), the Low Frequency Array
(LOFAR,
\citealt{haarlem,yata13}) and  the Murchison Wide-field Array
(MWA, \citealt{bowman13,tingay13})
which aim to measure the power spectrum   of the 21-cm radiation from  
the Epoch of Reionization (EoR, $6 \lsim z \lsim 13$). Future telescopes 
like the Square Kilometer Array (SKA1
LOW, \citealt{koopmans15}) 
and the Hydrogen Epoch of Reionization Array
(HERA, \citealt{neben16}) are planned 
to achieve even higher sensitivity for measuring  the EoR 21-cm power spectrum.
Several other upcoming experiments  like the Ooty Wide Field Array (OWFA;
\citealt{prasad,ali14}), the Canadian Hydrogen Intensity Mapping
Experiment (CHIME{\footnote{http://chime.phas.ubc.ca/};
  \citealt{bandura}), the Baryon Acoustic Oscillation Broadband,
  Broad Beam Array (BAOBAB{\footnote{http://bao.berkeley.edu/};
    \citealt{pober13a}) and  the Square Kilometre Array (SKA1 MID;
    \citealt{bull15}) target the post-Reionization 21-cm signal 
($0 <    z \lsim 6$).

Despite the  sensitive new instruments, the main challenge 
still arises from the fact that 
the cosmological 21-cm   signal is buried in astrophysical foregrounds 
which are $4-5$  orders  of magnitude brighter 
 \citep{shaver99,dmat1,santos05, ali08,paciga11,ghosh1,ghosh2}.
 A large variety of techniques have been proposed  to overcome 
this problem and estimate the  21-cm power spectrum. 
The different approaches may be broadly   divided into two classes
(1.) Foreground Removal, and (2.) Foreground Avoidance.
 
The idea in Foreground Removal is to model the foregrounds and 
subtract these out either directly from the  data 
(eg. \citealt{ali08}) or from the power spectrum estimator after 
correlating the data (eg.  \citealt{ghosh1,ghosh2}). 
 Foreground Removal is a topic of intense current research  
\citep{jelic08,bowman09,paciga11,chapman12,parsons12,liu12,trott12,pober13,paciga13,
parsons14,trott16}.  

Various studies (eg. \citealt{adatta10}) show that  the foreground 
contribution to the Cylindrical Power Spectrum  $P(k_{\perp},k_{\parallel})$ 
is expected to be restricted within  a wedge in the two dimensional (2D) 
$(k_{\perp},k_{\parallel})$ plane. The idea in Foreground Avoidance is to 
avoid the Fourier modes within the foreground wedge and only use the 
uncontaminated  modes outside the wedge to estimate the 21-cm power 
spectrum \citep{vedantham12,thyag13,pober14,liu14a,liu14b,dillon14,dillon15,ali15}.  
In a recent paper \citet{jacob16} have compared several power spectrum
estimation techniques in the context of MWA.

Point sources dominate the low frequency sky  at the
angular scales $\le  4^{\circ}$ \citep{ali08} which are relevant 
for  EoR 21-cm power spectrum  with the telescopes  like the GMRT, LOFAR and
the upcoming SKA.   It is difficult to model and subtract the point sources 
which are located at   the periphery of the telescope's field of view (FoV).
The antenna response  deviates from circular symmetry, and is   highly
frequency and time dependent at the outer parts of the telescope's FoV.
 The calibration also differs from the phase center due to ionospheric 
fluctuations. The residual point sources located far away from the phase
centre cause the signal to  oscillates along the frequency direction
\citep{ghosh1,ghosh2}. This poses a severe problem for foreground removal
techniques which assume a smooth behavior of the signal along the frequency direction. Equivalently, these distant point sources reduce the EoR window by
increasing the area under the foreground wedge in $(k_{\perp},k_{\parallel})$
space \citep{thyag15}. In a recent paper, \citet{pober16} showed that
correctly modelling and subtracting the distant point sources are important for
detecting the redshifted 21-cm signal. Point source subtraction is also
important for measuring the angular power spectrum of the diffuse Galactic
synchrotron radiation \citep{bernardi09,ghosh12,iacobelli13}. Apart 
from being an important foreground component for the EoR 21-cm signal, this is also interesting in its own right.

It is possible to suppress the contribution from the outer parts of the
telescope's FoV by tapering the sky response through a suitably
chosen window function. \citet{ghosh2} have analyzed $610 {\rm MHz}$ GMRT
data to show that it is possible to implement the tapering by convolving the
observed visibilities with the Fourier transform of the window function. It is
found that this reduces the amplitude of the oscillation along the frequency
direction. Our earlier work \citet{samir14} (hereafter Paper I) has introduced
the Tapered Gridded Estimator (TGE) which   
places the findings of \citet{ghosh2} on a sound theoretical
footing. Considering observations at a single frequency, the TGE estimates the 
angular power spectrum $C_{\ell}$ of the 2D sky signal directly from the
measured visibilities while simultaneously tapering the sky response. 
As a test-bed for the TGE, Paper I  considers a situation where 
the point sources have been identified and subtracted out so that 
the residual visibilities are dominated by the Galactic synchrotron 
radiation. This has been used to investigate how well the TGE  is  able
to recover   the angular power spectrum of the input model used to simulate 
the Galactic synchrotron emission at $150 \, {\rm MHz}$.  While  most of the 
analysis was  for the   GMRT, simulations for  LOFAR were also considered. 
These investigations show that the TGE  is able to recover  the input model
$C_{\ell}^M$ to a high level of precision provided the baselines have a
uniform $uv$ coverage. For the GMRT, which has a patchy $uv$ coverage,   
the $C_{\ell}$ is  slightly overestimated using TGE though the excess 
is largely within the $1\sigma$ errors.  This deviation  is found to be reduced 
in a situation with  a more uniform and denser baseline distribution , like
LOFAR. Paper I  also analyzes the effects of gain errors and the $w$-term. 

In a recent paper \citet{samir16a} (hereafter Paper II) we have further
developed  the simulations of Paper~I to include the point sources.  We have
used conventional radio astronomical techniques to model and subtract the
point sources from the central region of the primary beam. As detailed in
Paper~II, it is difficult to do the same for the sources which are far away
from the phase center, and these persist as residuals in the visibility
data. We find that these residual point sources dominate the  $C_{\ell}$
estimated at large baselines. We also show that it is possible to suppress the
contribution from these residual sources located at  the periphery of the
FoV by using TGE with a suitably chosen window function.  

Removing the  noise bias  is an important issue for any power spectrum
estimator. As demonstrated in Paper II, the TGE internally estimates the actual
noise bias from the data and subtracts this out to give an unbiased
estimate of the power spectrum.

In the present work we report the progress on two counts. First, our earlier
implementation of the TGE assumed a uniform and dense baseline $uv$ coverage 
to calculate the normalization coefficient which relates  visibility
correlations to the estimated angular power spectrum $C_{\ell}$. We, however, 
found (Paper I) that this leads to an overestimate of $C_{\ell}$ for
instruments like the GMRT which have a sparse and patchy $uv$ coverage. In
Section 2 of this paper we present an improved TGE which overcomes this
problem by using 
simulations to estimate the normalization coefficient. Second, the entire
analysis of Papers I and II has been restricted to observations at a single
frequency wherein the relevant issue is to quantify the 2D angular
fluctuations of the sky signal.  This, however, is inadequate for the three
dimensional (3D) redshifted HI 21-cm signal  where it is necessary to
also simultaneously quantify the fluctuations along the frequency
direction. In Section 3 of this paper we have generalized the TGE to quantify
the 3D 21-cm signal  and estimate the spatial power spectrum of the 21-cm  
brightness temperature fluctuations $P(\k)$.  We discuss two different
binning schemes which respectively yield the  spherically-averaged (1D) power
spectrum $P(k)$ and the cylindrically-averaged (2D) power spectrum
$P(\kpm,\kp)$, and present theoretical expressions for predicting the expected
variance.  We have validated the estimator and its variance predictions using
simulations which are described in Section 4 and for which the results are
presented in Section 5.  Sections 6 presents the summary and conclusions.

In this paper, we have used cosmological parameters from the (Planck +
WMAP) best-fit $\Lambda$CDM cosmology (\citealt{ade15}).

\section{$C_{\ell}$ estimation}
\label{v2ps}

\subsection{An Improved TGE}
In this section we restrict our attention  to a single  frequency channel $\nu_a$ which we do not  show explicitly in any of the subsequent equations.  The measured visibilities $\V_i$   can be decomposed into two contributions, 
\begin{equation}
\V_i=\S(\u_i)+\N_i
\label{eq:a1}
\end{equation}
the  sky signal and system noise respectively, and $\u_i$ is the baseline corresponding to the  $i$-th   visibility. 
The signal contribution $\S(\u_i)$ records the
Fourier transform of the product of the telescope's primary beam pattern
${\mathcal A}(\th)$ and the specific intensity fluctuation on the sky $\delta I(\th)$. Expressing the signal in terms of brightness
 temperature fluctuations $\delta T(\th)$ we have 
\begin{equation}
\S(\u_i)= \left(\frac{\partial B}{\partial T}\right)  \int d^2 \theta \,  e^{2 \pi i \u_i
 \cdot \th} {\mathcal A}(\th) \delta T(\th),
\label{eq:a2a}
\end{equation}
where 
$B=2k_B T/\lambda^2$  is the Planck function in the Raleigh-Jeans limit which
is valid at the 
frequencies of our interest. In terms of Fourier components we have 
\begin{equation}
\S(\u_i)= \left( \frac{\partial B}{\partial T} \right)  \int \, d^2 U  \,
  \tilde{a}\left(\u_i - \u\right)\, 
 \, \Delta \tilde{T}(\u),   
\label{eq:a2}
\end{equation}
where $\Delta \tilde{ T}(\u)$ and $\tilde{a}\,(\u)$ are the
Fourier transforms of $\delta T(\th) $ and ${\cal A}(\th)$
respectively.  Here we assume that  $\delta T(\th) $ is a
particular realization of a statistically homogeneous and isotropic
Gaussian random process on the sky. Its  statistical properties  are 
completely characterized by the angular power spectrum of the
 brightness temperature fluctuations $C_{\ell}$ defined  through 
\begin{equation}
 \langle \Delta \tilde{T}(\u ) \Delta \tilde{T}^{*}(\u') \rangle = 
\delta_{D}^{2}(\u-\u') \, C_{2 \pi U}\
\label{eq:a3}
\end{equation}
where $\delta_{D}^{2}(\u-\u')$ is a two dimensional Dirac delta
function and $2 \pi U= \ell$, is the angular multipole. The angular brackets $\langle ... \rangle$ here denote an
ensemble average over different realizations of the stochastic
temperature fluctuations on the sky.

The noise  in the different visibilities is uncorrelated,
and we have  
\begin{equation}
\langle \V_i \V_j \rangle = \langle \S_i \S_j \rangle + \langle \mid \N_i \mid^2  \rangle  \delta_{i,j}
\label{eq:a4}
\end{equation}

where  $\langle \mid \N_i \mid^2  \rangle$ is the noise variance of the visibilities, $\delta_{i,j}$ is a 
Kronecker delta and 
\begin{equation}
\langle \S_i \S_j \rangle =
 \left( \frac{\partial B}{\partial T} \right)^2
 \int d^2 U \, \tilde{a}(\u_i-\u) \, \tilde{a}^{*}(\u_j-\u) \, C_{2 \pi U_i} \,
\label{eq:a5}
\end{equation}
This  convolution can be approximated by  a multiplicating factor if $C_{2 \pi U}$ is nearly constant across the
width of   $\tilde{a}(\u_i-\u) $,  which is the situation at large baselines where the antenna separation is large 
compared to the telescope diameter (Paper I), and we have
\begin{equation}
\langle \mid \V_i \mid^2 \rangle =V_0 C \, _{2 \pi U_i} +  \langle \mid \N_i \mid^2  \rangle \,
\label{eq:a6}
\end{equation}
where 
\begin{equation}
V_0= \left( \frac{\partial B}{\partial T} \right)^2
 \int d^2 U \, \mid \tilde{a}(\u_i-\u) \mid^2 \,.
\label{eq:a6a}
\end{equation}
We see that the correlation of a visibility with itself provides an estimate of the angular power 
spectrum, except for the terms $ \langle \mid \N_i \mid^2  \rangle$ which introduce  a positive noise bias.  

It is possible to control the sidelobe response of the telescope's beam pattern
${\mathcal A}(\th)$
by  tapering the sky response
through a frequency independent window function ${\cal W}(\theta)$.  
In this work we use a Gaussian ${\cal W}(\theta)=e^{-\theta^2/\theta^2_w}$  
with $\theta_w$ chosen  so that the window function  cuts off the sky response 
well  before the first null of ${\mathcal A}(\th)$.  This tapering is
achieved by convolving  the measured visibilities with the Fourier 
transform of ${\cal W}(\theta)$.  
We choose a rectangular grid
in the $uv$ plane and consider the convolved visibilities  
\begin{equation}
\V_{cg} = \sum_{i}\tilde{w}(\u_g-\u_i) \, \V_i
\label{eq:a7}
\end{equation}
where  $\tilde{w}(\u)=\pi\theta_w^2e^{-\pi^2U^2\theta_w^2}$ is the Fourier 
transform of ${\cal W}(\theta)$ and $\u_g$ refers to the different grid points.  As shown in Paper I, gridding reduces the computation in comparison to an estimator that uses pairs of visibilities to estimate the power spectrum. We now focus our attention 
on $\S_{cg}$  which is the sky
signal contribution to $\V_{cg}$. This can be written as 
\begin{equation}
\S_{cg}= \left( \frac{\partial B}{\partial T} \right)  \int \, d^2 U  \, 
 \tilde{K}\left(\u_g - \u\right)\,  \, \Delta \tilde{T}(\u),  
\label{eq:a8}
\end{equation}
where 
\begin{equation}
\tilde{K}\left(\u_g - \u\right)= \int d^2 U^{'} 
\tilde{w}(\u_g-\u^{'}) B(\u^{'}) \tilde{a}\left(\u^{'} - \u\right) 
\label{eq:a9}
\end{equation}
is an effective  ``gridding kernel'', and 
\begin{equation}
{\rm B}(\u)=\sum_i \delta^2_D(\u-\u_i)
\label{eq:a10}
\end{equation}
 is the  baseline sampling function of the measured visibilities. 

Proceeding in exactly the same way as we did for eq. (\ref{eq:a6}) we have 
\begin{equation}
\langle \mid \V_{cg} \mid^2 \rangle =V_{1g} C_{2 \pi U_g} + 
\sum_i \mid \tilde{w}(\u_g-\u_i) \mid^2 \langle \mid \N_i \mid^2  \rangle \, , 
\label{eq:a11}
\end{equation}
where 
\begin{equation}
V_{1g}= \left( \frac{\partial B}{\partial T} \right)^2
 \int d^2 U \, \mid \tilde{K}(\u_i-\u) \mid^2 \,.
\label{eq:a10a}
\end{equation}
Here again we see that the correlation of the tapered gridded visibility with
itself provides an estimate of the angular power  
spectrum, except for the terms $ \langle \mid \N_i \mid^2  \rangle$ which
introduces  a positive noise bias.   

Combining equations (\ref{eq:a6}) and (\ref{eq:a11}) we have 

\begin{equation}
\langle \left( \mid \V_{cg} \mid^2 -\sum_i \mid \tilde{w}(\u_g-\u_i) \mid^2
\mid \V_i \mid^2 \right)   \rangle 
=M_g C_{2 \pi U_g}
\label{eq:a11a}
\end{equation}
where 
\begin{equation}
M_g=V_{1g} - \sum_i \mid \tilde{w}(\u_g-\u_i) \mid^2 V_0
\label{eq:a12}
\end{equation}

This allows us to define the Tapered Gridded Estimator (TGE) as 
\begin{equation}
{\hat E}_g= M_g^{-1} \, \left( \mid \V_{cg} \mid^2 -\sum_i \mid
\tilde{w}(\u_g-\u_i) \mid^2  \mid \V_i \mid^2 \right) \,. 
\label{eq:a13}
\end{equation}

The TGE defined here (eq. \ref{eq:a13}) incorporates three novel
  features which are highlighted below. First, the estimator uses the gridded 
visibilities to estimate $C_{\ell}$, this is computationally much faster
 than individually correlating the visibilities. Second, the correlation
 of the gridded visibilities is used to estimate $C_{\ell}$. A 
positive   noise bias is removed by subtracting 
the auto-correlation of the  visibilities. Third, the estimator 
allows us to taper the FoV so as to restrict the contribution from the 
sources in the outer regions and the sidelobes.
 It is, however, necessary to 
note that this comes at a cost which we now discuss. First,  we lose 
information  at the largest angular scales due to the reduced FoV. 
This restricts the smallest $\ell$ value at which it is possible to estimate
the power spectrum. Second,  the reduced FoV results in a larger cosmic 
variance for the smaller angular modes which are within the tapered FoV.

The TGE provides an unbiased estimate of $C_{\ell_g}$ at the angular multipole
$\ell_g=2 \pi U_g$ {\it i.e.} 
\begin{equation}
\langle {\hat E}_g \rangle = C_{\ell_g}
\label{eq:a14}
\end{equation}

 We  use this to define the binned Tapered Gridded Estimator for bin $a$ 
\begin{equation}
{\hat E}_G(a) = \frac{\sum_g w_g  {\hat E}_g}{\sum_g w_g } \,.
\label{eq:a15}
\end{equation}
where $w_g$ refers to the weight assigned to the contribution from any particular 
grid point. In the entire subsequent analysis we have used the weight
$w_g=1$  which assigns equal weightage to all the  grid points which are sampled 
by the baselines.

The binned estimator  has an expectation value 
\begin{equation}
\bar{C}_{\bar{\ell}_a}  = \frac{ \sum_g w_g C_{\ell_g}}{ \sum_g w_g}
\label{eq:ga16}
\end{equation}
where $ \bar{C}_{\bar{\ell}_a}$ is the average  angular power spectrum  at 
 \begin{equation}
\bar{\ell}_a =
\frac{ \sum_g w_g \ell_g}{ \sum_g w_g}
\label{eq:a18}
\end{equation}
which is the   effective angular multipole  for bin $a$. 

\subsection{Calculating $M_g$}
The discussion, till now, has not addressed  how to calculate  $M_g$ which  
is the normalization constant for the TGE (eq. \ref{eq:a13}).  The values of
$M_g$ (eq. \ref{eq:a12}) depend on the baseline distribution
(eq. \ref{eq:a10}) and the form of the tapering  function
${\cal W}(\theta)$, and it is necessary to calculate $M_g$ at every grid
point in the $uv$ plane. 
Our earlier work (Paper I)  presents an analytic  approximation using
which it is possible to estimate $M_g$. 
While this has been found to work very well in a situation where the baselines
have a nearly uniform and dense $uv$ coverage (Fig. 7 of Paper I),  
it leads to an overestimate of $C_{\ell}$ if we have a sparse and non-uniform 
$uv$ coverage. Here we present a different  method to estimate   
$M_g$ which, as we show later, works very well even if we have a   sparse and
non-uniform  $uv$ coverage. 

 We proceed by   calculating simulated  visibilities $[\V_i]_{\rm UAPS}$
 corresponding to an unit angular power spectrum  (UAPS) which has
 $C_{\ell}=1$  
with exactly  the same baseline distribution as the actual observed visibilities. 
We then have (eq. \ref{eq:a11a}) 
\begin{equation}
M_g=\langle \left( \mid \V_{cg} \mid^2 -\sum_i \mid \tilde{w}(\u_g-\u_i) \mid^2 \langle \mid \V_i \mid^2 \right)  \rangle_{\rm UPAS} 
\label{eq:m1}
\end{equation}
which allows us to estimate $M_g$. We average over $N_u$ independent
realizations of the UPAS to reduce the statistical uncertainty
$(\delta M_g/M_g \sim 1/\sqrt{N_u})$ in the estimated $M_g$.

\begin{figure*}
\begin{center}
\psfrag{x}[t][t][1.5][0]{$u$ ($\lambda$)}
\psfrag{y}[b][c][1.5][0]{$v$ ($\lambda$)}
\psfrag{mg}[t][t][1.][0]{log(Mg)}
\includegraphics[width=80mm,angle=-90]{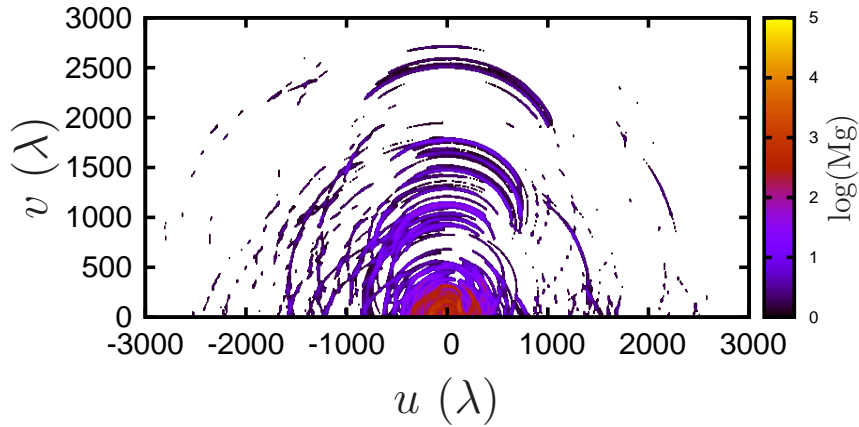}
\caption{ This shows $M_g$ for a fixed value of $f=0.6$. Note that, the baselines in the lower half of the $uv$ plane have been folded on to the upper half.}
\label{fig:fig01}
\end{center}
\end{figure*}

\subsection{Validating the estimator}
\label{sec:clvar}
We have tested the entire method of analysis using simulations of $8$ hours of 
$150 \, {\rm MHz}$ GMRT observations targeted on an arbitrarily selected  field
 located at RA=$10{\rm h} \, 46{\rm  m} \, 00{\rm s}$ and DEC=$59^{\circ} \, 00^{'} \, 59^{''}$.  
The simulations only incorporate the  diffuse Galactic
 synchrotron radiation  for which we  use the measured 
angular power spectrum  \citep{ghosh12} 

\begin{equation}
C^M_{\ell}=A_{\rm 150}\times\left(\frac{1000}{\ell} \right)^{\beta}  \,
\label{eq:cl150}
\end{equation}
as the input model to generate  the brightness temperature fluctuations  on
the sky. Here   $A_{\rm 150}=513 \, {\rm  mK}^2$ and  $\beta=2.34 $ 
\citep{ghosh12}.   The simulation covers a  $\sim 26.4^{\circ} \times 
26.4^{\circ}$ region of the sky,  which is slightly more than ten times  the
FWHM of the GMRT primary beam $(\theta_{FWHM}=157^{'})$. 
The diffuse signal was simulated on a grid of resolution $\sim 0.5^{'}$, and
the entire analysis was restricted to baselines within  $U \le 3,000$. 
Our earlier work (Paper II), and also the discussion of this paper,  show
that the noise bias cancels out from the TGE, and  we have not
included the system noise in these simulations.   

We have modelled the tapering  window function as a 
Gaussian  ${\cal W}(\theta)=e^{-\theta^2/\theta^2_w}$ where we parameterize
$\theta_w=f \theta_0$ where  $\theta_0=0.6 \times \theta_{FWHM}$, 
and preferably $f \le 1$ so that ${\cal W}(\theta)$  cuts off the sky
response well before the first null of the  primary beam. After tapering, we
have an effective  beam pattern ${\mathcal A_W}(\th)={\cal W}(\theta)\,{\cal A}(\th, \nu)$ which is well approximated by a Gaussian $ {\cal A_W}(\theta)=e^{-\theta^2/\theta_1^2}$ with $\theta_1=f (1+f^2)^{-1/2} \theta_0$. The spacing of the  $uv$ grid required for TGE is
decided by $\tilde{a}_W(U)=\pi \theta_1^2 e^{-\pi^2 U^2 \theta_1^2} $ which is
the Fourier transform of ${\mathcal A_W}(\theta)$.  We have chosen a grid
spacing  $\Delta U=\sqrt{\ln2}/(2\pi\theta_1)$ which corresponds to one fourth
of the FWHM of $\tilde{a}_W(U)$.  The convolution in eq. (\ref{eq:a7})   was
restricted to the visibilities within  a disc of radius $12 \times \Delta U$
around each grid point.  The  function $\tilde{w}(\u_g-\u_i)$  falls of
rapidly and we do not expect the visibilities  beyond this to make a
significant contribution.  

We have considered three different values $f=10, 2$ and $0.6$ for the
tapering, here $f=10$ essentially corresponds to a situation with no
tapering, and the sky  response gets confined to a progressively smaller 
region  as the value of $f$ is reduced to $f=2.0$ and $0.6$ respectively
(see Figure 1 of Paper II).  We have
used $N_u=128$ independent realizations of the UAPS to 
estimate $M_g$ for each point in the $uv$ grid. It is necessary to separately
calculate $M_g$ for each value of $f$. 
 Figure \ref {fig:fig01} shows the values of $M_g$ for $f=0.6$. 
We see that this roughly traces out the $uv$ tracks of the baselines, 
the convolution with  $\tilde{w}(\u_g-\u_i)$ results in a thickening 
of the tracks. The values of $M_g$ are roughly proportional to 
$N_g^2 -N_g$, where $N_g$ is the number of visibilities that contribute 
to any particular grid point.

The estimator (eq. \ref{eq:a13}) was applied to the simulated
visibility data which was generated using the model angular power
spectrum (eq. \ref{eq:cl150}).  The estimated angular power spectrum was
binned into $20$ annular bins of equal logarithmic spacing.  We have
used $N_r=128$ independent realizations of the simulation to calculate
the mean and standard deviation of $C_{\ell}$ shown in the left panel
of Figure \ref {fig:fig1}. We see that the TGE is able to recover the
input model $C^M_{\ell}$ quite accurately.  As mentioned earlier, our
previous implementation of TGE (Paper I) had a problem in that the
estimated $C_{\ell}$ was in all cases in excess of the input model
$C^M_{\ell}$, though the deviations were within the $1\sigma$ error
bars throughout. The right panel of Figure \ref {fig:fig1} shows the
fractional deviation $(C_{\ell}-C^M_{\ell}) /C^M_{\ell}$ for the
improved TGE introduced in this paper for the three different values
of $f$ mentioned earlier. We see that for all the values of $f$ the
fractional deviation is less than $10\%$ for $\ell\ge500$.  This is a
considerable improvement over the results of Paper I where we had
$20\%$ to $50\%$ deviations. The fractional deviation is seen to
increase as we increase the tapering {\em i.e.} reduce the value of
$f$. We see that for $f=10$ and $2$, the fractional deviation is less
than $3\%$ for all values of $\ell$ except at the smallest bin.  The fractional deviation for
$f=0.6$ is less than $5\%$ except at the smallest value of  $\ell$  where
it becomes almost $40\%$. This is possibly an outcome of the fact that the 
width of the convolution window $\tilde{w}(\u_g-\u_i)$ increases  as the
value of $f$  is reduced, and the variation of the signal amplitude within
the width of $\tilde{w}(\u_g-\u_i)$ becomes important at small baselines 
where it is reflected as an overestimate of the value of $C_{\ell}$.
Theoretically, we expect the fractional deviation to have random, statistical
fluctuations of the order  $\sigma_{E_G}/\sqrt{N_r}C^M_{\ell}$, where
$\sigma_{E_G}$ is the standard deviation of the estimated angular power
spectrum. We have shown the statistical fluctuation expected  for $f=0.6$ 
as a shaded region in the right panel of Figure \ref {fig:fig1}. We see that
the fractional deviation is roughly consistent with statistical fluctuations
for $\ell \ge 500$. 

\begin{figure*}
\begin{center}
\psfrag{cl}[b][b][0.8][0]{$\ell (\ell+1) C_{\ell}/2 \pi \, [\rm K^2]$}
\psfrag{l}[c][c][0.8][0]{$\ell$}
\psfrag{model}[r][r][0.8][0]{Model}
\psfrag{tap10}[r][r][0.8][0]{f=10}
\psfrag{tap2}[r][r][0.8][0]{f=2}
\psfrag{tap0.6}[r][r][0.8][0]{f=0.6}
\psfrag{diff}[c][c][0.8][0]{$(C_{\ell}-C^M_{\ell}) /C^M_{\ell}$}
\includegraphics[width=75mm,angle=0]{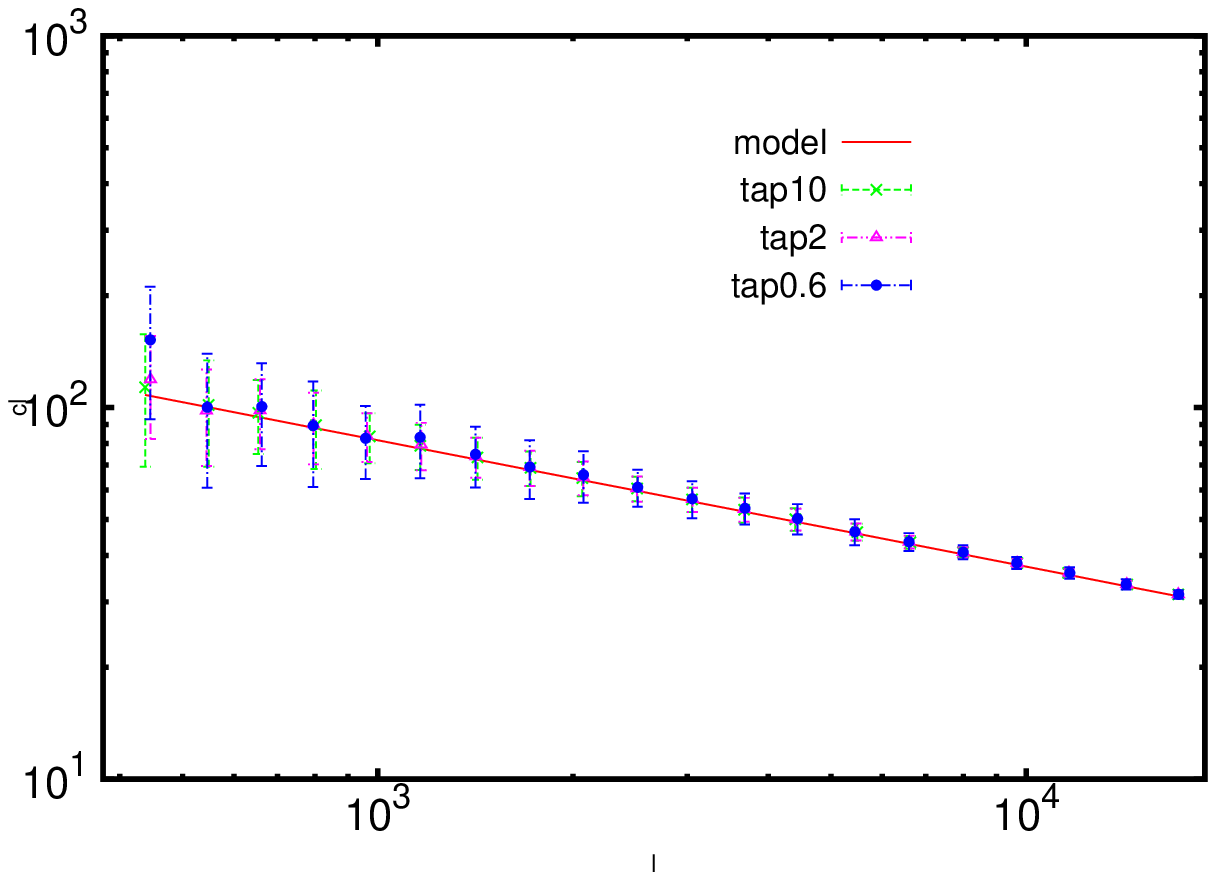}
\includegraphics[width=75mm,angle=0]{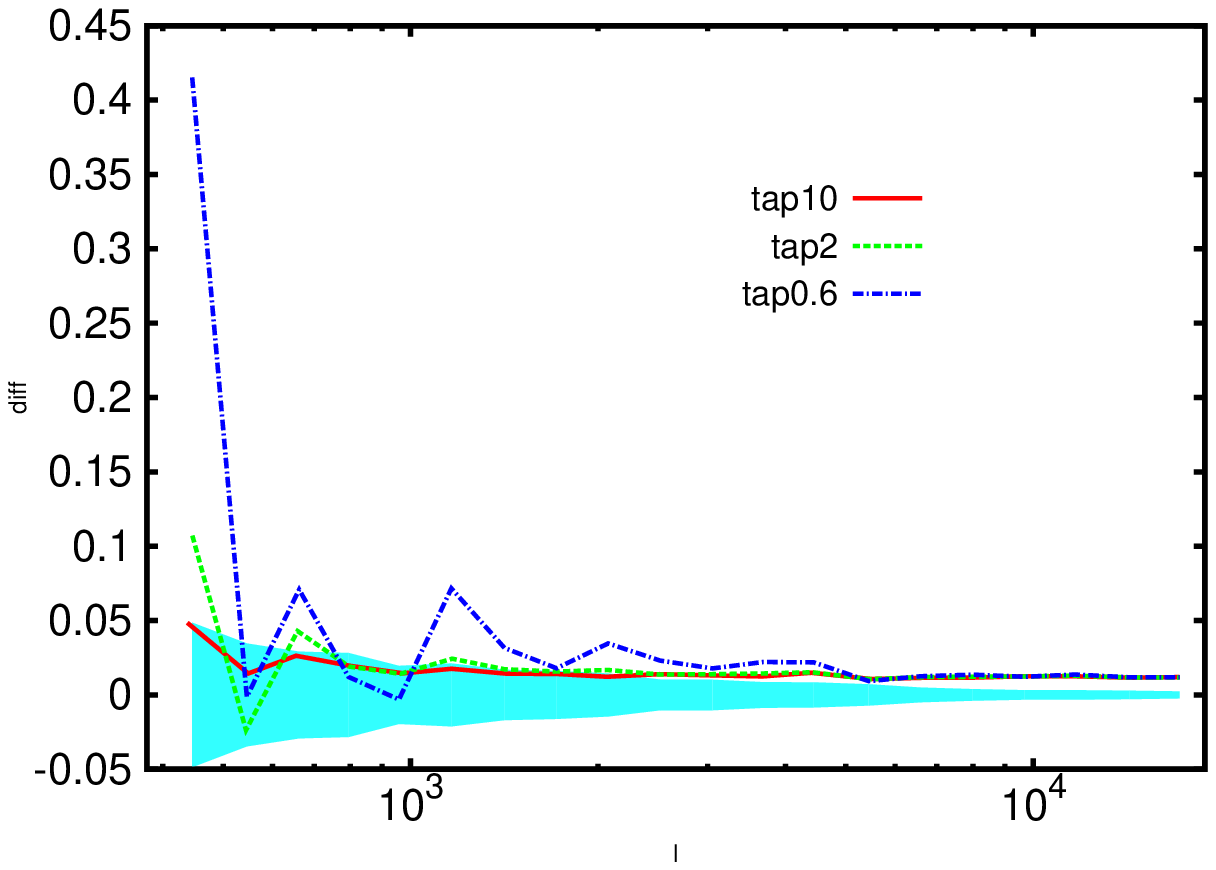}
\caption{The left panel shows a comparison of the input model and the values
  recovered from the simulated visibilities using the improved TGE for
  different tapering of values $f=10,2$ and   $0.6$,  with 1-$\sigma$ error bars
estimated   from   $N_r=128$ realizations of   the simulations. 
 The  right panel shows the fractional deviation of the estimated  $C_{\ell}$
with respect to the   input   model. Here the   shaded region shows 
 the expected statistical fluctuations   ($\sigma_{E_G}/\sqrt{N_r}C^M_{\ell}$)
 of the fractional deviation  for $f=0.6$.} 
\label{fig:fig1}
\end{center}
\end{figure*}

\subsection{Variance}
\label{sec:clvar1}
In the preceding discussion we have used several statistically independent
realizations of the signal to determine the variance of the estimated
binned  angular power spectrum. Such a procedure is, by and large, only
possible with simulated data. We usually  have access to only one 
statistically independent realizations  of the input signal, and the aim is to
use this to not only estimate the angular power spectrum but also estimate the 
uncertainty in the estimated angular power spectrum. In this subsection we
present  theoretical predictions for the variance  of the binned TGE
(eq. (\ref{eq:a15})) 
\begin{equation}
\sigma^2_{E_G}(a)=\langle \hat E^2_G(a) \rangle - \langle \hat E_G(a) \rangle^2\,
\label{eq:var1}
\end{equation} 
which can be used to estimate the uncertainty in the measured angular power
spectrum. 
Following Paper I, we ignore the term 
$\sum_i \mid \tilde{w}(\u_g-\u_i) \mid^2\mid \V_i \mid^2$  in
eq. (\ref{eq:a13})  
for calculating the variance. The signal contribution from this term to the 
estimator at the grid point $\u_g$ scales as $N_g$ which is the number of
visibilities that contribute to ${\hat E}_g$. In comparison to this, the
contribution from the term $\mid \V_{cg} \mid^2$ scales as $N_g^2$ which is
much larger when $N_g \gg 1$.  Assuming that this condition is satisfied  at 
every grid point which contributes to the binned TGE, 
it is  justified to drop the term  
$\sum_i \mid \tilde{w}(\u_g-\u_i) \mid^2\mid \V_i \mid^2$  for calculating the
variance. We then have 
\begin{equation}  
\sigma^2_{E_G}(a) = \frac{\sum_{g g^{'}} w_g w_{g^{'}}  M_g^{-1}
  M^{-1}_{g^{'}} \mid \langle \V_{cg}\V^{*}_{cg^{'}} \rangle \mid^2 }{[\sum_{g
    } w_g]^2} \,
\label{eq:var2}
\end{equation}
which is identical to eq. (41) of Paper I, except that we now have  the
normalization constant $M_g^{-1}$ instead  of $K^{-2}_{1g}/V_1$. 

It is necessary to model the correlation between the convolved visibilities at
two different grid points $ \langle \V_{cg}\V^{*}_{cg^{'}} \rangle$
in eq. (\ref{eq:var2})  in order
to make further progress. This correlation is a sum of two parts 
\begin{equation}
\langle \V_{cg}\V^{*}_{cg^{'}} \rangle = 
 \langle \S_{cg}\S^{*}_{cg^{'}} \rangle + 
\langle \N_{cg}\N^{*}_{cg^{'}} \rangle
\label{eq:var2a}
\end{equation}
the  signal  and the  noise correlation respectively. 

 Earlier studies (Paper I) show that we expect the
 signal correlation  $\langle \S_{cg}\S^{*}_{cg^{'}} \rangle$
to fall off as $e^{-\mid \Delta \u_{g g^{'}} \mid^2/\sigma_1^2}$
if  the grid separation is increased, here
$\sigma_1=f^{-1}\sqrt{1+f^2}\sigma_0$ 
where $\sigma_0=0.76/\theta_{\rm FWHM}$. We use this to approximate the 
signal correlation as
\begin{equation}
\langle \S_{cg}\S^{*}_{cg^{'}} \rangle= \sqrt{M_g M_{g^{'}}} e^{-\mid \Delta
  \u_{g g^{'}} \mid^2/\sigma_1^2} \, \bar{C}_{\bar{\ell}_a} 
\label{eq:var3}
\end{equation}
where $\bar{C}_{\bar{\ell}_a} $ refers to the  angular power spectrum
measured at the particular bin $a$ for which the variance $\sigma^2_{E_G}(a)$ 
is being calculated.   

The noise correlation
\begin{equation}
\langle \N_{cg}\N^{*}_{cg^{'}} \rangle= \sum_i  \tilde{w}(\u_g-\u_i)
\tilde{w}^{*}(\u_{g^{'}}-\u_i)  \langle \mid \N_i \mid^2  \rangle \,  
\label{eq:var4}
\end{equation}
also is expected to   fall off as the grid separation is increased, and we
have modeled this $\mid \Delta   \u_{g g^{'}} \mid$ dependence as 
\begin{equation}
\langle \N_{cg}\N^{*}_{cg^{'}} \rangle= \sqrt{K_{2gg}K_{2g^{'}g^{'}}}e^{-
  \mid \Delta \u_{g g^{'}} \mid^2/\sigma_2^2}(2\sigma_n^2)\, 
\label{eq:var5}
\end{equation}
where, $K_{2gg}=\sum_i \mid \tilde{w}(\u_g-\u_i) \mid^2 $, $\sigma_2=3\sigma_0f^{-1}$ and $\sigma_n^2$ is
the variance of the real (and also imaginary) part of $\N_i$. 

We have used eqs. (\ref{eq:var5}), (\ref{eq:var3})  and (\ref{eq:var2a}) in
eq. (\ref{eq:var2}) to calculate $\sigma^2_{E_G}(a)$,  the analytic prediction
for the variance of the estimated  binned angular power spectrum $
\bar{C}_{\bar{\ell}_a}$.

\begin{figure*}
\begin{center}
\psfrag{sig}[b][b][0.8][0]{$\ell (\ell+1) \sigma_{E_G}/2 \pi\, [\rm K^2] $}
\psfrag{l}[c][c][0.8][0]{$\ell$}
\psfrag{sim0.6}[r][r][0.8][0]{Simulation}
\psfrag{ana0.6}[r][r][0.8][0]{Analytic}
\psfrag{tap10}[r][r][0.8][0]{f=10}
\psfrag{tap2}[r][r][0.8][0]{f=2}
\psfrag{tap0.6}[r][r][0.8][0]{f=0.6}
\psfrag{f0.6}[r][r][0.8][0]{f=0.6}
\includegraphics[width=75mm,angle=0]{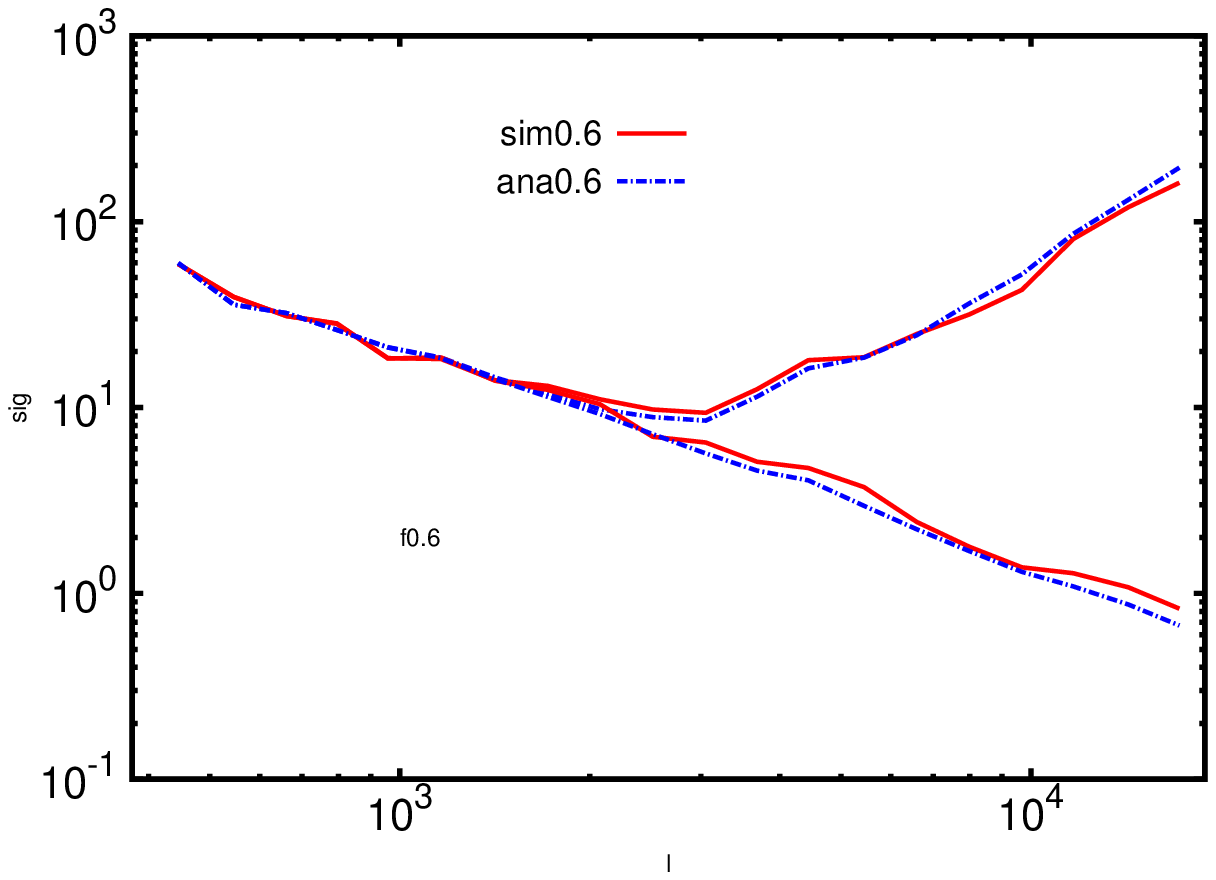}
\includegraphics[width=75mm,angle=0]{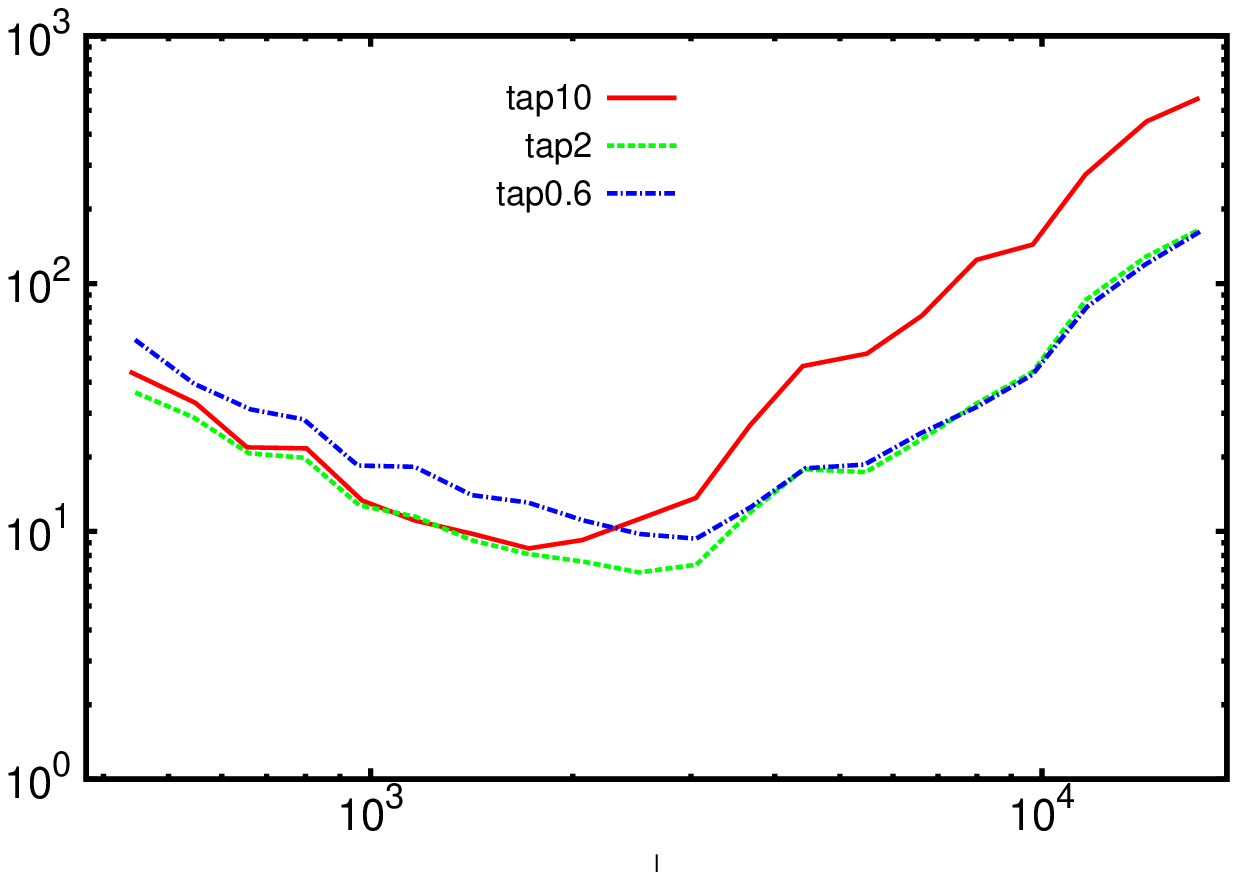}
\caption{In the  left panel  the analytic prediction for the
  variance (eq. \ref{eq:var2}) is compared with variance estimated from
  $N_r=128$ realizations of the simulated visibilities. Results are shown both 
  with (upper curves) and without (lower curves) the system noise contribution. Both match at small $\ell$
  where cosmic variance dominates, the system noise however is important at
  large $\ell$ where the two sets of results are different. 
The right panel shows how the variance with system noise obtained from simulations varies for different values of $f$.}
\label{fig:fig2}
\end{center}
\end{figure*} 

The left panel of Figure \ref {fig:fig2}  shows the analytic prediction for
the variance calculated using eq. (\ref{eq:var2}) for a fixed value of
$f=0.6$. For comparison we also show  the variance 
estimated from $N_r=128$ independent realizations of the simulated
visibilities. We have considered two situations, the first where the simulated
visibilities only have the signal corresponding to the input model
(eq. \ref{eq:cl150}) and no system noise,  and  the second situation where in
addition to the signal the visibilities also have a system noise
contribution  with $\sigma_n=1.03 \, {\rm Jy}$ which corresponds to $16 \, {\rm s}$
integration time and a  channel width of  $125 \, {\rm kHz}$.   
We see that the variance calculated from the simulations is dominated by
cosmic variance at small $\ell$ $(\le 2,000)$ where the
variance  does not  change irrespective of whether we include the system noise
or not. The variance calculated from the simulations is dominated by the  
system noise at large $\ell$ $(\ge 5,000)$. We see that the analytic
predictions  are in reasonably good  agreement with the values
obtained from the simulations over the entire $\ell$ range that we have
considered here. We have also considered  situations where
$f=2.0$ and  $10$ for which the comparison with the analytic results are not
shown here.  In all cases we find that analytic predictions are in
reasonably good agreement with the values obtained from the simulations.  

The right panel of Figure \ref{fig:fig2} shows how the variance obtained from
the simulations changes with $f$. We see that at low $\ell$ the variance
increases  if the value of $f$ is reduced. This is a consequence of the fact
that cosmic variance increases as the sky response is tapered by reducing
$f$. The same effect has also been discussed in detail in our
 earlier paper
(Paper I). We also see that at large $\ell$  the variance is 
considerably higher   
for  $f=10$ in comparison with $f=2$ and $0.6$. This $\ell$ range  is
dominated by the system noise contribution. The number of 
independent visibilities which are combined to estimate the power spectrum at
any grid point increases as $f$ is reduced, and this is reflected in a smaller
variance as $f$ is reduced. 

\section{3D $P(\kpr,\kp)$ estimation}
\subsection{3D TGE}
\label{sec:3dtge}
We now turn our attention to  
the  redshifted $21$-cm HI  brightness temperature fluctuations where it is
necessary to consider  different frequency channels for which 
eq. (\ref{eq:a1}) is generalized to 
\begin{equation}
\V_i(\nu_a)=\S(\u_i,\nu_a)+\N_i(\nu_a).
\label{eq:b1}
\end{equation}
Proceeding in exactly  the same manner as for a single frequency channel
(eq. \ref{eq:a2a}), we have 
\begin{equation}
\S(\u_i,\nu_a)= \left(\frac{\partial B}{\partial T}\right)_{\nu_a}  \int d^2
\theta \,  e^{2 \pi i \u_i  \cdot \th} {\mathcal A}(\th,\nu_a) \delta
T(\th,\nu_a), 
\label{eq:b2a}
\end{equation}
and the noise in the different visibility measurements at different frequency channels
are uncorrelated 
\begin{equation}
\langle \N_i(\nu_a) \N_j(\nu_b) \rangle =  \langle \mid \N_i(\nu_a) \mid^2  \rangle  \delta_{i,j}
\delta_{a,b} \,.
\label{eq:b3}
\end{equation}
Note that the baseline corresponding to a fixed antenna separation
$\u_i={\bf d}_i/\lambda$, the antenna beam pattern  ${\mathcal A}(\th,\nu_a)$ 
 and the factor $\left( \frac{\partial B}{\partial T} \right)_{\nu_a}$ all
vary with the frequency $\nu_a$ in eq. (\ref{eq:b2a}). However, for the present
analysis we only consider the frequency dependence of the HI signal  $\delta T(\th,\nu_a)$ which is assumed to vary much more rapidly with $\nu_a$ in comparison 
to the other terms which are expected to have a relatively slower  frequency
dependence which has been ignored here.  We then have 
\begin{equation}
\S(\u_i,\nu_a)= \left( \frac{\partial B}{\partial T} \right)  \int \, d^2 U  \,
  \tilde{a}\left(\u_i - \u\right)\, 
 \, \Delta \tilde{T}(\u,\nu_a),   
\label{eq:b2}
\end{equation}
which is similar to eq. (\ref{eq:a2}) introduced earlier. 

In eq. (\ref{eq:b2}),  we can express  $\Delta \tilde{T}(\u,\nu)$  in terms of  $\Delta T(\k)$ which refers to the three
dimensional (3D) Fourier decomposition of the HI brightness temperature
fluctuations in the region of space from which the redshifted 21 cm radiation
originated. We use 
equation (7) of  \citet{bharadwaj011} (or equivalently eq. (12) of
\citet{bharadwaj05}) to express $\S(\u_i,\nu)$ in terms of the three dimensional
brightness temperature fluctuations 
\begin{equation}
\S(\u_i,\nu)= \left( \frac{\partial B}{\partial T} \right)  \int \,
\frac{d^3 k}{(2 \pi)^3}  \,
  \tilde{a}\left(\u_i - \frac{\kpr r}{2 \pi}\right)\, e^{- i \kp r^{'} \nu}
 \, \Delta \tilde{T}(\k),   
\label{eq:b5}
\end{equation}
where $(\kpr,\kp)$ are the components of the comoving wave vector $\k$ respectively
perpendicular and parallel to the line of sight, $r$ is the comoving distance
corresponding to the redshifted 21-cm radiation at the observing frequency
$\nu$, $r^{'}=\mid dr/d \nu \mid$, and 
\begin{equation}
\langle \Delta \tilde{T}(\k) \, \Delta \tilde{T}^{*}(\k^{'}) \rangle = (2
\pi)^3 \delta^3_D(\k-\k^{'}) P(\kpr,\kp)
\label{eq:b6}
\end{equation}
defines $P(\kpr,\kp)$, the 3D power spectrum of HI brightness temperature
fluctuations.   $\nu$ here is measured with respect to the central frequency of
the observation, and   $r$ and $r^{'}$ are held fixed at the values
corresponding to the central frequency.  

We next consider observations with  $N_c$ discrete frequency channels $\nu_a$
with $a=0,1,2,...,N_c-1$, each channel of
width $\Delta \nu_c$ and the total spanning a frequency bandwidth ${\rm
  B_{bw}}$. This corresponds to a comoving spatial extent of  ($r^{'}\rm
  B_{bw}$) along the line of sight and $\kp$ now assumes discrete values 
\begin{equation}
\kp=\frac{2 \pi \tau_m}{r^{'}}
\label{eq:b7}
\end{equation}
where $\tau_m$ is the delay variable (\citealt{morales04,mac06}) which takes values $\tau_m=m/{\rm B_{bw}}$ with
$-N_c/2 < m \le Nc/2$. The $\kp$  integral  in eq.  (\ref{eq:b5})  is now 
replaced   by a discrete sum $\int \kp/(2 \pi) \rightarrow (r^{'}{\rm B_{bw}})^{-1}\sum_m$. It is   further convenient to use 
\begin{equation}
\kpr =\frac{2 \pi \u}{ r}
\label{eq:b8}
\end{equation}
whereby 
\begin{equation}
\S(\u_i,\nu_a)= \left( \frac{\partial B}{\partial T} \right)  \int \,
d^2 U \,  \tilde{a}\left(\u_i - \u \right)\, \sum_m e^{- 2\pi i \tau_m  \nu_a} 
 \, \frac{\Delta \tilde{T}(\u,\tau_m)}{{\rm B_{bw}}\, r^2 r^{'}}\,.   
\label{eq:b9}
\end{equation}
Note here that we can identify  $\tau_m$ as being the Fourier conjugate of
$\nu_a$.  

We now consider the Fourier transform along the frequency axis 
of the measured visibilities which gives the visibilities 
$v_i(\tau_m)$ in delay space
\begin{equation}
v_i(\tau_m)=(\Delta \nu_c) \sum_a e^{2 \pi i \tau_m \nu_a} \,  \V_i(\nu_a) \,.
\label{eq:b11}
\end{equation}
The subsequent analysis of this section is entirely based on the delay space
visibilities $v_i(\tau_m)$ defined in eq. (\ref{eq:b11}). 

Calculating $s(\u_i,\tau_m)$,  the HI signal contribution to $v_i(\tau_m)$ using 
eq. (\ref{eq:b9}),  we have 
\begin{equation}
s(\u_i,\tau_m)= \left( \frac{\partial B}{\partial T} \right)  \int \,
d^2 U\,  \tilde{a}\left(\u_i - \u \right)\, 
 \, \left[ \frac{\Delta \tilde{T}(\u,\tau_m)}{r^2 r^{'}} \right] \,,   
\label{eq:b12}
\end{equation}
and, rewriting eq. (\ref{eq:b6}) in terms of the new variables $\u$ and 
$\tau_m$ we have  
\begin{equation}
\langle \Delta \tilde{T}(\u,\tau_m) \,\Delta \tilde{T}^{*}(\u,\tau_n) \rangle
= \delta_D^2(\u-\u^{'}) \left[ \delta_{m,n} ({\rm B_{bw}}\, r^2 r^{'})
  P(\kpr,\kp) \right] \,. 
\label{eq:b13}
\end{equation}
It can be seen that the signals at two different delay channels are uncorrelated. It is also
straightforward to verify that the noise contribution $n_i(\tau_m)$ at
two different delay channels are uncorrelated.  

In summary of the calculations discussed till now in this section, we see that
the visibilities $v_i(\tau_m)$ 
at two  different delay channels are uncorrelated. It therefore suffices to
individually analyze each delay channel separately, and in  the subsequent
discussion  we restrict our attention to a fixed delay channel $\tau_m$.
Calculating  the correlation of a visibility with  itself, we have  
\begin{equation}
\langle \mid v_i(\tau_m) \mid^2  \rangle = V_0 \left[ \frac{{\rm 
      B_{bw}}}{r^2 r^{'}}   P(\kpr,\kp) \right]
 + (\Delta \nu_c)^2  \sum_a \langle \mid \N_i(\nu_a)  \mid^2  \rangle \,. 
\label{eq:b14}
\end{equation}
It is important to note that eqs. (\ref{eq:b12}), (\ref{eq:b13})  and 
(\ref{eq:b14}) which hold for a fixed delay channel
 are exactly analogous  to eqs. (\ref{eq:a2}), (\ref{eq:a3})  and
(\ref{eq:a6})   which hold for  a fixed frequency channel.  We define the 
convolved visibilities in exact analogy with eq. (\ref{eq:a7}) 
\begin{equation}
v_{cg}(\tau_m) = \sum_{i}\tilde{w}(\u_g-\u_i) \, v_i(\tau_m) \,,
\label{eq:b15}
\end{equation}
and we define the 3D TGE  in exact analogy with eq.  (\ref{eq:a13}). 
\begin{equation}
{\hat P}_g(\tau_m)= \left(\frac{M_g {\rm B_{bw}}}{r^2 r^{'}} \right) ^{-1} \,
\left( \mid v_{cg}(\tau_m) \mid^2 -\sum_i \mid \tilde{w}(\u_g-\u_i) \mid^2  
\mid v_i(\tau_m) \mid^2 \right) \,.  
\label{eq:a16}
\end{equation}

The 3D TGE is, by construction,   an unbiased estimator of the three
dimensional power spectrum $P(\kpr,\kp)$, and  we have  
\begin{equation}
\langle {\hat P}_g(\tau_m) \rangle = P({\kpr}_g,{\kp}_m)
\end{equation}
 where ${\kp}_m$ and ${\kpr}_g$ are related to $\tau_m$ and $\u_g$ through eqs.  
(\ref{eq:b7}) and (\ref{eq:b8}) respectively.

\subsection{Frequency Window Function}
\label{filter}
The discrete Fourier transform used to calculate $v_i(\tau_m)$ 
in eq. (\ref{eq:b11}) assumes that the measured visibilities $\V_i(\nu_a)$ are 
periodic across the frequency bandwidth ${\rm B_{bw}}$  ({\em i.e.}
$\V_i(\nu_a)=\V_i(\nu_a+ {\rm B_{bw}})$. In reality, the
measured visibilities are not periodic over the observational bandwidth,
and the discrete Fourier transform encounters 
a discontinuity at the edge of the band. It is possible to avoid
this problem by multiplying the measured visibilities with a frequency window
function 
$F(\nu_a)$ which smoothly falls to zero at the edges of the band.  This
effectively makes the product $F(\nu_a) \times  \V_i(\nu_a)$ periodic,   thereby
doing away with the discontinuity  at the edges of the band. 
This issue has been studied by   \citet{vedantham12} and  \citet{thyag13} 
who have proposed the Blackman-Nuttall \citep{nut81} window function 
\begin{equation}
F(a)=c_0-c_1{\rm cos}\big(\frac{2\pi a}{N_c-1}\big)+c_2{\rm cos}\big(\frac{4\pi a}{N_c-1}\big)-c_3{\rm cos}\big(\frac{6\pi a}{N_c-1}\big) \,
\label{eq:filter}
\end{equation} 
where $c_0=0.3635819,c_1=0.4891775,c_2=0.1365995$ and $c_3=0.0106411$.  In a
recent paper, \citet{chapman14}  
have compared different frequency window functions to  conclude that the
extended Blackman-Nuttall window is 
the best choice for recovering the HI power spectrum. For the present work we
have used the Blackman-Nuttall window as given by eq. (\ref{eq:filter})
above.  The left panel of
Figure \ref{fig:fig3} shows the frequency window function  for $256$ frequency
channels spanning a frequency bandwidth of ${\rm B_{bw}}=16 \, {\rm MHz}$
which corresponds to the values which we have used in our simulations (discussed later).

We now have 
\begin{equation}
v^f_i(\tau_m)=(\Delta \nu_c) \sum_a e^{2 \pi i \tau_m \nu_a} \,  F(\nu_a)
\V_i(\nu_a) \, 
\label{eq:c1}
\end{equation}
where $v^f_i(\tau_m)$ refer to  the delay space visibilities after
introducing the frequency window function.  The filtered delay space
visibilities  
$v^f_i(\tau_m)$ are related to the original delay space visibilities 
$v_i(\tau_m)$ (eq. (\ref{eq:b11})) through a convolution 
\begin{equation}
v^f_i(\tau_m)=\frac{1}{{\rm B_{bw}}} \sum_n   \tilde{f}(\tau_m-\tau_n)v_i(\tau_n) \,
\label{eq:c2}
\end{equation}
where $\tilde{f}(\tau)$ is  the Fourier transform of the frequency window
$F(\nu)$. Recollect that the delay space visibilities  $v_i(\tau_m)$ 
at the different $\tau_m$ are all independent and uncorrelated.
We however  see that this does not hold for the filtered delay space
visibilities $v^f_i(\tau_m)$  for which 
the different $\tau_m$ values are correlated,  the extent of this
correlation being  determined by the width of the function
$\tilde{f}(\tau_m-\tau_n)$ in eq. (\ref {eq:c2}). We now use this to calculate  
the correlation of $v^f_i(\tau_m)$  at two different values of $\tau_m$ for
which we have  
\begin{equation}
\langle v^f_i(\tau_m) v^{f*}_i(\tau_n)   \rangle = \frac{1}{{\rm B_{bw}}^2} \sum_a
\tilde{f}(\tau_m-\tau_a) \tilde{f}^{*}(\tau_n-\tau_a)  \langle 
\mid v_i(\tau_a) \mid^2  \rangle \,. 
\label{eq:c3a}
\end{equation}

This gives the self-correlation to be 
\begin{equation}
\langle \mid v^f_i(\tau_m) \mid^2  \rangle = \frac{1}{{\rm B_{bw}}^2} \sum_a
\mid \tilde{f}(\tau_m-\tau_a) \mid^2 \langle \mid v_i (\tau_a)\mid^2  \rangle \,.
\label{eq:c3}
\end{equation}

\begin{figure*}
\begin{center}
\psfrag{chan}[t][t][0.8][0]{Channel Number}
\psfrag{delay}[t][t][0.8][0]{Delay channel Number}
\psfrag{htau2}[b][b][0.8][0]{$\mid\tilde{f}(\tau)\mid^2$}
\psfrag{hnu}[c][c][0.8][0]{$F(\nu)$}
\includegraphics[width=75mm,angle=0]{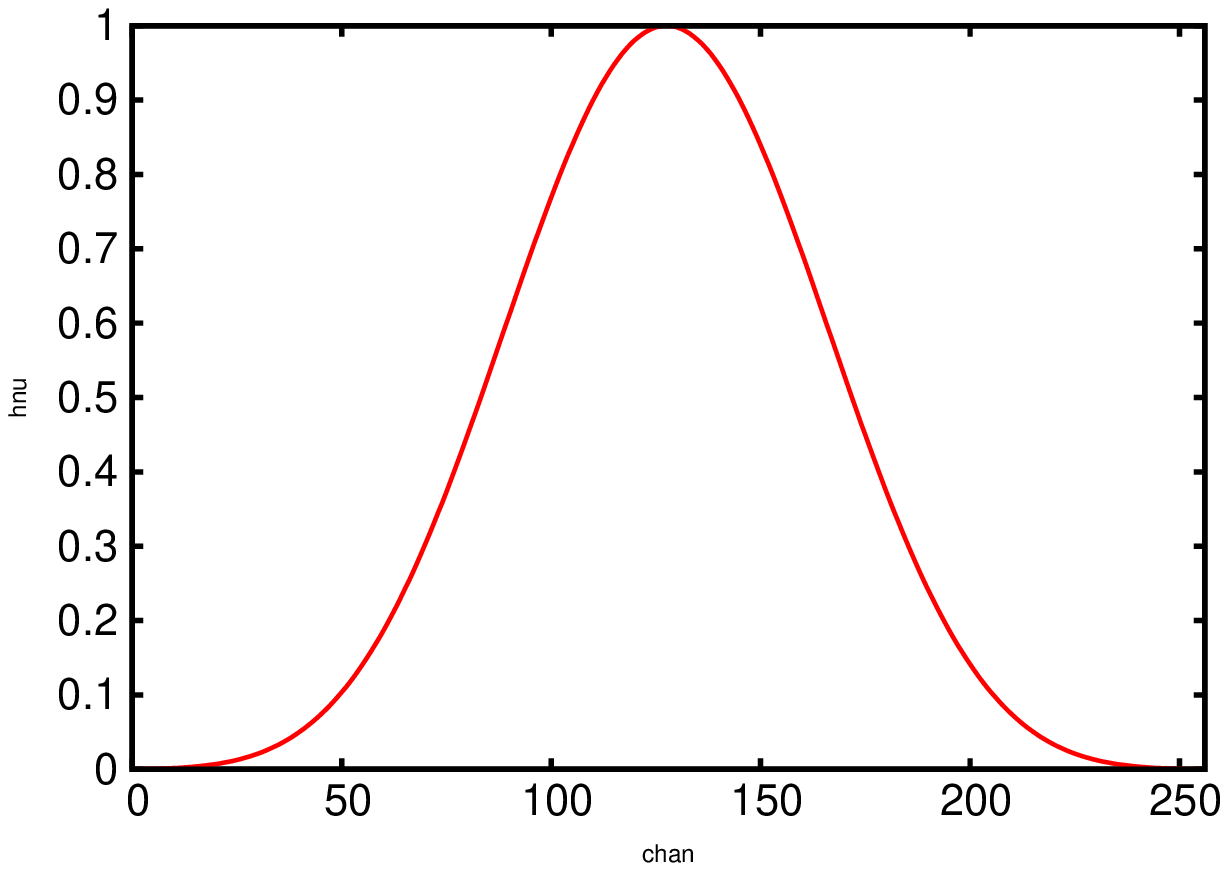}
\includegraphics[width=75mm,angle=0]{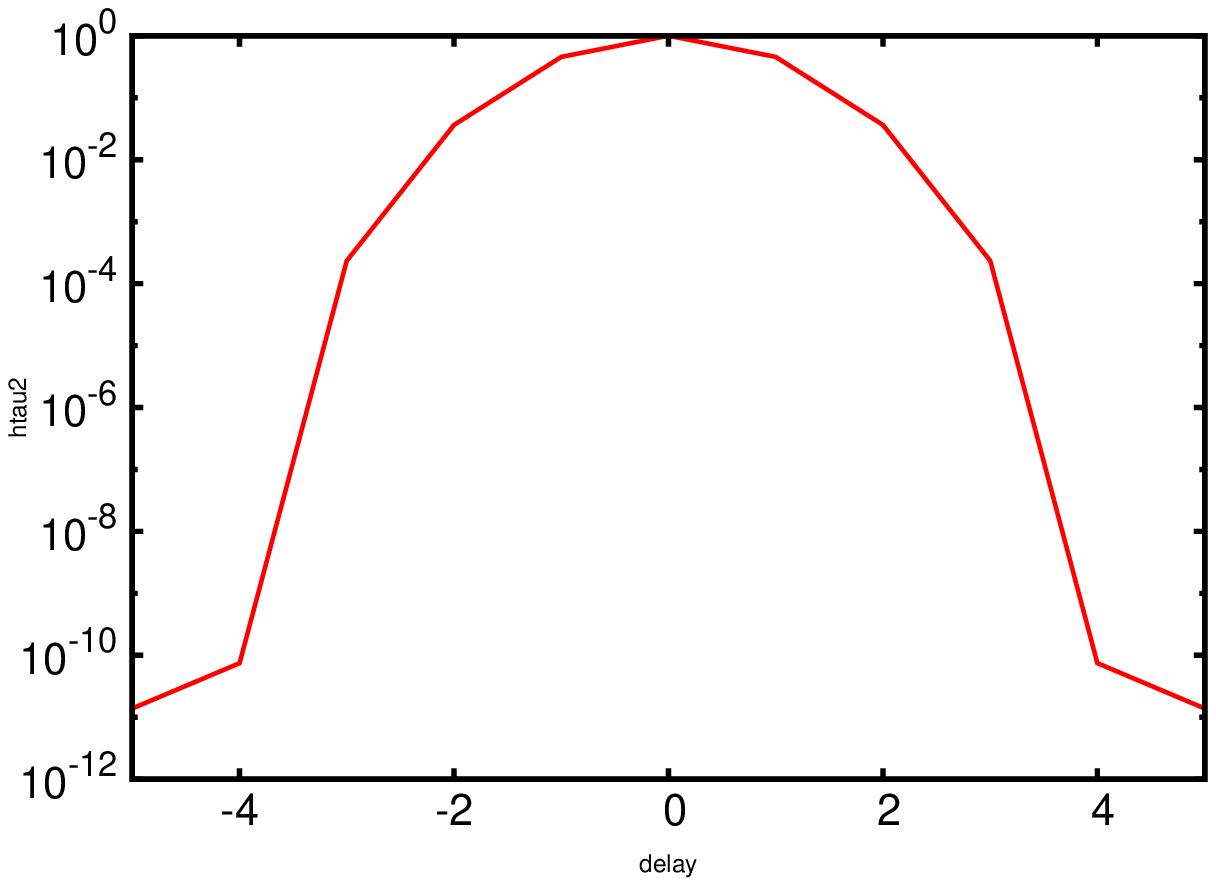}
\caption{The Blackman-Nuttall frequency window  $F(\nu)$ as a function of channel number is 
shown in the left panel. The right panel shows ($\mid\tilde{f}(\tau)\mid^2$)  which is 
the square of the Fourier transform of $F(\nu)$ . This is normalized to unity at the central delay channel.}  
\label{fig:fig3}
\end{center}
\end{figure*}

The right panel of Figure \ref {fig:fig3} show  $\mid\tilde{f}(\tau_m)\mid^2$
as a function of the delay channel number $m$. We see that
$\mid\tilde{f}(\tau_m)\mid^2$ has 
a    very narrow extent  in delay  space, implying that the visibilities 
 $v^f_i(\tau_m)$ in only three adjacent  delay channels are correlated, and  
 $v^f_i(\tau_m)$ are uncorrelated if the delay channel separation is larger
than this. This also allows us to approximate 
$\mid \tilde{f}(\tau_m-\tau_n) \mid^2$ using  a Kronecker delta function
$\approx {\rm B_{bw}^2} \, A_f(0) \, \delta_{m,n}$   where $A_f(0)= \frac{1}{\rm
  B_{bw}^2} \sum_n \mid \tilde{f}(\tau_n) \mid^2$. The convolution in
eq. (\ref{eq:c3}) now gives 
\begin{equation}
\langle \mid v^f_i(\tau_m) \mid^2  \rangle=A_f(0) \, \langle \mid v_i (\tau_m)\mid^2
\rangle \,. 
\label{eq:c4}
\end{equation}
We now generalize this to calculate the correlation for two different values
of $\tau_m$  which gives 
\begin{equation}
\langle v^f_i(\tau_m) v^{f*}_i(\tau_n)   \rangle =A_f(m-n) \, \langle \mid
v_i (\tau_m)\mid^2 \rangle 
\label{eq:c4a} 
\end{equation}
where 
\begin{equation}
A_f(m-n)=\frac{1}{\rm  B_{bw}^2} \sum_a  \tilde{f}(\tau_m-\tau_a) 
\tilde{f}^{*}(\tau_n-\tau_a)
\label{eq:c4b} 
\end{equation}
and $A_f(m-n)=A^{*}_f(n-m)$. We find that $A_f(m)$ has significant values only
for $m=0,1,2,3$ beyond which the values are rather small {\em i.e.} the
visibilities at only the three adjacent delay channels have significant
correlations, and the visibilities are uncorrelated beyond this separation. 
We have used the self-correlation (eq. \ref{eq:c4}) to calculate the power
spectrum estimator later in this subsection,  whereas the general expression for
the correlation (eq. \ref{eq:c4a}) comes in useful for calculating the
variance in a subsequent subsection.

Incorporating the frequency  window function in the  3D TGE introduces an 
additional factor of $A_f(0)$ in the normalization coefficient 
in  eq.  (\ref{eq:a16}).  We now have the final expression for the 3D TGE  
as
\begin{equation}
{\hat P}_g(\tau_m)= \left(\frac{M_g {\rm B_{bw}}\,A_f(0)}{r^2 r^{'}} \right)
^{-1} \, \left( \mid v^f_{cg}(\tau_m) \mid^2 -\sum_i \mid \tilde{w}(\u_g-\u_i)
\mid^2 \mid v^f_i(\tau_m) \mid^2 \right) \,.   
\label{eq:c5}
\end{equation}
As mentioned earlier,  ${\hat P}_g(\tau_m)$  gives an estimate of the power
spectrum  $P({\kpr}_g,{\kp}_m)$   where  ${\kp}_m$ 
and ${\kpr}_g$ are related to $\tau_m$ and $\u_g$ through eqs.   
(\ref{eq:b7}) and (\ref{eq:b8}) respectively.

\subsection{Binning and Variance}
\label{3dvar}
The  estimator ${\hat P}_g(\tau_m)$ presented in eq. (\ref{eq:c5}) provides an 
estimate of  the 3D power spectrum  $P({\kpr}_g,{\kp}_m)$  at an individual  
grid point $\k=({\kpr}_g,{\kp}_m)$ in  the three dimensional $\k$ space. 
Usually one would like to average  the estimated  power spectrum  over a bin
in $\k$ space  in order to increase the signal-to-noise ratio.   In this section we discuss the bin averaged 3D TGE and
obtain formulas for theoretically predicting the expected variance.

We  introduce   the binned 3D TGE which for the  bin labeled   $a$ is  
defined as   
\begin{equation}
{\hat P}_G (a)= \frac{\sum_{gm} w_{gm}  {\hat P}_g(\tau_m)}{\sum_{gm} w_{gm }}
\,
\label{eq:3dvar1}
\end{equation}
where the sum  is over all the $\k=({\kpr}_g,{\kp}_m)$  modes  or equivalently
the grid points ($\u_g$,$\tau_m$) included in the particular bin $a$, and   
$ w_{gm}$ is the weight assigned to  the contribution from any
particular grid point.  Earlier in this paper, in the discussion immediately
following eq. (\ref{eq:a15}),  we have introduced the weighing 
scheme $w_{g}=1$ in order to calculate $C_{\ell}$. Here we have adopted the 
same scheme $w_{gm}=1$  for estimating the 3D
power spectrum.  

The expectation value of the binned 3D TGE (eq.~\ref{eq:3dvar1}) 
\begin{equation}
\langle {\hat P}_G (a) \rangle = \bar{P}(\bar{k}_\perp,\bar{k}_\parallel)_a
\label{eq:3dvar1a}
\end{equation}
gives an
estimate of the bin averaged 3D power spectrum 
\begin{equation}
\bar{P}(\bar{k}_\perp,\bar{k}_\parallel)_a= \frac{\sum_{gm} w_{gm}  P({\kpr}_g,{\kp}_m)}{\sum_{gm} w_{gm }} \, 
\label{eq:3dvar2}
\end{equation}
at
\begin{equation}
(\bar{k}_\perp,\bar{k}_\parallel)_a=\Big(\frac{\sum_{gm} w_{gm}
    {\kpm}_g}{\sum_{gm} w_{gm }} ,\frac{\sum_{gm} w_{gm}  {\kp}_m}{\sum_{gm}
    w_{gm }}\,\Big). 
\label{eq:3dvar3}
\end{equation}
where for the particular bin $a$ the two components 
$(\bar{k}_\perp,\bar{k}_\parallel)_a$  refer to the  average  wave numbers 
respectively  perpendicular and parallel to the line of sight.  In this paper
we have considered two different binning schemes which we discuss later in this
sub-section. For the present, we turn our attention to calculate theoretical
predictions for the variance of the binned 3D TGE. 

The variance calculation closely follows the steps outlined in
section \ref{sec:clvar1}, and we  have the final expression 
\begin{equation}  
\sigma^2_{P_G} = \left(\frac{ {\rm B_{bw}}\,A_f(0)}{r^2 r^{'}} \right) ^{-2}
\frac{\sum_{gm,g^{'} m^{'}} \, w_{gm} w_{g^{'}m^{'}}  M_g^{-1} 
  M^{-1}_{g^{'}} \mid \langle v^f_{cg}(\tau_m) v^{f*}_{cg^{'}}(\tau_{m^{'}})
  \rangle \mid^2 }{[\sum_{gm} w_{gm}]^2} \,. 
\label{eq:3dvar4}
\end{equation}
which closely resembles eq. (\ref{eq:var2}) which we have used to calculate
the variance for $C_{\ell}$,  with the difference that we now 
have a 3D grid instead of the 2D grid encountered earlier for $C_{\ell}$. 

It is necessary to model the term 
$\langle v^f_{cg}(\tau_m) v^{f*}_{cg^{'}}(\tau_{m^{'}})
  \rangle $ in eq. (\ref{eq:3dvar4}) to make further progress.  
The correlation at two different $\tau_m$ values can be  expressed  
using eq. (\ref{eq:c4a}) as 
\begin{equation} 
 \langle v^f_{cg}(\tau_m) v^{f*}_{cg^{'}}(\tau_{m^{'}}) \rangle=A_f(m-m^{'})
 \langle v_{cg}(\tau_m) v^{*}_{cg^{'}}(\tau_{m}) \rangle \,.
\label{eq:3dvar5}
\end{equation}
Following eq. (\ref{eq:var2a}), we have decomposed 
the correlation $ \langle v_{cg}(\tau_m) v^{*}_{cg^{'}}(\tau_{m}) \rangle $
in eq. (\ref{eq:3dvar5})  into two parts 
\begin{equation}
\langle v_{cg}(\tau_m) v^{*}_{cg^{'}}(\tau_{m}) \rangle = \langle s_{cg}(\tau_m) s^{*}_{cg^{'}}(\tau_{m}) \rangle + \langle n_{cg}(\tau_m) n^{*}_{cg^{'}}(\tau_{m}) \rangle 
\label{eq:3dvar6}
\end{equation}
corresponding to the signal and the noise respectively. 

We have modeled  the signal correlation in exact analogy with eq. (\ref{eq:var3})
as 
\begin{equation}
\langle s_{cg}(\tau_m) s^{*}_{cg^{'}}(\tau_m) \rangle = \left(\frac{ {\rm
    B_{bw}}}{r^2 r^{'}} \right) \sqrt{M_g M_{g^{'}}} \, 
e^{-\mid \Delta  \u_{g g^{'}} \mid^2/\sigma_1^2}
\bar{P}(\bar{k}_\perp,\bar{k}_\parallel)_a  \,
\label{eq:3dvar7}
\end{equation}
and the noise correlation is similarly  modeled in exact analogy  with
eq. (\ref{eq:var5}) as 
\begin{equation}
\langle n_{cg}(\tau_m) n^{*}_{cg^{'}}(\tau_m) \rangle = (\Delta \nu_c){\rm
  B_{bw}}\sqrt{K_{2gg}K_{2g^{'}g^{'}}}e^{-   \mid \Delta \u_{g g^{'}}
  \mid^2/\sigma_2^2}(2\sigma_n^2)\,.  
\label{eq:3dvar8}
\end{equation}

We have used eqs. (\ref{eq:3dvar8}), (\ref{eq:3dvar7}), (\ref{eq:3dvar6}),
(\ref{eq:3dvar5}) and (\ref{eq:3dvar4}) to calculate the variance of the
binned 3D TGE. In the subsequent analysis  
we have considered two different binning schemes which we now present below. 
\begin{figure*}
\begin{center}
\psfrag{kx}[c][c][1.2][0]{${\bf k}_x$}
\psfrag{ky}[c][c][1.2][0]{${\bf k}_y$}
\psfrag{kpar}[c][c][1.2][0]{$\kp$}
\psfrag{ka}[c][c][1.2][0]{$k_a$}
\psfrag{dka}[c][c][1.2][0]{$\Delta {k}_a$}
\psfrag{kpara}[c][c][1.2][0]{${\kp}_a$}
\psfrag{dkpara}[c][c][1.2][0]{$\Delta {\kp}_a$}
\psfrag{kpera}[c][c][1.][0]{${\kpr}_a$}
\psfrag{dkpera}[r][r][1.][0]{$\Delta {\kpr}_a$}
\hspace*{-1.5cm}
\includegraphics[width=60mm,angle=0]{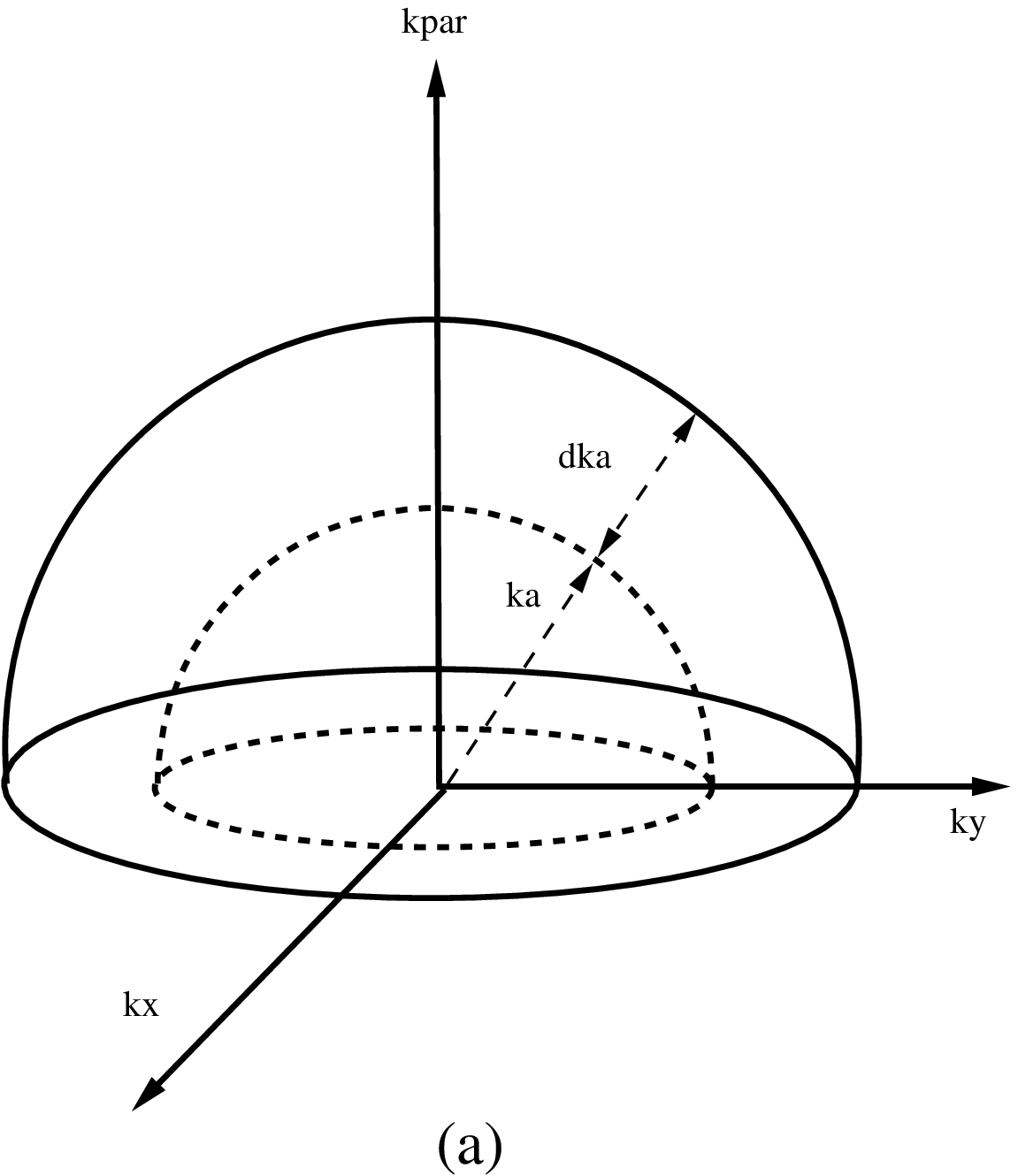}
\hspace*{2cm}
\includegraphics[width=60mm,angle=0]{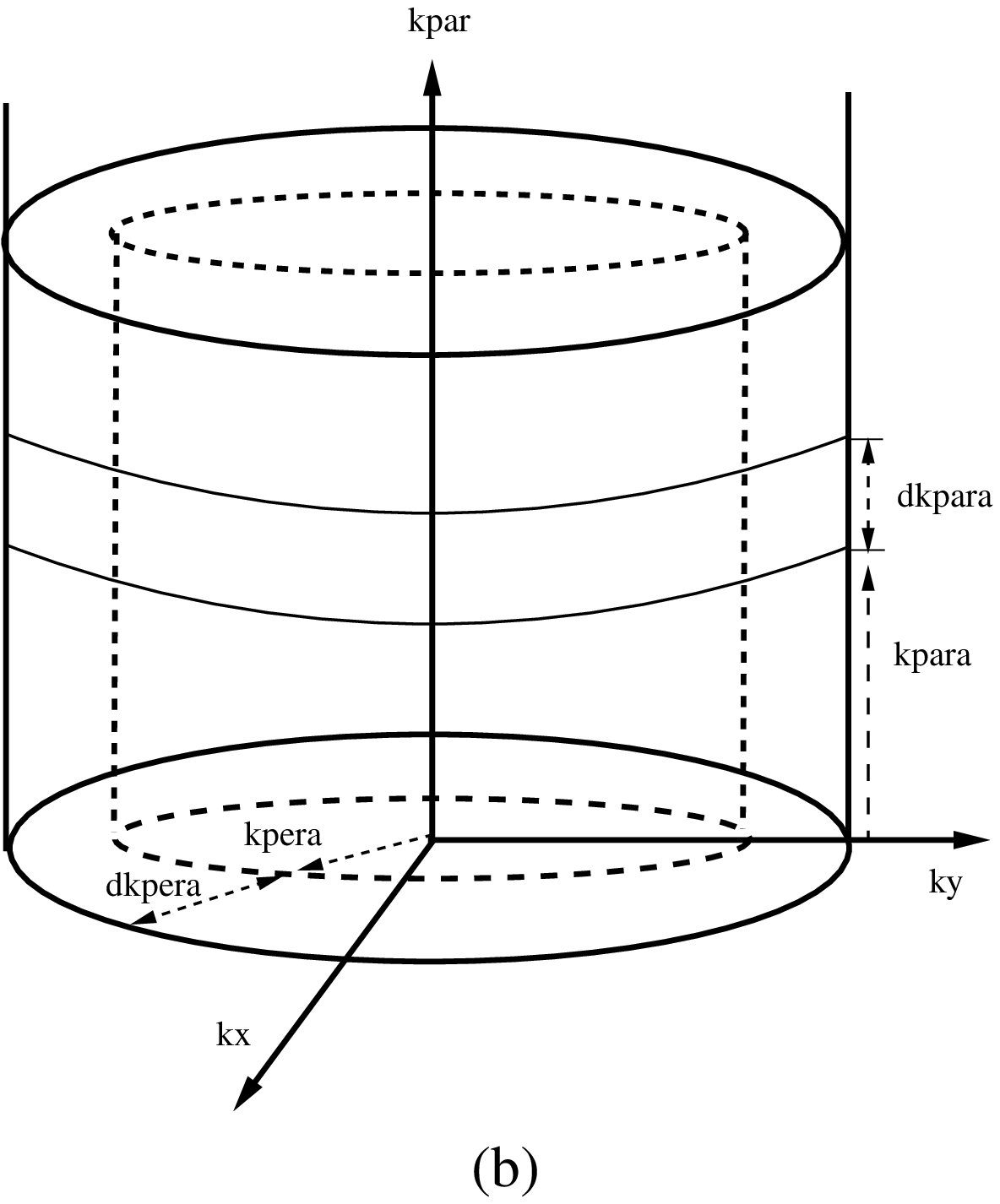}
\caption{This shows a typical bin for respectively calculating the Spherical Power Spectrum 
(left) and the Cylindrical Power Spectrum (right).}
\label{fig:bin}
\end{center}
\end{figure*}

\subsubsection{1D Spherical Power Spectrum}
The bins here are spherical shells of thickness $\Delta k_a$ as shown in the 
left panel of Figure \ref{fig:bin}, the shell thickness will in general vary
from bin to bin. The  Spherical Power Spectrum $\bar{P}(\bar{k}_a)$
is obtained by averaging the power spectrum $P(\k)$  over all the different
$\k$ modes which lie within the spherical shell corresponding to bin $a$ 
shown in the  left panel of Figure \ref{fig:bin}.  The  binning here
essentially averages out any anisotropy in the power spectrum, and yields the
bin averaged power spectrum as a function of the 1D bin averaged wave number
$\bar{k}_a$. While  
we use  eq. (\ref{eq:3dvar1}) to calculate the bin averaged power spectrum
$\bar{P}(\bar{k}_a)$, 
we have calculated the value of  $\bar{k}_a$ using 
\begin{equation}
 \bar{k}_a=\frac{\sum_{gm} w_{gm}    \sqrt{ {\kpm}_g^2 + {\kp}_m^2}}
     {\sum_{gm} w_{gm }} \,. 
\end{equation}

\subsubsection{2D Cylindrical Power Spectrum}
\label{sec:cylnbin}
Each bins here is, as shown in the  right panel of Figure \ref{fig:bin},   an
annulus of width $\Delta {\kpm}_a$ in the $\kpr \equiv (k_x,k_y)$ plane and it
subtends a thickness $\Delta {\kp}_a$ along the third direction $\kp$.  The
values of  $\Delta {\kpm}_a$ and $\Delta {\kp}_a$ will, in general, vary from bin
to bin. The bins here correspond to sections of a hollow cylinder, and the
resulting bin averaged power spectrum
$\bar{P}(\bar{k}_\perp,\bar{k}_\parallel)_a$ is referred to as the
Cylindrical Power Spectrum which   is defined on a 2D space 
$(\bar{k}_\perp,\bar{k}_\parallel)_a$  whose two components   refer to the
average  wave numbers  respectively  perpendicular and parallel to the line of
sight.  The binning of $P(\k)$ here does not assume that the signal is
statistically isotropic in the 3D space {\em i.e.} independent of the direction
of  $\k$. However, the signal  is assumed to be statistically
isotropic in the plane of the sky, and the binning in 
$\kpr$ is exactly identical to the binning that we
have used earlier for $C_{\ell}$. This distinction between $\kpm$ and $\kp$ is
useful to quantify the effect of redshift space distortion
\citep{bharadwaj01,bharadwaj011,bharadwaj044,barkana05,mao3,majumder13,jensen16} and also to distinguish the foregrounds
from the HI signal \citep{morales04}.  We have used  eq. (\ref{eq:3dvar1}) and
eq. (\ref{eq:3dvar3})   to calculate
$\bar{P}(\bar{k}_\perp,\bar{k}_\parallel)_a$ and
$(\bar{k}_\perp,\bar{k}_\parallel)_a$ respectively.

\section{Simulation}
\label{simu}
In this section we discuss the  simulations that we have used 
to validate the 3D power spectrum estimator (eq. \ref{eq:c5}). We start with
 an input model 3D power spectrum $P^M(k)$ of 
redshifted HI 21-cm brightness temperature fluctuations. 
The aim here is to test  how well the estimator is able to recover the
input model. For this purpose  the exact form of the input model power
spectrum  need not mimic the  expected cosmological HI signal, and we have
used a simple power law 
\begin{equation}
P^M(k)=\left(\frac{k}{k_0} \right)^n
\label{eq:pk1}
\end{equation}
which is arbitrarily normalized to unity at $k=k_0$, and has a power law index
$n$. In our analysis we have considered $n=-3$ and $-2$, and set 
$k_0=1 \, {\rm  Mpc}^{-1}$.  The quantity $\Delta_k^2=(2 \pi^2)^{-1} k^3 P(k)$
provides an estimate of the mean-square brightness  temperature 
fluctuations expected at different length-scales (or equivalently 
 wave numbers  $k$). We see that for $n=-3$ we have a constant 
 $\Delta_k^2= (2 \pi^2)^{-1}\,  {\rm K^2}$  across all  length-scales, whereas we have 
$\Delta_k^2= (2 \pi^2)^{-1} (k/1 \, {\rm Mpc}^{-1}) \, {\rm K^2}$ which increases
linearly with $k$ for $n=-2$. Note that we have used an isotropic  
input model where the power spectrum does not  depend on the direction of
$\k$ {\em i.e.}  $(P(\k) \equiv P(k))$ and  the 1D Spherical binning
and the 2D Cylindrical binning are expected to recover the same results.   

The simulations were carried out using a $N^3$ cubic grid of spacing $L$
covering a comoving volume $V$. We use the model power spectrum
(eq. \ref{eq:pk1}) to generate the Fourier components of the brightness
temperature fluctuations corresponding to this grid 
\begin{equation}
\Delta {\tilde T}(\mathbf{k}) = \sqrt{\frac{V P^M(k)}{2}} [a(\mathbf{k})+\mathit{i}
  b(\mathbf{k})] \,,
\label{eq:pk2}
\end{equation}
here $a(\mathbf{k})$ and $b(\mathbf{k})$ are two real valued independent
Gaussian random variable of unit variance.  The Fourier transform of  $\Delta
T(\mathbf{k})$ yields a single  realization of the brightness temperature
fluctuations $\delta T(\x)$ on the simulation grid. These fluctuations are, by
construction, a Gaussian random field with power spectrum $P^M(k)$. We
generate different statistically independent realizations of  $\delta T(\x)$
by using different sets of random variables  $a(\mathbf{k})$ and
$b(\mathbf{k})$ in eq. (\ref{eq:pk2}). 

The intention here is to simulate $150 \, {\rm MHz}$ GMRT observations with 
$N_c=256$ frequency channels of width $ (\Delta \nu_c)=62.5 \, {\rm kHz}$
covering a bandwidth of ${\rm  B_{bw}}=16 \, {\rm MHz}$. This corresponds to
HI at redshift  $z=8.47$ with  a comoving distance of 
$r = 9.28 \, {\rm  Gpc}$  and $r^{'}=\mid dr/d \nu \mid= 17.16 \,  {\rm Mpc
  \,  MHz}^{-1}$.  We have chosen the grid spacing  $L=1.073 \, {\rm Mpc}$  so
that it exactly matches the   channel width  $L =    r_{\nu}' \times (\Delta
\nu_c)$.  We have considered a $N^3=[2048]^3$ grid which corresponds to a
comoving volume of $[2197.5 \, {\rm Mpc}]^3$. The simulation volume is aligned
with the $z$ axis along the line of sight, and  the two transverse directions were
converted to angles relative to the box center
$(\theta_x,\theta_y)=(x/r,y/r)$.   The transverse extent of the 
simulation box covers an angular extent which is $\sim 5$ times the GMRT
$\theta_{FWHM}$.  The simulation volume corresponds to a frequency width $\sim
8 \times 16 \, {\rm MHz}$ along the line of sight. We have cut the box into $8$
equal segments along the line of sight to produce $8$ independent realizations
each subtending $16 \, {\rm MHz}$ along the line of sight. The grid index,
measured from the further boundary and increasing towards to observer along
the line of sight was directly converted to channel number $\nu_a$ with
$a=0,1,2,...,N_c-1$.  This procedure provides us with $\delta T(\th,\nu_a)$
the brightness temperature fluctuation on the sky at different frequency
channels $\nu_a$. 

We have considered $8$ hours of  GMRT observations with $16 \, {\rm s}$
integration time targeted on an arbitrarily selected  field  located at
RA=$10{\rm h} \, 46{\rm  m} \, 00{\rm s}$ and DEC=$59^{\circ} \,  00^{'} \,
59^{''}$.  Visibilities were calculated for the simulated baselines
corresponding to this observation, for which the $uv$ coverage is similar to the 
Figure 5 of Paper I. The signal contribution to the visibilities
$\S(\u,\nu_a)$  was calculated by taking the Fourier transform of the product
$\left(\frac{\partial B}{\partial T}\right)  \times 
{\mathcal A}(\th,\nu_a) \times \delta T(\th,\nu_a) $ as given by eq. (\ref{eq:b2a}).  
The simulations incorporate the fact that   the baseline corresponding to a
fixed antenna separation $\u_i={\bf d}_i/\lambda$, the antenna beam pattern
${\mathcal A}(\th,\nu_a)$    and the factor $\left( \frac{\partial B}{\partial
  T} \right)_{\nu_a}$ all vary with the frequency $\nu_a$ in
eq. (\ref{eq:b2a}). We have $\sigma_n= 1.45\, {\rm Jy}$ corresponding to a single
polarization, with $\Delta t=16 \, {\rm s}$ and $ (\Delta \nu_c)=62.5 \,
{\rm kHz}$. However, it is possible to reduce noise level by averaging independent data set
observed at different time. Here, we consider a situation where we average $9$ independent data sets 
to reduce the noise level by a factor of  $3$ to $\sigma_n= 0.48\, {\rm Jy}$. We have carried out  
the  simulations for 
two different cases, (i) no noise ($\sigma_n= 0\, {\rm Jy}$) and (ii) $\sigma_n= 0.48\, {\rm Jy}$.  We have  
carried out  $16$ independent  realization of the simulated visibilities to estimate the mean 
power spectrum and its statistical fluctuation (or standard deviation $\sige$)
presented in the next section.

\section{Results}
\label{result}
 The left panels of Figures \ref{fig:fig5} and \ref{fig:fig6} show
 $\Delta_k^2=(2 \pi^2)^{-1} k^3 P(k)$ for the spherically-averaged
 power spectrum for the power law index values $n=-3$ and $-2$
 respectively.    The results are shown for the three
 values $f=10 ,2$ and $0.6$ to demonstrate the effect of varying the tapering.
 The simulations here do not include the system noise contribution.
 For both $n=-3$ and $-2$, and for all the values of $f$ we find that
 $\Delta_k^2$ estimated using the 3D TGE is within the
 $1-\sigma_{P_G}$ error bars of the model prediction for the entire
 $k$ range considered here. The right panels of Figures \ref{fig:fig5}
 and \ref{fig:fig6} show the corresponding fractional deviations
 $(P(k)-P^M(k)) /P^M(k)$. For comparison, the relative statistical
 fluctuations, $\sigma_{P_G}/P^M(k)$ are also shown by shaded regions
 for different values of $f$. We find that for both cases $n=-3$ and
 $-2$, the fractional deviation is less than $4\%$ at $k > 0.2\,\rm
 {Mpc}^{-1}$. The fractional deviation increases as we go to lower
 $k$ bins. The fractional deviation also increases if the value of
 $f$ is reduced. The maximum fractional deviation has a value $\sim
 40\%$ and $\sim 20\%$ at the smallest $k$ bin for $n=-3$ and $-2$
 respectively. We find that the fractional deviation is within
 $\sigma_{P_G}/P^M(k)$ for $k \le 0.3\,\rm {Mpc}^{-1}$ and is slightly
 larger than $\sigma_{P_G}/P^M(k)$ for $k \ge 0.3\,\rm {Mpc}^{-1}$. Our results 
indicate  that the 3D TGE is able to recover the model power spectrum to a
reasonably good  level of accuracy ($ \le  20 \%)$ at  the $k$ modes 
$k \ge  0.1\,\rm {Mpc}^{-1}$. The fractional error at the smaller $k$ bins 
increases as the tapering is increased ($f$ is reduced). It may be noted that 
a similar behaviour was also found for $C_{\ell}$ (Figure \ref{fig:fig1}). 
As mentioned earlier,  we attribute this discrepancy to the variation of 
signal amplitude within the width of the convolving window 
$\tilde{w}(\u_g-\u_i)$. This explanation is further substantiated by the 
fact that 
the fractional deviation is found to be larger for $n=-3$, for which the power
spectrum is steeper compared to $n=-2$.

\begin{figure*}
\begin{center}
\psfrag{k}[b][b][0.8][0]{$k$ [$\rm {Mpc}^{-1}$]}
\psfrag{k3pk}[c][c][0.8][0]{$\Delta_k^2\,[\rm K^2]$}
\psfrag{model}[r][r][0.8][0]{Model}
\psfrag{tap10}[r][r][0.8][0]{f=10}
\psfrag{tap2}[r][r][0.8][0]{f=2}
\psfrag{tap0.6}[r][r][0.8][0]{f=0.6}
\psfrag{diffpk}[b][b][0.8][0]{$(P(k)-P^M(k)) /P^M(k),\sigma_{P_G}/P^M(k)$}
\includegraphics[width=75mm,angle=0]{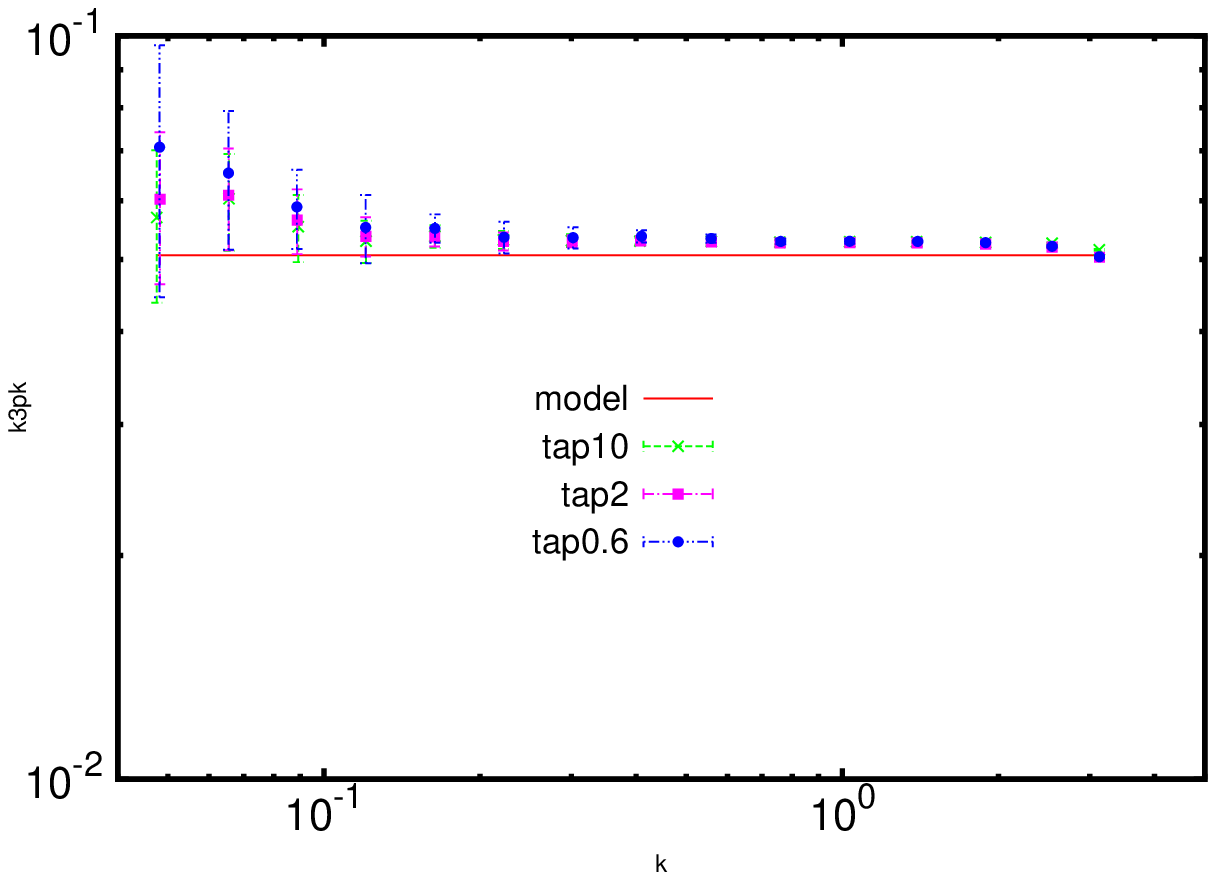}
\includegraphics[width=75mm,angle=0]{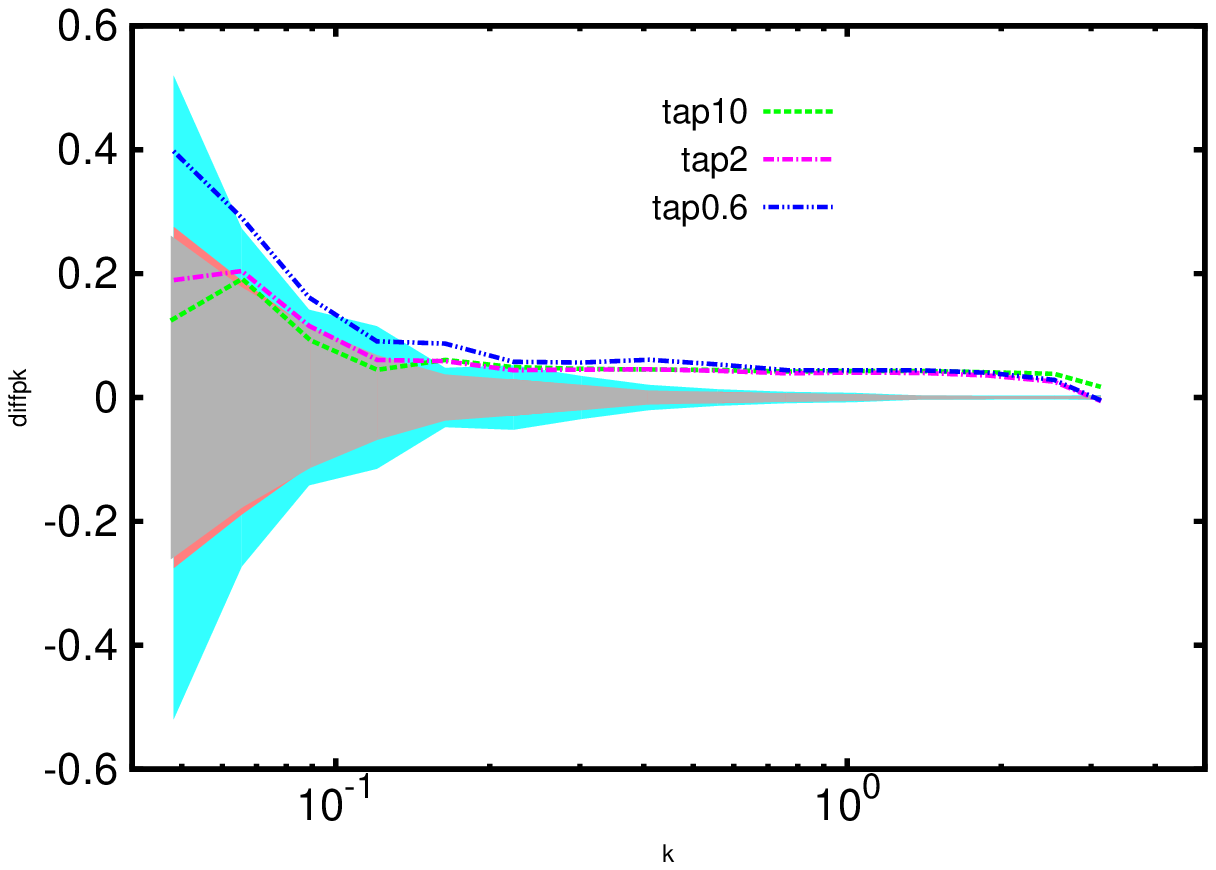}
\caption{The left panel shows the dimensionless power spectrum $\Delta_k^2$ for different values of $f$. The values obtained using the 3D TGE are compared with model power spectrum for $n=-3$ and $\sigma_n= 0 $. The 1-$\sigma_{P_G}$ error bars have been estimated using 16 different realizations of the simulated visibilities. The right panel shows the fractional deviation of estimated power spectrum, $(P(k)-P^M(k)) /P^M(k)$ relative to the input model  $P^M(k)$ for different values of $f$. The relative statistical fluctuations $\sigma_{P_G}/P^M(k)$ are also shown by shaded regions.}
\label{fig:fig5}
\end{center}
\end{figure*}
  
\begin{figure*}
\begin{center}
\psfrag{k}[b][b][0.8][0]{$k$ [$\rm {Mpc}^{-1}$]}
\psfrag{k3pk}[b][c][0.8][0]{$\Delta_k^2\,[\rm K^2]$}
\psfrag{model}[r][r][0.8][0]{Model}
\psfrag{tap10}[r][r][0.8][0]{f=10}
\psfrag{tap2}[r][r][0.8][0]{f=2}
\psfrag{tap0.6}[r][r][0.8][0]{f=0.6}
\psfrag{diffpk}[b][b][0.8][0]{$(P(k)-P^M(k)) /P^M(k),\sigma_{P_G}/P^M(k)$}
\includegraphics[width=75mm,angle=0]{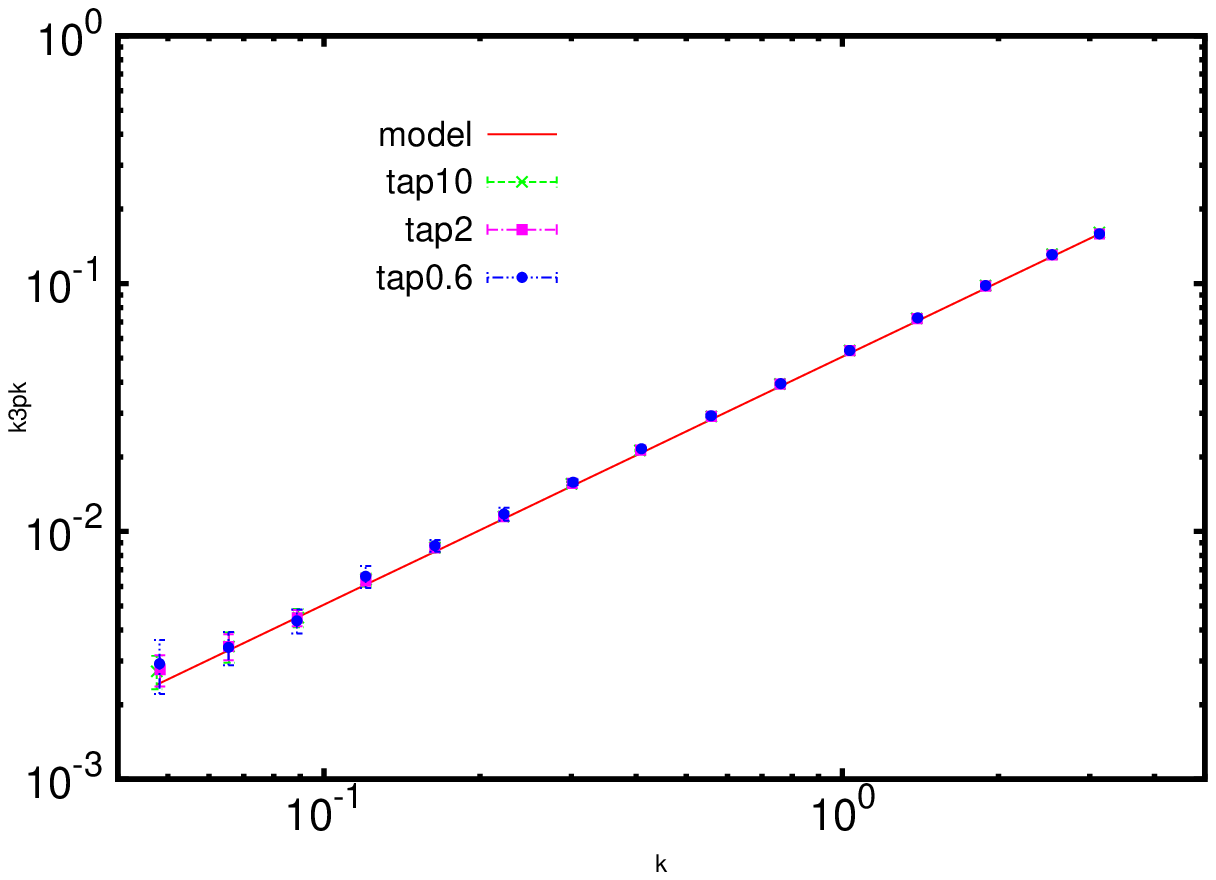}
\includegraphics[width=75mm,angle=0]{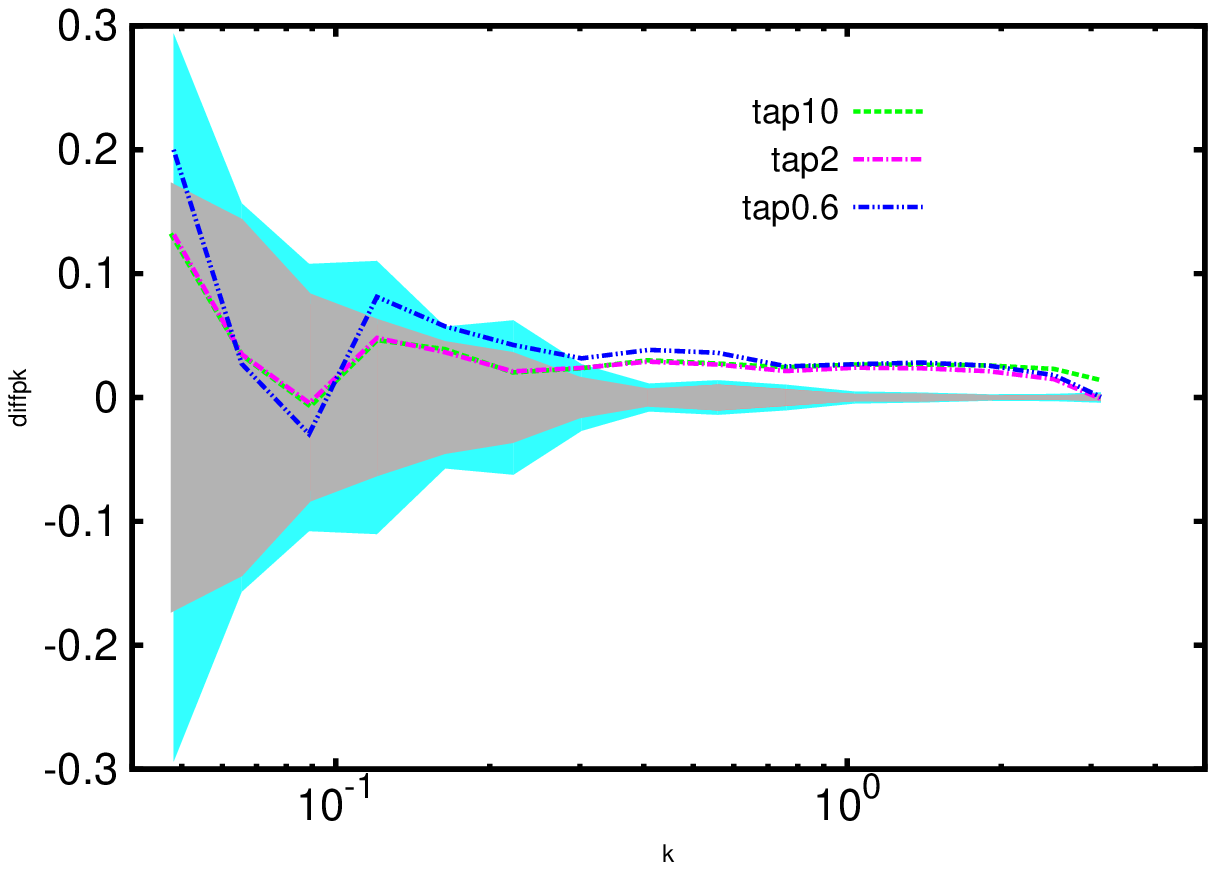}
\caption{Same as Figure~\ref{fig:fig5}, but with $n=-2$.}
\label{fig:fig6}
\end{center}
\end{figure*}

\begin{figure*}
\begin{center}
\psfrag{k}[b][b][0.8][0]{$k$ [$\rm {Mpc}^{-1}$]}
\psfrag{k3pk}[b][c][0.8][0]{$\Delta_k^2\,[\rm K^2]$}
\psfrag{model}[r][r][0.8][0]{Model}
\psfrag{nonse}[r][r][0.8][0]{No Noise}
\psfrag{nse}[r][r][0.8][0]{$\sigma_n=0.48\,{\rm Jy}$}
\psfrag{tap0.6}[r][r][0.8][0]{f=0.6}
\includegraphics[width=75mm,angle=0]{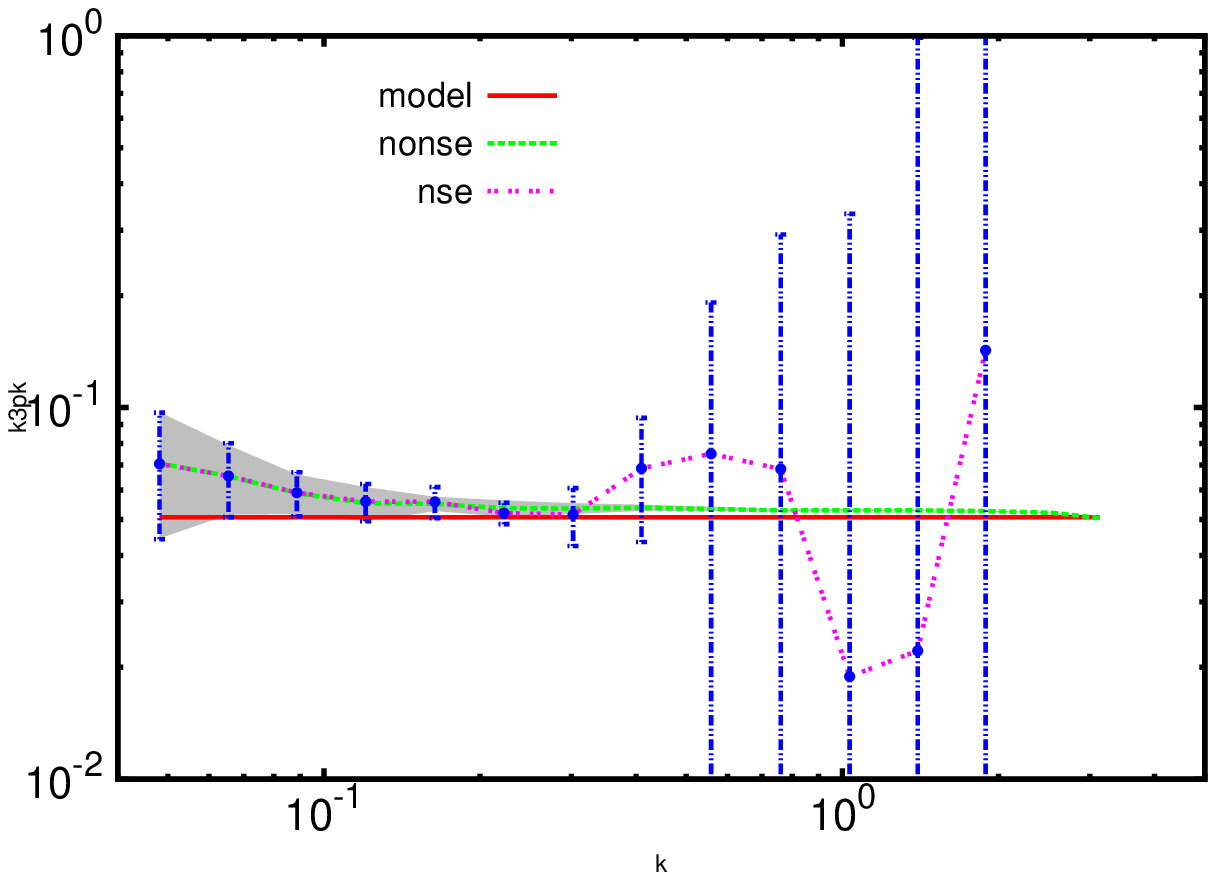}
\includegraphics[width=75mm,angle=0]{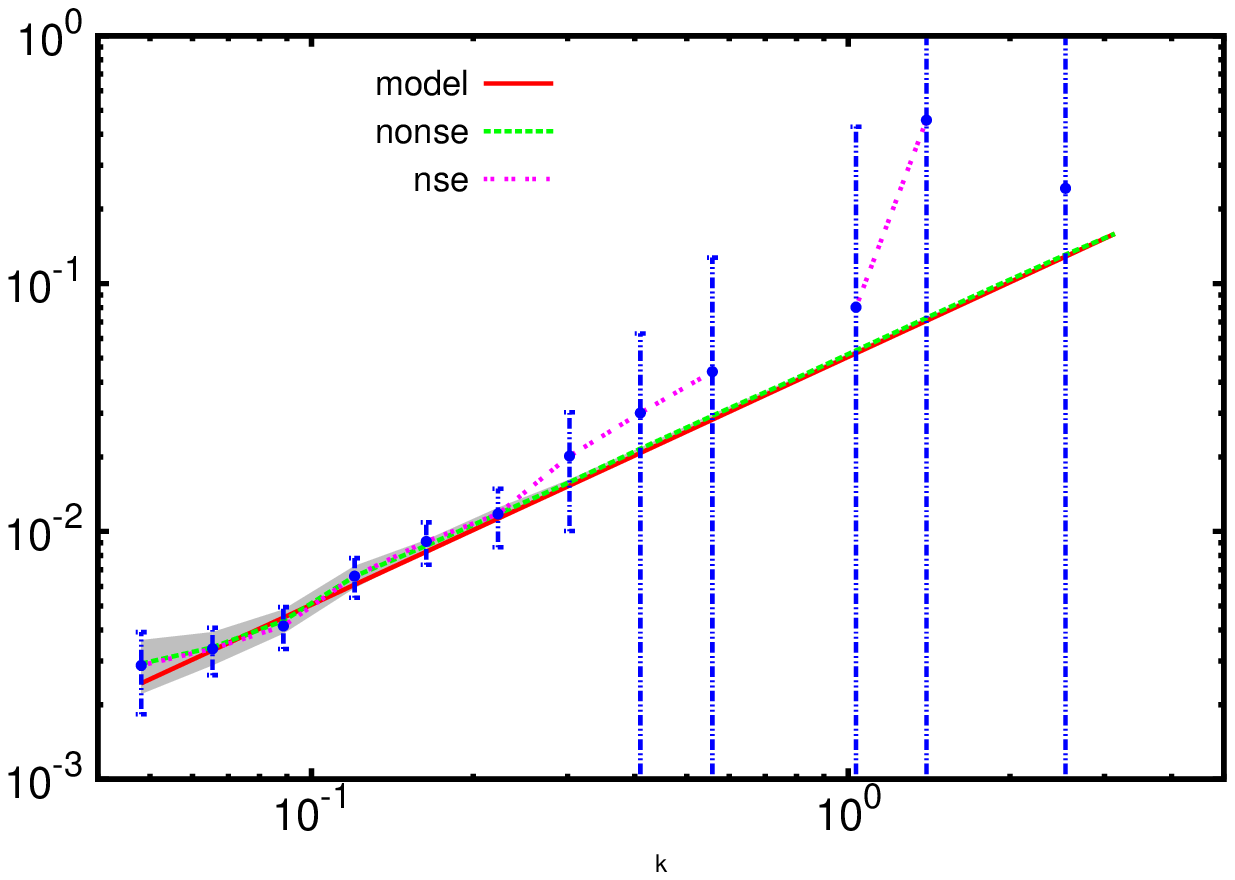}
\caption{The recovered dimensionless power spectrum $\Delta_k^2$ for
  $n=-3$ (left) and $n=-2$ (right), with and without noise for a fixed
  value $f=0.6$.  The statistical error (1-$\sige$) with (without)
  noise is shown with error bars (shaded region). Note that, the estimated
$\Delta_k^2$ has negative values at some of the $k$ values in the range where noise dominates the signal. These data points have not been displayed here.}
\label{fig:fig7}
\end{center}
\end{figure*}

The results until now have not considered the effect of system
noise. We now study how well the 3D TGE is able to recover the input
power spectrum in the presence of system noise. The left and right
panels of Figure \ref{fig:fig7} show the estimated $\Delta_k^2$ for
$n=-3$ and $-2$ respectively for the fixed value $f=0.6$. For
comparison, we also show the estimated $\Delta_k^2$ with $\sigma_n= 0
$. The statistical fluctuations with (without) noise are shown as error
bars (shaded region). We see that the error is dominated by the cosmic
variance at lower values of $k$ ($k < 0.2\,{\rm {Mpc}^{-1}}$) and the
system noise dominates at larger values of $k$. The statistical error
exceeds the model power spectrum at large $k$ and a statistically
significant estimate of the power spectrum is not possible in this $k$
range. We are able to recover the model power spectrum quite
accurately at low $k$ where $\sigma_{P_G} \le P^M(k)$.

We now investigate how well the analytic prediction
(eq. \ref{eq:3dvar4}) for $\sige$ compares with the values obtained
from the simulations (Figure \ref{fig:fig8} ) for different values of
$f$.  The number of grid points in each $k$ bin increase with the
value of $k$, and the computation time also increases with increasing
$k$.  We have restricted the $k$ range to $(k < 0.4\,{\rm{Mpc}^{-1}})$
in order to keep the computational requirements within manageable
limits. In the left panel we consider the situation where there is no
system noise. Here, the statistical fluctuations correspond to
the cosmic variance.  We see that the analytic predictions are in
reasonably good agreement with the simulation for both the values of
$f$. We find that the cosmic variance does not change if the value of
$f$ is changed from $2$ to $10$. As expected, the cosmic variance
increases as the sky tapering is increased.  The right panel shows the
statistical fluctuations with and without noise for the fixed value
$f=0.6$. The statistical fluctuations are dominated by the cosmic
variance at small values of $k$ $(k<0.2\,{\rm {Mpc}^{-1}})$, and the
system noise dominates at large $k$.  As mentioned earlier, the
statistical fluctuations are well modeled by the analytic predictions
in the cosmic variance dominated regime. We find that our analytic
prediction somewhat overestimates $\sige$ in the noise dominated
region.  This overestimate possibly originates from the noise
modelling in eq. (\ref{eq:3dvar4}), we plan to investigate this in
future work.

\begin{figure*}
\begin{center}
\psfrag{k}[b][b][0.8][0]{$k$ [$\rm {Mpc}^{-1}$]}
\psfrag{sigk}[b][c][0.8][0]{$k^{3}\sigma_{P_G}/2\pi^2\,[\rm K^2]$}
\psfrag{sim0.6}[r][r][0.8][0]{Simulation}
\psfrag{ana0.6}[r][r][0.8][0]{Analytic}
\psfrag{tap2}[r][r][0.8][0]{f=2, Simulation}
\psfrag{var2}[r][r][0.8][0]{Analytic}
\psfrag{tap0.6}[r][r][0.8][0]{f=0.6, Simulation}
\psfrag{var0.6}[r][r][0.8][0]{Analytic}
\psfrag{f0.6}[r][r][0.8][0]{f=0.6}
\includegraphics[width=75mm,angle=0]{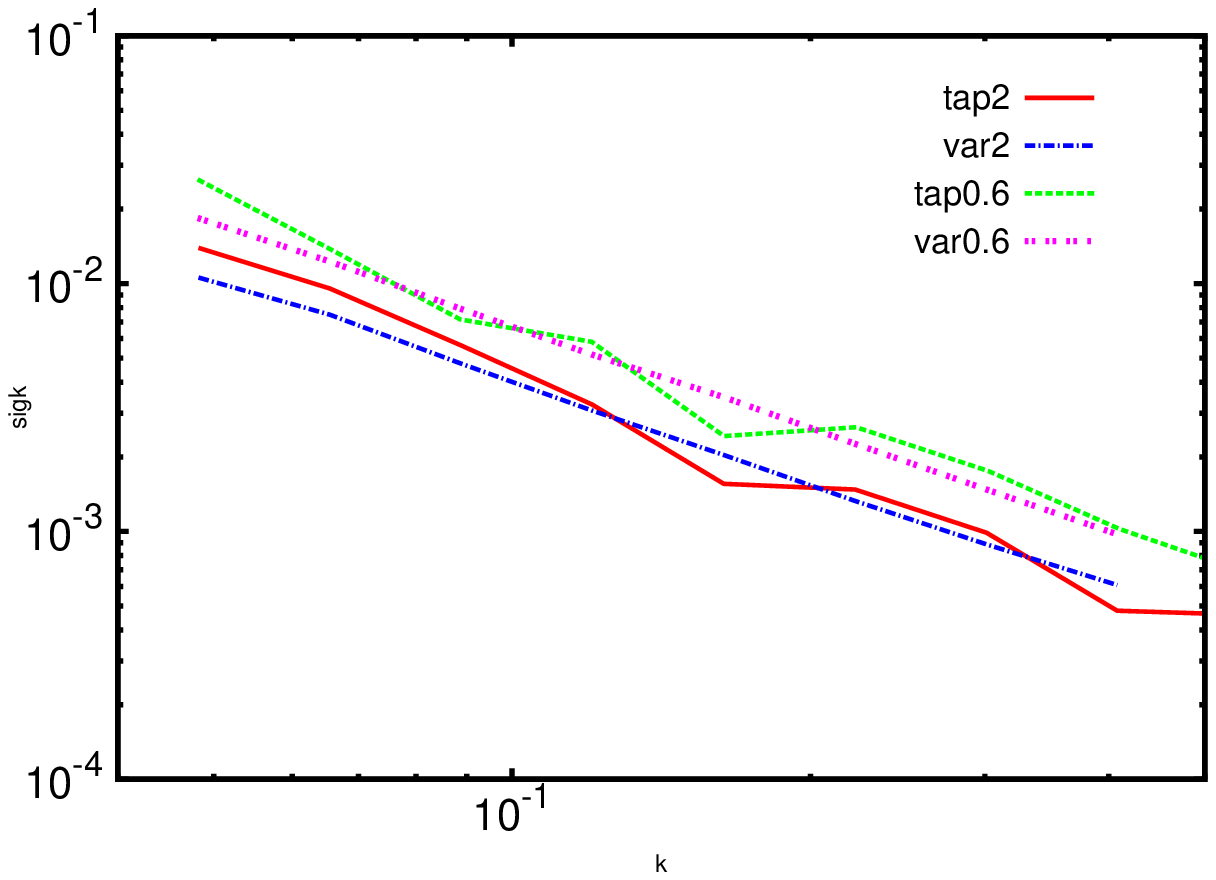}
\includegraphics[width=75mm,angle=0]{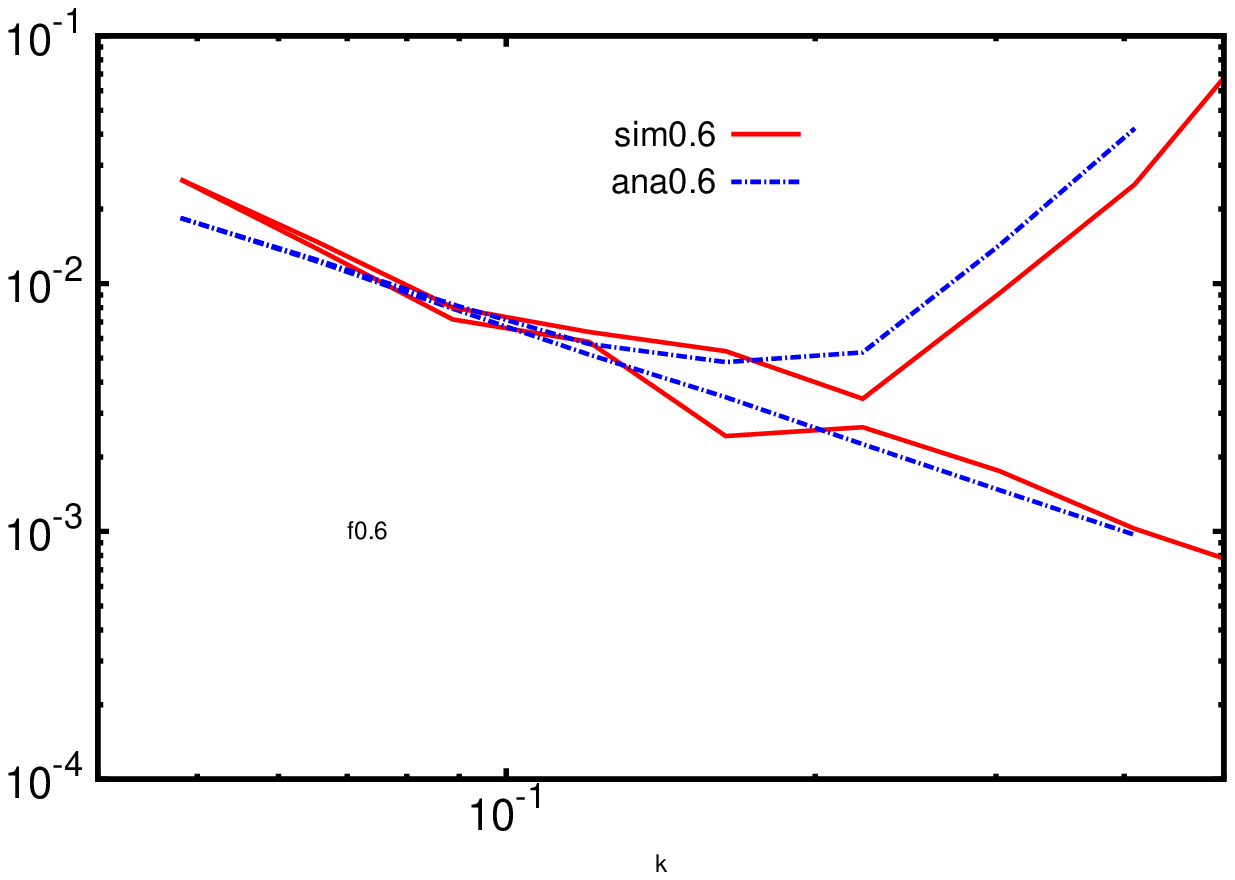}
\caption{The left panel shows a comparison of the analytic prediction
  for the statistical fluctuations of the power spectrum
  (eq. \ref{eq:3dvar4}) with the simulation for two different values
  of $f$, $n=-3$ and no system noise. The right panel shows the same
  comparison with (upper two curves) and without (lower two curves) noise for a fixed value $f=0.6$.}
\label{fig:fig8}
\end{center}
\end{figure*}

Till now we have discussed the results for the 1D Spherical Power
Spectrum, we now present the results for the 2D Cylindrical Power
Spectrum. We use $15$ equally spaced logarithmic bin in both $k_\perp$
and $k_\parallel$ direction to estimate the 2D Cylindrical Power
Spectrum. Figure \ref{fig:fig9} shows the 2D Cylindrical Power
Spectrum $P(k_\perp,k_\parallel)$ using 3D TGE. The left panel shows
the input model for $n=-3$. The middle and right panel respectively
show the estimated power spectrum with $f=0.6$ for situations where
the system noise is not included and included in the simulated
visibilities. The left and middle panels appear almost identical, 
indicating that the 3D TGE is able to recover the input model power
spectrum accurately across the entire $(k_\perp,k_\parallel)$ range.
We find that we are able to recover the model power spectrum in the
limited range $k_\perp \lsim \, 0.5\, \rm {Mpc}^{-1}$ and $k_\parallel
\lsim \, 0.5\, \rm {Mpc}^{-1}$ in presence of system noise.
  Figure \ref{fig:fig9a} shows the fractional deviation
  $(P^M(k_\perp,k_\parallel)-P(k_\perp,k_\parallel))/P(k_\perp,k_\parallel)$
 for $f=0.6$,  here the left and right panels show the results 
without and with system noise  respectively. From the left panel
 we see that the fractional 
  deviation is less than $14\%$ for the the entire $\mathbf{k}$ range
  when the system noise is not included in the simulation. We find that 
it is not possible to reliably recover the power spectrum at large 
$\mathbf{k}$ when the system noise is included. The right panel 
shows the fractional deviation only where it is within $30 \%$. The 
fractional deviation $100 \%$ at large $\mathbf{k}$, and these values have not been shown.

\begin{figure*}
\begin{center}
\psfrag{kpara}[h][h][0.8][0]{$\kp [\rm {Mpc}^{-1}]$}
\psfrag{kper}[b][b][0.8][0]{$\kpm [\rm {Mpc}^{-1}]$}
\psfrag{model}[r][r][0.8][0]{}
\psfrag{k2mpc3}[t][c][0.8][0]{log($P(k_\perp,k_\parallel)$)}
\hspace*{-3.0cm}
\includegraphics[width=75mm,height=60mm,angle=0]{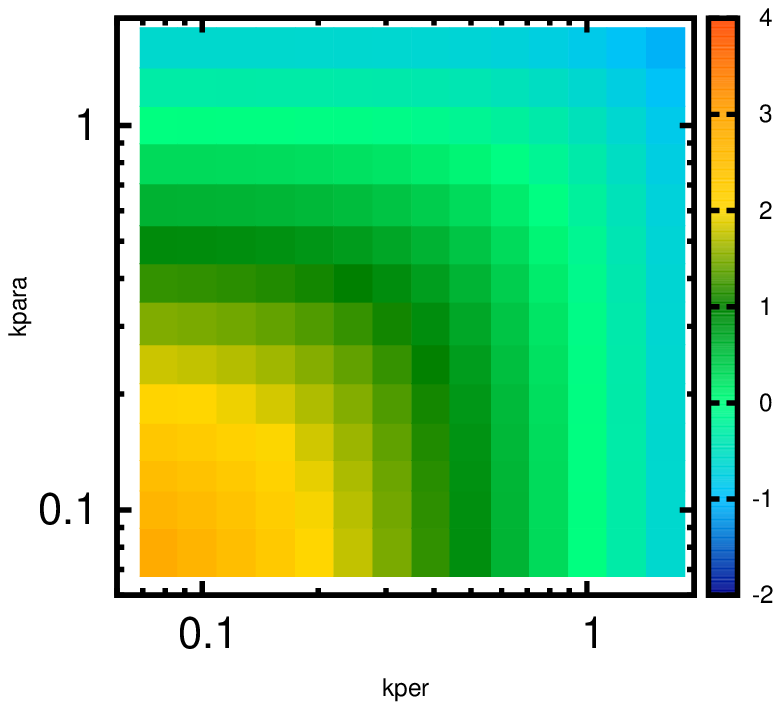}
\hspace*{-2.5cm}
\includegraphics[width=75mm,height=60mm,angle=0]{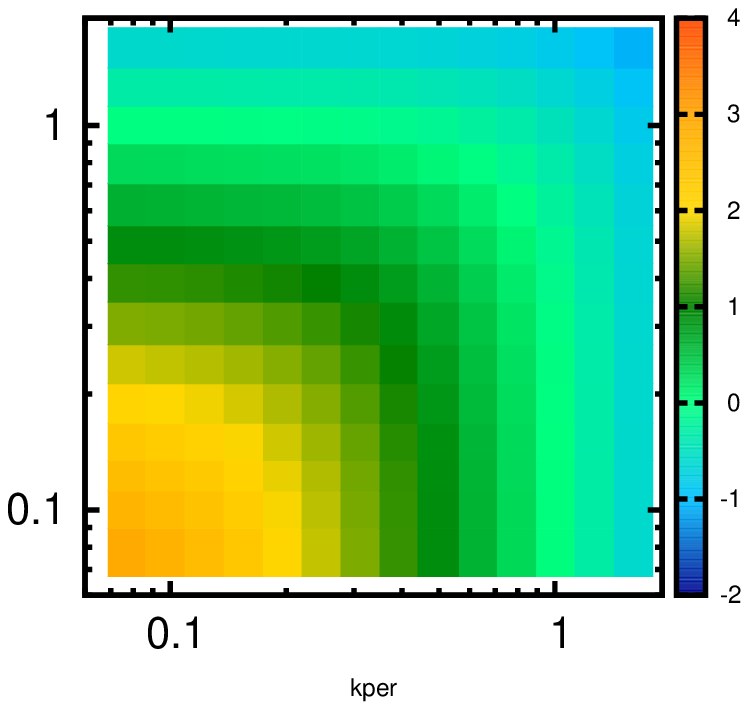}
\hspace*{-2.5cm}
\includegraphics[width=75mm,height=60mm,angle=0]{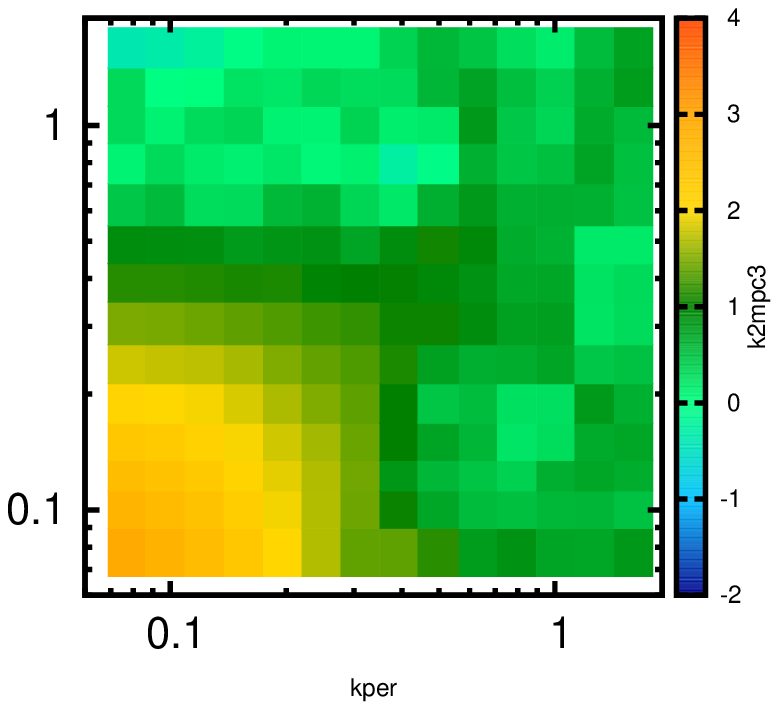}
\caption{This shows the 2D Cylindrical Power Spectrum for
  $n=-3$. The left panel shows the input model power spectrum.  The
  middle and right panels show the estimated power spectrum for
  $f=0.6$ without and with noise respectively.}
\label{fig:fig9}
\end{center}
\end{figure*}

\begin{figure*}
\begin{center}
\psfrag{kpara}[h][h][0.8][0]{$\kp [\rm {Mpc}^{-1}]$}
\psfrag{kper}[b][b][0.8][0]{$\kpm [\rm {Mpc}^{-1}]$}
\psfrag{k2mpc3}[t][t][0.8][0]{$\rm K^2{Mpc}^3$}
\hspace*{-3.0cm}
\includegraphics[width=80mm,height=60mm,angle=0]{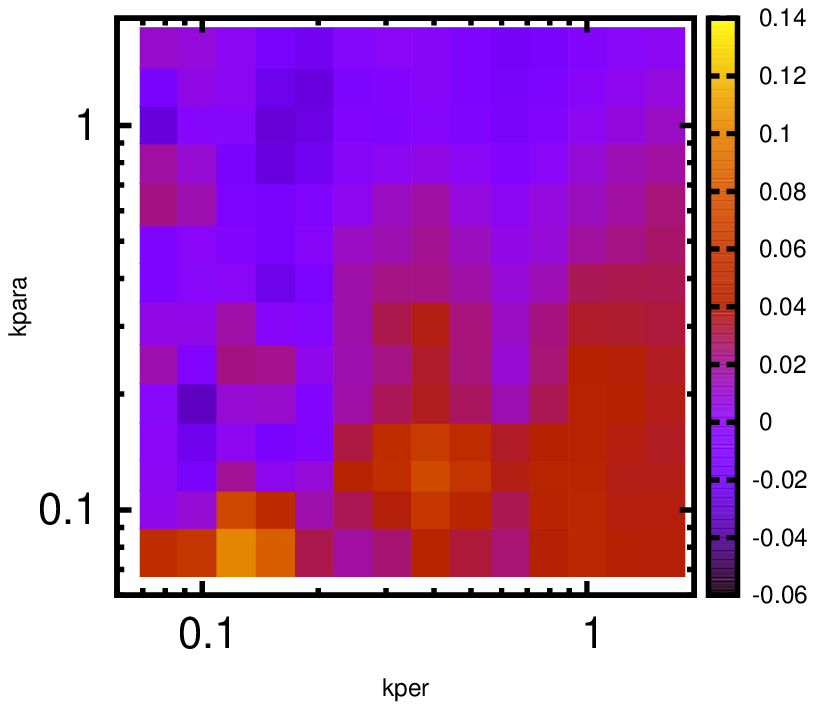}
\hspace*{-2.5cm}
\includegraphics[width=80mm,height=60mm,angle=0]{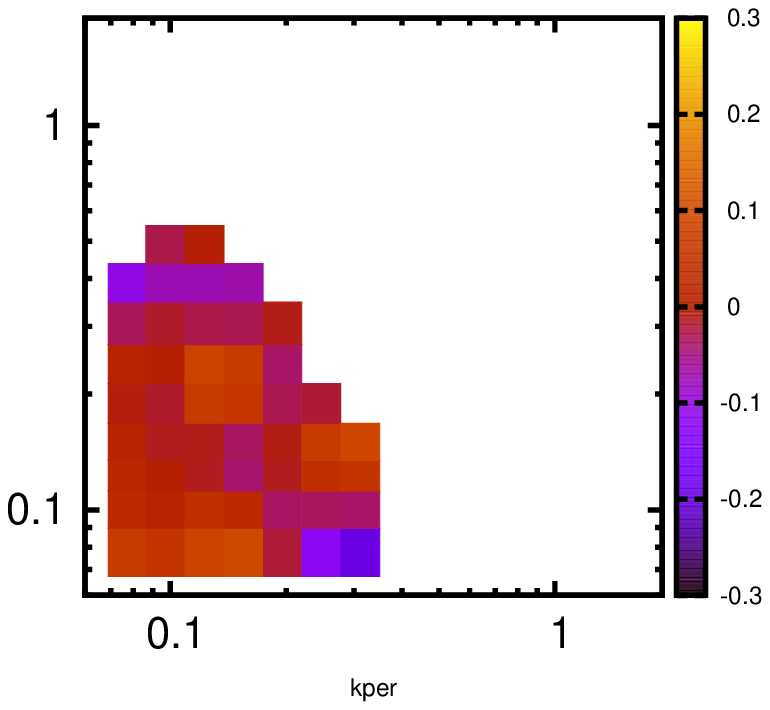}
\caption{The left and right panels show the fractional 
deviation $(P^M(k_\perp,k_\parallel)-P(k_\perp,k_\parallel))/P(k_\perp,k_\parallel)$ 
 without and with noise respectively for $n=-3$ and $f=0.6$.}
\label{fig:fig9a}
\end{center}
\end{figure*}

\begin{figure*}
\begin{center}
\psfrag{kpara}[h][h][0.8][0]{$\kp [\rm {Mpc}^{-1}]$}
\psfrag{kper}[b][b][0.8][0]{$\kpm [\rm {Mpc}^{-1}]$}
\psfrag{k2mpc3}[t][t][0.8][0]{log($\sige$)}
\includegraphics[width=75mm,angle=0]{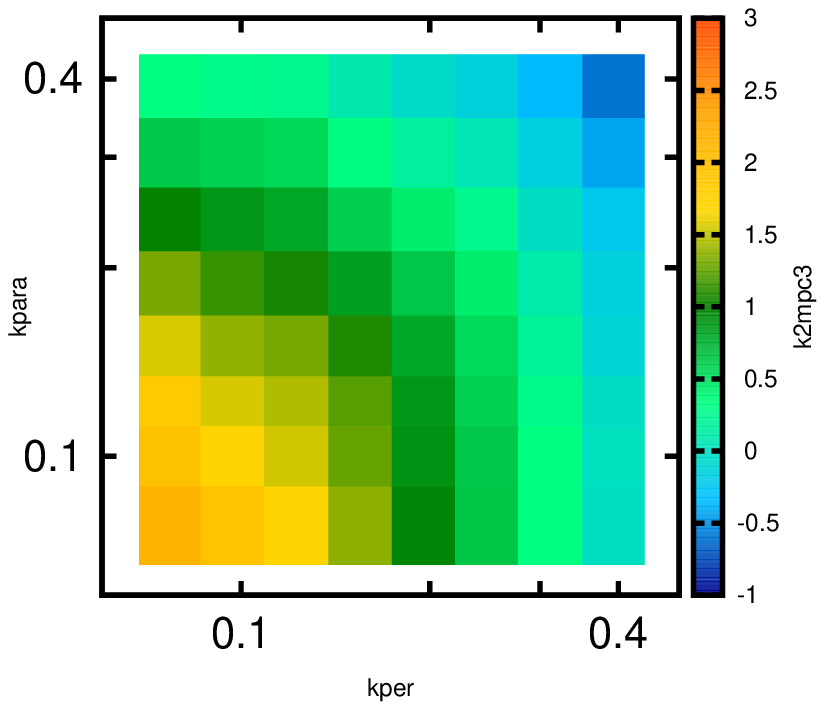}
\includegraphics[width=75mm,angle=0]{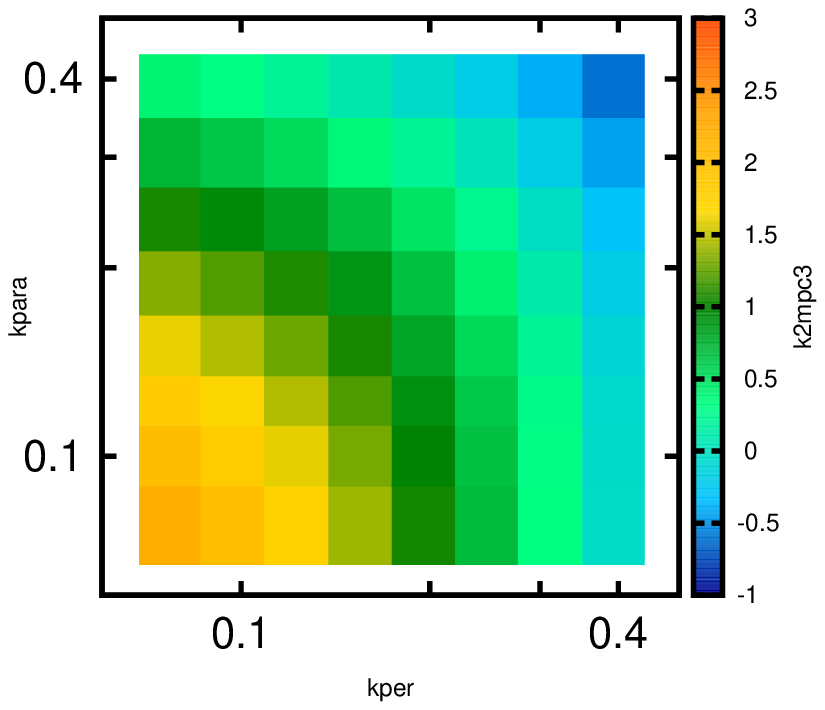}
\includegraphics[width=75mm,angle=0]{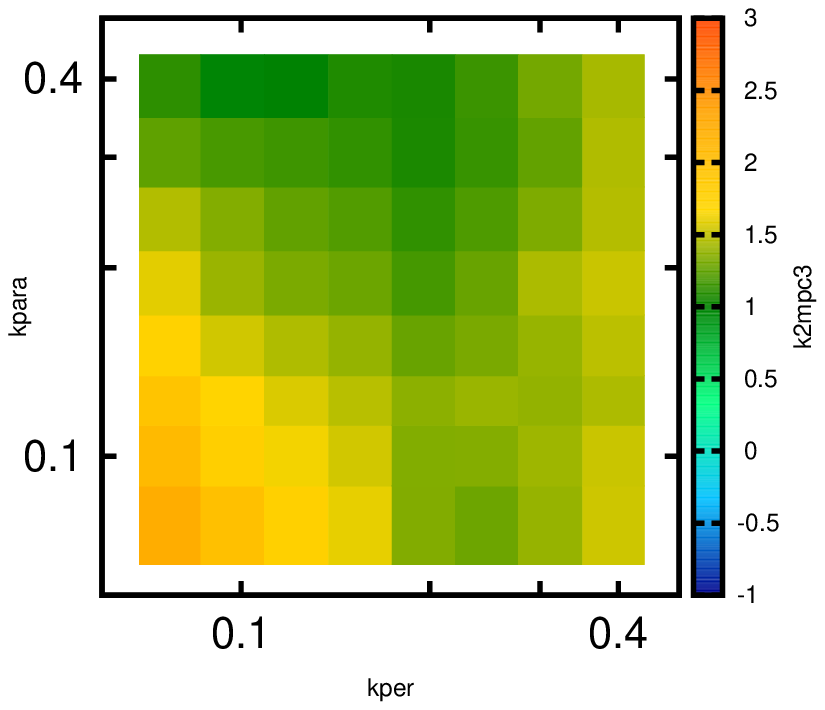}
\includegraphics[width=75mm,angle=0]{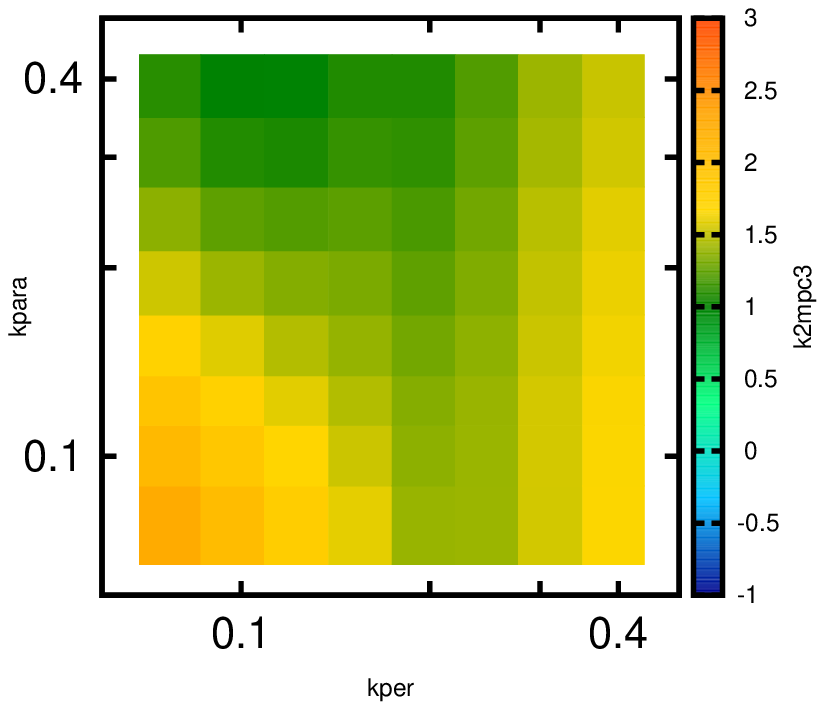}
\caption{This shows the statistical fluctuation ($\sige$) for
 the   2D Cylindrical Power Spectrum for $n=-3$ and $f=0.6$. The upper and
  lower panels show the results without and with system noise
  respectively, the left and right panels show the results from the
  simulations and the analytic prediction respectively.}
\label{fig:fig10}
\end{center}
\end{figure*}

\begin{figure*}
\begin{center}
\psfrag{kpara}[h][h][0.8][0]{$\kp [\rm {Mpc}^{-1}]$}
\psfrag{kper}[b][b][0.8][0]{$\kpm [\rm {Mpc}^{-1}]$}
\psfrag{k2mpc3}[t][t][0.8][0]{}
\includegraphics[width=75mm,angle=0]{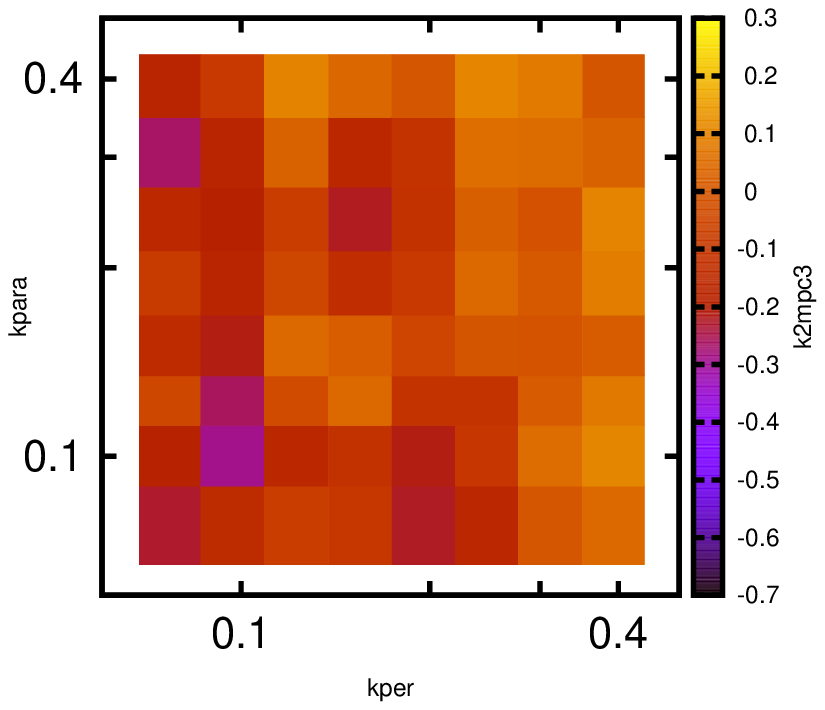}
\includegraphics[width=75mm,angle=0]{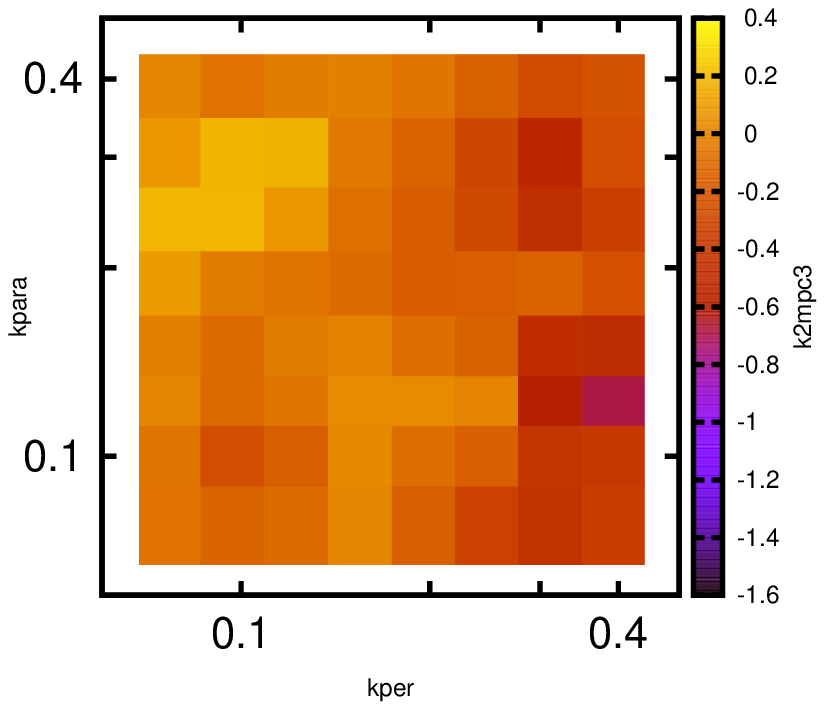}
\caption{The left and right panels  show
    the fractional deviation of $\sige$
    without and with system noise respectively.}
\label{fig:fig10a}
\end{center}
\end{figure*}

We now investigate how well the analytic prediction
(eq. \ref{eq:3dvar4}) for $\sige$ compares with the values obtained
from the simulations (Figure \ref{fig:fig10} ) for $f=0.6$. The two upper
panels consider the situation where there is no system noise for which
the left and right panels respectively show the simulated and the
analytic prediction for the statistical fluctuation $\sige$. We find
that the analytic predictions match quite well with the simulation for
the entire $\mathbf{k}$ range. The two lower panels consider the situation
where the system noise is included for which the left and right panels
respectively show the simulated and the analytic prediction for
$\sige$. The left and right panels of Figure \ref{fig:fig10a} show
  the fractional deviation between the simulated and analytic $\sige$ 
  without and with system noise respectively. We find that 
we have less than $20 \%$ fractional deviation   in $73 \%$ and $64 \%$ of 
the bins in $(k_\perp,k_\parallel)$ space without and with system noise 
respectively. The fractional deviation shows a larger spread in values 
when the system noise is included as compared to the situation without 
system noise. We do not, however,  find any obvious pattern in the distribution
 of the  bins that show a high fractional deviation.

\section{Summary and Conclusions}
Quantifying the statistical properties of the diffuse sky signal directly from
the visibilities measured in  low frequency radio-interferometric  observation
is an important issue.  In this paper we present a statistical estimator,
namely the Tapered Gridded Estimator (TGE), which has been  developed for this 
purpose. The measured visibilities are here  gridded in the $uv$ plane 
to reduce the complexity of the computation. The contribution from the
discrete sources in the periphery of the telescope's FoV,
particularly the sidelobes, pose a problem for power spectrum estimation. The
TGE suppresses the contribution from the outer regions by tapering the
sky response through a suitably chosen window function. The TGE also internally
estimates the noise bias from the input data, and subtracts this out to
give an unbiased estimate of the power spectrum. In addition to the
mathematical formalism for the estimator and its variance, we also present
simulations of  $150 \, {\rm MHz}$ GMRT observations which are used to
validate the estimator.

We have  first considered a situation where we have
observation at a single  frequency for which the 2D TGE provides an 
estimate of  the angular power spectrum $C_{\ell}$. The work here presents an
improvement over an earlier version of the 2D TGE presented in Paper I. 
 This is important in the context of the diffuse Galactic  synchrotron
 emission which is one of the 
major  foregrounds for the cosmological 21-cm signal. Apart from this, the
diffuse  Galactic synchrotron  emission is a probe of the cosmic
ray electrons and the magnetic fields in the ISM of our own Galaxy, and this
is an important study in its own right.

It is necessary to  also include the frequency variation of the sky signal in 
order to quantify the cosmological  21-cm signal. Here the 3D TGE provides an
estimate of $P(\k)$ the power spectrum of the 21-cm brightness temperature
fluctuations. We have considered two different binning schemes which provide
the 1D Spherical Power Spectrum $P(k)$ and the 2D Cylindrical Power Spectrum  
$P(k_\perp,k_\parallel)$  respectively. 
In all cases, we find that the TGE is able to accurately recover the input
model used for  the simulations.  The analytic predictions for the variance are
also found to be in reasonably good agreement with the simulations in most
situations.

Foregrounds are possibly   the biggest challenge for detecting the
cosmological 21-cm power spectrum. Various studies
(eg. \citealt{adatta10}) show that the foreground contribution to the
Cylindrical Power Spectrum $P(k_{\perp},k_{\parallel})$ is expected to
be restricted within a wedge in the $(k_{\perp},k_{\parallel})$ plane.
The extent of this ``foreground wedge'' is determined  by the angular extent 
 of the telescope's FoV. In principle,  it is  possible to 
limit the extent of the foreground wedge by tapering the telescope's FoV. 
In the context of estimating the angular power spectrum $C_{\ell}$, 
our earlier work (Paper II) has demonstrated 
that the 2D TGE is able to suppress the contribution from the outer parts 
and the sidelobes of the telescope's beam pattern. 
We have not explicitly considered the
foregrounds in our analysis of the 3D TGE presented in this paper. We however
expect the 3D TGE to suppress the contribution from the outer parts and the 
sidelobes of the telescopes beam pattern while estimating the power spectrum 
$P(k_{\perp},k_{\parallel})$, thereby reducing the area in the 
$(k_{\perp},k_{\parallel})$ plane  under the foreground wedge.

The 3D TGE holds the promise of allowing us to reduce the extent of the 
foreground wedge by tapering the sky response. It is, however, necessary to 
note that this comes at a cost which we now discuss. First,  we lose 
information  at the largest angular scales due to the reduced FoV. 
This restricts the smallest $k$ value at which it is possible to estimate
the power spectrum. Second,  the reduced FoV results in a larger cosmic 
variance for the smaller angular modes which are within the tapered FoV. 
The actual value of the tapering parameter $f$ that would be
 used to estimate $P(k_{\perp},k_{\parallel})$ will possibly be determined by 
optimising between the cosmic variance and the foreground contribution.  
A possible strategy would be to use different values of $f$  for different 
bins in the  $(k_{\perp},k_{\parallel})$ plane. 
It is also necessary to note that the effectiveness of the tapering 
proposed here depends on the actual baseline distribution, and a 
reasonably dense $uv$ coverage is required for a proper implementation 
of the TGE. We propose to include foregrounds in the simulations 
and address these issues in future work. 
We also plan to apply this estimator to $150 \, {\rm MHz}$ GMRT
data  in future.

%\clearpage{\pagestyle{empty}\cleardoublepage} %%%%%%%%%%%%%%%%%%%%
%\newpage
\setcounter{section}{0}
\setcounter{subsection}{0}
\setcounter{subsubsection}{2}
\setcounter{equation}{0}
%\pagenumbering{arabic}

\def\sige{\sigma_{P_G}}
\def\kpm{{k}_\perp}
\def\kp{k_\parallel}
\def\u{\vec{U}}
\def\S{{\mathcal S}}
\def\V{\mathcal{V}}
\def\N{{\mathcal N}}
\def\A{{\bf A}\,}
\def\Sc{S_2}

%-------------------------------------------
\chapter[Angular power spectrum for TGSS survey]{{\bf Measurement of Galactic Synchrotron emission using TGSS survey}\footnote{This chapter is adapted
   from the paper ``Measurement of Galactic Synchrotron emission using TGSS survey''
   by \citet{samir16d}}}
\label{chap:chap6}

\section{Introduction}
Observations of the redshifted 21-cm signal from the Epoch of
Reionization (EoR) contains a wealth of cosmological and astrophysical
information
\citep{bharadwaj05,furlanetto06,morales10,pritchard12}. The Giant
Metrewave Radio Telescope
(GMRT; \citealt{swarup})
is currently functioning at a frequency band corresponds to the 21-cm
signal from this epoch. Several ongoing and future experiment such as
the Donald C. Backer Precision Array to Probe the Epoch of
Reionization (PAPER,
\citealt{parsons10}), the Low Frequency Array
(LOFAR, \citealt{haarlem}), the
Murchison Wide-field Array
(MWA, \citealt{bowman13}), the
Square Kilometer Array (SKA1
LOW, \citealt{koopmans15})
and the Hydrogen Epoch of Reionization Array
(HERA, \citealt{neben16}) are
aiming to measure the EoR 21-cm signal. The EoR 21-cm signal is
overwhelmed by different foregrounds which are four to five orders of
magnitude stronger than the expected 21-cm signal
\citep{shaver99,ali08,ghosh1,ghosh2}. Accurately modelling and
subtracting the foregrounds from the data are the main challenges for
detecting the EoR 21-cm signal. The Galactic synchrotron emission is
expected to be the most dominant foreground at angular scale $\>10$
arcmin after point source subtraction at {\rm 10-20 mJy} level
\citep{bernardi09,ghosh12,iacobelli13}.  A precise characterization
and a detailed understanding of the Galactic synchrotron emission is
needed to reliably remove foregrounds in 21-cm experiments. In this
paper, we characterize the diffuse Galactic synchrotron emission at
arcminute angular scales which are relevant for the cosmological 21-cm signal studies.

The study of the diffuse Galactic synchrotron emission is also
important in its own right. The angular power spectrum ($C_{\ell}$) of
the diffuse Galactic synchrotron emission quantifies the fluctuations
in the magnetic field and in the electron
density of the turbulent interstellar medium (ISM) of our Galaxy
(e.g. \citealt{Waelkens,Lazarian,iacobelli13}).

There are several observations towards characterizing the diffuse
Galactic synchrotron emission spanning a wide range of frequency.
\citet{haslam82} have measured the all sky diffuse Galactic
synchrotron radiation at ${\rm 408 MHz}$. \citet{reich82} and
\citet{reich88} have presented the Galactic synchrotron maps at a
relatively higher frequency $({\rm 1.4~GHz})$. Using the ${\rm 2.3~GHz}$
Rhodes Survey, \citet{giardino01} have shown that the $C_{\ell}$ of
the diffuse Galactic synchrotron radiation behaves like a power law
$(C_{\ell}\propto\ell^{-\beta})$ where $\beta=2.43$ in the $\ell$
range $2\le\ell\le100$.  \citet{giardino02} have found that the value of
$\beta$ is $2.37$ for the ${\rm 2.4~GHz}$ Parkes Survey in the $\ell$
range $40\le\ell\le250$. The $C_{\ell}$ measured from the {\it
  Wilkinson Microwave Anisotropy Probe} (WMAP) data show a slightly
lower value of $\beta$ $(C_{\ell}\propto\ell^2)$ for $\ell<200$
\citep{bennett03}. \citet{bernardi09} have analysed ${\rm 150~MHz}$
Westerbork Synthesis Radio Telescope (WSRT) observations to
characterize the statistical properties of the diffuse Galactic
emission and find that
\begin{equation}
C_{\ell}=A\times\big(\frac{1000}{\ell}\big)^{\beta} {\rm mK^2}
\label{eq:eq1}
\end{equation}
where $A=253~{\rm mK^2}$ and $\beta=2.2$ for
$\ell\le900$. \citet{ghosh12} have used GMRT ${\rm 150~MHz}$
observations to characterize the
foregrounds for 21-cm experiments and find that
 $A=513~{\rm mK^2}$ and $\beta=2.34$
in the $\ell$ range $253\le\ell\le800$. Recently, \citet{iacobelli13}
present the first LOFAR detection of the Galactic diffuse synchrotron
emission around ${\rm 160~MHz}$.  They reported that the $C_{\ell}$ of
the foreground synchrotron fluctuations is approximately a power law
with a slope $\beta\approx1.8$ up to angular multipoles of ${\rm
  1300}$.

In this paper we study the statistical properties of the diffuse
Galactic synchrotron emission using two fields observed by the TIFR
GMRT Sky Survey
(TGSS{\footnote{http://tgss.ncra.tifr.res.in}}; \citealt{sirothia14}).
We have used the data which was calibrated and processed by
\citet{intema16}, who have
identified and subtracted all the point sources from the central
region of the telescope's filed of view (FoV). We have applied the
Tapered Gridded Estimator (TGE; \citealt{samir16b}) to the residual
data to measure the $C_{\ell}$ of the background sky signal after
point source subtraction. The TGE suppresses the contribution from the
residual point sources in the outer region of the telescope's FoV and
also internally subtracts out the noise bias to give an unbiased
estimate of $C_{\ell}$ \citep{samir16a}. For each field we are able to
identify an angular multipole range where the measured $C_{\ell}$ is
dominated by the Galactic synchrotron emission, and we present power
law fits for these.
\section{Data Analysis}
\label{analysis}
The TGSS survey
contains 2000 hours of observing time and is divided of 5336
individual pointings on an approximate hexagonal grid. The observing
time for each field is about ${\rm 15}$ minutes.  For the purpose of
this paper, we have used only two data sets for two fields located at Galactic
coordinates $(9^{\circ},+10^{\circ}$; {\bf Data1}) and
$(15^{\circ},-11^{\circ}$; {\bf Data2}). The central frequency of this
survey is ${\rm 147.5~MHz}$ with an instantaneous bandwidth of ${\rm
  16.7~MHz}$ which is divided into $256$ frequency channels. All the TGSS
raw data was analysed with a fully automated pipeline based on the
SPAM package \citep{intema09,intema099,intema14}. The operation of the
SPAM package is divided into two parts: (a){\it Pre-processing}
and (b) {\it Main pipeline}. The Pre-processing step calculates good-quality instrumental calibration from the best
available scans on one of the primary calibrators, and transfers these
to the target field. In the Main pipeline the direction independent
and direction dependent calibrations for each fields are calculated
and this finally converts the calibrated visibilities into a ``CLEANed''
deconvolved radio image. The left panel of Figure \ref {fig:fig01}
shows the deconvolved image for {\bf Data1}. Here, the pixels in the range
$8~{\rm mJy}$ to $130~{\rm mJy}$ are shown for clear visualization. The off
source rms noise ($\sigma_n$) for this
field is around $4.1~{\rm mJy}/{\rm Beam}$.  In the right panel of Figure
\ref {fig:fig01} we show the residual image of {\bf Data1} after
subtracting the point sources upto the $5\sigma_n$ level. For {\bf Data2}, the off
source rms noise in the continuum image is around $3.1~{\rm mJy}/{\rm Beam}$ and we have used same cut-off level ($5\sigma_n$) to subtract the point sources from the data.
\begin{figure}
\begin{center}
\includegraphics[width=75mm,angle=0]{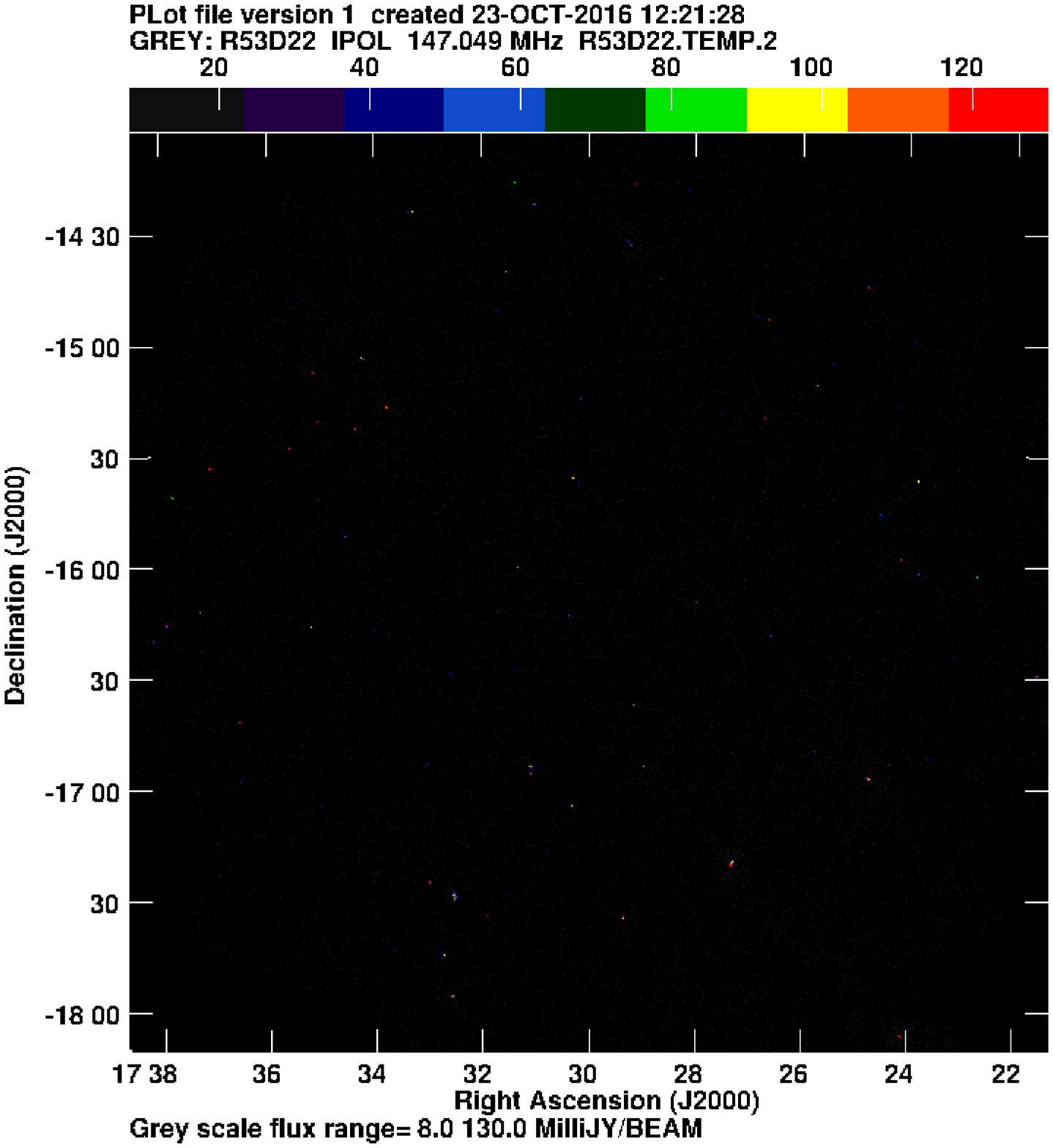}
\includegraphics[width=75mm,angle=0]{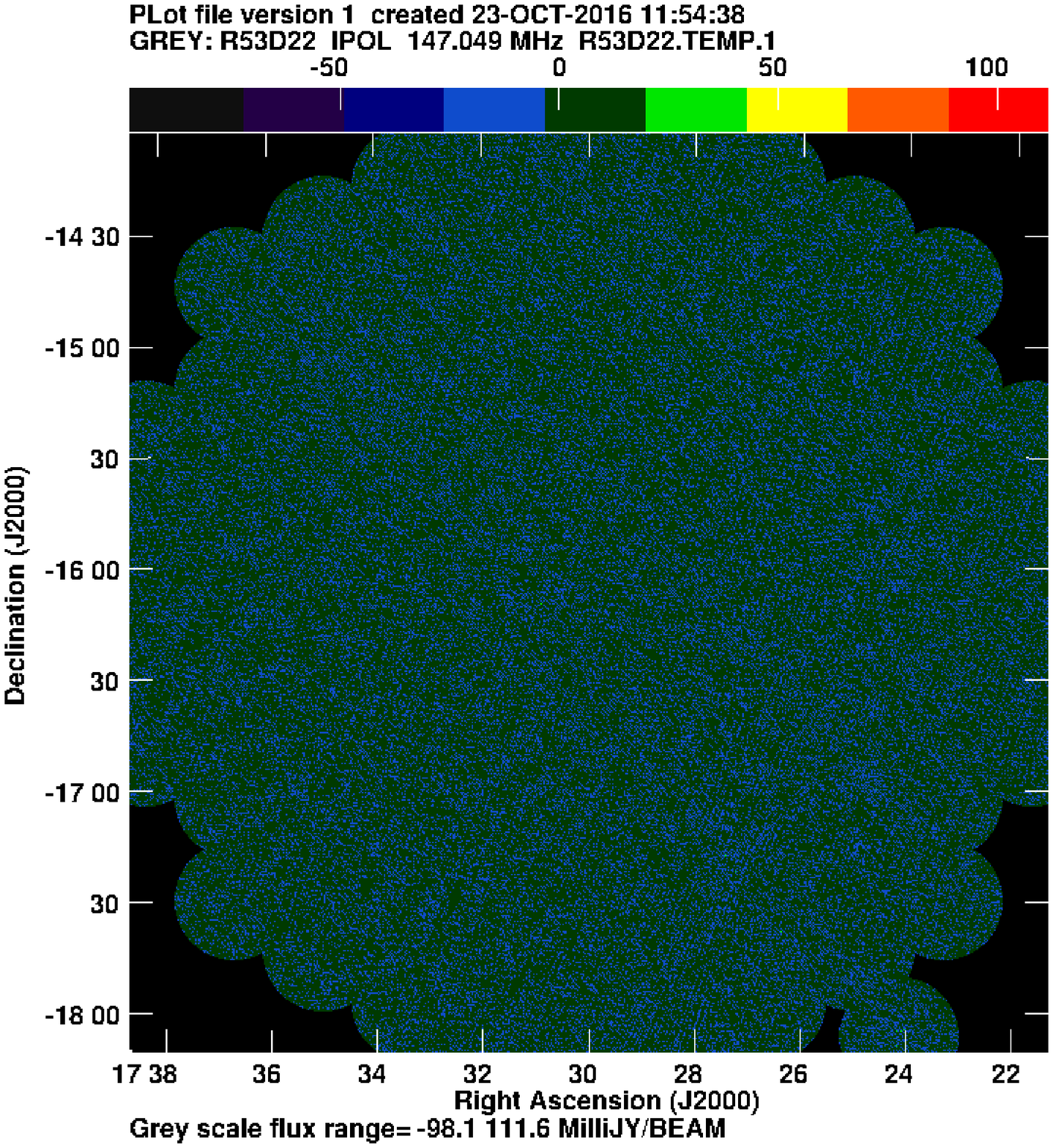}
\caption{This figure shows the deconvolved images of {\bf Data1} before (left panel) and after (right panel) point source subtraction. Here, we have shown 
the continuum images of bandwidth ${\rm 16.7~MHz}$. The total angular size  is $4.1^{\circ} \times 4.1^{\circ}$ and synthesized beam size is $25^{''} \times 25^{''}$. The off source rms
noise ($\sigma_n$) for this images are around $4.1~{\rm mJy}/{\rm
Beam}$. }
\label{fig:fig01}
\end{center}
\end{figure}

We have used the TGE to estimate the angular power spectrum $C_{\ell}$
both before and after point source subtraction. We have used $f=1$
for the tapering window.

\section{Results and Conclusions}
\label{results}
The left panel of Figure \ref {fig:fig1} shows the angular power
spectrum $C_{\ell}$ before and after point source subtraction for the
{\bf Data1}. The estimated $C_{\ell}$ before subtracting the point
sources is almost flat (upper red curve). This is mainly due to the
Poisson distribution of the point sources which dominates at all 
angular multipole $\ell$. The lower curve of this figure is for 
the estimated $C_{\ell}$ after subtracting the point sources from the
central region of the FoV. In this case the contribution from the
residual point sources dominates at $\ell_{max}\ge580$. We believe 
that the 
Galactic synchrotron emission has a significant contribution at lower 
values of $\ell$ $(\ell_{max}\le500)$. The right panel of Figure \ref {fig:fig1}
shows the same but for {\bf Data2}. Here, the value of $\ell_{max}$ is 440.
\begin{figure}
\begin{center}
\psfrag{cl}[b][t][1][0]{$C_{\ell} ~~{\rm mK}^2$}
\psfrag{l}[c][c][1][0]{$\ell$}
\psfrag{tot}[r][r][0.8][0]{Total}
\psfrag{res}[r][r][0.8][0]{Residual}
\includegraphics[width=75mm,angle=0]{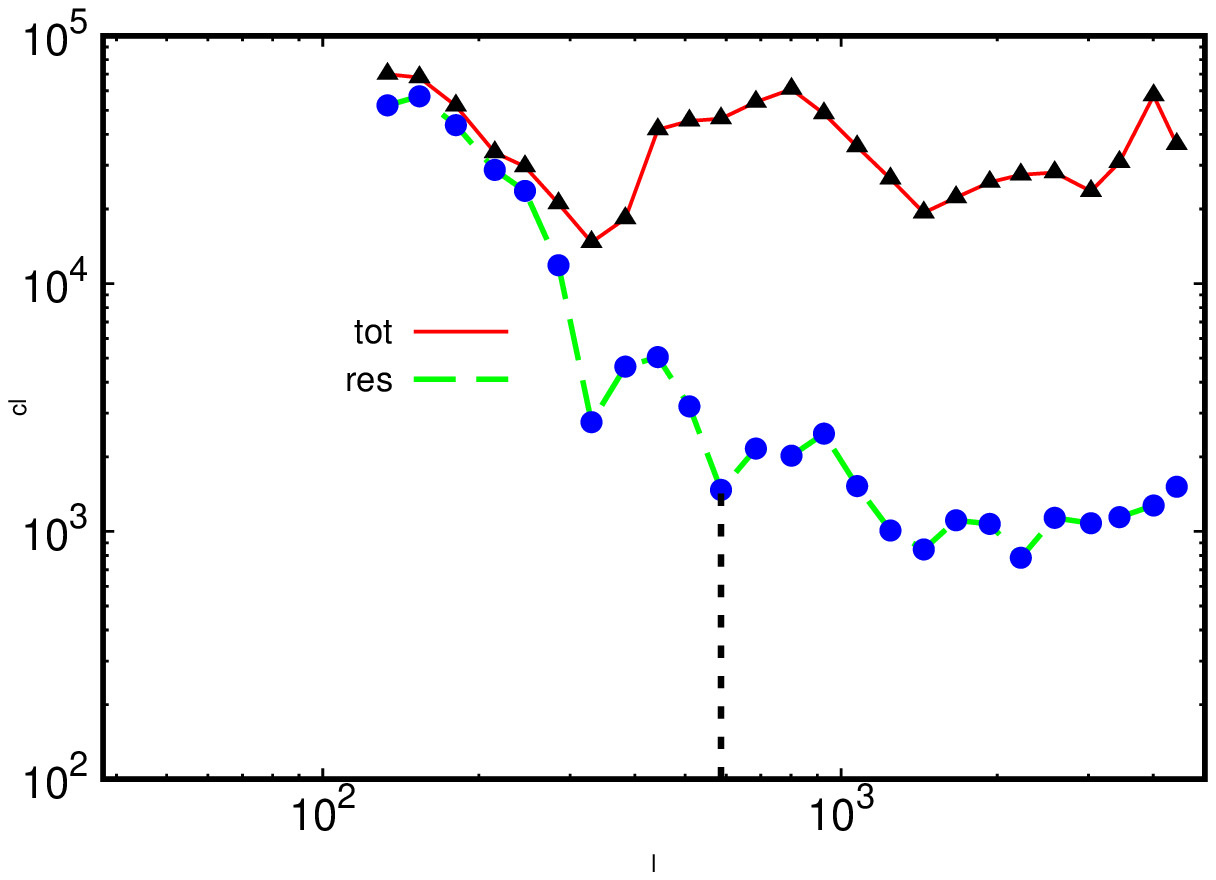}
\includegraphics[width=75mm,angle=0]{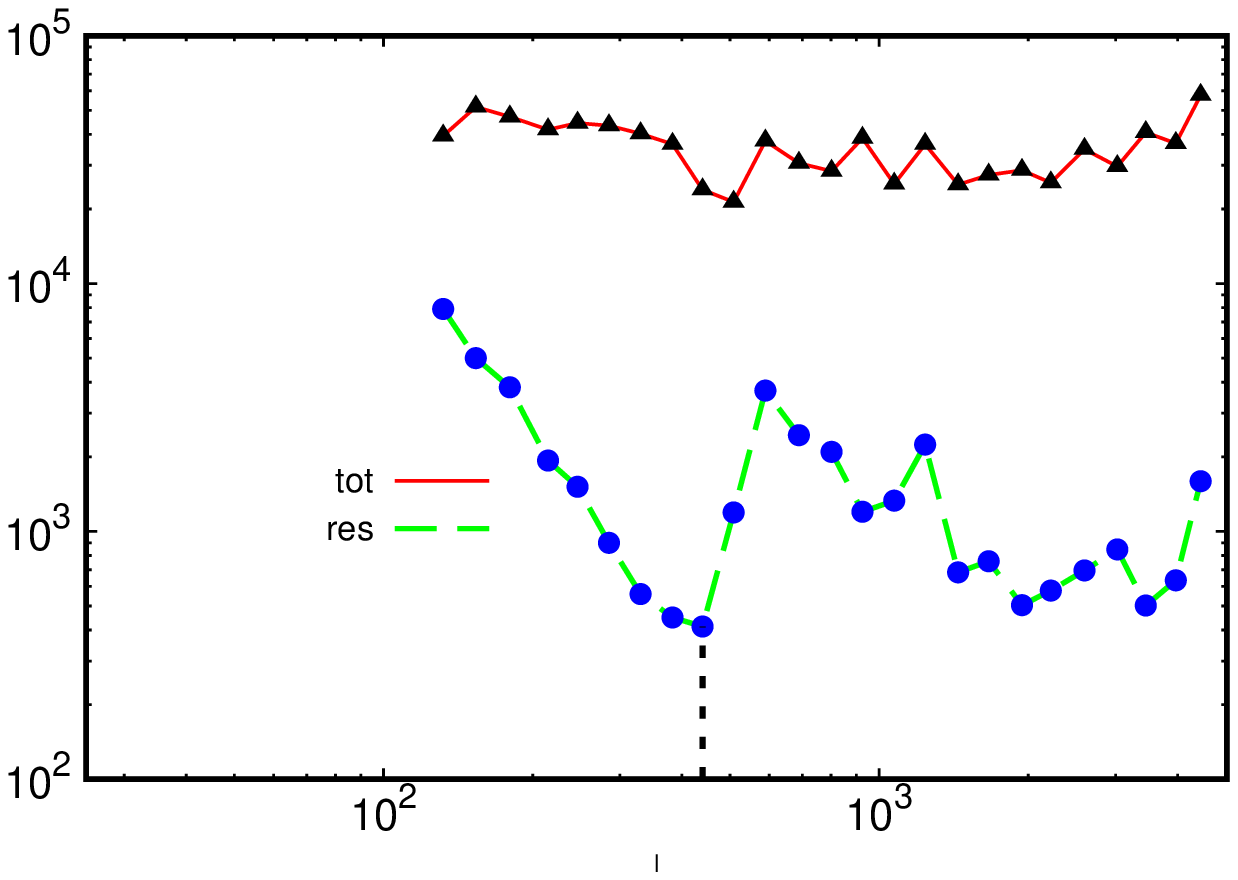}
\caption{The left (right) panel shows the $C_{\ell}$ before and after point source subtraction for {\bf Data1} ({\bf Data2}). The vertical dotted lines in both panels show $\ell_{max}$ after which the residual $C_{\ell}$ is dominated by unsubtracted point sources.}
\label{fig:fig1}
\end{center}
\end{figure}

We note that the convolution with the effective primary beam
(product of the primary beam and the taper window) affects the
estimated $C_{\ell}$ in the lower range of $\ell$ (Figure 3,
\citealt{samir14}). To identify the $\ell$ range upto which
the convolution is significant, we generate mock visibility data with same
observation parameters but with a known power law angular power
spectrum. Figure \ref {fig:fig2} shows the $C_{\ell}$ estimated from
the simulated data for two different power law indices $(\beta)$
$3$ and $1.5$. In this case we have used the same baseline
configuration as for {\bf Data1}. We see that the effect of the
convolution is important in the range  $\ell\le\ell_{min}=240$. In the
region $\ell\ge\ell_{min}$ we would be able to recover the model angular
power spectrum quite accurately.  We did the same analysis for {\bf
  Data2} for which the value of $\ell_{min}$ is almost same and we have
not shown this in the figure.
\begin{figure}
\begin{center}
\psfrag{cl}[b][t][1][0]{$C_{\ell}~~{\rm mK}^2$}
\psfrag{l}[c][c][1][0]{$\ell$}
\psfrag{mod3}[r][r][1][0]{$\beta=3.0$}
\psfrag{sim3}[r][r][1][0]{}
\psfrag{mod1.5}[r][r][1][0]{$\beta=1.5$}
\psfrag{sim1.5}[r][r][1][0]{}
\psfrag{data}[r][r][1][0]{Residual}
\psfrag{simu}[r][r][1][0]{Simulation}
\psfrag{model}[r][r][1][0]{$C^M_{\ell}$}
\includegraphics[width=80mm,angle=0]{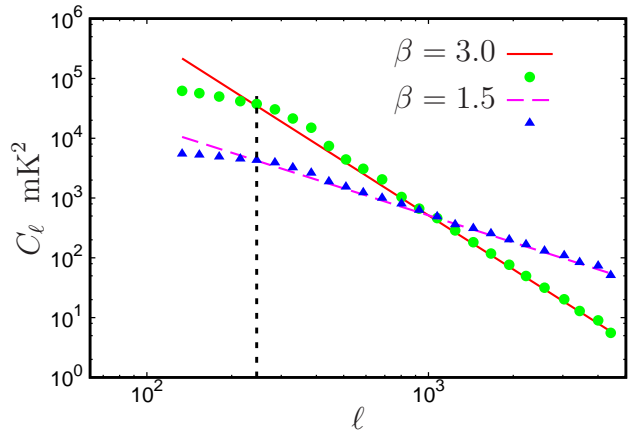}
\caption{This figure shows the region where the estimated $C_{\ell}$ is affected by the convolution with the effective primary beam. Here we see that the effect of the convolution is important in the range $\ell_{min}\le240$.}
\label{fig:fig2}
\end{center}
\end{figure}

We have used the residual visibilities after subtracting the point sources to estimate the angular power spectrum $C_{\ell}$. Figure \ref {fig:fig3} shows the $C_{\ell}$ estimated from the residual visibilities with $1-\sigma$ error bar for {\bf Data1} (left panel) and {\bf Data2} (right panel). We identify the region in the $\ell$ space $(\ell_{min}\le\ell\le_{max})$ which we expect to be dominated by the Galactic
synchrotron emission. In Figure \ref {fig:fig3} we show this region by drawing two vertical lines corresponding to $\ell_{min}$ and $\ell_{max}$ respectively. We see that the estimated $C_{\ell}$ in this region behaves as a power law. We fit equation~(\ref{eq:eq1}) to the measured $C_{\ell}$ in this $\ell$ range $(\ell_{min}\le\ell\le\ell_{max})$. The best fits values of $(A,\beta)$ are $(356.23\pm109.5,2.8\pm0.3)$ and $(54.6\pm26,2.2\pm0.4)$ for {\bf Data1} and {\bf Data2} respectively.  The $C_{\ell}$ using the best fit parameters 
are also shown in Figure \ref {fig:fig3}. The values of $\beta$ from this analysis are quite consistent with earlier measurements \citep{bernardi09,ghosh12,iacobelli13}. In Figure \ref {fig:fig3} we have also shown the $C_{\ell}$ using the simulated data. In this simulation we have used best fit values of $A$ and $\beta$. The $1-\sigma$ errors for the simulated $C_{\ell}$, estimated using $128$ independent realizations, are also shown by the shaded region. We mentioned earlier that the estimated $C_{\ell}$ for $\ell\ge\ell_{max}$ is due to the residual point sources which is almost flat in nature. We have shown the theoretical prediction of $C_{\ell}$ for the Poisson fluctuation of residual point sources in a situation where the all bright sources of flux density $S>50~{\rm mJy}$ has been subtracted from the data.

\begin{figure}
\begin{center}
\psfrag{cl}[b][t][1][0]{$C_{\ell}~~{\rm mK}^2$}
\psfrag{l}[c][c][1][0]{$\ell$}
\psfrag{data}[r][r][0.8][0]{Residual}
\psfrag{simu}[r][r][0.8][0]{Simulation}
\psfrag{model}[r][r][0.8][0]{$C^M_{\ell}$}
\psfrag{50mjy}[rt][rt][0.8][0]{$S_c=50{\rm mJy}$}
\includegraphics[width=75mm,angle=0]{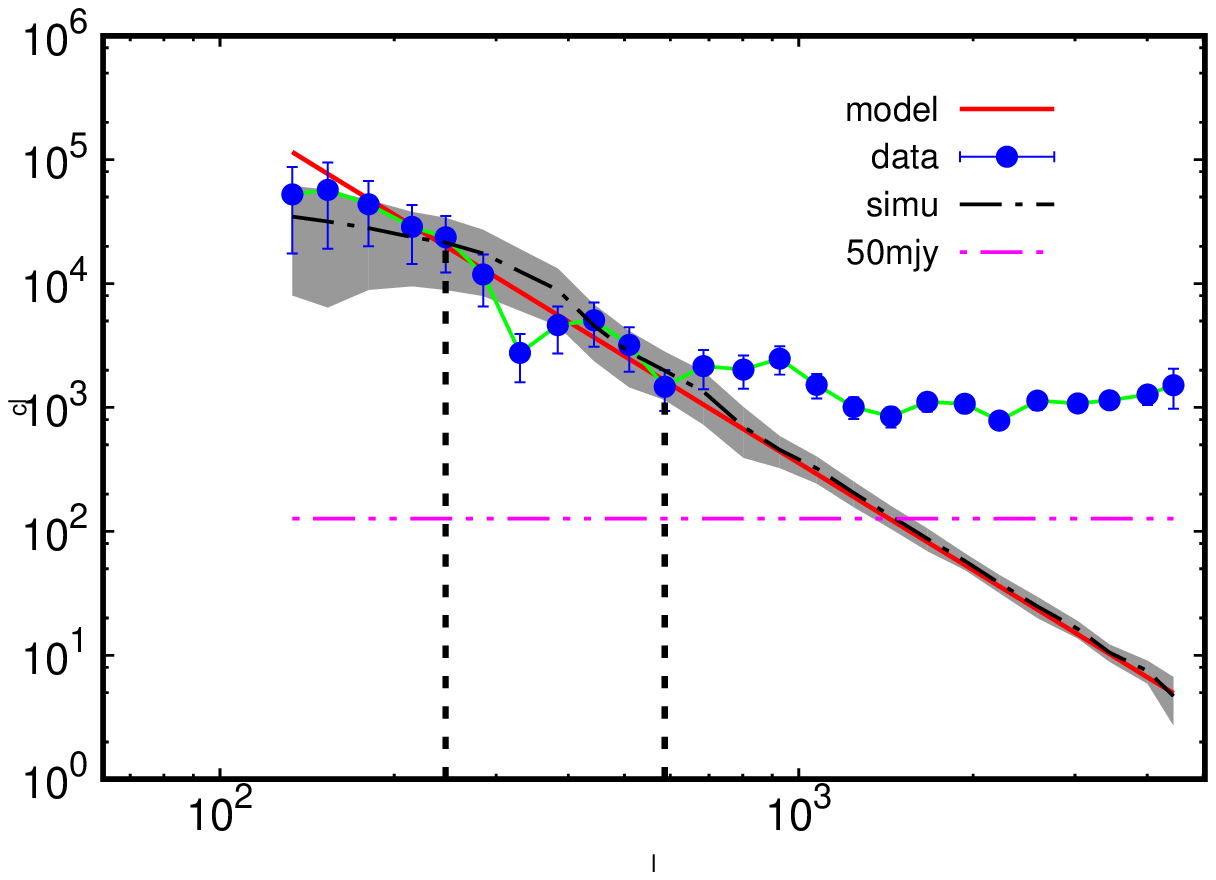}
\includegraphics[width=75mm,angle=0]{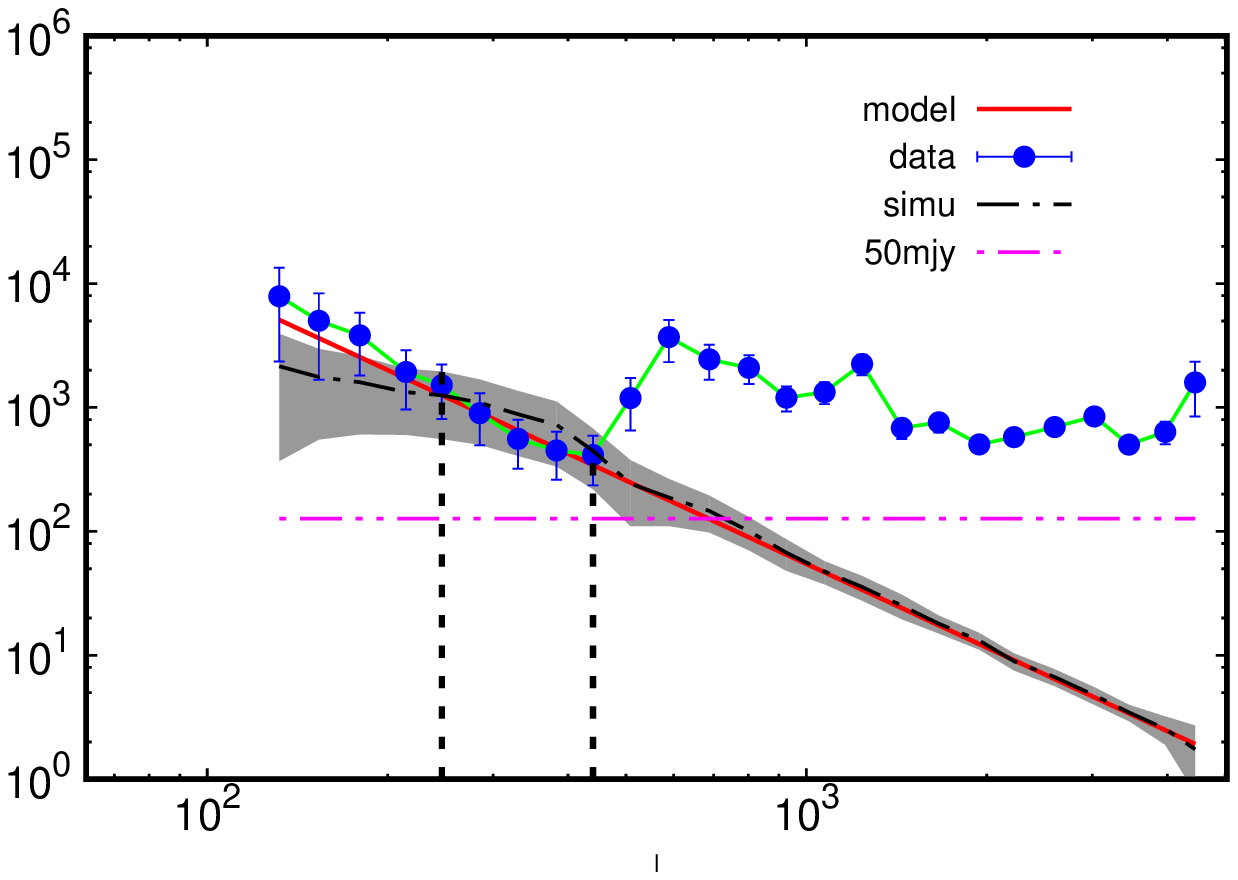}
\caption{This left panel shows the estimated $C_{\ell}$ from the residual data with $1\sigma$ error bar for {\bf Data1}. The solid line shows the $C_{\ell}$ using the best fit parameters. The dash-dot line shows the recovered $C_{\ell}$ using simulation where we have used the best fit parameters to generate the mock data. The $1-\sigma$ error in the recovered $C_{\ell}$ using $128$ independent realizations is also shown with shaded region. The theoretical
prediction of $C_{\ell}$ for the Poisson fluctuation of residual point sources upto flux density $50{\rm mJy}$ is shown by dot-dot-dash line. The right panel shows the same but for {\bf Data2}.}
\label{fig:fig3}
\end{center}
\end{figure}

We have estimated the angular power spectrum $C_{\ell}$ using two fields observed by TGSS in the $\ell$ range $150\le\ell\le4000$. The estimated $C_{\ell}$ is affected by the convolution with the effective primary beam in the range $\ell\le240$. The residual point sources have a significant contribution in the estimated $C_{\ell}$ at $\ell\ge450$. We identify the region in $\ell$ space $(240\le\ell\le450)$ which we expect to be dominated by the diffuse Galactic synchrotron emission. We present a power law fits (equation\ref{eq:eq1}) to the estimated $C_{\ell}$ over this $\ell$ range. The best fit values of the amplitude ($A$) and the power law index $(\beta)$ are $(356.23\pm109.5,2.8\pm0.3)$ and $(54.6\pm26,2.2\pm0.4)$ for two data sets observed by TGSS. We plan to extend this analysis for the whole sky using the full TGGS survey in future.

%\clearpage{\pagestyle{empty}\cleardoublepage} %%%%%%%%%%%%%%%%%%%%
\chapter[Summary and Future Scope of Study]{{\bf Summary and Future Scope of Study}}
\label{conclus}

\section{Summary of contributions}

Precise measurement of the power spectrum of the diffuse sky signal in the
presence of foregrounds is a topic of intense current research. In
this thesis we present the visibility based Tapered Gridded Estimator to
accurately measure the power spectrum of the diffuse sky signal from
low frequency radio interferometric observations.  The TGE
incorporates three novel features.  First, the estimator uses the
gridded visibilities to estimate the angular power spectrum
$(C_{\ell})$, this is computationally much faster than individually
correlating the visibilities. Second, a positive noise bias is removed
by subtracting the auto-correlation of the visibilities. Third, the
estimator allows us to taper the field of view (FoV) so as to restrict
the contribution from the sources in the outer regions and the
sidelobes of the telescope's primary beam. The mathematical formalism
of the TGE and its variances are presented in this thesis. The
estimator and its variance predictions are validated using realistic
simulations.

We also present the Bare Estimator which uses the individual
visibilities to estimate the $C_{\ell}$. The Bare estimator avoids the
self correlation of the visibilities which is responsible for noise
bias to give an unbiased estimate of the sky signal. The estimator and
the statistical error are presented mathematically and validated
using simulations. The simulations here include the Galactic diffuse
synchrotron emission and system noise for GMRT $150 {\rm MHz}$
observation. Our result show that the Bare estimator is very precise
for recovering the model power spectrum but computationally very
expensive. The TGE is relatively faster but gives an overestimate,
although it is within $1-\sigma$, for GMRT patchy $uv$ coverage. The
effect of the residual gain error and $w-term$ are also studied in the
estimated $C_{\ell}$. The estimated $C_{\ell}$ is exponentially
sensitive to the variance of the phase error but insensitive to the
amplitude error. But, the statistical uncertainties are affected both
by the amplitude and phase error. The $w-term$ does not have a
significant effect on the angular scale of our interest.

We have extended our earlier simulations by including discrete point
sources. We investigate different techniques to subtract point sources
from the central region of the primary beam. The TGE suppresses the
contribution from the outer region that's why we have not attempted to
subtract any point source from this region. Using simulation we have
shown that incomplete spectral modelling of the point sources leaves
some residual in the vicinity of the point sources which cause an
extra power at large angular multipole $\ell$. It is concluded that by
taking the source catalogue from other survey to choose the
``CLEANing'' region along with the accurate spectral modelling of the
point sources is the best strategy to subtract their contribution form
the multifrequency data and extract the $C_{\ell}$ for the underlying
diffuse signal.

We studied the effect of tapering the outer region on estimating the
$C_{\ell}$ of diffuse Galactic synchrotron emission using simulated
${\rm 150 MHz}$ GMRT observation. We have subtracted all the point
sources from the central part of the primary beam. It is really very
difficult to subtract the point sources form the outer region where
the primary beam is highly frequency dependent and also, calibration
differ from the central part. It is shown that the TGE very
effectively suppresses the contribution of the residual point sources
located at the periphery of the telescope's field of view.  We also
demonstrates that the TGE correctly estimates the noise bias from the
input visibilities and subtracts this out to give an unbiased estimate
of $C_{\ell}$.

We have further improved the 2D TGE where the overestimate due to
the patchy $uv$ coverage is corrected. Using simulated $150 {\rm MHz}$
GMRT observation, we have shown that the improved 2D TGE is able to
recover $C_{\ell}$ quite accurately. Here, the fractional deviation is
less than $5\%$ which is a considerable improvement over the earlier TGE
where the fractional deviation was $20\%$ to $50\%$. We have extended
the 2D TGE to the 3D TGE to estimate the power spectrum $(P(\k))$ of
the brightness temperature fluctuations of the redshifted 21-cm
signal. Our simulated results show that 3D TGE is also able to recover
both 1D Spherical Power Spectrum $(P(k))$ and 2D Cylindrical Power
Spectrum $P(k_\perp,k_\parallel)$ quite accurately and the analytic
predictions for the variance are in good agreement with the simulated
ones.

We have applied the 2D TGE to the data observed at $150 {\rm
  MHz}$ using GMRT. We find that the sky signal, after subtracting the
point sources, is dominated by the diffuse Galactic synchrotron
radiation across the angular multipole range $200 \le \ell \le
500$. We present power law fits,
$C_{\ell}=A\times\big(\frac{1000}{l}\big)^{\beta}$ to the measured
$C_{\ell}$ over this $\ell$ range. We find that the values of $\beta$
are in the range of 2 to 3 which are consistent with earlier
observations. The measured $C_{\ell}$ is dominated by the residual
point sources and artifacts at smaller angular scales ($\ell>500$).

\section{Future scope}
We plan to generalize the TGE to the Multi-frequency angular 
power spectrum (MAPS;\citet{kdatta07}) which quantifies the angular
and frequency dependence of the fluctuations in the sky plane.
The MAPS is relevant for separating the 21-cm signal from the 
foregrounds. The foregrounds are expected to behave smoothly 
as a function of frequency separation whereas the 21-cm signal 
decorrelates much faster. 

We have not explicitly considered the foregrounds in our analysis of
the 3D TGE presented here. We however expect the 3D TGE to suppress
the contribution from the outer parts and the sidelobes of the
telescopes beam pattern while estimating the power spectrum
$P(k_\perp,k_\parallel)$ thereby reducing the area in the
$(k_\perp,k_\parallel)$ plane under the foreground wedge. We plan to
include the foreground contribution in the simulation and to study the
effect of the tapering in the foreground wedge in details. We also
plan to apply 3D TGE in the real GMRT data to estimate the power spectrum in
$(k_\perp,k_\parallel)$ plane. 

We plan to extend the TGE to use multiple pointings (mosaic fields). This will enables us to recover the power spectrum at large angular scales and also simultaneously increase the SNR at smaller angular scales.

We plan to estimate the $C_{\ell}$ for the whole sky using TGSS survey
and to find out the variation of the amplitude and the power law index
of $C_{\ell}$ as a function of Galactic coordinate.

%\clearpage{\pagestyle{empty}\cleardoublepage} 

%\bibliographystyle{ifacconf}         %{klunamed}       %{acm}
%\bibliography{references}
\clearpage

%\backmatter
%\chaptermark{Appendix}
%\addcontentsline{toc}{chapter}{Appendix}
%\appendix
%\input{others/Appendix.tex}
%\clearpage{\pagestyle{empty}\cleardoublepage} %%%%%%%%%%%%%%%%%%%%
%\chaptermark{Appendix}
%\addcontentsline{toc}{chapter}{Appendix } 
%\appendix
%\input{others/Appendix.tex}
%\clearpage{\pagestyle{empty}\cleardoublepage} %%%%%%%%%%%%%%%%%%%%
%\chaptermark{Publications}
\addcontentsline{toc}{chapter}{Curriculam Vitae} 
\setcounter{section}{0}
\setcounter{subsection}{0}
\setcounter{subsubsection}{2}
\setcounter{equation}{0}
\begin{singlespacing}
\newcommand{\entry}[2]{ {{\bf #1}} & &  {#2} \\}
\newcommand{\eskip}{& & \\}

\chapter*{Curriculum Vitae}
\vspace{2.5cm}
%\centering
\begin{tabular}{rcl}
\entry{\large {Name : }}{\large {Samir Choudhuri}} 
\eskip
\entry{Affiliation : }{Department of Physics}
\entry{}{Indian Institute of Technology Kharagpur}
\entry{}{Kharagpur 721302, India}
\eskip
\entry{Date of Birth : }{12$^{th}$ July, 1989}    
\eskip
\entry{Email : }{samir.svc@gmail.com}
\entry{}{samir11@phy.iitkgp.ernet.in}
\eskip
\entry{Educational    }{{\bf Master of Science, Physics,} (2011)}
\entry{Qualifications : }{Jadavpur University, Kolkata.}
\entry{}{{\bf Bachelor of Science, Physics (Hons.),} (2009)}
\entry{}{Suri Vidyasagar College, Burdwan University}
\eskip
\entry{Research     }{Radio-interferometric observations,}
\entry{Interests : }{Diffuse radiation, HI Cosmology, Large scale structure.}
\end{tabular}
\end{singlespacing}
\newpage

\addcontentsline{toc}{chapter}{List of Publications} 
\renewcommand{\thefootnote}{\fnsymbol{footnote}}
\section*{List of Publications } 
\begin{singlespacing}
\subsection*{In Journals :}
\begin{itemize}
\item {\it Visibility-based angular power spectrum estimation in low-frequency radio interferometric observations}\\ {\bf Samir Choudhuri}, Somnath Bharadwaj, Abhik Ghosh, Sk. Saiyad Ali\\ 2014, MNRAS, 445, 4351

\item {\it Tapering the sky response for angular power spectrum estimation from low-frequency radio-interferometric data }\\ {\bf Samir Choudhuri}, Somnath Bharadwaj, Nirupam Roy, Abhik Ghosh, Sk. Saiyad Ali\\ 2016, MNRAS, 449, 151

\item {\it The visibility based Tapered Gridded Estimator (TGE) for the redshifted 21-cm power spectrum  }\\ {\bf Samir Choudhuri}, Somnath Bharadwaj,  Suman Chatterjee, Sk. Saiyad Ali, Nirupam Roy, Abhik Ghosh\\ 2016, MNRAS, 463, 4093 

\item {\it The angular power spectrum measurement of the Galactic synchrotron emission in two  fields of the TGSS survey  }\\ {\bf Samir Choudhuri}, Somnath Bharadwaj, Sk. Saiyad Ali, Nirupam Roy, Huib.~T.~Intema, Abhik Ghosh\\ 2017, MNRASL, 470, L11 

\item {\it Validating a novel angular power spectrum estimator using simulated low frequency radio-interferometric data }\\ {\bf Samir Choudhuri}, Nirupam Roy,  Somnath Bharadwaj,  Sk. Saiyad Ali, Abhik Ghosh, Prasun Dutta \\ 2017, NEW ASTRONOMY, 57, 94 

\end{itemize}

\subsection*{As a co-author}

\begin{itemize}

\item {\it The prospects of measuring the angular power spectrum of the diffuse Galactic synchrotron emission  with SKA1 Low}\\ Sk. Saiyad Ali, Somnath Bharadwaj, {\bf Samir Choudhuri}, Abhik Ghosh, Nirupam Roy    \\ 2016, JOAA, 37, 35

\item {\it Imaging the redshifted 21-cm pattern around the first sources during the cosmic dawn using the SKA  }\\ Raghunath Ghara,  T. Roy Choudhury, Kanan K. Datta, {\bf Samir Choudhuri}\\ 2017, MNRAS, 464, 2234

\end{itemize}

\end{singlespacing}
\newpage

%\chaptermark{Response to reviewers comments}
%\input{others/reviewer_reply.tex}
\end{document}